\documentstyle[12pt,rev]{report}

\begin{document}
\thispagestyle{empty}
~
\vspace{3cm}
\begin{center}
\begin{Huge}
\bf 
One-dimensional Fermi liquids
\end{Huge}\\
\rm
\bigskip
\bigskip
\bigskip
\Large
Johannes Voit\\
\bigskip
Bayreuther Institut f\"{u}r Makromolek\"{u}lforschung (BIMF) \\
and Theoretische Physik 1\\
Universit\"{a}t Bayreuth \\ 
D-95440 Bayreuth (Germany)\footnote{Present address}\\
\bigskip
and Institut Laue-Langevin \\
F-38042 Grenoble (France) \\
\vspace{2cm}
submitted to \em Reports on Progress in Physics \rm on \\
November 19, 1994 \\ 
~\\
last revision and update on \\
\today
\end{center}

\chapter*{Abstract}
\pagenumbering{roman}
\setcounter{page}{1}
We review the progress in the theory of one-dimensional (1D) Fermi liquids
which has 
occurred over the past decade. The usual Fermi liquid theory based on
a quasi-particle picture, breaks down in one dimension because of the
Peierls divergence in the particle-hole bubble producing anomalous
dimensions of operators, and because of charge-spin separation. Both
are related to the
importance of scattering processes transferring finite momentum.
A description of the low-energy properties of gapless one-dimensional
quantum systems can be based on the exactly solvable \lm\ which incorporates
these features, and whose correlation functions can be calculated.
Special properties of the eigenvalue spectrum, parameterized by 
one renormalized
velocity and one effective coupling constant per degree of freedom
fully describe the physics of this model. Other gapless 1D models share
these properties in a low-energy subspace. The concept of a ``\LL '' 
implies that their low-energy properties are described by
an effective \lm , and constitutes the universality class of these quantum
systems. Once the mapping on the \lm\ is achieved, one has an asymptotically
exact solution of the 1D many-body problem. 
Lattice models identified as \LL s include the 1D Hubbard model off 
half-filling, and variants such as the $t-J$- or the extended Hubbard model.
Also 1D electron-phonon systems or metals with impurities can be \LL s, as
well as the edge states in the quantum Hall effect. 

We discuss in detail various solutions of the \lm\ which emphasize different
aspects of the physics of 1D Fermi liquids. Correlation functions are 
calculated in detail using bosonization, and the relation of this method
to other approaches is discussed. The correlation functions decay as
non-universal power-laws, and scaling relations between their exponents
are parameterized by the effective coupling constant. Charge-spin separation
only shows up in dynamical correlations.
The \LL\ concept is developed from perturbations of the \lm . Mainly
specializing to the 1D Hubbard model, we review a variety of mappings for 
complicated models of interacting electrons onto \lm s, and thereby
obtain their correlation functions. 
We also discuss the generic behaviour
of systems not falling into the \LL\ universality class because of gaps
in their low-energy spectrum. The Mott transition provides an example
for the transition from Luttinger to non-Luttinger behaviour, and recent
results on this problem are summarized. Coupling chains by interactions
or tunneling allows transverse coherence to establish in the single- or 
two-particle dynamics, and drives the systems away from a \LL . We discuss
the influence of charge-spin separation and of the anomalous dimensions
on the transverse dynamics of the electrons. 
The edge states in the quantum
Hall effect provide a realization of a modified, chiral \LL\ whose detailed
properties differ from those of the standard model. 
The article closes with a summary of experiments which can be interpreted
in favour of \LL -correlations in the ``normal'' state of
quasi-1D organic conductors and superconductors, charge density wave systems,
and semiconductors in the quantum Hall regime.

\tableofcontents

\chapter{Introduction}
\label{chapone}
\pagenumbering{arabic}
\setcounter{page}{1}
\section{Motivation}
\label{intro}
\pagenumbering{arabic}
\setcounter{page}{1}

Strongly correlated fermions are an important problem in solid state
physics. Over the last one or two decades, experiments on many classes
of materials have provided evidence that strong correlations are a central
ingredient for the understanding of their physical properties. Among them
are the heavy fermion compounds, the \htcs , a variety of intimately
related organic metals, superconductors, and insulators, just to name
a few. Also in normal metals, the \em interactions \rm between
the electrons are rather strong, although the \em correlations \rm may
be much weaker than in the systems mentioned before.
The effective dimension of the electron gas plays an important
role in correlating interacting fermions, 
and the materials listed are essentially three-(3D), two-(2D), and 
one-dimensional (1D), respectively. Correlations are also very important
in semiconductor heterostructures and quantum wires, being two- or 
one-dimensional, including the Quantum Hall regime.

The theoretical description of strongly interacting electrons poses
a formidable problem. Exact solutions of specific models
usually are impossible, exception
made for certain one-dimensional models to be discussed later. 
Fortunately, such exact solutions are rarely required (and more rarely
even practical) when comparing with experiment. Most measurements,
in fact, only probe correlations on energy 
scales small compared to the Fermi
energy $E_F$ so that only the \em low-energy sector \rm of a given model is
of importance. Moreover, only at low energies can we hope to excite
only a few degrees of freedom, for which a meaningful comparison to
theoretical predictions can be attempted. 

Correlated fermions in three dimensions are a well studied problem. 
Their theoretical description,
by Fermi liquid theory, is approximate but well understood 
\cite{landau,Noz}.
It becomes  an asymptotically exact solution 
for low energies and small wavevectors ($E \rightarrow E_F, \;
{|{\bf k}|} \rightarrow k_F, \; T \rightarrow 0$).
The limitation to low energies is instrumental here
because, together with Fermi statistics, 
it implies that the phase space for excitations is severely restricted.
In one dimension, there is a variety of exactly solvable models, which
have been known for quite a time, but a deeper understanding of their
mutual relationships and their relevance for describing the generic 
low-energy physics, close to the 1D Fermi surface, has emerged only
rather recently. These relations as well as the properties of such
one-dimensional Fermi liquids, or following Haldane ``\LL s''
\cite{Haldane}, are the main subjects of this review article. Here
we shall use the terms ``one-dimensional Fermi liquids'' and 
``\LL s'' synonymously, although, as we show below, Fermi liquid
behaviour as it is established in 3D is not possible in 1D.

Fermi liquid theory is based on (but not exhausted by) a picture of
quasi-particles evolving out of the particles (holes) of a Fermi gas
upon adiabatically switching on interactions \cite{landau,Noz}. 
They are in one-to-one
correspondence with the bare particles and, specifically, carry the
same quantum numbers and obey Fermi-Dirac statistics. 
The free Fermi gas thus is the solvable model on which Fermi liquid theory
is built. The electron-electron interaction has three main effects:
(i) it renormalizes the kinematic parameters of the quasi-particles
such as the effective mass, and the thermodynamic properties (specific
heat, susceptibility), described by the Landau parameters $F_n^{a,s}$;
(ii) it gives them a finite lifetime diverging, however, 
as $\tau \sim (E-E_F)^{-2}$ as the Fermi surface is approached, so that
the quasi-particles
are robust against small displacements away from $E_F$; (iii) it
introduces new collective modes. The existence of quasi-particles 
formally shows up through a finite jump $z_{k_F}$ 
of the momentum distribution function $n(k)$ 
at the Fermi surface, corresponding to a finite residue of the
quasi-particle pole in the electron's Green function.

One-dimensional Fermi liquids are very special in that they retain
a Fermi surface (if defined as the set of points where the momentum
distribution or its derivatives have singularities) enclosing
the same $k$-space volume
as that of free fermions, in agreement with Luttinger's theorem
\cite{lt}. However,
there are \em no fermionic quasi-particles, \rm 
and their elementary excitations 
are rather bosonic collective
charge and spin fluctuations dispersing with different velocities. An 
incoming electron decays into such charge and spin excitations 
which then spatially separate with time (\em charge-spin separation). 
\rm The correlations between these
excitations are anomalous and show up as interaction-dependent
\em nonuniversal power-laws \rm in many
physical quantities where those of ordinary metals are characterized by
universal (interaction independent) powers.

To be more specific,
a list of salient properties of such 1D Fermi liquids includes: (i)
a continuous momentum distribution function $n(k)$, varying with
as $\mid k-k_F \mid^{\alpha}$ with an interaction-dependent
exponent $\alpha$, and a pseudogap in the single-particle
density of states $\propto \mid \omega \mid^{\alpha}$, consequences
of the non-existence of fermionic quasi-particles
(the quasi-particle residue vanishes as
$z_k \sim |k-k_F|^{\alpha}$ as $k \rightarrow
k_F$); (ii) similar 
power-law behaviour in all correlation functions, specifically 
in those for superconducting and spin or charge density wave fluctuations,
with \em universal scaling relations \rm
between the different nonuniversal exponents,
which depend only on one effective coupling constant per degree of freedom;
(iii) finite spin and charge response at small wavevectors, and a
finite Drude weight in the conductivity; (iv) 
charge-spin separation; (v) persistent currents quantized in units of
\tkf . All these properties can be described in terms of only two
effective parameters per degree of freedom which take over in 
1D the role of the Landau parameters familiar from Fermi liquid theory.

The reasons for these peculiar properties are found in the 
very special Fermi surface topology of 1D fermions producing both
singular particle-hole response and severe conservation laws. 
In a 1D chain, one 
has simply two Fermi ``points'' $\pm k_F$, and the Fermi surface of
an array of chains consists of two parallel sheets (in the absence of
interchain hopping). In both cases, one has perfect nesting, namely one
complete Fermi sheet can be translated onto the other by a single
wavevector $\pm 2k_F$. This produces a 
singular particle-hole response at \tkf , the well-known Peierls instability
\cite{Peierls}. This type of response is assumed finite in Fermi liquid
theory but, in 1D, is divergent for repulsive forward scattering
(or attractive backscattering, the case considered by Peierls), leading
to a breakdown of the Fermi liquid description. In addition, we have, as
in 3D, the BCS singularity for attractive interactions \cite{bcs}. 
On the other hand, the disjoint Fermi surface gives a well-defined 
dispersion, i.e. particle-like character, to the low-energy particle-hole
excitations. They now can be taken as the building blocks upon  which
to construct a description of the 1D low-energy physics.

These properties are generic for one-dimensional Fermi liquids but
particularly prominent in a 1D model of interacting fermions proposed
by Luttinger \cite{lut} 
and Tomonaga \cite{tom} and solved exactly by Mattis and Lieb \cite{matl}.
All correlation functions of the \lm\ can be computed exactly,
so that one has direct access to all physical properties of interest.
The notion of a ``\LL '' was coined by Haldane to describe these
universal low-energy properties of gapless 1D quantum systems, and to
emphasize that an asymptotic ($\omega \rightarrow 0, q \rightarrow 0$)
description can be based on
the Luttinger model in much the same way as the Fermi liquid theory
in 3D is based on the free Fermi gas. The basic ideas and procedures
had been discussed earlier by Efetov and Larkin \cite{efe} but
passed largely unnoticed. The name ``Tomonaga -- \LL '' might be more
appropriate to give credit to Tomonaga's important early contribution
but has not become widely popular today.

Despite this apparently very different set of physical properties,                                        
there are also similarities in the structures of Fermi and Luttinger
liquids. Some concepts
make these similarities particularly apparent: conformal field theory,
where we essentially exploit the fact that both Fermi and Luttinger
liquids (the former in 1D, of course) are critical, in the
language of the theory of phase transitions, and possess the same
central charge; a description of both
theories based on Ward identities (i.e. symmetries and conservation
laws), and the notion of a ``Landau-\LL '', where one formulates
a Fermi liquid picture for the pseudo-particles appearing in the
exact Bethe-Ansatz solution of models like the 1D Hubbard model.
Other methods, often more suitable for the practical calculations
required by a solid state physicist, like bosonization, more strongly
emphasize the differences between Fermi and Luttinger liquids. 

In two dimensions, the applicability of Fermi liquid theory,
specifically to the \htc\  problem, is quite controversial. In fact,
much of the recent interest in Luttinger liquids
is due to Anderson's observation that the normal state properties of the 2D 
\htcs\ are strikingly different from all known metals and 
cannot be reconciled with Fermi-liquid theory; 
they are more similar to properties of 1D models
\cite{ar}. Anderson proposed that the essential physics
be contained in the 2D Hubbard model 
and suggested a picture of a ``tomographic
Luttinger liquid'' for the ground state and the low-energy excitations of 
this model, building on Haldane's earlier work in 1D, to give a 
more systematic basis to these conjectured 
non-Fermi liquid properties of the \htcs. Arguments have been advanced,
however, also in favour of Fermi-liquid physics \cite{fl2d}. In addition 
a theory
somewhat intermediate between Fermi and \LL s, a ``marginal Fermi
liquid'' has been proposed \cite{mfl}, where the quasi-particle residue $z_k$
vanishes logarithmically as $k \rightarrow k_F$.
(We parenthetically note that there is no simple solvable model,
like the Fermi gas, or the \lm , onto 
which one could build the marginal Fermi liquid phenomenology
\cite{mfl}. Very recent work seems to indicate, however, that certain
impurity models do produce marginal Fermi liquid behaviour \cite{mflimp}.)

While the relevance of Anderson's ideas is still quite controversial and
no unambiguous formal justification has been published to date,
they have refocussed attention on 1D models as paradigms for the
breakdown of Fermi-liquid theory: there are few other instances where 
this has been established firmly. The main progress of the last years, 
to be reviewed here, is related to the realization that, in 1D,
a variety of models allows essentially exact calculations of the 
physical properties of ``exotic'' non-Fermi-liquid metals. Emphasis
has been directed in two main directions. (i) The relation of models
defined on a lattice, such as the 1D Hubbard model, to continuum
theories of the Tomonaga-Luttinger type. There had been a 
widespread opinion, that the lattice models would be appropriate to
model the limit of strong electron-electron interactions, while the
field theories would be better suited for weak-coupling situations.
It has now become clear that this is not so, and that the continuum theories
rather are the asymptotic low-energy limits of the lattice models even
at arbitrarily strong coupling. Moreover, this mapping has provided us with
several algorithms to extract the effective parameters of the 
continuum models from the (either Bethe Ansatz or numerical)
solution of the lattice models. It therefore provides an \em asymptotically
exact solution \rm to the 1D many-body problem. We now can 
compute essentially all correlation functions for lattice models, 
an impossible task if one wanted to use the lattice solution 
directly. (ii) The calculation of physical properties from the now
known correlation function allows to work out the distinctive 
difference of such \LL s  from the predictions of Fermi liquid theory
in higher dimensions, so as to get tools for the diagnosis of
non-Fermi liquid behaviour.

With the general excitement in the community over the spectacular physics
of the \htc superconductors, it has been somewhat forgotten that there
are many families of organic and inorganic quasi-1D metals \cite{jersch,icsm}
which do deviate strikingly from Fermi-liquid behaviour (at least from 
ordinary metals)
in their normal state, and undergo a variety of low-temperature phase 
transitions into, e.~g., charge or spin density wave (CDW/SDW) 
insulating phases or even become superconducting. The normal state 
properties of
these materials are often highly anisotropic
and justify application of 1D theory.
We therefore possess 
a laboratory playground where we can confront theoretical evaluations
of the distinctive properties of such ``Luttinger liquids'' with
experimental reality -- in a situation where the theoretical basis
(namely one-dimensionality) is quite firmly established from experiment.

There is thus at least a threefold motivation to study models of 1D 
interacting electrons: (i) The search for a coherent description  of 
the quasi-1D metals whose ``exotic'' properties have been studied over
nearly two decades  and which continue to be the focus of intense experimental
efforts. (ii) 1D models as a paradigm for ``metallic'' systems which are
not Fermi liquids. The detailed calculations possible here will hopefully
sharpen our understanding of critical requirements for the breakdown of
Fermi-liquid theory in general,
and how such scenarios translate into experimental
reality. There are only a few established examples of non-Fermi liquid metals
in higher dimension such as the multichannel Kondo problem, but even these
reduce, due to the spherical symmetry commonly assumed, 
to effective 1D problems
\cite{kondo}.
(iii) The possibility of finding exact solutions to nontrivial
many-body problems.

\section{Purpose and structure of this review}
This article will present a practical introduction to \LL s. It attempts
to combine a review of the progress of the last couple of years with
a self-contained and pedagogical presentation of the \lm , its
solution, and its properties (i.e. correlation functions)
and especially emphasize bosonization as a simple practical means
both to solve the model and to calculate correlation functions.
Based on this, we carry on to the notion of a \LL\ and a 
discussion of the various methods employed to map a complicated 1D
many-body problem onto the relatively simple \lm. 
For all models to be discussed, 
the emphasis will
be on their properties, i.e. correlation functions,
which ultimately can be compared with experiment. 
I also hope to demonstrate that the \LL\ often is a useful device 
if none of the
exact nonperturbative methods to compute correlation exponents works:
incorporating all essential features of the 1D Fermi liquid, it is
presumably the best possible starting point for a \rg\ analysis 
of the problem at hand. 

On the other hand, we shall be quite schematic concerning the methods
used to achieve an exact solution of the (lattice) models we are
interested in, such as the Bethe Ansatz or numerical diagonalization
techniques. We shall be more concerned with the various methods which have
been invented to extract effective Luttinger parameters given a certain
type of solution of the starting
models, to compare their virtues and drawbacks and emphasize their
complementarity.
Moreover, we do not attempt to present a complete overview
of the numerous 1D models used to describe strongly correlated electrons.
Rather, we shall concentrate on a few of them, most often the paradigmatic
Hubbard model. 
The methods discussed often can be
applied without significant changes to other models the reader might
be more interested in.  

The field of 1D Fermi liquids is not new. Stimulated by and stimulating
the research on quasi-1D organic conductors in the seventies and early
eighties, a number of useful review articles has been available
for some time. The meanwhile classical reviews of S\'{o}lyom \cite{solyom}
and Emery \cite{emrev} contain much material on the use of renormalization
group (with respect to a 1D Fermi gas) to treat the singularities generated
perturbatively by the 1D interactions. There is also material on the
solution of the Tomonaga-\lm\ either by bosonization (in a somewhat
approximate but for many purposes sufficient form) and through the use
of Ward identities. We are extremely brief on papers duely covered there.
Our presentation here will be limited to abelian
bosonization. Nonabelian bosonization, retaining manifestly the 
$SU(2)$-invariance in the spin sector, is reviewed by Affleck 
\cite{nonabbos}. Firsov, Prigodin, and Seidel \cite{firsov} and
Bourbonnais and Caron \cite{bourbonnais} more strongly emphasize phase
transitions in real materials made of coupled 1D chains, and especially
the paper by Bourbonnais and Caron gives a very modern presentation
combining functional integral representations with \rg . A classical
review on the earlier work on organic conductors is by J\'{e}r\^{o}me
and Schulz \cite{jersch} and the more recent developments have been
summarized by J\'{e}r\^{o}me \cite{jersci} and Williams \em et al. \rm
\cite{et}. A pedagogical overview of both experimental and theoretical
aspects can be found in the Proceedings of the 1986 NATO-ASI in Magog
(Canada) \cite{magog} while the latest progress on organic materials
is collected in the Proceedings of the biannual Synthetic Metals conferences
\cite{icsm}. Some subjects do not receive due coverage in this
article:
concerning the Bethe Ansatz, there are reviews by Sutherland \cite{suth},
Korepin et al.~\cite{korerev},
and Izyumov and Skryabin \cite{izsk}, and there is also a vast
literature on conformal field theory \cite{confft}. Other problems
of high current interest, and intimately related to our subject, 
could not be included for restrictions in space and time: spin chains,
which can be related by a variety of methods to 1D interacting fermions;
all developments starting from the Calogero-Sutherland model;
one-dimensional bosons; persistent currents in mesoscopic rings,
where interesting contributions originate from the study of 1D fermions
despite the 3D spherical Fermi surface in the real materials,
and many more. We hope that others take up the challenge to review these
active areas.

This article is structured as follows.
Chapter 2 will complement this brief introduction in that it discusses
the breakdown of the Fermi liquid on a more technical level and identifies
the relevant features of one-dimensionality. Specifically we show (i) how
the Peierls divergence produces an instability of the 1D Fermi gas in the
presence of repulsive interactions, (ii) how the breakdown of a quasi-particle
picture can be seen in a second-order perturbation calculation, and (iii)
where Landau's derivation of a transport equation crashes in 1D.

Chapter \ref{chaplm} will present a detailed discussion of the Luttinger
model from various angles. In Section \ref{lowenph}, we argue that
a universal low-energy description of 1D Fermi liquids can be based on
this model. We define the Hamiltonian and discuss its symmetries and
conservation laws which are essential for all solutions. In Section
\ref{hamdia}, we give a solution using a boson representation of
the Hamiltonian, before constructing an explicit operator identity 
between fermions and these bosons in Section \ref{secbos}. This
representation is fruitfully employed in Section \ref{secprop} for
a calculation of the \lm\ correlation functions. The manifestation
of charge-spin separation in dynamical correlations is the subject
of Section \ref{secdyn}. An alternative
method of solution based on the equations of motion of the Green functions,
and on Ward identities, is presented in Section \ref{altmet}. 
Another alternative for constructing a boson representation of the
fermions, conformal field theory, is introduced in Section \ref{cft}
and shown to be fully equivalent 
to the operator approach of the earlier sections.

The \lm\ is based on very strong restrictions on the dispersion and
interaction of the particles. In Chapter \ref{chapll} we shall pass beyond
these restrictions and show that the Luttinger physics still is conserved
in a low-energy subspace of more realistic models. This is conjectured
in Section \ref{llconj}, and then case studies are presented to its
support. With nonlinear band dispersion (Section \ref{nonldisp}), the
fermion operators acquire higher harmonics in $k_F$. Large-momentum
transfer scattering in (Section \ref{scatt}) lifts unrealistic
degeneracies of \lm\
correlation functions by logarithmic corrections. The correlations 
of various lattice models are evaluated in Section \ref{sec1d}. 
Electron-phonon systems or dirty 1D metals can also have \LL\ correlations,
and we touch upon these problems in Section \ref{phonie}. Section
\ref{lltrans} shows that rich transport phenomena occur as one goes away
from the simple \lm . Finally, in Section \ref{lalu}, we show
that the low-energy physics of the 1D Hubbard model is determined by
low-energy excitations in the charge-momentum- and 
spin-rapidity-distribution functions, and that in each sector, one
can therefore formulate a Fermi-liquid theory for its excitations. 

Not all 1D systems are \LL s. In some cases, there are gaps in either
the charge or spin excitation spectra, and phase separation can occur
(at least in \em models). \rm Chapter \ref{mottch} discusses
these cases. We expand especially on the Mott transition in Section
\ref{motttr} which has been studied in considerable detail during the
past years. 

In Chapter \ref{ext}, we go beyond the framework of the \LL\ outlined
before. We discuss multi-band models in Section \ref{multi}. An 
important issue has been the crossover from the 1D \LL\ to higher-dimensional
behaviour. We elaborate on this problem in Section \ref{twochains}. Finally,
we describe the modelling of edge excitations supporting the
transport in the quantum Hall effect in terms of a chiral \LL\ in Section
\ref{fqh}. While the general features are similar to the standard \LL ,
the edge excitations have irrational charges and constitute a new universality
class, described by a conformal field theory with central charge $c \neq 1$.

This review closes with a summary of experiments which provide (often
controversial) evidence for \LL\ correlations in several classes of materials. 
We discuss organic conductors and superconductors, inorganic charge density
wave systems, and semiconductors in the quantum Hall regime.

The general approach chosen here is to give some space to the discussion of
the basic methods used in this field in the last decade and to some 
selected examples. This necessarily requires selection, often biased
by the author's prejudices, and many important papers are discussed only
briefly. It is hoped that the general discussion will give tools, 
and that the overview on the current status give orientation to the
reader to locate and appreciate the original articles relevant to him. 
Although I have tried to incorporate a maximum of the published literature 
available to me, space restrictions did not allow to do so systematically.
I apologize to all those whose contributions have not
received due coverage.

\chapter{Fermi liquid theory and its failure in one dimension}
\label{chaptwo}
\section{The Fermi liquid}
\label{secfl}
Macroscopic properties of ordinary (3D) metals can be described remarkably
well by the model of a Fermi gas although the interactions are not
weak. Why is this possible? The answer is provided by Landau's theory
of the Fermi liquid \cite{landau,Noz}. 

The key observation is that macroscopic properties involve only
excitations of the system on energy scales (say temperatures) small
compared to the Fermi energy.
The state of the system can be specified in
terms of its ground state, i.e. its Fermi surface, and its low-lying 
elementary excitations -- a rarified gas of 
``quasi-particles''. These quasi-particles evolve continuously out of
the states of a free Fermi gas when 
interactions are switched on adiabatically, and are in one-to-one
correspondence with the bare particles (adiabatic continuity).
They possess the same quantum
numbers as the original particles, but their dynamical
properties are renormalized by the interactions. 
This scenario emerges because the phase space for scattering
of particles is severly restricted by Fermi statistics: at 
low temperatures, most particles are frozen inside the Fermi sea,
and only a fraction $T/T_F \ll 1$ participate in the scattering
processes. Apart those originating from the requirement of stability there 
are, however, no restrictions on the magnitude of the effective interactions 
between the quasi-particles, as measured by the Landau parameters. The
restriction to low-lying excitations implying low densities of excitations,
and Fermi statistics are enough to ensure Fermi liquid properties.

The ground state of a gas of free particles is fully described by its
momentum distribution function $n_0(\bk)$. For the interacting system,
it can be specified by the quasi-particle distribution function which
is the same as that of the bare particles in the free system.
Excitations are then determined by the deviations 
they produce in the momentum distribution with respect
to the ground state, $\delta n(\bk) = n(\bk) - n_0(\bk)$. So long as there
are few excitations, $\delta n(\bk)$ is small.
The change in energy $\delta E$ 
associated with quasi-particle excitations can
then be expanded in powers of $\delta n(\bk)$
\begin{equation}
\label{landhyp}
\delta E = \sum_{\bk} \left[ \varepsilon_0(\bk) - \mu \right]
\delta n(\bk) + \frac{1}{2}
\sum_{\bk,\bk'} \delta n(\bk) f(\bk,\bk') \delta n(\bk') + \ldots \;\;\;,
\end{equation}
where $f(\bk, \bk')$ is the quasi-particle interaction and $\mu$ is the
chemical potential. Although the single-particle term is of first order in
$\delta n(\bk)$ and the interaction term of second order, they are in fact
of equal importance and the second term cannot be neglected: the notion of
a quasi-particle making sense only in the neighbourhood of the Fermi surface,
$ \varepsilon_0(\bk) - \mu $ is small there and of the same sign as 
$\delta n(\bk)$.

On a more formal level, the Green function of an electron is 
\begin{equation}
\label{greenfl}
G(\bk,\omega) = \frac{1}{\varepsilon_0(\bk) - \omega -\Sigma(\bk,\omega)} 
\;\;\;,
\end{equation}
where $\varepsilon_0(\bk)$ is
the bare dispersion and $\Sigma(\bk,\omega)$ 
is the self-energy containing all the many-body
effects. The poles of the Green functions give the single-particle 
excitation energies, and the imaginary part of the self-energy provides
the damping of these excitations. 
$\Sigma(\bk,\omega)$ is, for fixed $\bk$, a smooth function of $\omega$
and continuous in $\bk$.
This guarantees solutions to the equation
\begin{equation}
\label{poles}
\varepsilon_0(\bk) - \omega - \Sigma(\bk,\omega) = 0 \;\;\;,
\end{equation}
determining the single particle excitation energies. One hopes
that there is only a single solution to this equation -- but this need
not be so. In fact, having a single solution -- the quasi-particle pole 
with finite residue \cite{lt}
\begin{equation}
\label{qpr}
z_{\bk} = \left( 1 - \frac{\partial {\rm Re}\Sigma(\bk,\omega)}{\partial 
\omega} \right)^{-1}_{\omega = \varepsilon(\bk)} \leq 1 
\end{equation}
-- implies a normal Fermi liquid. We shall see below that
the the breakdown of Fermi liquid theory in 1D is signalled by the appearance
of multiple solutions or vanishing of $z_{\bf k}$. 
The quasi-particle residue $z_{k_F}$ gives the magnitude of the jump 
of the momentum
distribution function of the \em bare \rm particles at the Fermi surface
\cite{lt}. 
Expanding the self-energy to second order, the Green function close to
the Fermi surface becomes
\begin{equation}
\label{grfl}
G(\bk,\omega) = G_{inc}(\bk,\omega) + 
\frac{z_{\bk}}{\omega - v(|\bk|-k_F) + i \: u \: 
{\rm sign}(|\bk|-k_F)(|\bk|-k_F)^2} \;\;\;.
\end{equation}
There is no damping of the
quasi-particles at the Fermi surface. They will exist off the Fermi surface
only to the extent that their damping is sufficiently small 
(their lifetime long enough) to make them behave like an eigenstate
over a reasonably long time scale. Damping of a quasi-particle with energy 
$\omega$
is provided by complex configurations of 
quasi-particle--quasi-hole excitations.
They also produce incoherent background Im$G_{inc}(\omega)$ in the 
spectral function which, interfering with the coherent part, gives  
Im$G(\omega) \propto \omega^2$ for $\omega \rightarrow 0$ at finite $k$. 
Eq. (\ref{grfl}) is to be compared to the 1D Green functions
derived in Chapter \ref{secprop}, and to that of the marginal Fermi
liquid whose quasi-particle residue vanishes as \cite{mfl}
\begin{equation}
\label{grmfl}
z_{\bk} \sim - 1 / \ln \mid \varepsilon ({\bk}) \mid \;\;\; {\rm for}
\;\;\; \varepsilon (\bk ) \rightarrow \mu = 0 \;\;\;.
\end{equation}

The quasi-particle is the central concept in the theory of the Fermi liquid.
From the quasi-particle picture, Landau derived, in his first paper, 
a Boltzmann-like transport equation for the Fermi liquid \cite{landau}. 
To this end, one assumes that spatially inhomogeneous excitations in the
system take place on a \em macroscopic \rm scale only, so that the
wavevector $\bk$ remains a good quantum number at least within a volume
of macroscopic size. One can then define a local distribution function
$\delta n(\bk,\br)$. The time evolution of this distribution is then
given by 
\begin{equation}
\label{boltz}
\frac{\partial \delta n(\bk , \br)}{\partial t} + {\bf v}_{\bk}
\cdot \nabla \delta n( \bk , \br) + \delta ( \varepsilon_{\bk} - \mu)
\sum_{\bk'} f( \bk , \bk' ) {\bf v}_{\bk} \cdot \nabla \delta 
n( \bk' , \br) = I[n] \;\;\;,
\end{equation}
where $I[n]$, the collision term, is a functional of $n ( \bk , \br)$ and
the velocity ${\bf v}_{\bk} = \nabla_{\bk} \varepsilon ( \bk )$. 
Since $\delta n$ and $\delta(\varepsilon_{\bk} - \mu)$ appear, it is
clear that this equation applies only close to the Fermi surface. 
Notice that the assumption of variation of $n( \bk , \br )$ over 
macroscopic length scales implies coarse graining any underlying 
microscopic theory over length scales at least of the order of the 
thermal de Broglie length
$\xi \sim v_F / \pi T$. $\xi$ measures the length over which the
quasi-particles loose their phase coherence.
Moreover, due to the collision term, (\ref{boltz}) contains
dissipation, produced by the elimination of degrees of freedom in the
coarse graining process.

Subsequently however, Landau was able
to derive the same equation from the general formalism of many-body theory
without making reference to the quasi-particle picture \cite{landau},
and one could conceive generalizations of the Fermi liquid theory based
on this equation.
For the one-dimensional Fermi liquid, however,
the analogon of the Landau-Boltzmann
transport equation has not yet been derived, and the usual derivation
fails in 1D. In the remainder of this article, we shall base our notion
of a Fermi liquid on the quasi-particle picture.

\section{Breakdown of Fermi liquid theory in one dimension}
\label{breakfl}
Adiabatic continuity is, a priori, a hypothesis which needs
verification: while it works for repulsive interactions in 3D,
it cannot be justified for attractive interactions where a transition
to superconductivity takes place -- but neither can it be justified
for repulsive interactions in 1D, the case of highest interest in the
present article.                                                       
Here, we discuss where Fermi liquid theory
breaks down in 1D. 
The first discussion is rather qualitative and handwaving. A 
second one computes the perturbation
corrections in the Green function of a 
1D Fermi gas due to some interactions and therefore probes
quasi-particles. The third part finally indicates where the derivation
of Landau's quasi-particle interactions and transport equation breaks down
and suggests that also the latter will have a new shape in 1D.

On the microscopic level, the central problem in the theory of 1D
interacting electrons is the Peierls instability \cite{Peierls},
Figure 2.1: 1D electrons spontaneously open a gap at the 
Fermi surface when they are coupled adiabatically to phonons with
wave vector \tkf.
The mechanism operates, however, also for electron-electron interactions. 
The particle-hole susceptibility in Figure 2.1 diverges as
$\ln [ \max( v_Fq, \omega)]$ if momentum $2k_F+q$ and frequency $\omega$
are transferred through the bubble. Its origin is the nesting property
of the 1D Fermi surface: one piece of the Fermi surface can be matched 
identically 
onto the other by a ``translation'' with $Q = \pm 2k_F$. 
(In higher dimensions,
in the generic case, a given $2 {\bf k}_F$ only matches two 
points -- a Fermi surface part of measure zero.)~ Summing
up a particle-hole ladder, i.e. doing a mean-field theory,
one would predict a (charge or spin)
density wave instability at some finite temperature for \em repulsive
\rm interactions -- implying that
\em there can be no Fermi liquid in 1D. 
\rm The finite transition temperature is,
of course, unphysical and an artefact of mean-field theory. It is removed
by realizing that, since the Peierls channel is as divergent as the
Cooper pairing channel, both types of instabilities interfere and one has
to solve at least a ``parquet'' of diagrams \cite{bychkov}. 
The Peierls--Cooper
interference conveys a marked non-mean-field character to this problem:
mean-field theories are constructed by selecting one important series of
diagrams. Here two of them interfere and compete!
The 1D Fermi gas is inherently unstable towards any finite interaction,
suggesting that it is not a good point of departure for analyzing 
interacting electrons in 1D. (Notwithstanding this statement, much 
progress in our understanding of 1D fermions is due 
perturbing the 1D Fermi gas by electron-electron interaction \cite{solyom}.)
There is thus urgent 
need for new low-energy phenomenology, similar in spirit to the Fermi
liquid picture, but adapted to the specific problems of 1D electrons. 

The breakdown of Fermi liquid theory in 1D is also visible in a
second order perturbation calculation, as we will demonstrate now.
We consider a simplified problem of 1D electrons with a 
density-density interaction parameterized by a coupling constant $g$.
We calculate the self-energy
$\Sigma_{rs}(q,\omega)$ in Eq. (\ref{greenfl}) in second order perturbation
theory. The relevant diagrams are shown in Figure 2.2.
Anticipating on the next Chapters, we limit ourselves to (forward) scattering
processes transferring only small momentum $q \ll k_F$,
and discuss the relevant processes 
separately in order
to avoid obscuring interferences. All arguments are robust, however.

We start with the process where all scattering partners are on the
same side $r = \pm$ of the Fermi surface, to be called $g_4$ hereafter
[the Hamiltonian is written out in Eq.~(\ref{h4})]. 
So long as $g_4$ is independent of momentum transfer, Hartree and 
Fock terms will cancel each other for scattering partners having the same
spin. If they have opposite spin ($g_{4\perp}$), (b) and (d) are absent,
and (a) only renormalizes the chemical potential. The 
self-energy (c) can be calculated and injected into (\ref{greenfl}). 
The pole of the Green function should give the energy for quasi-particle
excitations, but here we obtain
\em two solutions \rm
\begin{equation}
\label{cspole}
\omega = \left(v_F \pm \mid \frac{g_{4\perp}}{\sqrt{8} \pi} \mid \right)
\left( rk - k_F \right) \;\;\;!!!
\end{equation}
This violates the single-pole assumption at the origin of the Fermi liquid.
Anticipating Chapter \ref{chaplm} the meaning of the two poles is clear:
\em charge-spin separation. \rm 
The two poles 
are not converged into a single pole by higher order terms, which
generate more and more poles around
the two found in Eq.~(\ref{cspole}) and finally merge into a branch cut,
giving this model the specific spectral features discussed in detail in
Section \ref{secdyn}.

We now turn to forward scattering where both partners are on opposite
sides of the Fermi surface [labelled $g_2$ hereafter, cf.~Eq.~(\ref{h2})]. 
We drop spin indices since
one has only the Hartree diagrams (a) (renormalizing again
the chemical potential)
and (c) both for $g_{2\|}$ and
$g_{2\perp}$. Diagram (c) contains a counterpropagating electron-hole
pair at $\pm k_F$. This is precisely the Peierls bubble from Fig. 2.1
which gives a logarithmic dependence to $\Sigma (k, \omega)$. 
The pole in the Green function (\ref{greenfl}) now has a residue
$z_k \sim -1 / \ln | rk - k_F | \rightarrow 0 $ as $k \rightarrow k_F$.
Any quasi-particle character of the excitation fades away as we approach
the Fermi surface! Again, higher order terms
cannot restore the quasi-particle pole. They produce
higher powers of the logarithm which sum up to a power law.
These ubiquitous power laws have been mentioned in the Introduction and
will be discussed in more detail in Chapter \ref{chaplm}.

A complete and rigorous microscopic justification of the Landau theory
can be given \cite{Noz}. Here, we 
limit ourselves to a sketch of where these arguments break down in
1D. The quasi-particle interaction $f(k,k')$
defined via Landau's expansion of the total energy (\ref{landhyp})
is related through
\begin{equation}
\label{fkk}
f(k,k') = 2 \pi i z_k z_{k'} \lim_{\omega 
\rightarrow 0} \lim_{q \rightarrow 0}
\Gamma(k,E_F,k',E_F;q,\omega)
\end{equation}
to the complete particle-hole interaction vertex $\Gamma(k,E,k',E';q,\omega)$.
Notice that \em no \rm momentum transfer is involved in the 
qua\-si-par\-ti\-cle interaction.
The complete particle-hole interaction $\Gamma$ is related to the irreducible
one $I$ by the Bethe-Salpeter equation which we only display graphically
in Figure 2.3.
Singularities in $\Gamma$ are required for eventually destabilizing 
quasi-particles ($\Gamma$ determines the two-particle Green function
which is coupled, via the interaction, into the single-particle Green
function). So long as $I$ is nonsingular, singular $\Gamma$ can only arise
from the internal Green functions in the right 
diagram in Figure 2.3.
Physically, they represent that part of the effective interaction which
is mediated by propagating particles. 
There are, in fact,  such
singularities when the difference of (four-)momenta on the internal lines 
tends to zero. Due to the particular limit involved in Eq.~(\ref{fkk}),
the quasi-particle interaction is not sensitive to these singularities
and remains regular. The singularities matter, however, in opposite
(forward scattering) limit $\omega = 0, \; q \rightarrow 0$ when momentum
transfer is allowed. Then collective (zero sound) modes can be excited.
Their velocity, however, exceeds the Fermi velocity so that they do not
interfere with the quasi-particles.

Now consider one dimension. As we have seen in Figure 2.1
above there is a logarithmic
singularity in the (Peierls) particle-hole susceptibility at $q = 2k_F$.
It is clear from Figure 2.3 that the Peierls bubble 
gives an additional divergence in the Bethe-Salpeter equation when
the momentum of the internal Green functions differs by \tkf\
whereas 
the derivation of Landau theory assumes this vertex to be finite at $2k_F$. 
Moreover, the Peierls divergence is worse than that at $q=0$ in that
the internal momentum integrals sample the full bandwidth; at small $q$,
the pole structure of the singular part is such that one only integrates
over a slice of width $q$. This is why the singularity does not enter the
quasi-particle interaction. The two-particle Green function then
carries the singularity in $\Gamma$ 
into the single-particle Green function where it will ruin the 
quasi-particle pole. 

The quasi-particle interaction $f(k,k')$ in (\ref{fkk}) does not
involve momentum transfer between the quasi-particles. In other
words, the components of the interaction which do transfer momentum
are irrelevant in 3D. This is very different from 1D where, on dimensional
grounds, these interactions are marginal and cannot be neglected compared
to those which do not transfer momentum. This finally generates 
charge-spin separation. We shall give a more detailed argument in the
next chapter, at the end of Section \ref{symm}, 
after we have introduced the relevant Hamiltonian. A reduction of the
interactions to an effective quasi-particle interaction (\ref{fkk}) cannot
be operated in 1D. 

Any search for an extension or a replacement of Fermi liquid theory
in 1D must necessarily incorporate in a consistent manner the Peierls
divergence and momentum transfer in the interaction process. 
This is what the Luttinger liquid approach, to be discussed
in the next chapter, does. A valid though less satisfactory
alternative, starting from a 1D Fermi gas, is offered by either solving 
parquet equations or performing \rg\ as ``devices'' to sum up consistently
the offending divergences discussed here \cite{solyom,bychkov}.

\chapter{The Luttinger model}
\label{chaplm}
\section{Low-energy phenomenology in 1D -- the Luttinger model}
\label{lowenph}
\subsection{Ground state and elementary excitations of 1D fermions}
\label{gdstate}
We have seen in the previous chapters that both the Peierls singularity
and charge-spin separation,
both related to the small phase space in 1D, 
spoil a Fermi liquid description of 1D correlated fermions and require
new approaches. 
On the other hand, the rewarding feature of 1D physics is that
the particle-hole 
excitations acquire well-defined particle-like dispersion in
the long-wavelength limit $q \rightarrow 0$, Figure 3.1.
These collective density fluctuations obey approximately
bosonic commutation relations and can indeed be used to construct 
the new low-energy phenomenology called for.

To describe the low-energy physics, we need to know the ground state and
the elementary excitations. Consider a system with $N_{0}$ electrons in
a system of length $L$. In the absence of external fields,
the ground state of the free system is the
Fermi sea $|FS \rangle$ with $k_F = N_0 \pi / 2 L $. 
In general, the ground state may be different,
however. In a magnetic field, the number of up- and down-spins,  is 
different, $k_{F \uparrow} \neq k_{F \downarrow} $, 
and in an electric field, the number of right- and left-moving fermions
is different, $k_F^{(+)} \neq - k_F^{(-)}$, producing a net magnetization
and current, respectively. Varying the chemical potential changes all four
$k_F$.  With respect to the reference state given by
$k_F$, one can therefore introduce four numbers $N_{r,s}$
measuring the addition or removal of fermions, 
above or below the reference $k_F$, in the channel $(r,s)$,
where $r$ labels the dispersion branch close to $rk_F$ and $s$ the spin. 
The total charge and spin as
well as charge and spin currents with respect to the reference state
are obtained by linear combination.

What are the elementary excitations? For the
free system, one could add a fermion in a $k$-state with $ | k | > k_F$
and create a quasi-particle. However, we have seen in the preceding 
chapter that these
quasi-particles are not stable against turning on interactions. Next consider
particle-hole excitations $c^{\dag}_{k+q} c_k | FS \rangle$, 
Fig.~3.1 (left). Firstly, notice that
the electron
and hole created travel at the same group velocity and therefore form an
almost bound state which is certainly extremely susceptible to interactions, 
particularly in 1D. 
Secondly, for small $q$ where the dispersion is almost
linear, there is a huge degeneracy $L | q | / 2 \pi$ of these
excitations with energy $\omega(q)$. We can form ``particles'', 
corresponding to the linear dispersion branch $\omega(q) \propto |q|$
for $q \rightarrow 0$ in Fig.~3.1,  by coherently superposing particle-hole 
excitations with different $k$, Eq.~(\ref{rhoeq}) below.  
These fluctuations are also present in the Fermi liquid but there,
the low-energy spectral region ($\omega < v_F k_F, \; 0 < q < 2k_F$)
is filled in. The presence of these finite-q low-energy states allows
the decay of these excitations into their constituent quasi-particles and
therefore is responsible for a kind of ``chemical'' equilibrium between
quasi-particles and collective excitations. In 1D this decay into
quasi-particles is not possible and makes these charge-
(spin-) fluctuations stable elementary excitations of the system.
They have bosonic commutation properties. 

With respect to our reference state $|FS \rangle$, there are thus two
types of elementary excitations: (i) the charge and spin and their 
corresponding current excitations which change the Fermi wavevectors
$k_{F,s}^{(r)}$ and thus the number of fermions in the system \cite{Haldane},
and (ii) the collective bosonic charge and spin-density fluctuations. 
There are no stable quasiparticles, and the addition of a fermion 
generates both types of elementary excitations, cf.~Eq.~(\ref{bos}) below.

These features are generic to 1D gapless quantum systems but are particularly
prominent in the exactly solvable \lm. In the following, we describe and
solve this model before we turn, in the next chapter, to the reduction
of microscopic lattice models (e.g. Hubbard model) onto effective Luttinger 
Hamiltonians.

\subsection{Tomonaga-Luttinger Hamiltonian}
\label{lumod}
The \tlm\ describes 1D right- and left-moving fermions
through the Hamiltonian \cite{Haldane}, \cite{lut} -\cite{matl},
\cite{heiden}-\cite{everts}
\begin{eqnarray}
\label{hlutt}
H & = & H_0 + H_2 + H_4 \;\;\;, \\
\label{hfree}
H_0 & = & \sum_{r,k,s} v_F (rk - k_F) 
: c^{\dag}_{rks} c_{rks} : \;\;\;, \\
\label{h2}
H_2 & = & \frac{1}{L} \sum_{p,s,s'} \left[ g_{2\|}(p) \delta_{s,s'} +
g_{2\perp}(p) \delta_{s,-s'} \right] \rho_{+,s}(p) \rho_{-,s'}(-p) \;\;\;, \\
\label{h4}
H_4 & = & \frac{1}{2L} \sum_{r,p,s,s'} \left[ g_{4\|}(p) \delta_{s,s'} +
g_{4\perp}(p) \delta_{s,-s'} \right] : \rho_{r,s}(p) \rho_{r,s'}(-p) : 
\;\;\;. 
\end{eqnarray}
$c_{rks}$ describes fermions with momentum $k$ and spin $s$ on the
two branches ($r=\pm$) of the dispersion varying linearly [$\varepsilon_r(k)
= v_F (rk - k_F)$] about the two Fermi points $\pm k_F$. 
\begin{equation}
\label{rhoeq}
\rho_{r,s}(p) 
=  \sum_k : c^{\dag}_{r,k+p,s} c_{r,k,s} : 
=  \sum_k \left( c^{\dag}_{r,k+p,s} c_{r,k,s} -
\delta_{q,0} \langle c^{\dag}_{r,k,s} c_{r,k,s} \rangle_0 \right) 
\end{equation}
is the density fluctuation operator (describing the ``particles'' introduced
above), and
$: \ldots :$ denotes normal ordering, defined by the second equality. 
The Tomonaga and \lm s are distinguished by different cutoff prescriptions
on the dispersion. In the Tomonaga model \cite{tom} there is a finite
bandwidth cutoff $k_0$, i.e. the allowed $k$-space states for branch $r$
are $rk_F - k_0 < k < rk_F + k_0$. This simulates the finite bandwidth
of all real physical systems but, unfortunately, 
only allows an asymptotically
exact solution. In the \lm\ \cite{lut}, 
on the contrary, the dispersion extents to
infinity: $-\infty < k < \infty$ for both branches. In order to obtain
physically meaningful results, all the negative energy states have to
be occupied. The presence of these unphysical states is not expected to
affect the low-energy physics of the model ($| \omega | \ll E_F, \; 
| q | \ll k_F$). (A thorough discussion of various cutoff procedures
is given by S\'{o}lyom \cite{solyom} and Apostol \cite{apostol}.)
The normal ordering convention in Eqs.~(\ref{hfree})
and (\ref{rhoeq}) is necessary to
avoid reference to the infinite quantity $ \sum_k \langle
c^{\dag}_{rks} c_{rks} \rangle$, the total particle number, which is 
ill-defined. The coupling constants $g_2$ and
$g_4$ measure the strength of forward scattering  (momentum transfer
$| q | \ll k_F$) between particles on different or on the same
branch of the dispersion, respectively, Figure 3.2. 
They may
depend on the relative orientation of the spin of the scattering particles.
The interaction terms with $p = 0$ give the change
in Hartree-Fock energy of the system upon addition of particles, and
those with finite $p$ describe the scattering of the elementary excitations.

An exact solution of the \lm\ 
is possible \cite{Haldane,matl,heiden,dzya} if a cutoff $\Lambda$ is imposed
on the momentum transfer of the interactions \cite{Haldane,heiden}. 
The coupling ``constants'' $g_i(p)$ therefore depend on the momentum
transfer (below, we shall exhibit explicitly this momentum dependence
only where necessary).

\subsection{Symmetries and conservation laws}
\label{symm}

The possibility for an exact solution of the Luttinger model can be traced
back to severe conservation laws. The Hamiltonian not only conserves
the total charge and spin of the system 
\begin{equation}
\label{constot}
[N_{\rho}, H ] = 0 \;\;\;, \hspace{1cm} [N_{\sigma}, H] = 0
\end{equation}
but is does so separately on
each branch $r$
\begin{equation}
\label{conssep}
[N_{r,\rho}, H ] = 0 \;\;, \;\;\;\; [N_{r,\sigma}, H] = 0 \;\;,
\hspace{1cm} {\rm  or} \;\;\; [N_{r,s}, H ] = 0 \;\;\;.
\end{equation}
Clearly, this implies conservation of the charge and spin currents
\begin{equation}
\label{conscur}
[J_{\rho}, H] = 0 \;\;, \;\;\;\; [J_{\sigma}, H] = 0 \;\;, \hspace{1cm} 
{\rm or} \;\;\;
[J_s, H] = 0 \;\;\;.
\end{equation}
Consequently, the Hamiltonian is invariant under the gauge transformations
\begin{equation}
\label{gauge}
\Psi_{rs}(x) \rightarrow \exp(i \theta_r) \Psi_{rs}(x)
\end{equation}
for each branch separately. Expressed differently, the \lm\ possesses,
in addition to the usual gauge symmetry $\Psi_{rs}(x) \rightarrow 
\exp(i \theta) \Psi_{rs}(x)$, a chiral symmetry
\begin{equation}
\label{chiral}
\Psi_{rs}(x) \rightarrow \exp(i r \theta) \Psi_{rs}(x) \;\;\;.
\end{equation}
The physical origin of these conservation laws is the restriction of the
interaction Hamiltonian $H_2 + H_4$ to small momentum transfer (forward)
scattering: processes scattering particles across the Fermi surface are
excluded from the model.                                

For specific values of the interaction constants
\begin{equation}
\label{spinrot}
g_{2,\|} = g_{2,\perp} \;\;\; {\rm and} \;\;\; g_{4,\|} = g_{4,\perp} \;\;\;,
\end{equation}
the Hamiltonian is invariant under a spin-rotation
\begin{equation}
\label{spintf}
\Psi_{rs}(x) \rightarrow \sum_{s'} g_{ss'} \Psi_{rs'}(x) \;\;\;,
\end{equation}
where $g_{ss'} = (\exp[i {\bf \Omega} \cdot {\bf \sigma}])_{ss'}$ is a
SU(2)-matrix. As will be seen below, 
correlation functions are spin-rotation invariant also when only the
left equation in (\ref{spinrot}) is fulfilled. 

The linear dispersion and the normal ordering involved in the
density operators (\ref{rhoeq}) makes the model charge-conjugation symmetric
[$ \Psi_{rs}(x) \rightarrow \Psi_{rs}^{\dag}(x)$]. While a more complete model
need not be charge conjugation symmetric, linearizing the dispersion 
amounts to a constant-density-of-states approximation -- often employed also
in higher-dimensional systems. Finally, when $g_{2,s,s'} = g_{4,s,s'}$
the Luttinger model can be considered as the small momentum transfer limit
of a physical Hamiltonian involving only density-density interactions. 

Conservation of total charge and spin applies to most models commonly studied.
Their conservation separately on each branch is, however, a specific
property of the \lm\ and not shared by
more realistic 1D models. There, it holds in a low-energy subspace if 
interaction terms not commuting with the charge and spin currents are
irrelevant. If they are relevant, the low-energy physics
is characterized by different (possibly reduced) symmetries and cannot
be described by an effective Luttinger model. Two such interaction 
processes are depicted in  Figure 3.3.
$g_{1\perp}$ describes exchange scattering across the Fermi surface
and spoils spin current conservation but respects charge current conservation.
$g_{3\perp}$ is Umklapp scattering of two particles in the same direction
across the Fermi surface and destroys charge current conservation while
conserving the spin current. However, momentum conservation usually
inactivates $g_{3\perp}$, except for commensurate band fillings. Other
interaction processes violating both charge and spin current conservation are
possible, too.

Charge conjugation and spin rotation occur as separate symmetries 
because of the interactions. The free Hamiltonian has a higher symmetry
which, however, is broken by the interaction terms. The $g_2$-interaction
does not commute with the kinetic energy $[H_2 , H_0 ] \neq 0$. It therefore
can modify the ground state by exciting particle-hole pairs out of the
Fermi sea. On the other hand, $g_4$ commutes $[H_4 , H_0] = 0$.
With this term alone the Fermi sea remains the ground state.
Its influence is limited to removing degeneracies in the excitations,
as can a magnetic field or a hopping matrix element between chains.

The corresponding interactions are present also in higher
dimensions. Still, it seems that they play no role there. The reason is
that these interactions are marginal 
or scale invariant in 1D, in a \rg\ sense, while they are irrelevant in
$D > 1$ and drop out of the problem. 
Marginality means that the coupling constant
does not change under a change in the length (or energy) scale while
relevance or irrelevance imply an increase resp. decrease of the coupling
constant as the length (energy) scale is increased (decreased). 
The marginality of $g_2$ and $g_4$ can be seen by simple power counting. 
Taking the length scale to have canonical
dimension $[L] = 1$, the Hamiltonian has
$[H] = -1 $, the fermion operator has $[\Psi_r(x)] = -1/2$ and the 
density operator $[\rho_r(x)] = -1$. Consequently $[g_2] = [g_4] = 0$
i.e. $g_2$ and $g_4$ do not change with scale. The dimension of $\Psi_r(x)$
can be changed by the presence of marginal operators of $g_2$-type but
not by those of $g_4$-type. The dimension of $\rho_r(x)$ is not changed
by marginal interactions. Notice that the coupling
constants $g_i$ transfer momentum in the scattering process, 
and their marginality
implies that this momentum transfer cannot be neglected 
on any length (energy)
scale. In contrast, the momentum transfer of interactions in 3D can be 
neglected 
because the interaction is irrelevant (the explicit prefactor $L^{-3}$ 
in $H_{\rm int}$ gives it a dimension $-2$).

To see the consequences of this marginality, and in particular of $g_4$,
inject a particle into the second empty
plane wave state above the Fermi surface $| \varphi_1 \rangle
= c^{\dag}_{k_F+4 \pi / L, s} | FS \rangle$. 
Where can it be transferred
by $H_4$ which conserves the energy? The only allowed process 
relaxes it into the first empty state above the Fermi surface and excites the
last particle from the Fermi sea to the same empty state:
$|\varphi_2 \rangle = H_{4} | \varphi_1 \rangle = c^{\dag}_{k_F + 2\pi/L, -s} 
c_{k_F, -s} c^{\dag}_{k_F + 2\pi /L, s}$ $|FS \rangle$. Now, within this
two-state subsystem $\{ | \varphi_1 \rangle , | \varphi_2 \rangle \}$,
the Hamiltonian reduces to the matrix
\begin{equation}
H = \left( \begin{array}{ll} 
4 v_F \pi / L & 2 g_{4 \perp} / L \\
2 g_{4 \perp} / L & 4 v_F \pi / L
\end{array} \right) \;\;\;.
\end{equation} 
The diagonal terms come from the kinetic energy, and  the $1/L$-factor
comes in from the quantization of the $k$-vectors, and the off-diagonal
interaction terms are proportional to $1/L$ because of the explicit
normalization factor in the Hamiltonian $H_4$. Interaction and kinetic
energy both scale with $1/L$ and are of equal importance! Carrying 
through the same argument in 3D, the kinetic terms will continue to scale
with $1/L$ while the interaction terms scale with $1/L^3$ and therefore
can safely be neglected at finite momentum transfer
\cite{mechtha}. The new eigenvalues are $ (4 \pi /L) (v_F \pm g_{4\perp}
/2 \pi)$ suggesting that the particles have split into two objects propagating
at two different renormalized velocities 
$v_{\rho,\sigma} = v_F \pm g_{4\perp}/ 2 \pi$ -- charge-spin separation. 

The argument continues
to hold as one injects the particles at higher momenta $2 n \pi / L$
where $H_4$ couples it to $n-1$ other states of lower energy. It also
carries over to the $g_2$-interaction. Due to the non-conservation of 
energy, however, an infinite number of states are coupled to the particle
at any momentum. Momentum transfer scattering, therefore, can never be
neglected in 1D, and a reduction to a quasi-particle interaction
(\ref{fkk}) cannot be justified in any circumstances.
The marginality of forward scattering with finite momentum transfer
in 1D is at the origin both of the anomalous correlation exponents
and charge-spin separation which we shall discuss in more detail in the
subsequent sections. It is \em the \rm important difference to 
higher-dimensional systems.

\section{Boson solution of the Luttinger model}
\label{secham}
A variety of solutions for the \lm\ have been produced in the past.
Historically, the first solution involved a boson representation of
the Hamiltonian \cite{matl} and will be reviewed first. This solution
was ``completed'' by the construction of a boson representation for
the fermion operators \cite{lupe,mattis} 
which has been made rigorous by Haldane \cite{Haldane} and Heidenreich
et al.~\cite{heiden}. It emphasizes the differences between Fermi liquids
in one and in higher dimensions. Methods developed in Fermi liquid theory,
more strongly emphasizing similarities between 1D and 3D Fermi liquids,
have also been used to solve the \lm\ \cite{dzya,everts}
and are reviewed in Section \ref{altmet}. 

\subsection{Diagonalization of Hamiltonian}
\label{hamdia}

The \tlm\ (\ref{hlutt}), (although using Luttinger's version with infinite
bands and a momentum transfer cutoff in the interactions
throughout, we shall often attach Tomonaga's name to the model, too)
describes \em excitations \rm with respect to a ground state
described by the Fermi wave vector $k_F = (\pi / 2)(N_0/L)$
where $N_0$ is the number of physical electrons in a chain of length
$L$. Due to the unphysical negative energy states, 
$N_0 \neq \sum_{rks} \langle c^{\dag}_{rks} c_{rks} \rangle_0$; the 
left-hand side of this equation is finite, the right-hand side is infinite.
The infinitely extended dispersion introduces many more subtleties
into the model which are crucial to obtain a correct solution.
There are three important steps in achieving a complete solution of this
model: (i) the realization that due to the infinite dispersion, the 
$\rho_{r,s}(p)$ obey exact boson commutation relations \cite{matl}; 
(ii) a representation
of the free Hamiltonian (\ref{hfree}) as a bilinear in these boson operators
\cite{matl}; (iii) the explicit construction of a boson representation
for the fermion operators $\Psi_{rs}(x) = (1/\sqrt{L}) \sum_k c_{rks} 
\exp(ikx)$. 

``Normally'' (the precise meaning of this will become apparent below)
density operators commute $[\rho_s(p_1),
\rho_s(p_2)] = 0$ because their Fourier transforms $\rho_s(x) = 
\Psi_s^{\dag}(x) \Psi_s(x)$ are \em local \rm objects. This is no longer true
for Luttinger's density operators 
because of the fermion doubling
\begin{equation}
\Psi_s(x) \rightarrow \Psi_{r,s}(x) \;\;\;; \hspace{1cm}
c_{ks} = \sum_{r = \pm} \Theta(rk) c_{rks} \;\;\;.
\end{equation}
$\Theta(x)$ is the step function. There is now a \em
nonlocal \rm relation between the physical fermions $\Psi_s(x)$
and the right- and
left-moving $\Psi_{r,s}(x)$
\begin{equation}
\Psi^{\dag}_{s}(x) = \frac{1}{2 \pi i} \sum_r \int_{-L/2}^{L/2}
\! dy \: K(y) \: \Psi^{\dag}_{r,s}(x+y) \;\; {\rm with} \;\; 
K(y) = \frac{\pi}{L} \cot \left( \frac{\pi y}{L} \right) \;\;,
\end{equation}
and the density operators no longer commute.

The commutator of the density operators is
\begin{equation}
\label{comm}
[\rho_{r,s}(p), \rho_{r',s'}(-p') ]  = \delta_{r,r'} \delta_{s,s'}
\sum_k \left( c^{\dag}_{r,k+p,s} c_{r,k+p',s} - 
c^{\dag}_{r,k+p-p',s}c_{r,k,s} \right)
\;\;\;.
\end{equation}
In a finite band \em containing both $\pm k_F$ \rm ($c_{k,s}$
\em without \rm the subscript
$r$), 
it is permissible to change the summation variable
$k \rightarrow k+p'$ in the second term which makes the commutator
vanish. For the Tomonaga model (finite bands around $\pm k_F$), 
for $p \neq p'$ one has an operator acting on the states near the band
edges $rk_F \pm k_0$, and the approximate bosonic commutators of the
Tomonaga model are obtained by neglecting these band edge terms. For
$p=p'$ one measures the difference in the number of occupied states
at $k$ and $k+p$, i.e. $p$ -- making the commutator a finite number.
For infinite bands (\lm ), 
one manipulates the (ill-defined) difference of two infinite quantities,
and one must introduce normal ordered operators, Eq.~(\ref{rhoeq}),
into the right hand side of Eq. (\ref{comm}).
The problem of the band edge terms is then rigorously absent since there
is no band edge left, and for $p=p'$ the argument for the Tomonaga model
carries over:
\begin{eqnarray}
[\rho_{r,s}(p), \rho_{r',s'}(-p') ] & = &  \delta_{r,r'} \delta_{s,s'} 
\sum_k : c^{\dag}_{r,k+p,s} c_{r,k+p',s} - 
c^{\dag}_{r,k+p-p',s}c_{r,k,s} : \nonumber \\
& + & \delta_{r,r'} \delta_{s,s'}
\delta_{p,p'} \sum_k [ \langle n_{r,k+p,s} \rangle_0 
- \langle n_{rks} \rangle_0 ] \nonumber  \\
\label{tlmcom}
& = & - \delta_{r,r'} \delta_{s,s'} \delta_{p,p'} \frac{rpL}{2 \pi} \;\;\;.
\end{eqnarray}
One can safely change the summation variable in
the first line of (\ref{tlmcom}) because the operators are normal-ordered;
the two terms add up to zero, leaving the contribution of the second line.
In the Tomonaga model with finite
bands for right- and left-movers, the boson algebra obtains approximately
(for wave vectors far from the band edges) because one 
works with \em truncated \rm density operators. 
The algebra (\ref{tlmcom})
is known as the $U(1)$ Kac-Moody algebra in field theory, and the 
nonvanishing of the commutator (\ref{tlmcom}) 
due to the infinite number of negative energy states (\lm ) or the
cutoff procedure (Tomonaga model) is called an ``anomaly''.

Acting on the ground state of the free Hamiltonian $H_0$, the $\rho_{r,s}(p)$
behave either as creation or annihilation operators, depending on sign$(p)$
\begin{equation}
\rho_{+,s}(-p) |0 \rangle = \rho_{-,s}(p) |0 \rangle 
= 0 \;\; {\rm for} \;\; p>0 \;\;\;.
\end{equation}
To complete the algebra, 
it is necessary to construct a ladder operator $U_{rs}$ which
changes the fermion number without affecting the bosonic
excitations. This operator is necessary again because of the infinite
dispersion: since there are no upper and lower limits to the number of
particles, the number operator cannot be expressed in terms of raising
and lowering operators. Haldane and Heidenreich et al. 
have given such a construction in terms
of the bosons $\rho_{r,s}(p)$ and the fermions $\Psi_{rs}(x)$
\cite{Haldane,heiden}, cf.\ below.
              
There are several ways to see that the free fermion Hamiltonian $H_0$,
Eq.~(\ref{hfree}), is equivalent to an operator bilinear in the bosons
\rr. The simplest one \cite{matl} is to examine the commutator
\begin{equation}
[H_0, \rho_{r,s}(p) ] = v_F \: r \: p \:  \rho_{r,s}(p) 
\end{equation}
which is obviously compatible with
\begin{equation}
\label{hkr}
H_0 = \frac{\pi v_F }{L} \sum_{r,p \neq 0,s} : \rho_{r,s}(p) \rho_{r,s}(-p) : 
\: + \: {\rm const.} 
\end{equation}
The equivalence of the Hamiltonians (\ref{hfree}) and (\ref{hkr}) is
known as Kronig's identity \cite{kronig}, and is valid at fixed particle
number. If particles are added to the system , the ``$+$const.'' becomes
important, however, because one must add their kinetic energy
to the Hamiltonian. 
We put them into the lowest available states above the Fermi sea
(other states can be reached by acting with the boson operators).
The complete Hamiltonian then takes the form
\begin{eqnarray}
\label{h0bos}
H_0 & =  &
\frac{\pi v_s}{L} \sum_{r, p\neq 0, s} : \rho_{r,s}(p) \rho_{r,s}(-p) : \\
& + & \frac{\pi}{2L} \sum_s \left[ v_N (N_{+,s} + N_{-,s})^2 + v_J 
(N_{+,s}-N_{-,s})^2 \right] \;\;\;, \nonumber \\
(-1)^J_s & = & - (-1)^N_s \;\;\;, \hspace{1cm} (v_s = v_N = v_J = v_F) 
\;\;\;, \nonumber
\end{eqnarray}
where the $N_{r,s} \equiv \rho_{r,s}(p=0)$, Eq.~(\ref{rhoeq}), 
are taken relative to their (infinite) ground state value and therefore
measure excitations with respect to a given ground state charge.
The symmetric combination $N_{s} = \sum_r N_{r,s}$ measures charge
and the antisymmetric combination $J_{s} = \sum_r r N_{r,s}$ measures
current excitations, 
both carrying spin $s$. Total charge and spin, as well as charge
and spin currents, are obtained by the appropriate sums over $s$.
These quantities specify the number and the left-right asymmetry of the 
fermions added to the reference state ($N_{s} = N_0/2, \; J_s = 0$).
The equality of the three velocities in (\ref{h0bos}) to the bare one
only holds for the free model and is violated by interactions
(the Hartree-Fock energy of the added particles appears as the 
$q=0$ components of the interaction Hamiltonian). 
Including the charge and current excitations, 
(\ref{hfree}) and (\ref{h0bos}) possess the same spectrum, by construction.
That the multiplicities of the levels also are equal
can be proved by calculating the grand partition function both in the
fermion (\ref{hfree}) and in the boson (\ref{h0bos}) representation. 
Thus the fermionic and bosonic Hilbert spaces are identical. 

Why are such two different representations of the same Hamiltonian possible?
(i) Reconsider the elementary
particle--hole excitations in Figure 3.1.
They acquire a well-defined
particle-like character in 1D as $q \rightarrow 0$. 
In the \lm\, the low-$q$ branch of their 
dispersion is strictly linear in $q$. Decay of these excitations 
in the constituent
particles and holes is forbidden on account of 1D kinematics -- it would
involve states in the void low-frequency part of the spectrum. There should
thus be a representation of the Hamiltonian, which describes \em excitations, 
\rm in terms of these particles
alternative to the original fermionic one. Moreover, the absence of dispersion
implies that these excitations do not interact: one excitation with momentum
$q + q'$ has the same energy as two excitations with momenta $q$ and $q'$. 
Certainly, these collective modes also exist in higher
dimensions, but so does the electron-hole continuum which permits their
decay into quasi-particles and quasi-holes. 
(ii) An intimately related observation is that the particle and the hole
created in such an excitation, travel at the same group velocity
and therefore form an almost bound state which surely is extremely
susceptible to dramatic modification by interactions where, in any case,
momentum transfer cannot be neglected.
(iii) All states with even (odd) fermion charge 
$N- N_0$ have excitation energies that
are even (odd) multiples of $\pi v_F / L$. In other words, the spectrum
effectively becomes that of a harmonic oscillator. This fact again suggests
that an equivalent boson representation of $H_0$ should be possible.
(iv) The Kac-Moody algebra (\ref{tlmcom}) can be obtained either by
representing $\rho_{rs}(p)$ as a fermion bilinear (\ref{rhoeq}) or as
the gradient of true bosonic field \firs\ [Eq.~(\ref{phirs}) below].
Since the algebra is unique, the two representations must be equivalent.
While for the noninteracting problem, the two representations are true
alternatives, the success of bosonization is related to the fact that
the bosonic one becomes more ``natural'' once interactions are introduced.

Now the Luttinger Hamiltonian can
be diagonalized by a Bogoliubov transformation 
\cite{Haldane,matl,heiden}.
First transform 
to charge and spin variables
\begin{eqnarray}
\rho_{r}(p) & = & \frac{1}{\sqrt{2}} \left[\rho_{r,\uparrow}(p) +
\rho_{r,\downarrow}(p) \right] \;\;\;, \;\;\; N_{r,\rho} = \frac{1}{\sqrt{2}}
\left[ N_{r,\uparrow} + N_{r,\downarrow} \right] \;\;\;, \nonumber \\
\label{chsdens}
\sigma_{r}(p) & = & \frac{1}{\sqrt{2}} \left[\rho_{r,\uparrow}(p) -
\rho_{r,\downarrow}(p) \right] \;\;\;, \;\;\; N_{r,\sigma} = \frac{1}{\sqrt{2}}
\left[ N_{r,\uparrow} - N_{r,\downarrow} \right] \;\;\; .
\end{eqnarray}
We only include the $z$-component of the spin density operator working
within abelian bosonization. At this point, the SU(2)-spin transformation
properties of the fermions (\ref{spintf}) has been broken down to U(1)
[just like the gauge transformation (\ref{gauge}) for the charges], and
likewise for the symmetry of the Hamiltonian, even if (\ref{spinrot}) is
satisfied. One can keep the spin densities transforming explicitly 
according to SU(2)
\begin{equation}
\label{spinsu2}
{\bf S}_r(x) = \sum_{s,s'} \frac{1}{2} \Psi^{\dag}_{rs}(x) {\bf \sigma}_{ss'}
\Psi_{rs'}(x) 
\end{equation}
and represent the Hamiltonian in terms of the U(1)-$\rho_{r}$- and
SU(2)-${\bf S}_r$-fields. In this way, one can keep SU(2)-invariance
manifest at every stage of the calculation. The price to be paid is, however,
a significantly more complicated boson respresentation which will not
be reviewed here \cite{nonabbos,witten}. 

The interactions transform as
\begin{equation}
g_{i\rho} = \frac{1}{2} \left( g_{i\|} + g_{i\perp} \right) \;\;\; , \;\;\;
g_{i\sigma} = \frac{1}{2} \left( g_{i\|} - g_{i\perp} \right) \;\;\; .
\end{equation}
The Hamiltonian then becomes ($\nu = \rho, \sigma$ henceforth)
\begin{eqnarray}
\label{hnu}
H_0 & = & \frac{\pi v_F}{L} \sum_{\nu r p \neq 0} : \nu_r(p) \nu_r(-p) : \\
& + & \frac{\pi}{2L} \left[ v_{N\nu} (N_{+\nu} + N_{-\nu})^2
+ v_{J\nu} (N_{+\nu} - N_{-\nu})^2 \right] \;\;\;\;\;
(v_{N\nu} = v_{J\nu} = v_F) \;\;\;, \nonumber \\
H_2 & = & \frac{2}{L} \sum_{\nu p} g_{2\nu}(p) \nu_+(p) \nu_-(-p) \;\;\;, \\
H_4 & = & \frac{1}{L} \sum_{\nu r p} g_{4\nu}(p) : \nu_r(p) \nu_r(-p) :
\;\;\; .
\end{eqnarray}
We diagonalize by the canonical transformation
\begin{eqnarray}
\label{canon}
\tilde{H} & = & e^{iS_{\nu}} H e^{-iS_{\nu}} \;\;\;, \;\;\;
\tilde{\nu}_r(p)  =  e^{iS_{\nu}} \nu_r(p) e^{-iS_{\nu}} \nonumber \\
S_{\nu} & = & \frac{2 \pi i}{L} \sum_{p>0} \frac{\xi_{\nu}(p)}{p}
\left[ \nu_+(p) \nu_-(-p) - \nu_-(p) \nu_+(-p) \right] \;\;\;.
\end{eqnarray}
The $\nu_r$'s explicitly transform as
\begin{equation}
\tilde{\nu}_r(p) = v_r(p) \cosh [\xi_{\nu}(p) ] + 
\nu_{-r}(p) \sinh [\xi_{\nu}(p) ] \;\;\;,
\end{equation}
and $\tilde{H}$ is diagonal under the condition
\begin{equation}
\label{knu}
K_{\nu}(p) \equiv e^{2 \xi_{\nu}(p)} = 
\sqrt{\frac{\pi v_F + g_{4\nu}(p)-g_{2\nu}(p)}{\pi 
v_F+g_{4\nu}(p)+g_{2\nu}(p)}} \;\;\;.
\end{equation}
For repulsive interactions, $K_{\nu} < 1$ while for attraction 
$K_{\nu}>1$.
The diagonal form is then
\begin{eqnarray}
\label{hdiag}
\tilde{H} & = & \frac{\pi}{L} \sum_{r \nu p \neq 0 } 
v_{\nu}(p) : \tilde{\nu}_r(p) \tilde{\nu}_r(-p) : \\
& + & \frac{\pi}{2L} \left[ v_{N\nu} (N_{+\nu} + N_{-\nu})^2
+ v_{J\nu} (N_{+\nu} - N_{-\nu})^2 \right] \;\;\;, 
\nonumber \\
\label{speccoup}
& {\rm with} & v_{N\nu} v_{J\nu} = v_{\nu}^2 \;\;\;
 {\rm i.e.}  \;\;\; v_{N\nu} = v_{\nu} / K_{\nu} \;\;\;\; {\rm and} \;\;\;\;
v_{J\nu} = v_{\nu} K_{\nu} \;\;\;.
\end{eqnarray}
(The $N_{r\nu}$-operators are not changed by the canonical transformation.)~
The renormalized charge and spin fluctuation velocity is
\begin{equation}
\label{vnu}
v_{\nu}(p) = \sqrt{\left[ v_F + \frac{g_{4\nu}(p)}{\pi} \right]^2
- \left[\frac{g_{2\nu}(p)}{\pi}\right]^2} \;\;\;,
\end{equation}
and therefore
\begin{equation}
\label{nvj}
v_{N\nu} = v_F + g_{4\nu} + g_{2\nu} \;\;\;, \;\;\;
v_{J\nu} = v_F + g_{4\nu} - g_{2\nu} \;\;\;,
\end{equation}
and the limit $p \rightarrow 0$ is implied whenever $p$ is not exhibited
explicitly. Due to the momentum transfer cutoff $\Lambda$, we have
asymptotically
\begin{equation}
\label{asymp}
K_{\nu}(p) \rightarrow \left\{ \begin{array}{rll}
K_{\nu} & {\rm for} & p \ll \Lambda \\
1 & {\rm for} & p \gg \Lambda \end{array} \right.
\hspace{1cm} 
v_{\nu}(p) \rightarrow \left\{ \begin{array}{rll}
v_{\nu} & {\rm for} & p \ll \Lambda \\
v_F & {\rm for} & p \gg \Lambda \end{array}  \right. \;\;\;.
\end{equation}

Eqs.~(\ref{hdiag}) and (\ref{speccoup}) are the central constitutive
relations for the \lm\, and the \LL\ hypothesis discussed in the
next chapter will postulate that these relations continue to hold 
in a low-energy
subspace of all solvable gapless 1D models. 
The quantity $K_{\nu}(p \rightarrow 0) $, Eq.~(\ref{knu}), is the essential
renormalized coupling constant for each degree of freedom, and physically
plays the
role of a stiffness constant. $K_{\nu}$
governs the power-law
decay of most correlation functions. The two parameters
$v_{\nu}$ and $K_{\nu}$ completely describe the low-energy physics of
each degree of freedom $\nu$ of the model.
That there are just two such parameters is not surprising: 
the Hamiltonian has only two parameters $g_{2\nu}$ and
$g_{4\nu}$, and we just get back what we have put in. Important are
the following facts: (i) the three different velocities in the 
problem are all renormalized by the interactions,
Eq.~(\ref{speccoup}), and describe different physical processes. $v_{\nu}$,
the renormalized Fermi (or ``sound'') velocity governs the bosonic
excitations; $v_{N\nu}$ is related to the fermionic charge excitations,
i.e., for $\nu = \rho$ 
measures the shift in chemical potential upon varying the
Fermi wave vector $\delta \mu = v_{N\rho} \delta k_F$ and, for $\nu
= \sigma$ the relation of the magnetic field to the magnetization 
$M = v_{N\sigma} 
(k_{F\uparrow} - k_{F\downarrow})$. $v_{J\nu}$ finally measures the
energy necessary to create persistent charge or spin currents on the
periodic chain. (ii) All three velocities are properties of the spectrum
of the model. Spectra can, however, be calculated 
either exactly by Bethe Ansatz (e.g.~Hubbard model) or to 
high accuracy with numerical methods, and the velocities can 
then be determined. (iii) The three velocities
determine the renormalized coupling constant $K_{\nu}$ which in turn
determines all correlation functions of the \lm. It is now obvious
how Eq.~(\ref{speccoup}) turns the \lm\ into a very
useful device for accessing the correlation functions of all 1D gapless
models. 

One prominent property of the \lm\ -- charge-spin separation -- is 
manifest here: charge and spin fluctuations propagate with
different velocities and will therefore separate in time. 
In realistic models, charge-spin separation will be dynamically generated
close to their Fermi surface. In the \lm\, describing just this 
subspace, it has become a manifest property of the model.

Collective charge density and spin density fluctuations propagating 
with different velocities do also occur in higher dimensional models, 
and in particular in the Fermi liquid. This is not special to 1D.
Dramatic consequences in 1D arise, however, from the lack of robustness
of hypothetical quasi-particles with respect to these elementary excitations
separating in time. Quasi-particles do not exist in 1D systems with 
charge-spin separation. This may again be traced back to the lack of
a continuum of low-energy excitations for $0 \leq q \leq 2k_F$, (Figure
3.1): in 1D there is no way how these collective modes can decay into
the hypothetical constituent quasi-particles (holes) which therefore never
reappear once interactions have been introduced. The absence of quasi-particles
is most directly seen in the single-particle spectral function which 
cannot be written in a form similar to Eq.~(\ref{grfl}). Detailed results
can be found in Section \ref{secdyn}.

Labelling the charge-localization (Mott-Hubbard) transitions generated 
by strong Cou\-lomb interactions as ``charge-spin separation'' is 
somewhat misleading.
It happens in higher dimensions, too. At issue are the excitations out of
this state, and whether there are quasi-particles at low energies.
Of course, there may be borderline cases, where the quasi-particle 
residue in (\ref{grfl}) is small but finite, and most of the spectral weight 
resides in the collective modes.

Charge-spin separation is also visible in the many-particle correlation
functions. This is trivial, and also happens in higher dimension, for the
small-$q$ parts of density and spin density correlation functions. 
The novel
feature of the correlation functions in 1D is the appearance of two 
separate singularities close to $2k_F$ (and, partly, higher multiples thereof)
where a single one is expected in the absence of charge-spin separation.
This will also be discussed in Section \ref{secdyn}. Before, we need
to find a practical representation of the fermion operator in terms of
the bosons diagonalizing the Hamiltonian in order to be able to 
calculate correlation functions.

\subsection{Bosonization}
\label{secbos}
A completely satisfactory boson solution of the \lm\ also requires an
explicit representation of the fermion operators $\Psi_{rs}(x)$ in terms
of the bosons \rr . Then any correlation function
can be given an equivalent boson representation and, the 
diagonal Hamiltonian being a simple boson bilinear, the calculation
of any of these correlation functions becomes
almost trivial, reducing to Gaussian averages.

Pioneering work in this direction was performed by Luther and Peschel
\cite{lupe} and Mattis \cite{mattis}. 
They proposed a bosonization formula which allowed an
asymptotic calculation of correlation functions but was certainly not
an operator identity transforming between fermions and bosons.
The cutoff procedures and the interpretation of these cutoffs were
ambiguous \cite{apostol}. Field theorists proposed similar constructions
at the same time \cite{mandel}.

A precise formulation of such an operator identity was
given independently by Haldane \cite{Haldane} and Heidenreich et 
al.~\cite{heiden}, and involves the construction of the 
(unitary) ladder operator
$U_{r,s}$. This operator increases by unity the number of fermions
with spin $s$ on branch $r$ and must commute with the boson operators
with finite momentum. It is sufficient for that purpose to consider states
$| N_{r,s} \rangle$ 
where all states below a certain wave vector are filled and
above empty:
\begin{equation}
U_{r,s} |N_{r,s} \; N_{\bar{r},\bar{s}} \rangle 
= |N_{r,s}+1 \; N_{\bar{r},\bar{s}} \rangle
\;\;\;, \;\;\; [\rho_{r,s}(p \neq 0), U_{r',s'} ] = 0
\end{equation}
in evident notation.
A natural guess is to put the new fermion into the first free level above
the reference state, occupied up to $\mid k_F + 2 \pi N_{r,s} / L \mid $,
\begin{eqnarray}
\label{uprop}
U_{r,s} & = & \frac{1}{\sqrt{L}}\sum_k c^{\dag}_{rks} \:
\delta \left( rk - \left[k_F + 
\frac{(2 N_{r,s} + 1) \pi }{L} \right] \right) \\
& = & \frac{1}{\sqrt{L}} \int_0^L \! dx \: \Psi_{r,s}^{\dag}(x) \:
\exp \left( ir \left[k_F + 
\frac{(2 N_{r,s} + 1) \pi }{L} \right] x \right) \;\;\;. \nonumber
\end{eqnarray}
Its commutator with the bosons does, however, not vanish
\begin{equation}
[\rho_{r,s}(p), U_{r',s'} ] = \frac{\delta_{r,r'} \delta_{s,s'}}{\sqrt{L}}
\int_0^L \! dx \:e^{ipx} \: \Psi_{r,s}(x) \:
\exp \left( ir \left[k_F + 
\frac{(2 N_{r,s} + 1) \pi }{L} \right] x \right) \;\;\;.
\end{equation}
The idea now is to introduce, into Eq.~(\ref{uprop}), a 
bosonic field $\phi_{r,s}(x)$ whose commutators with 
$\rho_{r,s}(p)$ compensate the unwanted commutator from $\Psi_{r,s}$.
One then has 
\begin{eqnarray}
\label{udef}
U_{r,s} & = & \frac{1}{\sqrt{L}} \int_0^L \! dx e^{irk_Fx}
e^{-i \phi_{r,s}^{\dag}(x)} \: \Psi^{\dag}_{r,s}(x) \:
e^{-i \phi_{r,s}(x)} \\
& {\rm with} & \nonumber \\
\label{phirs}
\phi_{r,s}(x) & = & - \frac{\pi r x}{L} N_{r,s} +
\lim_{\alpha \rightarrow 0} \left( \frac{2 \pi i}{L} \sum_{p \neq 0}
\frac{ e^{- \alpha \mid p \mid / 2 - ipx} }{\mid p \mid} 
\Theta(rp) \rho_{r,s}(-p) \right) 
\end{eqnarray}
which is the desired operator. This expression can be inverted for
$\Psi_{r,s}(x)$, now given in terms of bosons and
the ladder operator, and compactified into 
\begin{equation}
\label{bos}
\Psi_{rs}(x) = \lim_{\alpha \rightarrow 0}
\frac{ e^{ir(k_F-\pi/L)x} }{ \sqrt{2 \pi \alpha} } 
U^{\dag}_{rs}
\exp \left( \frac{-i}{\sqrt{2}} 
\left[ r \Phi_{\rho}(x) - \Theta_{\rho}(x)
+ s \left\{ r \Phi_{\sigma}(x) - \Theta_{\sigma}(x) \right\} 
\right] \right) \;\;\;.
\end{equation}
The two phase fields are
\begin{equation}
\label{phi}
\Phi_{\nu}(x) = - \frac{i\pi}{L} 
\sum_{p \neq 0 } 
\frac{e^{-\alpha \mid p \mid / 2 -ipx}}{p} 
\left[\nu_+(p)+\nu_-(p) \right]
- (N_{+,\nu} + N_{-\nu})\frac{\pi x}{L} \;\;\;,
\end{equation}
and
\begin{equation}
\label{theta}
\Theta_{\nu}(x) = \frac{i\pi}{L} \sum_{p \neq 0 } \frac{e^{-\alpha
\mid p \mid / 2 -ipx}}{p} \left[\nu_+(p) - \nu_-(p) \right]
+ (N_{+,\nu} - N_{-\nu})\frac{\pi x}{L} \;\;\;
\end{equation}
and are constructed from the $\phi_{r,s}$ and $\phi_{r,s}^{\dag}$ plus
commutator terms.
The charge density operator is related to $\Phi_{\rho}$ by
\begin{equation}
\label{rhox}
\rho(x) = \sqrt{2} \left[ \rho_+(x) + \rho_-(x) \right] =
- \frac{\sqrt{2}}{\pi} \frac{\partial \Phi_{\rho}(x)}{\partial x} \;\;\;,
\end{equation}
where the factor $\sqrt{2}$ comes from (\ref{chsdens}), and there is an
analogous expression for the spin density.
$\Theta_{\nu}(x)$ is related to the momentum canonically conjugate to
$\Phi_{\nu}(x)$ 
\begin{equation}
\label{pi}
\Pi_{\nu}(x) = 
\frac{1}{L} \sum_{p \neq 0 } e^{-\alpha \mid p \mid / 2 -ipx} 
\left[\nu_+(p)-\nu_-(p) \right]
+ (N_{+,\nu} - N_{-,\nu})\frac{\pi}{L} 
\end{equation}
by
\begin{equation}
\Theta_{\nu}(x) = \pi \int^x_{-\infty} \! dz \: \Pi_{\nu}(z)
\end{equation}
and the commutation relations are
\begin{equation}
[\Phi_{\nu}(x), \Pi_{\nu'}(x')]  =  i \delta_{\nu,\nu'} \delta(x-x') 
\end{equation}
\begin{equation}
[\Phi_{\nu}(x), \Theta_{\nu'}(x')]  =  i \frac{\pi}{2}
\delta_{\nu,\nu'} \: {\rm sign}(x'-x) \;\;\;.
\end{equation}
We can rewrite the Hamiltonian in terms of these phase fields as
\begin{equation}
\label{phaseham}
H = \frac{1}{2 \pi} \sum_{\nu} \int \! dx \left\{ v_{J\nu} \: \pi^2
\Pi_{\nu}^2(x) + v_{N\nu} \left( \frac{\partial \Phi_{\nu}(x)}{
\partial x} \right)^2 \right\} \;\;\;,
\end{equation}
making obvious the equivalence to the Gaussian model of statistical mechanics.
Under the Bogoliubov transformation (\ref{canon}), the phase fields
transform as
\begin{equation}
\label{phastf}
\Phi_{\nu} \rightarrow \Phi_{\nu} \sqrt{K_{\nu}} \hspace{1cm} 
 {\rm and} \hspace{1cm} \Theta_{\nu} \rightarrow \Theta_{\nu} / \sqrt{K_{\nu}}
\end{equation}
if we neglect the momentum dependence of the interactions $g(p)$ so that
the $K_{\nu}$ can be taken outside the summations in (\ref{phi}) and
(\ref{theta}).
These expressions can now be employed for calculating arbitrary
correlation  functions. Examples are given in the following.

There is also a more physical way of arriving at the general boson structure
of the fermion operators \cite{halhfl,ijmp}. Define a boson field
$\Phi_{r,s}(x)$ by $\partial \Phi_{r,s}(x) / \partial x = - \pi \rho_{r,s}(x)$
where $\rho$ describes density \em fluctuations. \rm Introducing a
particle at site $x$ creates a kink of amplitude $\pi$ in the field
$\Phi_{r,s}$, i.e. the phases of all other particles have to shift to
accommodate the new particle. 
This field can be considered as a dynamical implementation of the Fermi
surface phase shifts appearing
in Anderson's arguments in favour of a Luttinger liquid in
2D \cite{ar}. 
Since displacement operators are exponentials of momentum 
operators, one could guess $\Psi_{r,s}(x) \sim \exp [ i \pi 
\int_{-\infty}^x dz \Pi_{r,s}(z) ]$
where $\Pi_{r,s}(z)$ is the momentum canonically conjugate to \firs.
This operator commutes, however, with itself. The required change of
sign at $x = x'$ is achieved by multiplying with $\exp[\pm i \Phi_{r,s}(x)]$
which yields
\begin{equation}
\label{lupeq}
\Psi_{r,s}(x) \approx \lim_{\alpha \rightarrow 0}
\frac{1}{2 \pi \alpha} \exp \left[ irk_Fx -ir \Phi_{r,s}(x) 
+ i \pi \int_{-\infty}^x \! dz \Pi_{r,s}(z) \right] \;\;\;.
\end{equation} 
This is essentially the Luther--Peschel--Mattis formula \cite{lupe,mattis} 
which contains all the important bosonic terms for the calculation of 
physical properties but does not have the status of an operator identity
in the full Hilbert space of the \lm. 

\subsubsection{Spinless fermions}
\label{secslf}
At various stages of this article, we will need spinless fermions,
either because of their physical relevance as elementary 
charge excitations (holons) in the Hubbard and related models, or just
for simplification. Here, we compile the most important formulae from
the preceding paragraphs for spinless fermions. 

The Hamiltonian is obtained by dropping the spin label and summations
in (\ref{hfree}) -- (\ref{h4}), i.e.
\begin{eqnarray}
\label{hslf}
H & = & \frac{\pi v_F}{L} \sum_{rp} : \rho_r(p) \rho_r(-p): 
+ \frac{\pi v_F}{2L} \left( N^2 + J^2 \right) \\
& + & \frac{1}{L} \sum_p \left( g_2(p) \rho_+(p) \rho_-(p) +
\frac{g_4(p)}{2} \sum_r : \rho_r(p) \rho_r(-p): \right)
\end{eqnarray}
with $N,J = N_+ \pm N_-$. The Hamiltonian is diagonalized as
\begin{equation}
\tilde{H} = \sum_{r} \frac{\pi v(p)}{L} : \tilde{\rho}_r(p) 
\tilde{\rho}_r (-p) : + \frac{\pi}{2L} (v_N N^2 + v_J J^2 ) + {\rm const.}
\end{equation}
with the velocities
\begin{equation}
v(p) = \sqrt{ \left[ v_F+ \frac{g_4(p)}{2\pi} \right]^2 - 
\left[ \frac{g_2(p)}{2 \pi} \right]^2 } \;, \;\;
v_N = v_F + \frac{g_4(0) + g_2(0)}{2\pi} \;, \;\;
v_J = v_F + \frac{g_4(0) - g_2(0)}{2\pi} \;,
\end{equation}
and the stiffness ``constant''
\begin{equation}
K(p) = \sqrt{\frac{2 \pi v_F + g_4(p) - g_2(p)}{2 \pi v_F + g_4(p) + g_2(p)}}
\;\;\;.
\end{equation}
Finally, the bosonization identity for spinless fermions is
\begin{equation}
\label{slf}
\Psi_r(x) = \lim_{\alpha \rightarrow 0} 
\frac{e^{ir(k_F-\pi/L)x}}{\sqrt{2 \pi \alpha}} U_r^{\dag}
\exp \left( - i \left[ r \Phi (x) - \Theta(x) \right] \right) \;\;\;.
\end{equation}
The fields $\Phi(x)$ and $\Theta(x)$ are given by the expressions
(\ref{phi}) and (\ref{theta}) for the charges, and the operators $\rho_r(p)$
now refer to spinless fermions. With these fields, the Hamiltonian the
has the following phase representation
\begin{equation}
\label{phasslf}
H = \frac{1}{2 \pi} \int \! dx \left\{ v_J \pi^2 \Pi^2(x) + v_N \left(
\frac{\partial \Phi(x) }{\partial x} \right)^2 \right\} \;\;\;.
\end{equation}

\section{Physical Properties of the Luttinger Model -- Thermodynamics
and Correlation Functions}
\label{secprop}

The machinery set up in the preceding section is extremely useful in
calculating correlation functions. A remarkable feature of the Luttinger
model is that all correlation functions can be calculated  exactly.
With the boson representation of an operator, all the expectation 
values reduce to Gaussian averages, as we shall show on some examples
here. The linear response of an operator $B$ in a system 
described by a Hamiltonian $H_0$, coupled to an external field $a(x,t)$
by the operators $A(x)$, i.e. $H = H_0 + \int \! dx \:a (x,t) \: A(x)$ 
is related, through the Kubo formulae, to correlation functions of the system
in the absence of the external field (in the interaction picture and
assuming translational invariance)
\begin{eqnarray}
\label{kubo}
\langle B(x,t) \rangle & = & \langle B^{(a=0)}(x,t) \rangle +
\int_{-\infty}^{\infty} \! \chi_{BA}(x-x',t-t')
a(x',t') dx'dt' \;\;\;, \nonumber \\
\chi_{BA}(x,t) & = & - i \Theta(t) \langle \left[ B(x,t) , A(0,0) 
\right] \rangle_{a=0} \;\;\;.
\end{eqnarray}
$\chi$ is called susceptibility, response function, retarded correlation
function, etc. There are many closely related functions and below, we shall
denote all of them as $R_{BA}$ or, if symmetric, simply as $R_A$, with
some exception for cases of special relevance.
Within linear response theory, the \lm\ can make predictions for all 
possible measurements.

\subsection{Thermodynamics and transport}
\label{themdy}
The Luttinger model has a specific heat linear in temperature
\begin{equation}
C(T) = \gamma T \;\;\;, \;\;\; \frac{\gamma}{\gamma_0} = \frac{1}{2}
\left( \frac{v_F}{v_{\rho}} + \frac{v_F}{v_{\sigma}} \right) \;\;\;.
\end{equation}
The linearity is both characteristic of the underlying fermions
(linear specific heat in any dimension) as well as of the bosonic
excitations (phonons in 1D also have a linear specific heat).
$\gamma_0$ is the coefficient of free electrons which can be calculated 
from both representations as
\begin{equation}
\gamma_0 = \frac{\pi^2 k_B^2}{3} N(E_F) = \frac{2 \pi k_B^2}{3 v_F} \;\;\;,
\end{equation}
the density of states of the free \lm\ being a constant $N(E) = 2/\pi v_F$
including spin and \em both \rm branches. 
The spin susceptibility and compressibility are
\begin{equation}
\label{susc}
\frac{1}{\chi} = \frac{1}{L} \frac{\partial^2 E_0(\sigma)}{\partial \sigma^2}
\;\;\; , \;\;\;
\frac{1}{\kappa} = \frac{1}{L} \frac{\partial^2 E_0(n)}{\partial n^2} \;\;\;,
\end{equation}
where $E_0$ is the ground state energy as a function of the particle
(spin)
density $n$ ($\sigma$). Throughout this article, we denote the average 
particle density (band filling factor) by $n$ and the density fluctuations
by $\rho(x)$. The susceptibilities
are renormalized by the interactions
\begin{equation}
\chi = \frac{2 K_{\sigma}}{\pi v_{\sigma}} = 
\frac{2}{\pi v_{N\sigma}}
\;\;\; {\rm and} \;\;\;
\kappa = \frac{2 K_{\rho}}{\pi v_{\rho}} =
\frac{2}{\pi v_{N\rho}} \;\;\;.
\end{equation} 
They are related to the renormalized velocities for the
charge (spin) excitations defined in Eq.~(\ref{nvj}). This is expected,
of course, because $v_{N\nu}$ measures the change in energy upon 
changing the number of electrons in the system, cf. (\ref{hdiag}).
As we shall see below, spin-rotation invariance requires $K_{\sigma} = 1$.

The electrical conductivity is determined from the current-current correlations
through the Kubo formula
\begin{equation}
\label{kubcon}
\sigma(\omega) = \frac{i}{\omega} \left[ \frac{D}{\pi} 
+ R_j^R(\omega) \right]
\end{equation}
where the first term is the diamagnetic part and the (second) paramagnetic
term is given in terms of
the retarded current-current correlation function
\begin{equation}
\label{curcor}
R_j^R(\omega) = - \frac{i}{L} \int_0^L \! dx \int_0^{\infty} \! dt
\langle [j(x,t),j(0,0)] \rangle e^{i \omega t} \;\;\;.
\end{equation}
The Drude weight $D$ in fact is a susceptibility and 
is related to the derivative of the ground
state energy with respect to an applied flux $\Phi$ \cite{flux}
\begin{equation}
\label{druwei}
D = \frac{\pi}{2L} \left. \frac{\partial^2 E_0 (\Phi)}{\partial \Phi^2}
\right|_{\Phi = 0} = 2 v_{J \rho} \;\;\;
\end{equation}
A flux creates a ``persistent'' current in a Luttinger ring, 
and the appearance of $v_{J\rho}$ here should not surprise from (\ref{hdiag})
again. $2 v_{J\rho} = 2 v_{\rho} K_{\rho}$ plays the role of the
plasma frequency in 1D \cite{giam1}. 

One has to be careful in the definition of the current operators. Naively,
one has $j(x) = \sqrt{2} v_F \left[ \rho_+(x) - \rho_-(x) \right] = 
\sqrt{2} v_F \Pi_{\rho}(x)$. A more careful evaluation via the continuity
equation $\partial_t \rho(xt) + \partial_x j(xt) = 0$ gives, however, 
\begin{equation}
\label{currop}
j(xt) = \frac{\sqrt{2}}{\pi} \frac{\partial}{\partial t} \Phi_{\rho}(xt)
= \sqrt{2} v_{\rho} K_{\rho} \Pi_{\rho}(xt) = \sqrt{2} v_{J\rho} 
\Pi_{\rho}(xt) \;\;\;.
\end{equation}
The difference is due to the fact that, in the \lm , $g_2$ may be
different from $g_4$ and the density does
not necessarily commute with the interaction Hamiltonian, as it does
in  a well-defined lattice model. Notice, however, that $v_{\rho} K_{\rho}
= v_{J \rho}$; if the \lm\ is derived from a well-defined lattice model
with density-density interactions only, $g_{2\nu} = g_{4\nu}$ is required
and, using Eq. (\ref{nvj}), one obtains $v_{J \rho} = v_F$, i.e. the current
operators are not renormalized by interactions. This applies to
galilean invariant models in general where, in the limit $q \rightarrow 0$,
the current becomes proportional to momentum which is conserved by the
interactions \cite{giamil}. Consequently, as can be shown by two
partial integrations on Eq.~(\ref{curcor}) producing
\begin{equation}
\label{conint}
\sigma(\omega) = - \lim_{q \rightarrow 0} {\rm Re} \: (i \omega)^{-3}
\int_0^{\infty} \! dt e^{i \omega t} \langle \left[ \: [H,j(qt)] \: , \:
[H, j(q,0)] \: \right] \rangle \;\;\;,
\end{equation}
one has
$R_J^R(\omega) \equiv 0$
\cite{giamil}, so that the conductivity reduces to a pure Drude peak 
\begin{equation}
\label{condlm}
\sigma(\omega) = 2 v_{J \rho} \delta(\omega) = 2 v_F 
\delta(\omega)
\end{equation}
with
an interaction-independent strength. This relation has been derived by
a number of people \cite{giam1,metzner,shankar,schulz}.

There are a few most remarkable facts about these unspectacular formulae. 
(i) The finiteness of the susceptibilities characterizes the
system as a ``normal metal''. It is highly nontrivial in view of the
ubiquitous divergences we shall encounter in the following sections.
The physical origin lies in the strong conservation laws of the 1D phase
space in the absence of backscattering \cite{metzner}. (ii) These quantities
can be calculated both in a fermion representation where one considers
the charge excitations $N_{\nu}$ and uses (\ref{susc}), 
and from the $q \rightarrow 0$ limit
of bosonic correlation functions which we shall compute in Section 
\ref{altmet}. The result is the same. The reason will be given below
\cite{shankar}. (iii) The boson representation gives the 
susceptibilities in absolute magnitude for lattice models provided parameters 
are identified correctly \cite{shankar}. 
One can invert the procedure  and use these relations to
identify the parameters of a low-energy boson theory for lattice models
from (\ref{susc}) \cite{efe,schulz,Haldprl}. 
(iv) They can be obtained from the
energy alone which can be calculated either from an exact solution or
accurately with numerical diagonalization where correlation functions 
are not readily available \cite{schulz,Haldprl}.

\subsection{Single- and two-particle correlation functions}
The thermodynamic properties do not differ from the Fermi liquid.
There, compressibility and susceptibility are renormalized by interactions,
too, and the renormalization is given by the Landau parameters
$F_0^s$ and $F_0^a$. We neither see the anomalous power-laws not 
the effects of charge-spin separation highlighted earlier. To this end,
we carry on to the space- and/or time-dependent correlation functions 
\begin{equation}
R_O(x,t) = -i \Theta(t) \left\langle \left[ O(x,t) , O^{\dag}(0,0) 
\right]_{\pm} 
\right\rangle
\end{equation}
of various operators $O$ of interest. $[ \ldots ]_{\pm}$ denotes commutator
or anticommutator for bosons and fermions, respectively. 
Other, e.g.
time-ordered, correlation functions are obtained in a similar way. 

We give a rather explicit calculation for the 
single-electron Green function 
\begin{equation}
\label{gf}
G_{rs}(xt) = -i \Theta(t) \langle \{ \Psi_{rs}(xt) , \Psi^{\dag}_{rs}(00) \}
\rangle \equiv -i \Theta(t) \left( \tilde{G}(xt) + \tilde{G}
(-x-t) \right)
\end{equation}
where \prs\ has been defined in Eq.~(\ref{bos}), to sketch how such a 
calculation works in practice, and to give some formulae useful for the
work with boson operators. In (\ref{gf}), we have incorporated that $G$
is diagonal in $r$ and $s$. 
Using the bosonization identity (\ref{bos}) in (\ref{gf}), we first
commute the $U_{rs}^{\dag}$-operator from \prs\ at the left of the
expression through the exponentials until it arrives at the right, where 
$\Psi_{rs}^{\dag}(0)$ has a $U_{rs}$. Being unitary, we have $U_{rs}^{\dag}
U_{rs} = 1$. What are the terms we pick up during the commutations?
$U_{rs}$ commutes with $\rho_{rs}(p \neq0)$ by construction, so that the
only nonvanishing terms come from the operators $N_{rs}$ measuring the charge
excitations 
in the phase fields $\Phi_{\nu}$ and $\Theta_{\nu}$. These terms, however, 
involve prefactors $1/L$ so that their contribution vanishes in the limit
$L \rightarrow \infty$. If we are interested in this 
thermodynamic limit, we can neglect
both the $U_{rs}$- and $N_{rs}$-operators altogether. We see that \rm the
Luther-Peschel-Mattis formula (\ref{lupeq}) (neglecting the $U_{r,s}$-
and $N_{r,s}$-operators) gives the exact asymptotic behaviour of
the Green function! 
\rm (This statement is not completely true for the
many-particle functions: $U_{rs}$ anticommutes with $U_{r's'}$ if 
at least one index is different. When one considers operators $O$
pairing \prs\ with different indices, as we do in almost all two-particle
functions below, the $U_{rs}$ will produce phase factors. The exponent
of the power-law is unaffected by these phase factors but logarithmic
or prefactor corrections crucially depend on them \cite{ehm}.)

After diagonalizing the Hamiltonian, the Green function becomes 
(dropping the indices $r$ and $s$)
\begin{eqnarray}
\tilde{G}(xt) & = & \lim_{\alpha \rightarrow 0} \frac{e^{irk_Fx}}{2 \pi 
\alpha}
\left\langle \exp \left( - \frac{i}{\sqrt{2}} \left[ r \tilde{\Phi}_{\rho}
(xt) - \tilde{\Theta}_{\rho}(xt) \right] \right)
\exp \left(  \frac{i}{\sqrt{2}} \left[ r \tilde{\Phi}_{\rho}
(00) - \tilde{\Theta}_{\rho}(00) \right] \right) \right\rangle_{\rho}
\times \nonumber \\
& \times &
\left\langle \exp \left( - \frac{i}{\sqrt{2}} \left[ r \tilde{\Phi}_{\sigma}
(xt) - \tilde{\Theta}_{\sigma}(xt) \right] \right)
\exp \left(  \frac{i}{\sqrt{2}} \left[ r \tilde{\Phi}_{\sigma}
(00) - \tilde{\Theta}_{\sigma}(00) \right] \right) \right\rangle_{\sigma}
\;\;.
\end{eqnarray}
However, we keep the momentum
dependence of the interactions and do not use (\ref{phastf}); we denote
the transformed fields by $\tilde{\Phi}_{\nu}$ and $\tilde{\Theta}_{\nu}$.
Moreover, since the $\rho$-phase 
fields commute with the $\sigma$-fields, we have
separated the exponentials into products involving only $\rho$ and 
$\sigma$ separately. $\langle \ldots \rangle_{\nu}$ denotes the expectation
value with the $\nu$-part of the Hamiltonian. Next we use the
important relation 
\begin{equation}
\label{expuni}
e^A e^B = e^{A+B} e^{[A,B]/2} \hspace{0.5cm} {\rm valid \; if} \;\;
[A,B] \in C \!\!\!\! C
\end{equation}
to merge all $\nu$-phase fields into one exponential. 
The commutators contribute $\exp [ C_{\nu}(xt) ]$ with
\begin{equation}
C_{\nu}(xt) = - \frac{\pi}{L} \sum_{p \neq 0} \frac{e^{- \alpha |p|}
e^{-ipx}}{p}
\left( \left[ K_{\nu}(p) + \frac{1}{K_{\nu}(p)} \right] i \sin \left[ v_{\nu}(p) 
p t \right] + 2 r \cos \left[ v_{\nu}(p) t \right] \right) \;\;\;.
\end{equation}
The expectation value  $\langle e^{A+B} \rangle \equiv \exp 
[ D_{\nu}(xt) ]$ is evaluated using
\begin{equation}
\label{gaussav}
\langle \exp A \rangle = \exp \left( \frac{1}{2} \langle A^2 \rangle \right)
\end{equation}
valid for a linear form in boson operators whose exponential is
averaged with a harmonic oscillator (Gaussian) Hamiltonian. We find
\begin{eqnarray}
\lefteqn{D_{\nu}(xt) = \frac{\pi^2}{4L^2} \sum_{p,p'\neq0} 
\frac{e^{- \alpha(|p| + |p'|)/2 } }{p p'}  \times} \\
& & \times \sum_{R= \pm} \langle \nu_R(p) \nu_R(p') \rangle \prod_{P=p,p'} 
\left( r 
\sqrt{K_{\nu}(P)} + R \frac{1}{\sqrt{K_{\nu}(P)}} \right) 
\left( e^{-iP(x-R v_{\nu}(P)t)} - 1 \right) \nonumber
\end{eqnarray}
with 
\begin{equation}
\label{expect}
\langle \nu_r(p) \nu_r(p') \rangle = 
\delta_{p,-p'} \frac{L |p|}{2 \pi}
\left(
\frac{\Theta(rp)}{ \exp \left[ v_{\nu}(p)p/2kT \right] - 1}  + 
\Theta(-rp) \left\{ 
1 + \frac{1}{\exp \left[ v_{\nu}(p)p/2kT \right] - 1} 
\right\} 
\right) \;.
\end{equation}
For $T=0$, to be treated first, the Bose-Einstein distribution
vanishes for $p \neq 0$, and we can rearrange
$e^{C_{\nu}} e^{D_{\nu}}$ so that
\begin{eqnarray}
C_{\nu}(xt) + D_{\nu}(xt) & = & -\frac{1}{2} \left[ V_+^{\nu}(xt) + V_-^{\nu}
(xt) - 2 r V_0^{\nu}(xt) \right] \;\;\;, \\
\label{vpm}
V_{\pm}^{\nu}(xt) & = & \frac{1}{2} \int_0^{\infty} \! \frac{dp}{p}
e^{- \alpha p} K_{\nu}^{\pm 1}(p)
\left[ 1 - \cos(px) e^{- i v_{\nu}(p) p t} \right] \;\;\;, \\
\label{vzero}
V_0^{\nu}(xt) & = & \frac{i}{2} \int_0^{\infty} \! \frac{dp}{p} e^{- \alpha p}
\sin(px) e^{- i v_{\nu}(p) p t} \;\;\;.
\end{eqnarray} 
All correlation functions can be expressed in terms of 
$V_{\pm}$ and $V_0$. 

Now remember that in
the \lm\ a momentum transfer cutoff must be imposed on the interactions,
and the asymptotic values of $K_{\nu}$ and $v_{\nu}$ are given by 
(\ref{asymp}). Taking e.g. $V_+$, we split the integral in two terms
by adding and subtracting $[ \ldots ]$ on the right-hand side 
\cite{suzu,schucou}. Taking
together $(K_{\nu}-1)[\ldots]$, the important contributions will come
from $p \ll 1/\Lambda$, and we can replace $v_{\nu}(p) \rightarrow v_{\nu}$
there. The contribution of this term to $V_+$ then becomes
\begin{equation}
\label{contr}
\frac{K_{\nu}-1}{4} \ln \left( 
\frac{\Lambda_+^{\nu} + i v_{\nu}t + ix }{\Lambda_+^{\nu}} 
\right) +
\frac{K_{\nu}-1}{4} \ln \left( \frac{\Lambda_+^{\nu} 
+ i v_{\nu}t - ix }{\Lambda_+^{\nu}} 
\right)
\end{equation}
where the cutoff $\Lambda_+^{\nu}$ is given by
\begin{equation}
\label{cutoff}
\ln \left( \frac{\Lambda_+^{\nu}}{\Lambda_0} \right) = - \frac{1}{K_{\nu}-1}
\int_0^{\infty} \!  \frac{dp}{p} \left[ K_{\nu}(p) - 1 
- \left( K_{\nu} - 1 \right) e^{- \Lambda_0 p} \right] \;\;\;.
\end{equation}
$\Lambda_0$ is arbitrary but finite. Since (\ref{contr}) would be obtained by 
taking an exponential cutoff on $K_{\nu}(p) -1$, (\ref{cutoff}) amounts to 
finding an equivalent exponential cutoff to the cutoff of arbitrary form
contained in $K_{\nu}(p)$. In the following, we assume that there
is a single cutoff $\Lambda$ in the problem independent of the indices $\pm$
and $\nu$. In the second term from $V_+$, the only interaction-dependent
quantity is $v_{\nu}(p)$. Here it is simplest to use the fact that 
(\ref{contr}) can be obtained with an exponential cutoff $\Lambda$,
add and subtract $\exp(-\Lambda p)$ and use $v_{\nu}$ resp. $v_F$ for
the integrals weighted at small or large momentum. $V_0$ is treated in
the same way. The final result is then \cite{solyom,suzu,schucou,spec2}
(approximate expressions have been given by many others)
\begin{equation}
\label{green} 
\tilde{G}(x,t)  =  \frac{1}{2 \pi} e^{irk_Fx} \lim_{\alpha \rightarrow 0}
\frac{\Lambda + i (v_Ft-rx)}{\alpha + i (v_Ft-rx)} 
\prod_{\nu = \rho,\sigma} \frac{1}{\sqrt{\Lambda + 
i(v_{\nu}t-rx)}} \left( \frac{\Lambda^2}{(\Lambda + i v_{\nu}t)^2
+ x^2}\right)^{\gamma_{\nu}} 
\;\;\;.
\end{equation}
The exponent is
\begin{equation}
\label{gammanu}
\gamma_{\nu} = \frac{1}{8} \left( K_{\nu} + \frac{1}{K_{\nu}} - 2 \right)
\geq 0
\;\;\;.
\end{equation}
Eq.~(\ref{green}) gives the \em universal \rm behaviour of the Green function,
which is independent of detailed cutoff forms. Nonuniversal contributions which
have been eliminated by the trick of adding, subtracting and recombining 
terms above, can also be evaluated \cite{schome}. The spinless fermion result
can be obtained by putting formally $K_{\rho}=K_{\sigma}=K$ and $v_{\rho} =
v_{\sigma}=v$ \cite{lupe,theu}. 

For $t=0$, the Green function decays as
\begin{equation}
\label{gree}
G_{rs}(x) \sim x^{-1-\alpha} \;\;\;, \hspace{1cm} \alpha = 
2 \sum_{\nu} \gamma_{\nu}  \geq 0 \;\;\;. 
\end{equation}
The exponent $\alpha$ appears in all single-particle properties.
$\alpha/2$ is the ``anomalous dimension'' of the fermion operators.
[It has become  customary to use $\alpha$ both for the exponent of
the Green function and for the infinitesimal in the bosonization
identity (\ref{bos}); the context usually identifies clearly which $\alpha$ is
referred to, and confusion seems unlikely.] 
From $G_{rs}$, one can derive the momentum distribution function
\cite{spec2,gutsch,brech}
\begin{equation}
\label{nkut}
n(k) \sim \frac{1}{2} - C_1 {\rm sign} (k-k_F) \mid k-k_F \mid^{\alpha} 
- C_2 (k-k_F)
\end{equation}
which does not have a jump at $k_F$ but rather a continuous power-law
variation.
An exact calculation of the prefactors is also possible \cite{spec2}.
In (\ref{nkut}) the breakdown of Fermi liquid theory and the absence of 
quasi-particles are evident. 
Fermi liquids have a jump discontinuity of amplitude $z_{k_F} \leq 1$
at $k_F$ where $z_k$ is the wave-function
renormalization constant, Eq.~(\ref{qpr}). 
However, the velocities do not enter and charge-spin separation
does not manifest itself, and only the absence of quasi-particles due to
the Peierls-type coupling of the two Fermi points is probed.
The single-particle density of states $N(\omega)$
varies as a power-law
\begin{equation}
\label{nomega}
N(\omega) \sim \mid \omega \mid^{\alpha}
\end{equation}
with the same exponent $\alpha$. Again, exact but lengthy expressions are
available \cite{spec2}.

The density-density correlation function consists out of several pieces,
corresponding to the wave vectors $q \approx 0$ ($\rho$), $q \approx 
\pm 2k_F$ (CDW), and $q \approx \pm 4k_F$ (\fkf -CDW), and, in principle,
higher multiples
\begin{eqnarray}
\label{rho0}
O_{\rho}(x) & = & \sum_{r,s} \rho_{r,s}(x) = \sqrt{2} \sum_r \rho_r(x) 
= - \frac{\sqrt{2}}{\pi} \frac{\partial \Phi_{\rho}(x) }{\partial x}\\ 
O_{CDW}(x) & = & \sum_{s} \Psi^{\dagger}_{+,s} (x) \Psi_{-,s}(x)
\nonumber \\
\label{cdw}
& = & \frac{1}{2 \pi \alpha} \sum_{s} U_{+,s} U_{-,s}^{\dagger}
\exp \left\{ -2ik_Fx +
\sqrt{2} i  \left[ \Phi_{\rho}(x) + s \Phi_{\sigma} (x)
\right] \right\} \\
& \approx & \frac{1  }{\pi \alpha} 
\exp \left\{-2ik_Fx + \sqrt{2} i \Phi_{\rho}(x)
\right\}
\cos[\sqrt{2} \Phi_{\sigma}(x) ] \;\;\;. \nonumber \\
O_{4k_F}(x) & = & \sum_{s} \Psi_{+,s}^{\dagger}(x) \Psi_{+,-s}^{\dagger}(x)
\Psi_{-,-s}(x) \Psi_{-,s}(x) \nonumber \\
\label{cdw4}
& = & \frac{2}{(2 \pi \alpha)^2} 
\exp \left\{ -4ik_Fx +\sqrt{8} i\Phi_{\rho}(x) \right\} \;\;\;.
\end{eqnarray}
In the \lm, the $U_{rs}$-ladder operators give only contributions vanishing
in the thermodynamic limit $L\rightarrow \infty$ and have been dropped after
(\ref{cdw})
[see however the remark after Eq.~(\ref{gf})]. Notice also that  $O_{4k_F}$
involves four fermions in the \lm\ but, as will be explained below, is part 
of the (two-particle) density operator in lattice models. Moreover, the
wavelength $\lambda = 2 \pi / 4k_F = (N_0/L)^{-1}$ equals the inverse particle
density. Establishment of \fkf -CDW long-range order therefore corresponds
to the formation of a Wigner crystal, and we shall be interested in this
possibility as well as its short-range ordered variant below.
The expectation
values are evaluated exactly as in the case of the Green function above,
so that we only give the asymptotic decay laws
\begin{eqnarray}
\label{rroh}
R_{\rho}(x) & = & \frac{K_{\rho}}{(\pi x)^2} \;\;\;, \\
\label{rcdw}
R_{CDW}(x) & \sim & \cos(2k_F x) \: x^{-2+\alpha_{CDW}} \;\;\; , \;\;\;
\alpha_{CDW} = 2 - K_{\rho} - K_{\sigma} \;\;\;, \\
\label{r4kf}
R_{4k_F}(x) & \sim & \cos(4k_Fx) \: x^{-2 + \alpha_{4k_F}} \;\;\; , \;\;\;
\alpha_{4k_F} = 2 - 4 K_{\rho} \;\;\;,
\end{eqnarray}
We see that the \fkf -correlations decay very fast at weak-coupling but
become competitive with the \tkf\ ones when $K_{\rho}$ decreases
\cite{emery}. For $K_{\sigma} = 1$, \fkf\ correlations dominate over \tkf\
for $K_{\rho}  \leq 1/3$. 

The other correlation functions follow similar power-laws. 
Long wavelength spin fluctuations follow (\ref{rroh}) with $K_{\rho}
\rightarrow K_{\sigma}$. For later use, we also give the operators
for the $x,y,z$-components of the SDW correlations
\begin{eqnarray}
O_{SDW,x}(x) & = 
&  \sum_{s} \Psi^{\dagger}_{+,s}(x) \Psi_{-,-s}(x) \nonumber \\
\label{sdwx}
& = & \frac{1}{\pi \alpha} 
\exp \left[ -2ik_Fx + \sqrt{2} i \Phi_{\rho}(x) \right]
\cos \left[ \sqrt{2} \Theta_{\sigma}(x) \right] \\
O_{SDW,y}(x) & = & -i  \sum_{s} s
\Psi^{\dagger}_{+,s}(x) \Psi_{-,-s}(x) \nonumber \\
\label{sdwy}
& = & \frac{1}{\pi \alpha} 
\exp \left[ -2ik_Fx + \sqrt{2} i \Phi_{\rho}(x) \right]
\sin \left[ \sqrt{2} \Theta_{\sigma}(x) \right] \\
O_{SDW,z}(x) & = & \sum_{s} s \Psi^{\dagger}_{+,s}(x) \Psi_{-,s}(x) \nonumber
\\
\label{sdwz}
& = & \frac{i}{\pi \alpha} \exp \left[ -2ik_Fx + \sqrt{2} i \Phi_{\rho}(x) 
\right] \sin \left[ \sqrt{2} \Phi_{\sigma}(x) \right] \;\;\;.
\end{eqnarray}
The correlation functions decay as
\begin{eqnarray}
R_{SDW}(x) & \sim & \cos (2k_F x) x^{-2 + \alpha_{SDW}} \;\;\;, \\
\label{asdw}
\alpha_{SDW_x}  & = & \alpha_{SDW_y} = 
2 - K_{\rho} - K_{\sigma}^{-1} \;\;\;, \hspace{1cm}
\alpha_{SDW_z} = 2 - K_{\rho} - K_{\sigma} \;\;\;.
\end{eqnarray}
Singlet (SS) and triplet (TS)  superconducting correlations do not oscillate
and decay with exponents
\begin{eqnarray}
\label{ass}
\alpha_{SS} & = & 2 - K_{\rho}^{-1} - K_{\sigma} \;\;\;, \\
\label{alts}
\alpha_{TS0} & = & 2 - K_{\rho}^{-1} - K_{\sigma} \;\;\;, \hspace{1cm}
\alpha_{TS\pm 1} = 2 - K_{\rho}^{-1} - K_{\sigma}^{-1} \;\;\;.
\end{eqnarray}
Each correlation function has its proper special combination of the
two parameters $K_{\nu}$ in the power-law exponent which therefore 
parameterize completely the scaling laws between the exponents. 
Remember also that $K_{\nu}$ relates the three velocities
for each degree of freedom, i.e. the spectrum of low-lying eigenvalues
(\ref{speccoup}).
Different is only the correlation function of the long-wavelength charge or
spin fluctuations. The operator $\nu(x) \nu(0)$ is marginal with a scaling
dimension $-2$ and does not acquire an anomalous dimension. 
Also its correlation function does not depend on a cutoff (in the other
expressions, it has simply been suppressed), as has been discussed
in Section \ref{themdy}. 

The three components of the spin density and triplet superconductivity
operators have very different representations in terms of the phase
fields $\Phi_{\sigma}(x)$ and $\Theta_{\sigma}(x)$, and their correlation
functions differ (at least formally) even in the exponents. 
This is so because our abelian bosonization scheme treats $\sigma_z$
on a special footing and breaks the spin-rotation symmetry $SU(2)$ down to
$U(1)$. In the absence of external magnetic fields or spin-anisotropic
interactions, the correlation functions must be spin-rotation invariant.
We see that this requires $K_{\sigma} = 1$. We shall assume this to be
the case throughout this article except when stated to the contrary.
Again, nonabelian bosonization would allow 
to keep the spin-rotation invariance manifest at every stage of the 
calculation.

The Green function's $\alpha$
is invariant under $K_{\rho} \rightarrow 1/K_{\rho}$ and therefore
does not depend in an important way on the sign of the interaction.
It is positive and, had one only $g_2$, would be symmetric in
attraction and repulsion. It is only $g_4$ which slightly changes 
the modulus of $\alpha$ when $g_i \rightarrow - g_i$.
$\alpha = 0$ is possible only
when the fluctuations on all branches are free -- the system may still be
interacting, though, if $g_4 \neq 0 $. 
On the other hand, the many-particle correlations
do depend on the sign of the interactions: $K_{\rho} < 1$ for 
repulsion, and $K_{\rho} > 1$ for attraction.
Consequently, for repulsive interactions, 
the \tkf\ density wave correlations decay more slowly than for
free fermions ($\sim x^{-2}$), while for attractive coupling, the 
superconducting correlations decay slowest.
At first sight surprising will be the fact that the correlation
functions of density and spin density  (as well as those for singlet
and triplet superconductivity) are strictly degenerate in the 
spin-rotation invariant \lm . This is quite counterintuitive,
and nature is certainly richer than such simple-minded results. 
On the other hand, the \LL\ hypothesis requires that this degeneracy
of exponents carries over to more realistic models. The resolution of
this puzzle will be postponed to Chapter \ref{chapll}.

If one is interested in finite temperatures, there are several possibilities.
(i) One can use the conformal invariance of the \lm\ to map the $T=0$
correlation functions onto those at $T \neq 0$. This will be demonstrated
in the next section, Eq.~(\ref{corfexp}). 
(ii) One can introduce Matsubara frequencies 
and calculate the boson propagators $\sim D_{\nu}(x \tau)$ at imaginary times.
(iii) One can simply use the Bose-Einstein distribution $n_{BE}(p)$ at
finite temperature in (\ref{expect}). This will decorate the integrals 
appearing in (\ref{vpm}) and partly those in (\ref{vzero}) with factors
$\coth(\beta v_{\nu} p /2)$. The integrals can still be evaluated in 
terms of logarithms of Gamma functions which, for small temperatures
essentially add terms $\ln \{ \pi [x \pm i v_{\nu}t] / \sinh(\pi 
[x \pm i v_{\nu} t] / v_{\nu} \beta) \}$ to (\ref{contr}). If $x \gg
v_{\nu} \beta$, the hyperbolic sine will grow exponentially, and 
correlation functions like (\ref{rcdw})
will therefore decay exponentially on a 
scale set by the thermal coherence length $\xi_T = \pi v_F / T$. 
If $\xi_T \gg 1/\Lambda$, the power-laws discussed before will still
show up in the window in between.

Transforming to $k$-space, one has to distinguish between the 
instantaneous and static correlation functions. 
Given a correlation function in $x$-space
\begin{equation}
R_i (x) \sim \cos(nk_F x) x^{-2+\alpha_i} \;\;\;, \hspace{1cm} R(t) \sim
t^{-2+\alpha} \;\;\;,
\end{equation}
the instantaneous and static correlations behave as 
\begin{equation}
\label{rq}
R_i (k, t=0) \sim (k - n k_F)^{1 - \alpha_i} \;\;\;, \hspace{1cm}
R_i (k, \omega = 0) \sim (k - n k_F)^{- \alpha_i} \;\;\;,
\end{equation}
respectively, with equivalent formulae for the $\omega$-dependent
local and $q=0$-functions. For free fermions, the static correlations
have a logarithmic divergence which is changed into a power law divergence
by even weak interactions. On the other hand, the instantaneous correlations
are nonsingular usually (though possibly enhanced), 
and singularities can only be brought up by rather strong interactions. 
Divergences of this kind have been observed both in computer simulations
and in X-ray scattering on quasi-1D materials, and will be discussed below.

We have not discussed charge-spin separation in detail yet. While it is
contained in the full expression for the Green function (\ref{green}),
it does not influence the long-distance or time properties of the correlation
functions. It is clear that
this subtle feature of 1D interacting fermions can only be probed
in dynamic, $q$- and $\omega$-resolved correlation functions.

\section{Dynamical correlations: the spectral properties 
of Luttinger liquids}
\label{secdyn}
Fermi liquid theory breaks down in 1D for two reasons: (i) the anomalous
dimensions of the fermion operators, giving rise to the nonuniversal
power laws discussed in the preceding section, and (ii) charge-spin
separation. Either of them is sufficient to kill all quasi-particles
in the neighbourhood of the Fermi surface, and both together will
certainly cooperate. However, all correlation functions
of the previous section are affected only by the anomalous
dimensions. Much effort has been devoted to the study of these functions
over the last decade.

On the other hand, for a long time much less has been known about the dynamical 
($x- \; and \;t-$ resp.~$q- \; and \; \omega-$dependent) correlations. 
Also, how to measure charge-spin separation? Since this phenomenon is
characterized by different propagation velocities for charge and
spin fluctuations, fully dynamical correlation
functions are needed to put it into evidence. 
The single-particle spectral function $\rho_{rs}(q,\omega)$ is defined as
\begin{equation}
\label{rhors}
\rho_{rs}(q,\omega) = - \frac{1}{\pi} {\rm Im} G_{rs}(rk+q, \mu + \omega)
\end{equation}
where $G_{rs}$ is the retarded Green function (\ref{gf}), (\ref{green}).
There is no principal
difficulty in computing this quantity. All we need to do is Fourier transform. 
This can be done quite easily for spinless fermions or for the one-branch \LL\
($g_2 = 0$) but is laborious for the full model for $s=1/2$-fermions
we are most interested in. 

With spinless fermions we can single out the influence of
the anomalous fermion dimensions. 
This is the generic structure \cite{lupe,spec2,schome,spec1,ms}: 
At $q=0$~(i.e. $k=k_F$), $\rho(0,\omega) \sim
\mid \omega \mid^{\alpha-1}$, i.e. a power-law divergence (or cusp-singularity
for $\alpha>1$) instead of the 
$\delta$-function in Fermi liquid theory. Clearly,
as the 1D correlations increase from zero, spectral weight is pushed away
from the Fermi surface by the virtual particle-hole excitations generated
by $g_2$. Let us increase $q$. In a Fermi liquid, the $\delta$-function
would disperse with $q$ and broaden but essentially conserve its
shape. In the Luttinger liquid, $\rho(q,\omega)$ strongly deforms: There
is a power law singularity 
$\rho(q,\omega) \sim \Theta(\omega - v q)
( \omega - v  q )^{\gamma_{0}-1} $ at positive frequencies (for $q>0$)
and a weaker singularity $\sim \Theta(-\omega -vq) 
(-\omega - v q)^{\gamma_0}$ at negative frequencies. In the positive
frequency contribution -- particle creation above the Fermi surface --
spectral weight of an incoming particle is boosted to
higher energies by the particle-hole excitations on both branches. The negative
frequency contribution describes the destruction of 
particles above the Fermi surface present in the ground state as a
result of particle-hole excitations. As $q$ 
increases, the negative frequency part is exponentially suppressed and
all the spectral weight is transferred to positive frequencies.

For the ``one-branch'' Luttinger liquid ($g_2 =0$, charge-spin
separation only), one has finite spectral weight only
at positive frequencies (for $q>0$) between $v_{\sigma}q$ and $v_{\rho}q$
with inverse-square-root divergences at the edges \cite{spec2,spec1,fogedby}. 
At $k_F$, the spectral function reduces to
$\delta(\omega)$
and the momentum distribution is a step function with
a jump of unity at $k_F$, in agreement with Luttinger's theorem \cite{lt}. 
Although this seems to imply a Fermi liquid it is clear that the
physical picture is quite different and that the notion of a quasi-particle
does not make sense because the $\delta$-function does not survive the
slightest displacement from the Fermi surface.
The incident electron decays into multiple particle--hole-like
charge and spin fluctuations which all live on the same branch
as the incoming fermion. 
It is immediately apparent that $n(k)$ and, more
generally, any quantity depending on $k$ or $\omega$ alone will 
fail to detect
charge-spin separation. It can be seen {\em only}
in quantities depending on {\em both} $q$ {\em and} $\omega$.

We now turn to the spectral properties of the $s=1/2$-Luttinger liquid
\cite{spec2,schome,spec1,ms,spec3}.
We limit ourselves to the spin-rotation invariant case 
($\gamma_{\sigma}=0$). Fig. 3.6 displays the dispersion of
$\rho(q,\omega)$ for small $q$ and $\alpha=0.125$. 
It is apparent that the spectral function carries features both from
the spinless fermions (synonymous with ``anomalous fermion dimensions'')
and the one-branch problem (``charge-spin separation''). 
At very small $q$, on the scale of the Figure, $\rho$ looks pretty much
like the spinless fermions' function. 
As $q$ increases, the negative frequency weight (very small anyway) is 
transferred to positive frequency but, most importantly, the generic
two-peak structure of the spectral function becomes apparent. The exponent
of the singularity at $v_{\sigma} q$ is $2\gamma_{\rho}-1/2$ while it
is $\gamma_{\rho}-1/2$ at the $v_{\rho}q$-singularity and $\gamma_{\rho}$
at $-v_{\rho}q$. Since $\gamma_{\rho} = 1/16$ here, the correction to
the one-branch case is quite insignificant here and the charge-spin
separation aspect is clearly dominant at finite $q$. The weight above/below
$\pm v_{\rho}q$ originating from the anomalous dimensions is barely
visible. As $\alpha$ increases, the various power-law divergences weaken and
finally transform into cusp-singularities. At the same time, the spectral
function becomes much less structured, and spectral weight is shifted
by the electronic correlations both to above/below $ \pm v_{\rho}q$, more
reminiscent of the spinless fermion problem. As the correlations increase,
the features originating from charge-spin separation are more and more
obscured by transfers of spectral weight over significant energy scales.
The important scale here is the energy 
of the charge fluctuations $\pm v_{\rho}q$.

The spectral function in Figure 3.6 obeys to the sum rule 
\begin{equation}
\label{sumrule}
\int_{-\infty}^{\infty} \! d \omega \rho_{rs}(q,\omega) = 1 \;\; 
{\rm \; for \; all} \;\; q
\;\;\;.
\end{equation}
The single-particle density of states $N(\omega)$ has already been discussed
above. 
A local sum rule is not satisfied 
by $N(\omega)$ unless $g_{4\|} = 0$ \cite{suzu} as is the
case for local interactions; in general (long-range interactions), one has
\cite{spec2,spec3,schoensum}
\begin{equation}
\label{locsum}
\int_0^{\infty} \! d \omega \left[ N_{rs}
(\omega) - N_0(\omega) \right]
= - \frac{1}{4 \pi v_F} \int_{-\infty}^{\infty} \! \frac{dk}{2\pi} g_{4\|}(k)
\;\;\;,
\end{equation}
$N_0(\omega) = 1/2 \pi v_F$ being the noninteracting density of states.
It \em is \rm satisfied, however, by the Tomonaga model  with a finite
bandwidth cutoff \cite{schoensum}. Here, as usual, $\int_0^{\infty} \! 
d\omega N(\omega) = n$, the particle density which is not changed by the 
interactions. The failure of the local sum rule in the
\lm\ is certainly due to the introduction of the unphysical negative-energy
states which are sampled in the frequency integral. 

The many-particle spectral functions display similar features. 
Fig. 3.7 displays the charge [$S(q,\omega)$] and spin [$\chi(q,\omega)$]
structure factors at \tkf\ and the charge factor at \fkf\ [$S_4(q,\omega)$]
\cite{spec2,seoul}. Again there are power-law singularities at 
$\omega = \pm v_{\sigma}q$ and $\pm v_{\rho}q$ but the functions now are
symmetric because the CDW and SDW operators mix left- and right-moving
particles. At weak coupling, there are cusps, and only as $K_{\rho}
< 1/2$ do  they turn into divergences. Further interesting is the fact 
that the \tkf\ CDW and SDW fluctuations are sensitive to charge-spin
separation but the \fkf -CDW is not. This is easy to understand from
the boson representation of these operators (\ref{cdw}) -- (\ref{sdwz}): 
the \tkf -operators
necessarily involve the $\Phi_{\sigma}$ or $\Theta_{\sigma}$-fields
in addition to $\Phi_{\rho}$.
The only divergent \fkf -operator,
however, only depends on the charge field $\Phi_{\rho}$. \fkf -operators
involving the spin degrees of freedom are never divergent.

\section{Alternative methods}
\label{altmet}
\subsection{Green function methods}

There are alternative routes for solving the \tlm , based on
diagrammatic methods or equations of motion for the Green
function. They provide an interpretation of the novel physics
of the \LL\ from the standpoint of conventional many-body theory, and
therefore stress the formal similarities of Fermi and Luttinger liquids
while the bosonization approach 
more strongly emphasizes their differences.
Moreover, the connection between symmetries, conservation laws and the 
low-energy structure of 1D Fermi liquids may become more apparent in this
approach which we outline now. It has been pioneered by 
Dzyaloshinsk\u{i}i and Larkin for a spinless variant of the model,
Eq. (\ref{hslf}) \cite{dzya} and followed and extended
by others \cite{solyom,everts,metzner,peso}.

The power of the 1D conservation laws can be gauged from the fact that our
arguments for the breakdown of Fermi liquid theory in 1D in Section
\ref{secfl} were based on 
divergences encountered in a perturbation treatment of the self-energy
corrections to the 1D Green
functions. As a consequence of Ward identities, vertex and self-energy
correction cancel exactly in some quantities (such as density-density
correlation functions) and to such a large extent in others that 
meaningful answers are obtained and all results of the bosonization approach
reproduced. 

What are Ward identities? They are specific relations between the vertex 
operators and (single or $n$-particle)
Green functions of a theory, translating its conservation
laws i.e. its symmetries, into a Green function formalism which
describes the dynamics of the excitations. 
Vertex operators couple the charges and currents of a system to external
fields. They involve the corresponding density operator (e.g. $\rho(p), \;
j_{\rho}(p)$) plus two (more generally $2n$) fermions. The equation of
motion for the vertex operator in general produces Green functions 
involving even more particles. If the charge is conserved, however,
$\rho(p)$ obeys the continuity equation. Combining it with the equation
of motion of the simple vertex described, yields just the difference of
the two single-particle propagators involved, instead of complicated
objects involving intermediate excitations. The
principle is easy: use the continuity equation associated with the conserved
charge to
reduce the equations of motion for the object under consideration, then
Fourier transform the resulting expression to recover an algebraic
relation. This is particularly transparent for density-density
response which we study now before carrying on to the single-particle Green 
function.

In Section \ref{symm}, we had studied the conservation of charge and spin          
separately on each branch of the dispersion. 
This generates
the following continuity 
equations for the charge and spin densities and currents from the Heisenberg
equations of motion
\begin{eqnarray}
\label{cont1}
i \frac{\partial \nu(p,t) }{\partial t} & = & [ \nu(p,t), H ] = 
- v_{J\nu} \: p \: j_{\nu}(p,t) \\
\label{cont2}
i \frac{\partial \tilde{\nu}(p,t) }{\partial t} & = & 
[ \tilde{\nu}(p,t), H ] = 
- v_{N\nu}\:  p \: \tilde{j}_{\nu}(p,t) 
\end{eqnarray}
with the total charge ($\nu = \rho$) and spin ($\nu = \sigma$) densities
and currents
\begin{equation}
\label{nudef}
\nu (p) = \sum_r \nu_r (p) \;\;\; {\rm and} 
\;\;\; j_{\nu}(p) = \sum_r r \nu_r (p) 
\;\;\; .
\end{equation}
In the Green function approach, it is 
the physical charge and spin densities which enter the various 
operators. For this reason, we define in this section, and only in this
section 
\begin{equation}
\nu_r(p) = \rho_{r\uparrow} (p) \pm \rho_{r\downarrow} (p)
\end{equation}
at variance with the remainder of this paper. This will avoid a confusing
proliferation of factors $\sqrt{2}$ due to the different definition in
(\ref{chsdens}).
The ``axial'' charge and spin densities and currents (named after similar
constructions appearing in field theory) are
\begin{equation}
\label{axial}
\tilde{\nu} (p) = \sum_r r \nu_r (p) \;\;\; {\rm and} \;\;\;
\tilde{j}_{\nu}(p) = \sum_r  \nu_r (p) \;\;\; 
\end{equation}
and are identical to the usual currents
and charges, Eq.~(\ref{nudef}), respectively.
$v_{J\nu}$ and $v_{N\nu}$ are the velocities for charge and current
excitations \cite{Haldane} defined earlier (\ref{speccoup}). 

Notice in passing
that both equations can be put together to produce the
equations of motion of a harmonic oscillator \cite{metzner}
\begin{equation}
\frac{\partial^2 \nu(p,t)}{\partial t ^2} + v_{J\nu} v_{N\nu} \: p^2 \:
\nu(p,t) = 0 \;\;\;,
\end{equation}
indicating that the charge (spin) 
density fluctuations are the elementary excitations
of the systems which propagate
with an effective (sound) velocity $\sqrt{v_{J\nu} v_{N\nu}}$. 
The conservation laws thus completely
determine the dynamics of our system. 

For illustration, we investigate the density-density correlation
function 
\begin{equation}
R_{\rho\rho} (q,\omega) = \int_{-\infty}^{\infty} \!
dt e^{i \omega t} R_{\rho\rho} (q,t) \;\;\;, \hspace{1cm}
R_{\rho\rho}(q,t) = - \frac{i}{L} \langle T \rho (q,t) \rho (-q,0) \rangle 
\;\;\;.
\end{equation}
$T$ is the time ordering operator.
Applying $i \partial_t$ to this equation and using (\ref{cont1}), we have
\begin{equation}
\label{wi1}
i \frac{\partial R_{\rho}(q,t) }{\partial t} 
= \frac{i}{L} v_{J\rho} q \langle T j_{\rho} (q,t)
\rho (-q, 0) \rangle + \frac{1}{L} \delta(t) \langle 
\left[ \rho(q), \rho (-q) \right] \rangle
\end{equation}
where the last term originates from taking the time derivatives of the 
step functions implied by time ordering but vanishes on account of the
commutator algebra Eq.~(\ref{tlmcom}). A first Ward identity is obtained
from the Fourier transform
\begin{equation}
\omega R_{\rho\rho}(q, \omega) - v_{J\rho} q R_{j_{\rho}\rho} (q , \omega)
= 0 \;\;\;.
\end{equation}
Similar Ward identities can be derived for $R_{j_{\rho}\rho}(q,\omega)$
(with the difference that $[j_{\rho}(q) , \rho(q)]$ $ \neq 0$) and for
the axial charges and currents (\ref{axial}). The second derivative of
the density-density correlation function is then
\begin{equation}
\label{wii}
- \frac{\partial^2 R_{\rho\rho}(q,t)}{\partial t^2}  = 
v_{J\rho} v_{N\rho} q^2 R_{\rho\rho} (q,t) + \frac{2 v_{J\rho} q^2 }{\pi}
\delta(t) \;\;\; ,
\end{equation}
which is Fourier transformed into
\begin{equation}
\label{rrhio}
R_{\rho\rho}(q, \omega) = \frac{2}{\pi} 
\frac{v_{J\rho}q^2}{\omega^2 - v_{\rho}^2
q^2} \;\;\; {\rm with} \;\;\; v_{\rho} = \sqrt{v_{J\rho} v_{N\rho}} \;\;\; .
\end{equation}
Eq.~(\ref{wi1}) 
is an example of a very simple -- yet manifestly powerful -- Ward identity.
Metzner and Di Castro \cite{metzner} give many more.

Now consider the single-particle Green function
\begin{equation}
G_{rs}(k,t) = -i \langle T c_{rs} (k,t) c^{\dag}_{rs}(k,0) \rangle
\end{equation}
which obeys the equation of motion \cite{everts}
\begin{eqnarray}
(\omega - r v_F k) G_{rs} (k, \omega)  = 1 
+ i \sum_{\nu,q} \int \! \frac{ d \Omega}{2 \pi}
\left[ g_{2\nu} (1 - 2 \delta_{\nu,\sigma} \delta_{s,\downarrow})
F^{\nu}_{-rrs} (k,\omega;k+q, \omega + \Omega; q, \Omega) \right. & &
\nonumber \\
\label{eqmot}
\left.  \hspace{3cm} + g_{4\nu} 
(1 - 2 \delta_{\nu,\sigma} \delta_{s,\downarrow})
F^{\nu}_{rrs}(k, \omega; k+q, \omega+\Omega; q, \Omega) \right] \;\;\; .
\hspace{2cm}
\end{eqnarray}
To get (\ref{eqmot}), take
$ i \partial_t G_{rs}(k,t)$ using the Heisenberg equation of motion
and Fourier transform; deriving the $T$-operator
gives the $1$, taking the commutator with $H_0$ gives $r v_F k G_{rs}$,
and the commutators with $H_2$ and $H_4$ and using (\ref{chsdens}) to go from
the $\rho_{rs}$ to the $\nu_r$, gives the vertex functions
\begin{equation}
\label{tpges}
F_{r'rs}^{\nu} (k_1,t_1 ; k_2, t_2 ; q t)  = - \langle T \nu_{r'}(qt)
c_{rs}(k_2,t_2) 
c^{\dag}_{rs}(k_1,t_1) \rangle
\end{equation}
Continuing now without using 
(\ref{cont1}) would lead to a hopeless  hierarchy of equations.
However, (\ref{tpges}) obeys a remarkable Ward identity 
\begin{equation}
\label{wivert}
q F_{r'rs}^{\nu} (k, \omega; k+q, \omega + \Omega; q , \Omega) =
r \pi (1 - 2 \delta_{\nu,\sigma} \delta_{s,\downarrow}) 
R_{\nu_{r'} \nu_r}( q, \Omega) \left[ G_{rs} (k,\omega) -
G_{rs}(k+q, \omega+\Omega) \right]
\end{equation}
which helps to
simplify the problem.
The one-branch density-density correlation function
\begin{equation}
\label{obdens}
R_{\nu_{r'} \nu_r} (q,\omega) = - \frac{i}{L} \int \! dt
e^{i \omega t} \langle T \nu_{r'}(qt)
\nu_r(-q0) \rangle = \left\{ \begin{array}{ll}
\frac{rq}{\pi} \frac{\omega + r (v_{N\nu} + v_{J\nu}) q /2}{
\omega^2 - (v_{\nu} q)^2} & {\rm for} \; r = r' \\ 
\frac{1}{\pi} \frac{(v_{J\nu} - v_{N\nu})q^2/2 }{ \omega^2 -
(v_{\nu} q)^2} & {\rm for} \; r = - r'
\end{array} \right.
\end{equation}
can itself be derived from (\ref{wi1}) and related Ward identities.
One can now eliminate the vertex function from (\ref{eqmot}) and close
the equation of motion for the Green function. 
The resulting integral equation is then solved by Fourier transforming
back to a real space differential equation and taking into account boundary
and analyticity conditions. The result agrees with the
expression (\ref{green}) up to details of cutoff procedures.
Notice that the Ward identity for $F_{r'rs}^{\nu}$ (\ref{wivert}) involves
the chiral (charge and spin) density operators $\nu_r$. It therefore is
the consequence of two separate Ward identities, one for the density
$\sum_r \nu_r$ which is present also in the many-body problem in higher
dimensions, and a new one involving the axial density $\sum_r r \nu_r$
which is new and related to the disconnected 1D Fermi ``surface'' and
the absence of backscattering in the \lm .

In these two examples, we have rederived results via Ward identities
which are also quite easy to derive from bosonization. There are others
where the derivation via Ward identities are easier than with bosonization.
An example is the intra- or inter-branch
polarization bubble $\Pi^{\rho}_{rr'}(q,\omega)$
which is related to the density correlation function (\ref{obdens})
by Dyson's equation
\begin{equation}
\label{dyseq}
R_{\rho_r \rho_{r'}}(q, \omega) = \Pi^{\rho}_{rr'}(q, \omega)
+ \sum_{tt'} \Pi^{\rho}_{rt}(q,\omega) g_{tt'\rho} R_{\rho_{t'}
\rho_{r'}} (q, \omega) \;\;\;,
\end{equation}
represented graphically in Figure 3.4. $g_{rr'\rho}$ denotes $g_{2\rho}$
or $g_{4\rho}$. The polarization $\Pi^{\rho}_{rr'}$ is given by
the irreducible vertex $\Lambda^{\rho}_{rr's}$ and the exact single-particle
Green functions $G_{r's}$
\begin{equation}
\label{piro}
\Pi^{\rho}_{rr'}(q \omega) = - i \sum_s \int \! \frac{dk d\Omega}{(2 \pi)^2}
\Lambda^{\rho}_{rr's} (k,\Omega ; k+q, \omega + \Omega; q, \omega)
G_{r's}(k,\Omega) G_{r's}(k+q, \Omega+\omega)
\end{equation}
as shown in Fig.~3.5.
$\Lambda$ is obtained from $F$ by amputating the external fermion legs,
i.e. dividing by the product of the two Green functions involved
in (\ref{wivert}) and taking only the interaction-irreducible part of $F$.
$\Pi$ must be a wildly divergent function because the Green functions
have divergences and the vertex corrections certainly have divergences, 
too! This is not true, however, and with the Ward identity (\ref{wivert}),
converted into one for $\Lambda$, one obtains the simple, finite results
\begin{eqnarray}
\Pi^{\rho}_{r,-r} (q, \omega) & = & 0 \;\;\;,  \\
\Pi^{\rho}_{rr} (q, \omega) & = & \frac{-i}{\omega - r  v_F q} \sum_s 
\int \! \frac{dk d\Omega}{(2 \pi)^2} \left[
G_{rs}(k,\Omega) - G_{rs}(k+q, \Omega+\omega) \right] \;\;\; 
\nonumber \\
& = & \frac{r}{\pi} \frac{q}{\omega - v_F q} \equiv \Pi^{\rho(0)}_{rr}(q,
\omega) \;\;\;.
\end{eqnarray}
This result is remarkable: all vertex and self-energy corrections have
cancelled out as a consequence of the Ward identities, and the
polarization is identical to $\Pi^{\rho(0)}$ of free fermions. 
Eq.~(\ref{dyseq}) then reduces to a standard RPA summation, showing 
that RPA is exact for the density-density correlation functions. 
Moreover, the charge and spin susceptibilities 
$\lim_{q \rightarrow 0} \lim_{\omega \rightarrow 0} R_{\nu\nu}
(q,\omega)$ are finite, in agreement with (\ref{susc}).
The \LL s therefore 
are ``normal'' metals. This is entirely due to the conservation
laws and Ward identities which enforce the cancellation of all divergences
which would occur in a diagrammatic development.

Of course, one can also compute all the many-particle Green functions,
and construct the same picture as in the preceding sections using
the standard many-body formalism. We reemphasize that in the exact solutions
we had found, both the Ward identities related to the charges (currents) and
to the axial charges (currents) were essential. It is the latter one
that gives the one-dimensional Fermi liquids their special properties.

Similar results can also be obtained by more diagram-based techniques
\cite{solyom,dzya,peso}. In this case, the Ward identities are expressed by
the theorem that closed fermion loops with more than two fermion
lines vanish (equivalent in the vanishing of the transverse current
in quantum electrodynamics).  
The limitation to forward scattering only
in the \lm\ implies that a closed fermion line has all of its parts on
a definite branch $r$. 

Moreover, one can use the Ward identities to construct a field-theoretical
\rg\ formulation of the \lm\ with respect to the free Fermi gas 
\cite{metzner,metzcdc}. This verifies that all couplings are dimensionless,
and that consequently, the beta-function 
\begin{equation}
\beta(g) = \Lambda \left( \frac{\partial g}{\partial \Lambda} \right)
\equiv 0
\end{equation}
at the \LL\ fixed point. The density operators do not acquire anomalous
dimensions, and the coupling constants are \rg\ invariants. It also
verifies the correctness of the earlier scaling Ansatz \cite{solyom}.

\subsection{Other bosonic schemes}
In Section \ref{secbos}, 
we have solved the \lm\ via a boson representation of the
Hamiltonian and of the fermion operators. Other bosonic approaches, based
on functional integrals and a Hubbard-Stratonovich decoupling have been
developed in the past \cite{fogedby,leechen}. 
They are closer to the methods used in quantum field theory than Haldane's
operator approach.
They also provide 
an exact solution of the model,  and reproduce all the results
obtained by the two methods presented above. Which one to use is rather a
matter of taste and background than of the specific nature of the problem
at hand. 

A bosonic scheme widely used for strongly
correlated fermions are ``slave
bosons'' and one may naturally wonder if there is any relation to the
bosonization discussed above. 
Slave bosons are usually applied to problems where double occupancy of
lattice sites is dynamically forbidden because of strong electronic repulsion.
One tries to circumvent the difficult treatment of inequality constraints
(such as $\langle n_i\rangle 
\leq 1$) by introducing additional particles into an enlarged
Hilbert space whereby the inequality constraint translates into an equality
constraint which can be solved by Lagrange multipliers. Properties
are then obtained by projecting back onto the physical Hilbert space.
From these remarks, it is quite clear that slave bosons and the 
Tomonaga-Luttinger bosons are two distinct entities. While the latter are
the elementary excitations of the 1D Fermi liquid, the former are, in the
first place, a bookkeeping device to obtain good approximations to fermionic
properties. Still, slave bosons have been used successfully, together
with standard bosonization, to obtain low-energy properties of, e.g., the
$U=\infty$ Hubbard model \cite{schmeltz}. A deeper knowledge
of differences and similarities of both types of bosons is, however, 
just beginning to emerge \cite{mudry}.

\section{Conformal field theory and bosonization}
\label{cft}

In the language of the theory of phase transitions, 
one-dimensional Fermi liquids are critical at $T=0$. An arbitrary system,
close to a second order phase transition, exhibits strong precursor 
fluctuations
of the ordered phase, whose typical size is measured by the correlation 
length $\xi \sim | ( T - T_c ) / T_c |^{-\nu}$ which diverges as
the critical point $T_c$ is approached. Thermodynamic properties (specific
heat, magnetization, etc.) exhibit similar divergences whose sole origin
is the divergence in $\xi$. Therefore, their critical 
exponents can be related by
scaling relations to $\nu$ and the dimension of space. 
These scaling relations only 
depend on the symmetry of the theory (universality). At the critical
point, correlation functions decay as power-laws of distance and time
with some critical exponents which generally can be calculated from the
model under consideration \cite{phastr}. The power-law correlations
of one-dimensional Fermi liquids found in Section
\ref{secprop}, show explicitly that we have a $T=0$ quantum critical point.

\subsection{Conformal invariance at a critical point}

Conformal field theory is a powerful means of characterizing universality
classes of critical systems in 2D statistical mechanics or 1D quantum
field theories [time playing the role of a second dimension, these
theories in fact are (1+1)D] in terms of a single dimensionless
number, the central charge $c$ of the underlying Virasoro algebra 
\cite{confft}. The critical exponents are the scaling dimensions of 
the various operators in a conformally invariant theory and, generically,
are fully determined by $c$. A notable exception are theories with
central charge $c=1$ such as the Gaussian model, of particular relevance
to the problems considered here, where the exponents (scaling dimensions)
depend on a single effective coupling constant of the model. 
Both the central charge and
the scaling dimensions can be computed from the finite-size scaling 
properties of the ground state energy and the low-lying excitations
\cite{confft,bloeaff}. This is important because these quantities can
be computed accurately either by Bethe Ansatz (for models solvable by
the technique) or, in any case, by numerical diagonalization. 

What are the symmetries of systems at a critical point? It is certainly
translationally and rotationally invariant. Quantum field theories,
in addition are Lorentz invariant but in (1+1)D, Lorentz invariance
reduces to rotations in the $\bx = (x,t)$-plane. 
As we have seen above, a system
at criticality, in addition is characterized by scale invariance,
\begin{equation}
{\bf x} \rightarrow \lambda {\bf x} \;\;\;.
\end{equation}
It turns out that the combined rotational and scale invariance implies
that the system is invariant under a wider symmetry group, the global
conformal group.
On a classical level, conformal transformations are general coordinate
transformations which leave the angles between two vectors invariant.
In dimension $D>2$, the global conformal group is 
finite-dimensional, and so is the associated
Lie algebra of its generators. 
There is a finite number of constraints, and these allow
for an evaluation of the two-point and three-point correlation functions, 
but not for the higher ones. 

The situation is different in two dimensions, where all correlation functions
can be determined. Consider a general coordinate transformation
\begin{equation}
{\bf x} \rightarrow \bx ' = {\bf x} + {\bf \xi} (\bx) \;\;\;.
\end{equation}
For this transformation to be conformal, ${\bf \xi}$ must satisfy certain
constraints which can be expressed in a differential
equation (Killing-Cartan equation). In general
dimension $D$, this leaves for ${\bf \xi}(\bx)$ a polynomial of second degree
in $\bx$ (with tensor coefficients). 
In two dimensions, however, the Killing-Cartan equation reduces
to the Cauchy-Riemann equation, and therefore all analytic functions
are allowed for conformal transformations.
This group of transformations, 
called local conformal group, is much wider than the global conformal group
encountered before.
It is then natural to switch to complex variables 
$z, \bar{z} = x_1 \pm i x_2$,
so that we have
\begin{equation}
\label{locconft}
z \rightarrow z + \xi^z(z) = f(z)\;\;\;, \hspace{1cm} 
\bar{z} \rightarrow \bar{z} + \bar{\xi}^{\bar{z}} (\bar{z}) = 
\bar{f} (\bar{z}) \;\;\;.
\end{equation}
To determine the algebra corresponding to the local conformal group,
we need the commutation relations of the generators of the transformations.
Since $\xi^z(z)$ and $f(z)$ are analytic, they can be expanded in a
Laurent series
\begin{equation}
\xi^{z}(z) = \sum_{n = -\infty}^{\infty} \xi_n z^{n+1}
\end{equation}
[and a similar equation for $\bar{\xi}(\bar{z})$], and
we find the generators of the local conformal transformations
\begin{equation}
\ell_n (z) = - z^{n+1} \partial_z \;\;\;, \hspace{1cm} \bar{\ell}_n (\bar{z})
= - \bar{z}^{n+1} \partial_{\bar{z}} \;\;\;, \hspace{1cm} 
n \in Z \!\!\! Z \;\;.
\end{equation}
These generators obey the local conformal algebra
\begin{equation}
\label{viraclass}
[ \ell_m , \ell_n ] = (m-n) \ell_{m+n} \;\;, \hspace{0.8cm}
[ \bar{\ell}_m, \bar{\ell}_n ] = (m-n) \bar{\ell}_{m+n} \;\;,
\hspace{0.8cm} [\ell_m, \bar{\ell}_n ] = 0 \;\;.
\end{equation}
This infinite dimensional algebra is called the classical Virasoro algebra.
(The global conformal algebra is generated 
by $\{\ell_{-1}, \ell_0, \ell_1\}$.)
Since the two algebras are independent, one may take
$z$ and $\bar{z}$ as independent, corresponding to the natural variables
for left- and right-moving objects; the physical theory then lives on
$\bar{z} = z^{\star}$.

We now go to the quantum (or statistical mechanics) case.
How do fields and correlation functions of a quantum field
theory transform under conformal 
transformations? In general, an infinitesimal symmetry variation in a field 
$\phi$ is generated by $\delta_{\xi} \phi = \xi [Q, \phi]$ where $Q$
is the conserved charge associated with the symmetry.
Local coordinate transformations are generated by the charges constructed
from the \set\ $T_{ij}$. Rotational invariance constrains $T_{ij}$ to
be symmetric, and scale invariance requires its trace to vanish; then
conformal invariance does not impose additional constraints
showing that it is implied by rotational and dilatational 
invariance. Translating
these conditions into the complex variables $z$ and $\bz$, one can show
that only the diagonal components 
\begin{equation}
T(z) \equiv T_{zz} (z) \hspace{0.5cm} {\rm and} \hspace{0.5cm} 
\bar{T}(\bar{z}) \equiv \bar{T}_{\bar{z}\bar{z}} (\bar{z})
\end{equation}
do not vanish. In the radial quantization scheme, the conserved charge
then becomes        
\begin{equation}
Q = \frac{1}{2 \pi i} \oint \left[ dz T(z) \xi(z) + d\bar{z} \bar{T}(\bar{z})
\bxi (\bz ) \right] \;\;\; ,
\end{equation}
which generates a field variation 
\begin{equation}
\delta_{\xi, \bxi} \phi(w, \bw) = \frac{1}{2 \pi i} \int \! \left\{
dz \left[ T(z) \xi(z) , \phi (w, \bw ) \right] + 
d\bar{z} \left[ \bar{T}(\bar{z}) \bar{\xi}(\bz) , \phi( w ,\bw ) \right] 
\right\} \;\;\;.
\end{equation}
In general, it is difficult at this point to proceed further without
having explicit expressions at hand. There is, however, a distinctive
class of fields, to be called primary fields, for which
\begin{equation}
\label{primftf}
\delta_{\xi, \bxi} \phi(w, \bw)  =  \left( h \partial_z \xi^z(z) +
\xi^z(z) \partial_z +  
\bar{h} \partial_{\bar{z}} \bxi^{\bar{z}}(\bz) +
\bxi^{\bz}(\bz) \partial_{\bz} \right) \phi ( w, \bw )
\end{equation}
which can be recognized as the infinitesimal version of
\begin{equation}
\label{fieldtf}
\phi (w, \bw) \rightarrow \left(\frac{\partial f}{\partial w} \right)^h
\left( \frac{\partial \bar{f}}{\partial \bw } \right)^{\bar{h}}
\phi( f(w), \bar{f}(\bw) ) \;\;\;.
\end{equation}
All other fields are called secondary fields.
$h$ and $\bar{h}$ are two real numbers, the conformal weights of the
field $\phi$. The combinations $\Delta = h + \bar{h}$ and $s = h - \bar{h}$
are the scaling dimension and spin of the field $\phi$, respectively
[if one works in a basis of eigenstates of $L_0$ and $\bar{L}_0$,
the combinations $L_0 + \bar{L}_0$ and $i (\L_0 - \bar{L}_0)$ are generators
of dilations and rotations, respectively].  Eq. (\ref{fieldtf}) is the
transformation law of a complex tensor of rank $h, \bar{h}$. Normally,
such a tensor transforms with integer powers of $\partial f /\partial z$
and $\partial \bar{f} / \partial \bar{z}$ which are the number of 
$z$ and $\bar{z}$ indices; here, however, one could conceive also noninteger
exponents. They are called anomalous dimensions.
As a consequence, the scaling dimension of the field $\phi$ also can
become anomalous. We have seen examples in the \lm\ in the preceding section.

One reason for the special status of primary fields is that
one can derive (in fact
in any dimension) some of their correlation functions 
from the transformation property (\ref{primftf}).
The two-point function
$G^{(2)} = \langle \phi_1 (z_1, \bz_1) \phi_2 (z_2 , \bz_2) \rangle$
must be invariant under a conformal transformation (\ref{locconft})
\begin{equation}
\delta_{\xi , \bxi} G^{(2)} ( z_i, \bar{z}_i ) =
\langle ( \delta_{\xi , \bxi} \phi_1 ) \phi_2 \rangle
+ \langle \phi_1 \delta_{\xi , \bxi} \phi_2 \rangle = 0  \;\;\;.
\end{equation}
Using the transformation law (\ref{primftf}), one can derive a differential
equation for $G^{(2)}$ which can be solved to yield
\begin{equation}
\label{twoptfct}
G^{(2)} (z_i, \bz_i ) = \frac{C_{12}}{z_{12}^{2h} \bar{z}_{12}^{2 \bar{h}}}
\;\;\;, 
\end{equation}
where $z_{ij} = z_i - z_j $ and $C_{12} \propto \delta_{\Delta_1,\Delta_2}$ 
is a constant. The three-point
function $G^{(3)}$ can be determined in a similar manner, but the 
four-point function, at the present stage of development, can only
be determined up to a function of the cross-ratio $z_{12} z_{34} / 
z_{13} z_{24} $.

Not all fields are primary fields.
For the primary fields to transform according to (\ref{primftf}), 
the operator product expansion (OPE)
of the \set\ with $\phi$ for short distances
must go as
\begin{equation}
\label{ope}
T(z) \phi(w,\bw ) = \frac{h}{(z-w)^2} \phi(w, \bw )
+ \frac{1}{z-w} \partial_w \phi ( w, \bw ) + \ldots \;\;\;,
\end{equation}
where radial ordering is implied, 
and there is an equivalent
equation for the anti-holomorphic (left-moving) piece of the \set .
(In the following, we always imply the existence of such equivalent
equations for the anti-holomorphic dependences.)
A secondary field has a higher than double-pole singularity in its
OPE with $T(z)$. The most prominent representative is $T(z)$ itself
\begin{equation}
\label{opeset}
T(z) T(w) = \frac{c/2}{(z-w)^4} + \frac{2}{(z-w)^2} T(w) + \frac{1}{z-w}
\partial T(z) \;\;\;.
\end{equation}
The coefficient $c \; (= \bar{c} \geq 0)$ 
is called the central charge. It cannot be determined
by the requirement of conformal invariance alone, and will depend on the
theory studied. Different values of $c$ will imply different universality
classes.

The nonvanishing of $c$ represents an anomaly which often occurs in problems
with local symmetries. It means that a classical symmetry cannot be
implemented quantum-mechanically due to renormalization effects. 
Therefore not all fields but only the primary fields transform according
to (\ref{primftf}). As will be seen below, $T(z)$ determines the change in
action under a local coordinate transformation. In a path-integral
formalism, the anomaly in $T$ then implies that the complete measure cannot
be made conformally invariant. The anomaly is also called Schwinger term.
As examples, for a free boson $\phi(z)$, $T(z) = :[\partial_z \phi(z)]^2: /2$
and $c=1$; free real (Majorana) fermions $\psi(z)$, 
relevant for the 2D Ising model, have $T(z) = : \psi(z) \partial_z \psi(z) :
/2$ and $c=1/2$; finally, free complex (Dirac) fermions $\Psi(z)$, 
relevant for
the \lm , have $T(z) = i : [\partial_z \Psi^{\dag}(z) ] \Psi(z) -
\Psi^{\dag}(z) \partial_z \Psi(z) : /2$ and $c=1$ like the bosons.

This anomaly has important consequences for the algebra of the generators
of the local conformal transformations on the quantum level.
Just as above on the classical level,  
one can derive the algebra of the generators from a Laurent expansion 
of the \set\
\begin{equation}
\label{modeset}
T(z) = \sum_{n = -\infty}^{\infty} L_n z^{-n-2} \;\;\;.
\end{equation}
Using (\ref{opeset}), we obtain the Virasoro algebra with central
extension $c$
\begin{eqnarray}
[ L_n, L_m ] & = & (n-m) L_{n+m} + \frac{c}{12}(n^3 -n) \delta_{n+m,0}
\;\;\;, \nonumber \\
\label{vira}
[ \bar{L}_n, \bar{L}_m ] & = & (n-m) \bar{L}_{n+m} 
+ \frac{\bar{c}}{12}(n^3 -n) \delta_{n+m,0} \;\;\;, \\
\left[ L_n , \bar{L}_m \right] & = & 0 \;\;\;.  \nonumber
\end{eqnarray}
The classical Virasoro algebra is recovered for $c=0$.
Every conformal quantum field theory defines a representation of (\ref{vira})
with some central charge $c, \bar{c}$. The $L_n$ are the generators
of transformations of quantum fields associated with the monomial of
degree $n+1$ in $z$. For $\xi^z(z) = - \xi_n z^{n+1}$, we have
\begin{equation}
\label{ntf}
\delta \phi (z, \bz ) = -\xi_n [ L_n, \phi(z, \bz) ]\;\;\;.
\end{equation}
Unitarity constrains the generators to satisfy
\begin{equation}
L_m^{\dag} = L_{-m}
\end{equation}
and regularity of the \set\ at the origin implies
\begin{equation}
\label{genvac}
L_m |0 \rangle = 0 \;\;, \;\;\;\;m \geq -1 
\hspace{0.5cm} {\rm and} \hspace{0.5cm}
L_m^{\dag} |0\rangle = 0 \;\;, \;\;\;\; m \leq -1 
\end{equation}
in their action on the vacuum $|0 \rangle$.

There are two more important properties of the \set . Under a local
conformal transformation to $z' = f(z)$, it transforms as
\begin{eqnarray}
\label{settf}
T(z) & \rightarrow & T'(z) = \left( \frac{dz'}{dz} \right)^2 T(z')
+ \frac{c}{12} \{ z' , z \} \;\;\;, \\
\label{schw}
\{z' , z\} & = & \frac{\partial^3_z z'}{\partial_z z'} - 
\frac{3 (\partial^2_z z')^2}{2 (\partial_z z')^2} \;\;\;.
\end{eqnarray}
The first term in (\ref{settf}) translates the fact that $T(z)$ is
a field of conformal weight $(2,0)$ in agreement with (\ref{opeset}) above,
while the second term contains the conformal anomaly. 
(\ref{schw}) is known as the Schwarzian derivative.

We now turn to the representations of the Virasoro algebra, i.e. the
states of our Hilbert space. In general, the representations of symmetry
groups are constructed from highest weight vectors (states). Such a
highest weight state $|h \rangle $ is created by the action of a holomorphic
primary field $\phi$ on the vacuum, at the origin
\begin{equation}
\label{hwst}
|h \rangle = \phi(0) |0 \rangle
\;\;\;, \hspace{0.5cm}
L_0 |h \rangle = h |h \rangle \;\;\;, \hspace{0.5cm} L_n |h \rangle 
=0 \;\;\;, \;\;\; n>0 \;\;\;.
\end{equation}
$|h \rangle $ is thus eigenstate of $L_0$. The $L_n, \; n>0$ are the lowering
operators annihilating $|h\rangle$. The corresponding raising operators are
$L_{-n}, \; n>0$ and, acting on $|h \rangle $, generate the descendant states
\begin{equation}
\label{desc}
L_{-n_1} \ldots L_{-n_k} |0 \rangle \neq 0 \;\;\;, \hspace{1cm}
1 \leq n_1 \leq \ldots \leq n_k \;\;\;.
\end{equation}
They form a basis for the representation vector space. The eigenvalue
of $L_0$ on the state (\ref{desc}) is $h + n_1 + \ldots + n_k$. The
highest weight state $|h \rangle$ 
has the lowest eigenvalue among all the states
that can be created out of it by acting with the raising operators. 
It is the ground state in a given sector of the theory. The descendants
are the excited states. The $L_n$ ($n>0$) act as an infinite number
of harmonic oscillator annihilation operators, and the $L_n^{\dag} = L_{-n}$
then are the creation operators. The 
level of the state (\ref{desc}) is $\sum_{i=1}^k n_k$, and the level
associated with an operator $L_{-n}$ is $n$. The number of basis vectors
on a given level $N$ is $P(N)$, the number of partitions of $N$. The 
conformal weight of all the descendant states on level $N$ is $h+N$.
The vector space generated from $|h \rangle $ is called Verma module.

All states (and fields) in a conformal field theory can be grouped
into conformal families (towers). 
They consist of a highest weight state $|h \rangle $
and all the descendant states generated by the application
of the raising operators $L_{-n}$. The different highest weight states
are obtained from the action of the different primary fields $\phi_n
(z)$ [or, more generally,  $\phi_n(z, \bz )$] on the vacuum according to
(\ref{hwst}). The conformal families offer
a very convenient way to classify the excitations in the system
and the spectrum of the scaling dimensions.

All correlation functions involving secondary fields
can be calculated from those containing primary fields only, by
acting on them with a differential operator obtained from the transformation
property (\ref{ntf}). From global conformal invariance, we 
also know the two-point correlation
functions of the primary fields, Eq. (\ref{twoptfct}), and can construct
the three-point function according to the same scheme. What about the
$n$-point function for primary fields? It may happen that on a given
level of the theory, say $k$, 
the states are not linearly independent but that
there is a combination of states that vanishes (the family is then said
to be degenerate at level $k$). The equation describing this degeneracy 
can then be transformed into a differential equation to be satisfied
for an arbitrary correlation function of primary fields, if at least one
of them is degenerate. In this way, it is possible to obtain, for
conformal field theories with degenerate families, all correlation 
functions.

Unitary representations of the Virasoro algebra only exist for certain
values of $c$ and $h$
\begin{eqnarray}
\label{unit}
c  \geq 1 & , &  h \geq  0 \\
\label{discrete}
{\rm or} \; c & = & 1 - \frac{6}{m(m+1)} \;\;\;, \;\;\; 
h_{p,q} (m) = \frac{[(m+1)p - mq]^2 -1}{4m(m+1)} \;\;\; \\
{\rm with} \;\; m & = & 3,4,\ldots \;\;\;, \;\;\;
1 \leq p \leq m-1 \;\;\;, \;\;\; 1 \leq q \leq p \;\;\;, \nonumber
\end{eqnarray}
and at least  the discrete series (\ref{discrete}) does indeed 
have degenerate families. The models belonging to this discrete series
have quantized critical exponents \cite{friedan} contained in
(\ref{discrete}). The most famous among 
them is the 2D Ising model with $c=1/2$. 

Up to now, we have implicitly assumed that our fields are defined in
the infinite $z$-plane. What happens when we consider finite systems?
From Eq. (\ref{genvac}), we deduce that
\begin{equation}
\label{setzero}
\langle T(z) \rangle = \sum_{m=-\infty}^{\infty} \left\langle 0
\left| \frac{L_m}{z^{m+2}} \right| 0 \right\rangle = 0
\end{equation}
in the infinite complex $z$ plane. Now use the exponential transformation
\begin{equation}
\label{exptf}
z = \exp \left( \frac{2 \pi i }{L} u \right) \;\;\;, \hspace{1cm}
u = \frac{L}{2 \pi i} \log z
\end{equation}
to map the infinite $z$-plane onto a strip ($u$) of width $L$
with periodic boundary conditions. Observe that under this transformation,
the mode expansion (\ref{modeset}) simply becomes a Fourier transformation,
and the Virasoro generators $L_n$ become the Fourier coefficients of the
\set . We obtain
\begin{equation}
\label{setfin}
T_{\rm strip}(u) = \left( T_{\rm plane}(z) - \frac{c}{12} \{u,z \}
\right) \left( \frac{d z}{d u} \right)^2
\end{equation} 
and with (\ref{setzero}) therefore 
\begin{equation}
\label{setexp}
\langle T_{\rm strip} (u) \rangle = \frac{c}{24} \left( \frac{2 \pi}{L}
\right)^2     \;\;\;.
\end{equation}
The \set\ measures the cost of energy 
[change in the action $\delta S = (-1/2 \pi) $ $\times $ $ 
\int T_{ij} \partial_i \xi_j d^2 r$] 
of a change in metric. One 
can now calculate the change in energy associated with another 
(nonconformal) transformation,
a horizontal dilatation of the $u$-strip [$u_1' = (1+ \varepsilon) u_1,
\; u_2' = u_2$] which changes the length of the system, and integrate
to find 
\begin{equation}
\label{freen}
E(L) - E(\infty) = \frac{c \pi}{6L} \;\;\;,
\end{equation}
where $E(L)$ is the energy per unit length \cite{bloeaff}.
This formula is extremely important because it allows us to determine
the value of the central charge from calculations on finite systems!
Moreover, it
suggests an interpretation of the anomaly (\ref{schw})
as a Casimir effect, i.e.
a shift in the energy due to the finite geometry of the system. The 
mathematical reason is that the local conformal transformations 
(with the exception of the global ones) are (i)
usually not defined in all points of the complex plane and (ii) are not
one-to-one mappings of the complex plane on itself.

The exponential transformation (\ref{exptf})
is also important to obtain the scaling 
dimensions of primary fields from finite-size calculations \cite{cardy}. 
The two-point correlation function of a primary operator $\phi(z,\bz)$ 
with conformal weights $h,\bh$ 
transforms under a conformal transformation (\ref{exptf}) into
\begin{equation}
\label{corfexp}
\langle \phi(u, \bu) \phi (u' ,\bu') \rangle =
\frac{(\pi/L)^{2 \Delta}}{ 
\left( \sinh[ \pi (u - u') / L] \right)^{2h} 
\left( \sinh[ \pi (\bu - \bu') / L] \right)^{2\bh} } \;\;\;. 
\end{equation}
[Notice in passing: correlation functions at finite temperature $1/\beta$ must
satisfy periodic boundary conditions in the Matsubara time $\tau = it$.
(\ref{corfexp}) then gives directly the finite temperature expressions
for the correlation functions of the preceding section if we put
$L = 2 v \beta$, as suggested in Section \ref{secprop}.]
Writing $u=u_1 + i u_2$ and going on the physical surface $\bu = u^{\star}$, 
this can be expanded as
\begin{eqnarray}
\langle \phi(u, \bu) \phi (u' ,\bu') \rangle & = &
\left( \frac{2 \pi}{L} \right)^{2 \Delta} \sum_{N, \bar{N} = 0}^{\infty}
a_N a_{\bar{N}} \exp [ - 2 \pi (\Delta + N + \bar{N} )(u_1 - u_1')/L ] 
\nonumber \\
\label{corsca}
& \times & \exp [ 2 \pi i (s + N - \bar{N} ) (u_2 - u_2')/L ] \;\;\;.
\end{eqnarray}
Here, $\Delta$ and $s$ are scaling dimension and spin of the operator,
respectively. The correlation function can also be calculated using
operators $\hat{\phi}(u_2)$ which act on the states $|n,k \rangle$
of a Hilbert space
\begin{equation}
\label{corhil}
\langle \phi(u) \phi(u') \rangle = \sum_n \langle 0 | \hat{\phi}(u_2)
| n,k \rangle e^{-(E_n - E_0)(u_1 - u_1')} \langle n,k | \hat{\phi}(u_2')
|0 \rangle \;\;\;,
\end{equation}
where the matrix elements depend on $u_2$ as $\exp(iku_2)$ with momentum
$k$. Comparing (\ref{corsca}) with (\ref{corhil}), we find that
energies and momenta scale as
\begin{eqnarray}
\label{finen}
E_n^{(L)}(N,\bar{N}) & = & E_0^{(L)} 
+ 2 \pi v (\Delta + N + \bar{N} ) / L \;\;\;, \\
\label{finmom}
k^{(L)}(N,\bar{N}) & = & k^{(\infty)} + 2 \pi (s + N - \bar{N}) / L \;\;\;.
\end{eqnarray}
Here, the energies are taken on the system of length $L$, and $v$ is
the velocity of the excitations. 
In (\ref{finmom}), we have allowed for a finite momentum
$k^{(\infty)}$ of the highest weight state in the conformal tower built by
the primary field $\phi$, extrapolated to the infinite system.
Eqs.~(\ref{finen}), (\ref{finmom}) show that on a finite strip,
each primary operator generates
the whole spectrum of scaling dimensions and momenta
of the conformal tower, i.e. also those of the descendant operators
at level $(N,\bar{N})$. 

There are thus two ways to obtain the scaling dimensions and correlation
functions of a conformal field theory; (i) one can construct the \set ;
from its transformation properties (\ref{settf}) or its OPE with itself
(\ref{opeset}),
one can deduce $c$, and from its OPE with other fields one obtains
their scaling dimensions (\ref{ope}). (ii) one can study the finite size
scaling behaviour of the ground state and  a set of excited states, and
obtain the central charge from Eq. (\ref{freen}). 
Going back now from $z, \bz = z^{\star}$ to $(x,t)$ and
extrapolating $L \rightarrow \infty$, the correlation functions
of a primary $(h,\bh)$ field $\phi(xt)$ are then given 
from (\ref{twoptfct})
\begin{equation}
\label{crophys}
\langle \phi(xt) \phi(00) \rangle = \frac{e^{i k^{(\infty)}x}
e^{i \bar{k}^{(\infty)} x}}{(x 
+ i v t)^{2 h} (x - i v t)^{2 \bh}} 
\end{equation}
and those of the descendant fields have the same structure with
the corresponding $(h +N , \bh + \bar{N})$.

\subsection{The Gaussian model}

The first problem beyond the discrete classification scheme (\ref{discrete})
-- also the most important in the context of the present article --
is given by theories with central charge $c=1$ and realized by
free bosons, precisely the spinless \lm , or 
the Gaussian model of statistical mechanics. Here, we 
give a conformal field theory analysis. 
The action for free bosons is given by
\begin{equation}
\label{acbos}
S = \frac{1}{2 \pi} \int \! dz \: d\bz (\partial_z \Phi)(\partial_{\bz} \Phi) 
\;\;\;, \hspace{1cm} \Phi \equiv \Phi(z, \bz ) \;\;\;,
\end{equation}
and the equivalence to the \lm\ is clear when represented in terms of
phase fields (\ref{phaseham}) or (\ref{phasslf}).
The model is critical and manifestly conformally invariant, 
and remains so even upon introducing
a dimensionless coupling constant $g$ as a prefactor in $S$, for all
values of $g$ [cf. below after Eq. (\ref{marac})].
The solution of the equations of motion can be given in terms
of left- and right-moving (holomorphic and anti-holomorphic) fields
$\Phi(z ,\bz ) = [\phi(z) + \bar{\phi}( \bz ) ]/2$ where 
$z, \bz = x \pm i y$. Their correlation functions are
\begin{equation}
\label{boscor}
\langle \phi(z) \phi(w) \rangle = - \log (z-w) \;\;\;, \hspace{1cm}
\langle \bar{\phi}(\bz) \bar{\phi}(\bar{w}) \rangle = - 
\log (\bz - \bar{w}) \;\;\;.
\end{equation}
The fields $\phi(z)$ are not conformal fields, but their derivatives
are. To show this, we need the \set\ which can be identified from the
change in the action $\delta S = (-1/2 \pi) \int T_{ij} \partial_i
\xi_j d^2 r$ under a conformal transformation $\br \rightarrow \br +
{\bf \xi}$ of the fields $\phi$ as (after going to the complex $z$-plane)
\begin{equation}
T(z) = - \frac{1}{2} : [\partial \phi (z) ]^2 : \equiv - 
\frac{1}{2} \lim_{d \rightarrow 0}
\left[ \partial \phi \left(z+\frac{d}{2} \right) 
\partial \phi \left( z -  \frac{d}{2} \right) - 
\frac{1}{d^2} \right] \;\; ,
\end{equation}
where the identity defines the normal-ordering convention.
From the OPE of $\phi$ with the \set , we find
\begin{equation}
T(z) \partial \phi (w) = \frac{\partial \phi(w)}{(z-w)^2} +
\frac{1}{z-w} \partial^2 \phi(w) + \ldots     \;\;\;,
\end{equation}
which, comparing with Eq. (\ref{ope}), indeed identifies $\partial \phi$ as
a primary field with conformal weights $(1,0)$. 
It is now possible to write down a mode expansion (Laurent series)
for the field $\partial \phi(z)$
\begin{equation}
\label{mode}
J(z) \equiv i \partial \phi(z) = \sum_{n= -\infty}^{\infty} 
\frac{J_n}{z^{n+1}} \;\;\;, \hspace{1cm} J_n = \oint \frac{dz}{2 \pi i}
z^n J(z) \;\;\;.
\end{equation}
Again, on a cylinder, i.e. periodic
boundary conditions for a (1+1)D field theory, the $J_n$ simply are the
Fourier components of the current $J(z)$.
Their algebra from (\ref{mode}) is
\begin{equation}
\label{kacmood}
[J_n , J_m ] = n \delta_{n+m,0} \;\;\;,
\end{equation}
the $U(1)$-Kac-Moody algebra. We immediately note that the algebra
of the $J_n$, (i) up to the factor $n$ which can be absorbed into a
redefinition of $J_n$, is a bosonic one, and (ii) is identical to
the algebra satisfied by the Luttinger density operators $\rho_{rs}$,
Eq. (\ref{tlmcom}). Physically, this identifies the $J_n$ as 
(chiral) density fluctuation modes or currents, which is the same because we
look at a single branch only. The modes with $n<0$ are creation operators,
and those with $n>0$ are annihilation operators
\begin{equation}
\label{kacmolow}
J_n | 0 \rangle = 0 \hspace{1cm} {\rm for} \hspace{1cm} n > 0 \;\;\;.
\end{equation}
The correlation function of the currents
$J(z) = i \partial \phi(z)$ is 
\begin{equation}
\langle J(z) J(w) \rangle = \frac{1}{(z-w)^2 }
\end{equation}
from (\ref{twoptfct}) and the conformal weights $(1,0)$. 
Of course, for free bosons, these results can also be obtained directly
by simply calculating a Gaussian integral. 

Eq.~(\ref{kacmood}) is a special case of a more general expression
\begin{equation}
\label{kacgen}
[ J_n^a , J_m^b ] = i f^{abc} J^c_{n+m} + k n \delta^{ab} \delta_{n+m,0} 
\end{equation}
satisfied by the current generators $J_n^a$ of a more general Lie group,
e.g. $SU(2)$ as appears in nonabelian bosonization schemes \cite{nonabbos}. 
$f^{abc}$ are its structure constants, and the integer $k$
is the level of the Kac-Moody algebra. The central charge of the 
associated Virasoro algebra is related to the level $k$ by $c = 3 k /
(k+2)$. We do not go into further details here, although we shall 
encounter an example of a $U(1)$-Kac-Moody algebra with $c \neq 1$ 
in Section \ref{fqh} when we discuss the chiral \LL s formed by
the edge excitations in the fractional quantum Hall effect.

One can also
consider ``vertex operators'' $:\exp [ i \alpha \phi(z) ]:$ which,
from their OPE with $T(z)$, are identified as primary fields with
weights $(\alpha^2/2, 0)$. This determines the decay of their correlation
functions, which, like all Gaussian model correlations, can also be
evaluated explicitly
\begin{equation}
\label{vercor}
\langle :e^{i \alpha \phi(z)} : : e^{- i \alpha \phi(w)} : \rangle
= e^{\alpha^2 \langle \phi(z) \phi(w) \rangle } =
\frac{1}{(z-w)^{\alpha^2}} \;\;\;.
\end{equation}
The first equality is (\ref{gaussav}) and the
second equality has been obtained with (\ref{boscor}). 

Up to now, we have been silent on a parameter of the theory -- the
compactification radius $R$. This is a parameter of the theory,
and one can either fix it from certain constraints on the vertex operators,
or give it from the outset and then determine the operators which are
well-behaved. 
Single-valuedness of the vertex operators implies that
the fields $\phi(z)$ must be compactified with a compactification
radius $R$ (obey periodic boundary conditions on a circle with radius $R$)
\begin{equation}
\label{pbc}
\phi + 2 \pi R = \phi \;\;\;, \hspace{1cm} R = n/\alpha \;\;\;.
\end{equation}
In general, the vertex operators $:\exp[ i \alpha \phi(z) ]:$ have
weird commutation relations. 
We can require, however, the object
\begin{equation}
\label{bosfera}
\Psi_{\alpha}^{\dag} (z) = : \exp [ i \alpha \phi(x) ] :
\end{equation}
to obey fermionic anticommutation relations with $\Psi(z)$
and with its antiholomorphic
counterparts $\bar{\Psi}^{\dag}(\bz), \bar{\Psi}(\bz)$, and to
commute with the currents \cite{mandel}
\begin{eqnarray}
\{ \Psi(z) , \Psi^{\dag}(z') \} & = & \delta(z-z') \;\;\;\;,
\hspace{0.5cm} \{ \Psi(z) , \Psi(z') \} = 0 \;\;\;, \nonumber \\
\{ \Psi(z) , \bar{\Psi}^{\dag}(\bz') \} & = & \{ \Psi(z) , 
\bar{\Psi}(\bz') \} = 0 \;\;\;, \\
\left[ J(z) , \Psi(z') \right] & = & - \delta(z-z') \Psi(z) \;\;\;.  \nonumber
\end{eqnarray}
This imposes $\alpha = 1$ for free fermions. 
$: \exp [ i \phi(z)]:$ then creates a $(1,0)$ state from the vacuum. 
One can also prove that the current $J_f(z) = : \Psi^{\dag}(z) \Psi(z):$
written as a fermion bilinear is identical to the bosonic current
$J_b(z) = i \partial \phi(z)$ where the field $\phi$ is precisely the
one appearing in the exponential in (\ref{bosfera}), making the 
identity of the fermion and boson representations complete. 

The Kac-Moody generators $J_m$ are extremely useful in classifying the
excitations in our model. The Hamiltonian is related to 
the \set\ through [$x$ is the spatial coordinate of the
image of $z$ on the cylinder]
\begin{equation}
\label{set1df}
H = \frac{1}{2 \pi} \int \! \left[ T(x) + \bar{T}(x) \right] dx \;\;\;,
\hspace{1cm} T(x) = \frac{1}{2} : J(x) J(x) : +
\frac{1}{24} \left( \frac{2 \pi}{L} \right)^2 \;\;\;.
\end{equation}  
The constant shows that our model has a central charge $c=1$. 
Putting together the mode expansions (\ref{modeset}), (\ref{mode}) and the
exponential transform (\ref{exptf}), and realizing that the transformation
to the cylinder generates an anomaly similar to (\ref{settf})
in $:J(z) J(z):$, we find
the generators of the Virasoro algebra 
\begin{equation}
\label{viragen}
\frac{2 \pi}{L} L_m = \frac{1}{2 \pi} \int_0^{\infty} \! dx
\: e^{i \frac{2 \pi}{L} m x} \: T(x) = 
\frac{ \pi}{L} \sum_{n = - \infty}^{\infty} : J_n J_{m-n} : - 
\delta_{m,0} \frac{c\pi}{12L}
\;\;\;.
\end{equation}
The $L_m$ satisfy the Virasoro algebra (\ref{vira}) 
with central charge $c=1$ among them, and
\begin{equation}
\label{virakac}
[L_n , J_m ] = - m J_{n+m}
\end{equation}
with the generators of the Kac-Moody algebra. 
To every Kac-Moody algebra, there is an 
associated Virasoro algebra, and the present construction of the 
Virasoro generators from the Kac-Moody ones is due to Sugawara.
Specializing (\ref{virakac}) to $n=0$ yields (taking only one chiral
component)
\begin{equation}
[H , J_m ] = - \frac{2 \pi}{L} m J_m \;\;\;.
\end{equation}
showing that the $J_m$ 
act as raising ($m<0$) and lowering ($m>0$) operators of a harmonic
spectrum.
The spectrum is harmonic because of the linearized dispersion and the
equal $k$-spacing.
Moreover, Eq. (\ref{kacmolow}) implies that the state $|0 \rangle$ 
is annihilated
by the Virasoro generators (\ref{viragen}) with $n>0$, and therefore
qualifies as a highest-weight state.

The $J_m$ can be used to algebraically generate
this spectrum and its conformal tower (family) of descendant states
from a reference state $|0 \rangle $. We suppose this state completely
filled up to some energy, call it Fermi energy, $E_F = 0$. By applying
the $J_m$ with $m<0$, we make a particle-hole excitation with energy
$-m$ by raising a particle from below the Fermi energy into an 
unoccupied state above, $-m$ levels higher. 
The energy level structure of the Hamiltonian carries over to 
the descendants created from the reference state $| 0 \rangle$
\begin{equation} 
\label{kamodesc}
J_{-n_1} \ldots J_{-n_k} |0 \rangle \neq 0 \;\;\;, \hspace{1cm}
1 \leq n_1 \leq \ldots \leq n_k \;\;\;,
\end{equation}
as in Eq. (\ref{desc}) for the Virasoro generators. The level of the 
state (\ref{kamodesc}) is $\sum_{i=1}^k n_i,$ ${(n_i > 0)}$, 
and the level associated with a single
generator $J_{-m}$ is $m \; (>0) $. 
The energy of the state equals its level in units of $2 \pi / L$.
Of course, such a description is redundant:
a state at level $N$ can be generated in $P(N)$ ways; $P(N)$ is the
number of partitions of $N$. 
At level $N$, we have an N-particle--N-hole
excitation. The Kac-Moody algebra can therefore be used to classify
the particle-hole excitations from a reference state. 

There are states which the currents cannot create from our Fermi sea
$|0\rangle$: those with additional particles, i.e. a total charge $Q$
with respect to $|0 \rangle$,
and specifically the lowest-energy
states $|Q \rangle $ where the $Q$ particles occupy the first $Q$ states above
the Fermi level. We anticipate that there is an infinity of such 
states, and each of them will be the highest 
weight state in the sector with total  charge $Q$ of the theory.
From our discussion, it is clear that the vertex operators (with $\alpha
=1$ for a free theory) 
\begin{equation}
\label{bosfer}
\Psi^{\dag} (z) = : \exp [ i \phi(z) ] : 
\end{equation}
creates a chiral fermion. The chiral state 
$| Q \rangle$ then is created out of $|0 \rangle$ by
\begin{equation}
\label{qcrea}
| Q \rangle = : \exp [ i Q \phi (z) ] : |0 \rangle \;\;\;.
\end{equation}
The energy of a state $| Q \rangle$ is
\begin{equation}
E(Q) - E(0) = v_F \frac{ \pi Q^2}{L} \;\;\;.
\end{equation}
Acting on $| Q \rangle$, the Kac-Moody generators $J_{-m}$ will again
create the full spectrum of particle-hole excitations in the sector
$| Q \rangle$.

There are more general operators than (\ref{bosfer}). One can
build them by combining fields of both chiralities, writing e.g.
\begin{equation}
\label{psimn}
\Psi_{mn}^{\dag}(z,\bz) =
: \exp \left( i \left[ \alpha_{mn} \phi(z)
+ \bar{\alpha}_{mn} \bar{\phi} (\bz) \right] \right) : \;\;\;.
\end{equation}
Its scaling dimension is $(\alpha_{mn}^2 + \bar{\alpha}_{mn}^2)/2$, 
and $\alpha$
and $\bar{\alpha}$ are related to the compactification radius $R$ by
\begin{equation}
\label{alalbar}
\alpha_{mn} = \frac{1}{2} \left( \frac{m}{R} + \frac{nR}{2} \right) \;\;\;,
\hspace{1cm} \bar{\alpha}_{mn} = \frac{1}{2} \left( \frac{m}{R} - 
\frac{nR}{2} \right) \;\;\;. 
\end{equation}
Its physical meaning is quite
clear. Increasing $m$ means increasing the charge for both
chiralities $\Psi_{m0}^{\dag} | Q, \bar{Q} \rangle
= |Q+m, \bar{Q}+m \rangle$. 
Increasing $n$ transfers charge from one
chirality to the other, i.e. creates a finite persistent current
$\Psi_{0n}^{\dag} |Q ,\bar{Q} \rangle = |Q+n, \bar{Q}-n \rangle$
[this is an example of a primary operator
with $k^{(\infty)} \neq 0$ in (\ref{finmom})]. $\Psi_{mn}^{\dag}$ creates
charge or current excitations, or combinations thereof. 
As for $\Psi_{\alpha}(z)$, the particles making up these charge and 
current excitations are fermions only at the particular
compactification radius $R = 1$. Else, they are more general objects.

The Luttinger or Gaussian model has operators with continuously varying
exponents. Responsible for this are dimension $(1,1)$ marginal operators
which take the model along a whole critical line. The simplest of these
operators is 
\begin{equation}
\label{marac}
S' = \frac{g-1}{2 \pi} \int \! dz \: d\bz (\partial_z \Phi) (\partial_{\bz}
\Phi) \;\;\;, 
\end{equation}
which is proportional to the free action (\ref{acbos}) and whose only
effect is to introduce the coupling constant $g$ as a prefactor into $S$.
The effect of this interaction can simply be absorbed by 
redefining the fields $\phi \rightarrow \phi / \sqrt{g} $. With this
redefinition, the effective $\alpha \rightarrow \alpha/\sqrt{g}$ 
in the vertex operators changes
accordingly, and consequently both their compactification radius and
their conformal weight. When the compactification radius $R \neq 1$,
$\Psi_{\alpha}^{\dag}$ no longer 
describes a fermion but a more complicated object.
In the Luttinger model, the $g_2$-interaction
is such a marginal operator, coupling currents of both chiralities.
From Eq.~(\ref{vercor}) it is obvious then that $S'$ generates continuously
varying exponents in the correlation functions, and varying $g$ sweeps
the model over a whole critical line.

By transforming the phase-field representation of the spinless \lm\
\ref{phasslf} into an
imaginary time ($\tau = it$) 
action, one obtains the Gaussian model with an effective
coupling constant $g = K$. Importantly, the coupling constant $g$ depends
essentially on $g_2$. Finite $g_4$ only renormalizes the effective $g_2$
but, alone, is not able to give $g \neq 1$. This is
quite easy to understand because it only changes the Fermi velocity of the
\lm\ and is therefore absorbed when going to the second spatial coordinate
$y = v \tau$.

More interesting is the case when the interactions are not 
of current-current type, or when the theory is formulated on a lattice
so that the identification of the conformal operators is far from
obvious. Mironov and Zabrodin have given a simple application of the methods 
discussed above to interacting spinless fermions (or bosons) \cite{mirozab}
\begin{equation}
H = \int_0^L \! dx \: \partial_x \psi^{\dag}(x) \: \partial_x \psi(x)
+ \frac{g}{2} \int_0^L \! dx \: dy \: \psi^{\dag}(x) \: \psi^{\dag}(y) \: V(x-y)
\: \psi_(y) \: \phi(x) \;\;\;.
\end{equation}
$V(x)$ is a repulsive pair interaction of rather general form. 
The density of particles is $n = N/L$ and $k_F = \pi n$. 
The model
can be solved by Bethe Ansatz for $V(x) = \delta(x)$ but most of the 
results are expected to carry over for reasonable longer-range potentials.

From the finite-size scaling formula for the ground state energy
(\ref{freen}), one finds $c=1$ which puts it in the Gaussian 
(\LL ) universality class.
Now we want to find the correlation function of some local operator $O(x)$.
To this end, one must compute up to order $1/L$ 
the energy of the lowest excited state
$| \varphi \rangle$ whose matrix element $\langle 0 | 
O(x) | \varphi \rangle \neq 0 $ for $L \rightarrow \infty$. Then, one
can use Eq.~(\ref{crophys}) if its scaling dimension and spin is
known from (\ref{finen}) and (\ref{finmom}). As an example,
one can make particle-hole excitations [$\rho(x)$]
with momentum $2m \pi/ L$ ($m$ small). From their excitation energies
linear in $m$, one can deduce the renormalized sound velocity $v$ and
$\Delta = \pm s = 1$. Of course, this is in agreement with our earlier
discussion where we found that the currents of the Gaussian model are
$(1,0)$ or $(0,1)$ fields. Next, make an excitation at constant particle
number with $k^{(\infty)}_{0n} = 2 n k_F$ where the particle-hole
spectrum goes to zero. In a free system, this would correspond to applying 
the operator $\Psi^{\dag}_{0n}$ to the Fermi sea. The Bethe Ansatz
gives the energy change $\delta E_{0n} = (2 \pi / L) (2 k_F n^2 / v)$
and, using (\ref{finen}), the scaling dimension $\Delta_{0n} 
= 2k_F n^2 / v$. 
Comparing with (\ref{alalbar}), one finds a compactification
radius $R^2 = 2 k_F / v$. One can also add pairs of particles 
or a single particle, where a selection rule enforces half-integer $n$.
The energy shift is given by (\ref{finen}) with a scaling dimension
(\ref{alalbar}) with $m=1,n=1/2$ and the $R$ found above. The Green
function and the CDW correlation functions  (corresponding to the low-energy
excitation at \tkf\ discussed before) are then
\begin{equation}
G(x) \sim \cos(k_Fx) x^{-1/2R^2 - R^2/2} \;\;, \;\;\;\;
R_{CDW}(x) \sim \cos(2 k_F x) x^{- 2 R^2} \;\;, \;\;\;\;
R^2 = 2 k_F / v = K \;\;.
\end{equation}
The exponents satisfy the scaling relations of the spinless Luttinger
model, and the correlation exponent $K$ has been identified in the last
equality. (The factor 4 difference in $R^2$
to Mironov and Zabrodin \cite{mirozab}
must be due to a different prefactor [$2/\pi$] of the Gaussian action.)

The close correspondence of conformal field theory and the operator
approach to bosonization should be apparent now, at least for the
case $c=1$ of the Luttinger and Gaussian models. Density fluctuations
(currents), charge and current excitations, and fermion raising and
lowering operators all appear either in an operator approach based 
on a Hilbert space, or in an algebraic formalism based on the 
$U(1)$-Kac-Moody algebra, which is  satisfied by the currents as a consequence
of the $U(1)$-gauge symmetry (\ref{gauge}) corresponding to the conservation 
of left- and right-moving particles separately. The main problem
in the operator approach is the explicit identification of the
coupling constant of the \lm\ which then determines all correlation
functions. In conformal field theory, we must determine the scaling
dimensions of the various primary operators. As will be discussed
in the next chapter, in both cases the spectrum of
low-lying eigenvalues is sufficient for that purpose, showing once
more the full equivalence between bosonization and conformal field
theory. Which one to use is a matter of taste.

The preceding analysis of the interacting spinless fermions foreshadows the
application of conformal field theory to rather general models of interacting
electrons. We shall discuss this topic in the next chapter, where 
some other methods for extracting the low-energy physics will also be
presented.

\chapter{The Luttinger Liquid}
\label{chapll}
\section{The conjecture}
\label{llconj}

The \lm\ can be solved exactly at any interaction strength, except for
too strong attraction where the model becomes unstable towards phase
separation (formation of electron droplets; Section 
\ref{phassep}). However,
it contains drastic approximations with respect to a realistic many-body
problem: (i) its dispersion is strictly linear, and (ii) the electron-electron
interaction is limited to forward scattering only. The possibility of
an exact solution is precisely related to these two approximations.

One may therefore wonder if these approximations are essential in the sense
that the Luttinger physics is lost in any different model or if,
on the contrary, this physics
is robust. In this case, only parameters ($K_{\nu}, \; v_{\nu}$)
would be renormalized close to the Fermi surface, but the structure of 
the low-energy theory would be identical to the \lm . Of course, further
away from the Fermi surface, new phenomena such as boson-boson interactions
or lifetime effects could occur but it would be guaranteed that they fade
away as $(k-k_F, \; \omega, \; T) \rightarrow 0$. This is what happens in
the Fermi liquid.
The \lm\ would then represent the generic behaviour of gapless 1D quantum
systems, and one could build upon it
the universal low-energy phenomenology for all 1D
metals (the \LL ) called for by the breakdown of the Fermi liquid 
in 1D (Chapter \ref{chaptwo}). 

Haldane, in the early 1980's, conjectured that this was indeed possible
and supported this conjecture with a series of case studies of models
solvable by Bethe Ansatz \cite{Haldprl,halphlet}.
He also demonstrated that certain
features mapped away when passing to the Luttinger model, such as curvature
in the dispersion, only introduce nonsingular, perturbative interactions
among the bosons \cite{Haldane} which disappear as one goes to the
long-wavelength or low-frequency limit. Haldane's conjecture
has meanwhile been verified in an impressive number of instances
some of which will be discussed below
and, to my knowledge, no counterexample has been discovered yet.

What is the content of this ``Luttinger liquid conjecture''? Given any
1D model of correlated quantum particles (in 1D not even necessarily fermions)
and let there be a branch of gapless excitations: then the Luttinger model is 
the stable low-energy fixed point of the original model (or at least its 
gapless degrees of freedom). In other words, the asymptotic
low-energy properties of the degree of freedom
associated with this branch are described by an effective (renormalized)
Luttinger model, in particular with one renormalized Fermi velocity 
$v_{\nu}$ and one
renormalized effective coupling (stiffness) constant
$K_{\nu}$, up to perturbative
boson--boson interactions. All properties found for
the Luttinger model: (i) absence of fermionic quasi-particles in the 
vicinity of the Fermi surface, (ii) anomalous dimensions of the fermion
operators producing nonuniversal power-law correlations, (iii) charge-spin
separation, (iv) the universal relations among the nonuniversal exponents
of the correlation functions, among the velocities for sound, charge, and
current excitations and between velocities and the effective renormalized
coupling constants, carry over to \em the low-energy sector \rm of
the model under consideration. 
The nontrivial task  remaining then is to determine the two central
quantities of the \LL, the renormalized velocity $v_{\nu}$ and the renormalized
effective coupling constant $K_{\nu}$, from the original model. 
Once this is achieved, one has an \em asymptotically exact solution \rm
of the 1D many-body problem. 
This is what most of this chapter will be about.

A word of caution is required for systems with several degrees of freedom.
If all degrees of freedom remain gapless, the low-energy fixed point will
be a \LL\ in the sense described. If some degrees of freedom become
gapped while others do not, the physics \em within \rm the gapless degrees of
freedom can be described as a \LL\ while all quantities involving gapped
degrees of freedom will deviate qualitatively from the \LL . 
In Chapter \ref{mottch}, we will give examples for this kind of situation.

When mapping a more realistic model of interacting electrons
in 1D onto the Luttinger model, complications arise from two 
sources: (i) the dispersion of these models is not linear;
(ii) the interactions generally do not only contain the forward scattering
processes included in, and solved by the \lm\, but also the large momentum 
transfer backward (spin exchange)
and Umklapp scattering (for commensurate band filling) depicted in Fig.~3.3. 
Moreover if the interactions
are not weak, states far from the Fermi surface are coupled to the Fermi
surface states; curvature and interaction could then conspire to
invalidate the \LL\ picture. That none of this in fact happens, was 
demonstrated by Haldane \cite{Haldane,Haldprl}.

There are two basic ways of proceeding. One can start from a \lm\ and
extend it by various interactions and other features, and then study
the stability and renormalization of the \lm\ solution. The alternative
is to start directly from a realistic model of 1D correlated fermions, 
possibly on a lattice, and search for either \LL -correlations or
the specific \LL\ properties of the spectrum of charge, current and
sound excitations, Eq.~(\ref{speccoup}). The organization of the
chapter will follow roughly this order. 

\section{Luttinger model with nonlinear dispersion -- the emergence of higher
harmonics}
\label{nonldisp}

Haldane extended the Luttinger Hamiltonian, Eqs.~(\ref{hlutt})--(\ref{h4}),
by terms modelling nonlinear
dispersion
\begin{eqnarray}
\varepsilon_r(k) & = & v_F (rk - k_F) \nonumber \\
\label{disp}
& \rightarrow &  v_F (rk - k_F) + \frac{1}{2 m} (rk-k_F)^2
+ \frac{\lambda}{12 m^2 v_F} (rk - k_F)^3 \;\;\;. 
\end{eqnarray}
The third order term is necessary to ensure stability, and $\lambda > 3/4$
is required then. 
(\ref{disp}) can be bosonized, and one obtains quadratic terms of the
usual structure (\ref{h0bos}), with parameters renormalized
by $m$ and $\lambda$, but also cubic and quartic boson terms.
The quadratic form can be diagonalized as usual and the remaining terms
are written as (for spinless fermions for simplicity \cite{Haldane})
\begin{equation}
\label{hnl}
\delta H = \sum_r \frac{1}{2 \pi} \int_0^L \: dx
: \frac{1}{6m} \tilde{\Phi}_r^3(x) + \frac{\lambda}{48 m^2 v_F} 
\tilde{\Phi}_r^4(x) : \;\;\;,
\end{equation}
where the fields $\Phi_r(x) = [r\Phi_{\rho}(x) - \Theta_{\rho}(x)]/\sqrt{2}$
are given in terms of the phase fields of 
Eqs.~(\ref{phi}) and (\ref{theta}) and
the tilde implies that the Bogoliubov transformed fields (\ref{phastf}) enter.
$\delta H$ describes
boson-boson interactions. Their appearance is quite clear physically 
because, for a curved dispersion, two fluctuations with wave vector $q$ and
$q'$ will have an energy different from a single one with $q+q'$:
fluctuations interact. 
Fortunately, $\delta H$ is harmless:
one can either (i) argue, using \rg , that these
higher order boson terms are irrelevant and therefore do not influence
the fixed point physics described by the quadratic ones, or
(ii) as did Haldane, perform a systematic $1/m$-expansion to find that
the model still obeys the constitutive \LL\ relations 
(\ref{speccoup}) between velocities and 
coupling constant, and that the relation of $K$ to the correlation function
exponents, Eqs.~(\ref{gree}) -- (\ref{alts}), also remains unchanged.
Only the \em values \rm of $K$ and the velocities change. 

There is, however, another important effect caused by the nonlinear 
dispersion: the appearance, in the physical fermion operators $\Psi(x)$,
of higher harmonics in the chiral fermion operators $\Psi_r(x)$
which generates components with $3k_F, \; 5k_F, \ldots$ in the fermion
fields and with $4k_F, \; 6k_F, \ldots$ in the pair fields. 
Consequently, the single-particle Green function acquires components
at $3k_F, \; 5k_F \ldots$, and e.g. the density-density correlations
get oscillations at $4k_F, \; 6k_F \ldots$ in addition to the
usual ones at small $q$ and at \tkf . These components
are not present in the \lm\ but may appear in any more general model
with nonlinear dispersion $\varepsilon(k)$. They are necessary for
constructing \em local \rm density and fermion operators. 

This becomes apparent when one attempts to construct a representation
for excitations on all length scales from the long-wavelength ones which
dominate the low-energy physics \cite{halhfl}. Recall from our bosonization
procedure in Section \ref{secbos} that the field $\Phi(x)$ had a kink
of amplitude $\pi$ at the location of each particle. 
The location of the $l$th particle then is given by $\Phi(x) = l \pi$.
When going from the
smeared, long-wavelength density operators $\rho_r(x)$
to a physical local operator $\rho(x)$, it is sufficient to multiply
by a delta function $\sum_l \delta[ \Phi(x) - l \pi]$ to locate
the individual particles. The local density then is written as
\begin{equation}
\label{densloc}
\rho(x) = \left[ n + \Xi(x) \right] \sum_{m = -\infty}^{\infty}
\exp \left[ 2 i m \left\{\Phi(x) + k_F x \right\} \right] \;\;\;,
\end{equation}
where $n = N/L$ is the average density, and the field
$\Xi(x)$ describes the long-wavelength fluctuations. 
The fermion field essentially is the square-root
\begin{equation}
\label{fermloc}
\Psi(x) \sim \sqrt{n + \Xi(x)} \exp \left[ i \Theta(x) \right]
\sum_{m = -\infty}^{\infty} \exp \left[ (2 m + 1) i 
\left\{ \Phi(x) + k_F x \right\} \right] \;\;\;,
\end{equation}
but the sum must contain only odd terms to ensure
the anticommutation property.  
$\Theta(x)$ and $\Phi(x)$ have been introduced in (\ref{phi}) and
(\ref{theta}), and $\Xi(x)$ commutes with $\Theta(x)$ as 
$[ \Theta(x) , \Xi(x') ] = i \delta(x-x')$. Moreover, describing charge
density fluctuations, $\Xi(x)$ is 
related to $\Phi(x)$ by $\partial_x \Phi(x) = - \pi [ n + \Xi(x)]$. 
An equation equivalent to (\ref{fermloc})
for spin-1/2 fermions can be constructed easily. The correlation
exponents associated with these higher harmonics are then
deduced with the methods of Section \ref{secprop} \cite{Haldane}.

In the \lm\ with its linear dispersion, 
only the components with $m=0, \; \pm 1$ are present in
(\ref{densloc}) and those with $m = -1, \; 0$ in (\ref{fermloc}). 
This is related to the fact that in the \lm , the mean current
is [from the continuity equation (\ref{cont1}) for $q \rightarrow 0$]
$j = v_j J / L$, and is strictly conserved because $J$ is a good
quantum number \cite{Haldane}. With $\delta H$ [Eq.~(\ref{hnl})] 
containing the nonlinearity, the current operator as determined from
the continuity equation contains higher-order boson terms, and a
simple relation to the quantum number $J$ only obtains close to the
Fermi surface. In order to allow for a fermionic representation of
this complex current operator, the physical 
fermions must contain the higher harmonics in the chiral fermions.

\section{Backward and Umklapp scattering}
\label{scatt}
We now turn to the problem of non-Luttinger interactions. The Luttinger
model includes only the forward scattering interactions $g_2$ and $g_4$,
Eqs.~(\ref{h2}) and (\ref{h4}). This is certainly very restrictive since
any realistic model, say with an interaction 
\begin{equation}
H_{int} = \frac{1}{L} \sum_{k,k',q,s,s'} V(q) c^{\dag}_{k+q,s} 
c^{\dag}_{k'-q,s'} c_{k',s'} c_{k,s}  
\end{equation}
will also contain components of $V(q)$ with $q$ large, specifically
$q \approx 2k_F$. 
The restriction to forward scattering is, however,
absolutely essential in guaranteeing the exact solvability of the 
Luttinger model.

In any realistic theory, these offending interactions will be there.
The most important processes are depicted in Figure 3.3.
The contributions to the Hamiltonian are
\begin{eqnarray}
\label{hperp}
H_{1\perp} & = & g_{1\perp} \sum_s \int_0^L \: dx : \Psi_{+,s}^{\dag}(x)
\Psi_{-,s}(x) \Psi_{-,-s}^{\dag}(x) \Psi_{+,-s}(x) : \\
& = & \frac{2 g_{1\perp}}{(2 \pi \alpha)^2} \int \: dx
\cos [\sqrt{8} \Phi_{\sigma}(x) ] \;\;\;, \\
\label{h3perp}
H_{3\perp} & = & \frac{g_{3\perp}}{2} 
\sum_s \int_0^L \: dx : \Psi_{+,s}^{\dag}(x) \Psi_{+,-s}^{\dagger}(x)
\Psi_{-,-s}(x) \Psi_{-,s}(x) + {\rm H.c.} : \\
& = & \frac{2 g_{3\perp}}{(2 \pi \alpha)^2} \int \: dx
\cos [\sqrt{8} \Phi_{\rho}(x) ] \;\;\;.
\end{eqnarray}
The first term represents exchange scattering of 
two counterpropagating particles with opposite spin
across the Fermi surface (momentum transfer $\approx 2k_F$) and violates
spin-current conservation. [SU(2)-invariant backscattering would imply the
presence of a $g_{1\|}$-term in the Hamiltonian which, after bosonization,
can however be absorbed into $g_{2\nu} \rightarrow g_{2\nu} - g_{1\|} / 2$.]~
The second term is Umklapp scattering of two
particles moving in the same direction. 
The product of four fermion fields in Eq. (\ref{h3perp})
contains a factor $\exp(4ik_Fx)$ which, generally, oscillates rapidly and
suppresses contributions from this term. For half-filled bands, however,
\fkf\ equals a reciprocal lattice vector $\pm 2 \pi/a$, and Umklapp 
scattering becomes important. Charge-spin separation
is re\-spec\-ted here. This is not generic. However all purely electronic 
processes 
coupling charge and spin, i.e. arising from four fermion operators, 
are less relevant than (\ref{hperp}) and (\ref{h3perp}) \cite{ehm}. 

For the \LL\ phenomenology  to survive 
one must demonstrate that, although these interactions 
certainly renormalize velocities and stiffness
constant, they do not destroy the universal relations among them
nor those between $K$ and the correlation function exponents. 
To map the low-energy
physics onto a Luttinger model, 
one then has to (i) check that there are (how many?)
branches with gapless excitations; (ii) for each branch determine 
the relevant renormalized velocity and coupling constant; (iii) insert
these into the Luttinger liquid expressions for the quantities of interest.
If this works, it is proven that the originally offending interactions
are irrelevant at the \LL\ fixed point and that their only effect was
a quantitative renormalization of the \LL\ parameters.

This is the spirit of all approaches to a Luttinger liquid description
of interacting 1D electrons. It is most explicit in the \rg\
method if one accepts the limitation to weak coupling, and we shall 
treat $H_{1\perp}$ in this way. While more powerful mappings
of lattice models onto the \tlm\ are available now,
in conjunction with renormalization group, such ``direct''
extensions of the \tlm\ give a clear and simple idea of how the renormalized
effective parameters in the \LL\ are generated. Moreover, many
problems beyond the 1D electron-electron-interaction models, such as
coupling to phonons or scattering by impurities, only become tractable
with this approach. Also \rg\ allows to determine corrections to
the simple power-law decay (\ref{rcdw}) -- (\ref{alts})
of correlation functions of the \lm\ which are absolutely 
essential to obtain a correct picture of the physics of more complicated 
models. Finally, much of the early understanding of what is now 
called ``\LL '' was based on continuum models \cite{solyom,emrev}, by that 
time most often running under the label ``g-ology'', and \rg\ was the most
important tool for their understanding.

We derive \rg\ equations
for $H_{1\perp}$ following a method described by Chui and Lee \cite{chui}. 
There are other ways to formulate the \rg ; they have been reviewed in
detail elsewhere \cite{solyom,emrev,bourbonnais}. 
First diagonalize
the Luttinger part of $H_{\sigma}$ (shorthand for all terms containing
$\sigma$-operators). Then compute the partition
function $Z_{\sigma} = \langle \exp - \beta H_{\sigma} \rangle$ 
in the Matsubara formalism of imaginary times $\tau = it$.
$Z_{\sigma}$ can be expanded
in $H_{1\perp}$ and the expectation value be evaluated with
respect to the diagonal part
\begin{equation}
Z_{\sigma} = \sum_n \frac{1}{(n !)^2} 
\left( \frac{g_{1\perp}}{(2 \pi )^2} \right)^{2n}
\int \left( \prod_i^{2n} \frac{d^2 r}{\alpha^2} 
\right) \exp \left[ 2 K_{\sigma}
\sum_{i>j} q_i q_j \ln \left( 
\frac{\mid {\bf r}_i - {\bf r}_j \mid^2}{\alpha^2} \right) \right] \;\;\;.
\end{equation}
The 2D vector ${\bf r} = (x , v_{\sigma} \tau)$ and $q_i = 1$ for $i=1 
\ldots n$ and $-1$ else. 
$Z_{\sigma}$ is now identified as the partition function of a classical
2D Coulomb gas with charges $q_i$, at a fictitious temperature $\beta_{CG}
= 4 K_{\sigma}$ and a fugacity $g_{1\perp} / (2\pi)^2$. For this problem
Kosterlitz and Thouless \cite{kothou}
derived a set of \rg\ equations which translate into
\begin{equation}
\label{kot}
\frac{ d K_{\sigma}}{d \ell} = - \frac{1}{2} K_{\sigma}^2 
\left( \frac{g_{1\perp}}{\pi v_{\sigma}} \right)^2 \;\;\; , \;\;\;
\frac{ d g_{1\perp}}{d \ell} = g_{1\perp} ( 2 - 2 K_{\sigma}) \;\;\; .
\end{equation}
They describe the flow of the effective coupling constants
$g_{1\perp}$ and $K_{\sigma}$, shown in Figure 4.1,
when short-distance degrees of freedom
(between $\alpha$ and $\alpha e^{\ell}$) are integrated out. 
Here $\alpha$ is reinterpreted as a short-distance cutoff parameter
which may be of the order of a lattice constant. The
coupling constants must be rescaled so as to maintain
the Fermi surface physics and the asymptotic correlations invariant.
There are two different types of flow. (i) Assume 
$K_{\sigma}$ sufficiently large so that $|g_{1\perp}|$ decreases with 
increasing $\ell$ (lower right part of Fig.~4.1). 
If this remains so even for $\ell \rightarrow \infty$,
the \rg\ trajectory will flow into a fixed point $g_{1\perp}^{\star} = 0$
and $K_{\sigma} \rightarrow K_{\sigma}^{\star}$. $g_{1\perp}$ has dropped
out of the problem, i.e. at long distances the model behaves effectively
as a Luttinger model with a renormalized $ K_{\sigma}^{\star}$. 
The fixed point is spin-rotation invariant if it turns out that
$K_{\sigma}^{\star}=1$. Then the flow is precisely along the separatrix.
This is one example of a \LL .
[Even then, during intermediate stages of the
calculation, one may have $K_{\sigma}(\ell) \neq 1$;
this apparent breaking and final restoration of $SU(2)$-invariance is 
typical of abelian bosonization.]~(ii) 
If the bare $K_{\sigma}$ is not large enough
compared to $\mid g_{1\perp} \mid $, $K_{\sigma}$ will flow towards $0$ but
more importantly $\mid g_{1\perp} \mid$ will increase.
Derived from a perturbation expansion, the \rg\ manifestly 
looses its sense. It is clear that the system flows away from
the \LL\ fixed line, and the diverging $\mid g_{1\perp} \mid$ signals
an instability of the model towards a different ground state 
whose accurate description must, however,
be based on different methods. This regime will be the subject of 
Section \ref{spingaps}.
 
So long as the system is not half-filled, the charge exponent $K_{\rho}$
is not renormalized. 
At half-filling, the situation in the charge degrees of freedom is
isomorphic to the spin part discussed here. It is sufficient to 
change $g_{1\perp} \rightarrow g_{3\perp}, \; K_{\sigma} \rightarrow
K_{\rho}$ and carry over the Equations (\ref{kot}). Also more complicated
models where charge-spin coupling is important can be treated in this
way \cite{ehm}. The application to phonons and impurities will be
discussed below.

The essential weakness of the \rg\ approach is its limitation to weak 
coupling, being derived from
perturbation expansions. This limitation has been overcome
by several methods which will be discussed in the subsequent sections.

Before, however, we discuss in more details the correlation functions
of such a \LL\ where all non-Luttinger interactions have become irrelevant. 
A first idea about the correlations
is obtained by inserting the fixed point value $K_{\nu}^{\star}$ into
the correlation functions of Section \ref{secprop}. This is the standard
procedure in the \rg\ treatment of critical points \cite{phastr}.
In particular, we
would then find a degeneracy of exponents between SDW and CDW, and 
SS and TS, no matter what the precise fixed point values \knus .
Anticipating that the non-half-filled repulsive Hubbard model can
be described, at least for small $U$, by (\ref{hlutt})+(\ref{hperp}), it is 
clear that this cannot be the whole story. 

Let us consider the \tkf -$SDW_z$ correlation function for definiteness.
The $SDW_z$-operator is
\begin{equation}
\label{osdwz}
O^{\dag}_{SDW_z}(x) = \sum_{s} s \Psi^{\dag}_{-,s}(x) \Psi_{+,s}(x) =
\frac{-i}{ \pi \alpha} \exp \left[ 2ik_F x - \sqrt{2} i \Phi_{\rho}(x) \right]
\sin \left[ \sqrt{2} i \Phi_{\sigma}(x) \right] \;\;\;.
\end{equation}
Now consider the time-ordered
correlation function (again in imaginary-time formalism)
\begin{eqnarray}
- R_{SDW_z}( \br ) & = & \left\langle T_{\tau} O_{SDW_z}( \br ) 
O^{\dag}_{SDW_z} ({\bf 0 } ) \right\rangle \nonumber \\
& = &
\frac{1}{Z_{\sigma}} {\rm Tr} \left( T_{\tau} O_{SDW_z}( \br ) 
O^{\dag}_{SDW_z}
( {\bf 0 } ) \exp \left[ \int_0^{\beta \rightarrow \infty} d \tau H(\tau)
\right] \right) \;\;\; ,
\end{eqnarray}
where $H = H_{\rm Lutt} + H_{1\perp}$ and the trace (Tr) is performed over
$\sigma$ and $\rho$. The charge part is trivial and gives the Luttinger
result $| \br |^{- K_{\rho}}$. In the spin part, use Wick's theorem to
expand the exponential in $H_{1\perp}$. This will generate nonvanishing
contributions at all even orders which are essentially those contained
in the partition function $Z_{\sigma}$ multiplied by $OO^{\dag}$. 
In addition to these terms there will, however, be important new terms in odd 
orders of $H_{1\perp}$ not present in the partition sum \cite{pref,giam}. They
arise from contracting the $\sigma$-part of the $SDW_z$-operators
with $H_{1\perp}$
\begin{eqnarray}
\lefteqn{
\left\langle \sin \left[ \sqrt{2 K_{\sigma}} \Phi_{\sigma} ( \br ) \right]
\sin \left[ \sqrt{2 K_{\sigma}} \Phi_{\sigma} ( {\bf 0} ) \right]
\cos \left[ \sqrt{8 K_{\sigma}} \Phi_{\sigma} ( \br_1 ) \right] \right\rangle
} \nonumber \\
& &
= - \frac{1}{8} 
\left\langle \exp \sqrt{2 K_{\sigma}} \left[ \Phi_{\sigma}( \br )
+ \Phi_{\sigma} ( {\bf 0} ) - 2 \Phi_{\sigma}( \br_1) \right] + 
{\rm H.c.} \right\rangle \;\;\;.
\end{eqnarray}
These expectation values do not vanish because the prefactor of the 
$\Phi_{\sigma}$-field in the correlation function is half of that in
the perturbation operator or, in other words, because the 
$\Psi_{-,s}^{\dag}\Psi_{+,s}$-components of the $SDW_z$-operator also occur
as factors in $H_{1\perp}$. In the Coulomb gas language, this is equivalent
to saying that one considers the screening of two test charges $q/2$ by
charges $-q$. The terms up to second order 
can be reexponentiated in the spirit of a cumulant expansion. 
Now it is important to integrate up the correlation function
along the whole \rg\ trajectory
\cite{pref,giam}. 
The spin-part of the correlation function then becomes
\begin{equation}
R^{(\sigma)}_{SDW_z}( \br ;\alpha ) = 
\exp \left( - K_{\sigma} \ln \frac{| \br | }{\alpha} 
+ \int_0^{\ell} \! \frac{g_{1\perp}(\ell')}{\pi v_{\sigma}} d \ell'
+ \frac{1}{2} \int_0^{\ell} \left[ \frac{g_{1\perp}(\ell')}{\pi v_{\sigma}} 
\right]^2 \ln \frac{| \br |}{\alpha} d \ell' \right) \;\;\;.
\end{equation}
If scaling goes to weak coupling, the integrals can 
be extended to infinity and the usual expressions involving the fixed-point
exponent $K_{\sigma}^{\star}$ follow. Notice, however, that an ultimate
cutoff is provided by the observation scale $| \br |$ (if not by temperature
or system size) so that the integration cannot go beyond
$\ell^{\star} = \ln | \br | / \alpha$. The correlation function 
then decays as
\begin{equation}
\label{sdwlog}
R_{SDW_z} (\br ) = \left( \frac{| \br |}{\alpha} \right)^{-K_{\rho} -
K_{\sigma}^{\star}} \sqrt{\ln \frac{| \br | }{\alpha}} \;\;\;,
\hspace{1cm} K_{\sigma}^{\star} = 1 \;\;\;,
\end{equation}
where we have reintroduced the contribution from the charge density 
fluctuations. In doing the integrals, we have used explicitly the fact that
we scale along critical line (the separatrix in Fig.~4.1) so that the
logarithmic corrections only obtain in the spin-rotation invariant case.
In this case, one recovers expressions identical to (\ref{sdwlog}) for the
$x$- and $y$-components of the $SDW$ correlation function although, involving
$\Theta_{\sigma}$-fields, the intermediate expressions are quite different.
The charge density wave and superconducting correlations decay as
\begin{equation}
\label{corrlog}
R_{CDW} (\br ) \sim | \br |^{-K_{\rho} - 1} \ln^{-3/2} | \br | \;, \;\;\;
R_{SS} (\br ) \sim | \br |^{-K_{\rho}^{-1} - 1} \ln^{-3/2} | \br | \;, \;\;\;
R_{TS} (\br ) \sim | \br |^{-K_{\rho}^{-1} - 1} \ln^{1/2} | \br | \;.
\end{equation}
There is no logarithmic correction to the \fkf -CDW function because it
does not involve spin fluctuations. 
It is remarkable that at this level, the degeneracy of the CDW and SDW
correlation functions and between SS and TS is lifted: they have the same
exponents, correctly given by the \lm\ but the correlations are logarithmically
stronger for SDW and TS. 
For repulsive interactions, magnetic correlations must dominate! If we
have attractive backscattering $g_{1\perp} \rightarrow -g_{1\perp}$ with
$K_{\sigma}$ left unchanged, CDW and SS will be logarithmically enhanced
over SDW and TS [just exchange the log-exponents in (\ref{sdwlog})
and (\ref{corrlog})]. 
Finally, if spin-rotation invariance is broken and there is an easy plane
anisotropy, $g_{1\perp}$ scales to zero faster. In this case, the integration
along the trajectory only gives prefactor corrections to the power-law
correlations \cite{giam}. These results can be transposed straightforwardly
to commensurate systems when Umklapp scattering is irrelevant \cite{ehm}.

The phase diagram in the $g_{1\perp} - K_{\rho}$-plane
obtained in the absence of Umklapp scattering is 
displayed in Figure 4.2. At $g_{1\perp} > 0$, the dominant divergences
are SDW for $K_{\rho} < 1$ and TS for $K_{\rho} > 1$. Subdominant
fluctuations are indicated in parenthesis, and the preceding discussion
shows that CDW and SS have the same exponents as SDW and TS but are
disfavoured by their logarithmic corrections. We have assumed
the system to be spin-rotation invariant, and consequently, the fixed-point
$K_{\sigma}^{\star} = 1$. For $g_{1\perp} < 0$, a spin gap opens through
a Kosterlitz-Thouless transition, and formally $K_{\sigma}^{\star} = 0$.
Here, CDW and SS have the strongest divergences for $K_{\rho} < 1$ and
$K_{\rho} > 1$, respectively. They also diverge in the regimes $1 \leq 
K_{\rho} \leq 2$ and $1/2 \leq K_{\rho} \leq 1$, respectively, though
with a weaker power than the dominant fluctuations. 

Logarithmic corrections to the free energy of statistical models whose
fermionic description contains a marginally irrelevant Umklapp operator
and which are related to the singularities found here in the correlation
functions, had been discovered earlier by Black and Emery \cite{black}.

\section{Lattice models: Hubbard \& Co.}
\label{sec1d}

A variety of nontrivial lattice models can be solved exactly in 1D, 
for which no exact solution exists in higher dimension. 
A non-exhaustive list contains
the Heisenberg model \cite{heis}, the Hubbard model \cite{hubbard,liebwu}
and various long-range, supersymmetric or degenerate
extensions, the supersymmetric $t-J$-model \cite{babla}-\cite{schlotj}, 
and others. Solvable continuum models include, apart 
the Tomonaga-Luttinger model discussed above and the Luther-Emery model
reviewed in Section 5.1, 
the massive and massless Thirring model \cite{bergtha} and the
interacting Bose gas \cite{bosgas}.
Exact solutions are due to a large extent to very strong conservation
laws arising from the restricted phase space for 1D fermions.

We briefly discuss some important lattice models. A central role is played
by the Hubbard model, and our treatment of \LL\ correlations in lattice models
will be centered on this model. We therefore also 
present a short summary of important
Bethe-Ansatz results for this model to make this section more self-contained.

\subsection{Models}
\label{modls}

The Hubbard model \cite{hubbard} is described by the Hamiltonian
\begin{equation}
\label{hubbard}
H_{\rm Hub} = - t \sum_{<i,j>s} c^{\dag}_{i,s} c_{j,s} + \frac{U}{2} 
\sum_{i,s}
(n_{i,s} - 1/2)(n_{i,-s} - 1/2) - \mu \sum_{i,s} n_{i,s} \;\;\;,
\end{equation}
where $c_{i,s}$ describes fermions with spin $s$
in Wannier orbitals at site $i$, $n_{i,s} = c^{\dag}_{i,s} c_{i,s}$,
$U$ is the repulsion of two electrons on the same site and $\mu$ the
chemical potential. One can also fix the band filling to $n = N_{\rm electrons}
/N_{\rm sites}$. 
$<i,j>$ restricts the sum to nearest neighbours. 
This model is the simplest 
approximation for strongly correlated
electrons in a crystal lattice. 
The model is exactly solvable, cf.~Section \ref{bethans}.
For a long time, it was believed that
the \hm\ describes the strong-coupling limit of the 1D Fermi liquid
while the \tlm\ rather would represent the weak-coupling case. To show
that this is \em not \rm the case, and that both are closely related,
is a major purpose of Section \ref{sec1d}.

Various more realistic
extensions can be considered. 
In some cases, it is necessary to add longer-range interactions between
the electrons. The extended Hubbard model \cite{ehm,Hubtcnq}
\begin{equation}
\label{hehm}
H_{\rm EHM} = H_{\rm Hub} + V \sum_i n_i n_{i+1} 
\end{equation}
includes interactions between neighbouring sites, but one may obviously
go to longer interaction range [such as $1/r$ \cite{schuwi} or a Yukawa form
$\exp(-r)/r$]. In contrast to the Hubbard model, this
Hamiltonian is no longer exactly solvable. Also ``off-diagonal'' terms,
i.e. interactions coupling charge densities on site to those on bonds,
can be added \cite{kssh}-\cite{hirsch}
\begin{equation}
\label{hx}
H = H_{\rm Hub} + X \sum_{i,s} \left( c^{\dag}_{i=1,s} c_{i,s} + {\rm H.c.} 
\right) \left(n_{i,-s} + n_{i+1,-s} \right) \;\;\;.
\end{equation}
An important feature here is the breaking of charge-conjugation symmetry
generated by $X$. This term goes beyond the zero-differential-overlap 
approximation. A critical discussion of the approximations 
involved in going from a realistic correlation problem to the Hubbard
model in a 1D context has been given by Painelli and Girlando \cite{pai}
and Campbell et al.~\cite{cmp}.

At $U>0$ and half-filled band, the Hubbard model has an insulating ground
state whose spin fluctuations are described by an effective Heisenberg
model \cite{heis} with an (antiferromagnetic) exchange integral $J = 4t^2/U$. 
For a nearly half-filled Hubbard model, 
it is more convenient to think in terms of a few
holes doped into such an antiferromagnetic Heisenberg system. For large $U$,
double occupancy of lattice sites is dynamically forbidden, and
the energy scales for charge fluctuations ($\sim t$) and for spin
fluctuations ($ \sim J \ll t$) are well separated. One can then simplify
the problem by projecting out the states in the Hilbert space involving
double occupancies. In a restricted Hilbert space containing only 
singly-occupied and empty sites, one finds in second order
in $t/U$ the following Hamiltonian
($t-J$-model, \cite{babla}-\cite{schlotj})
\begin{equation}
\label{hamtj}
H_{t-J} = - t \sum_i \left[ (1-n_{i,-s}) c_{i,s}^{\dagger}
c_{i+1,s} (1-n_{i+1,-s}) + {\rm H.c.} \right] + J \sum_i \left[ {\bf S}_i
\cdot {\bf S}_{i+1} - \frac{1}{4} n_i n_{i+1} \right] \;\;\;.
\end{equation}
The fermions $c_{i,s}$ now behave as spinless fermions, and ${\bf S}_i
= \sum_{s,s'} c_{i,s}^{\dagger} ({\bf \sigma})_{s,s'} c_{i,s'}$ are spin
operators. This model can be solved in two limits. For $J = 0$, it reduces
to the $U = \infty$-Hubbard model which describes free spinless fermions, and
for $J/t=2$, it possesses an additional supersymmetry and can be solved
by Bethe Ansatz \cite{babla}-\cite{schlotj}. 
The $t-J$-model approximates the strong-coupling limit
of the Hubbard model only for $J \ll t$. Models with other interactions 
or more bands can, however, be approximated in a low-energy subspace
by a $t-J$-model with sizable $J$ \cite{pafo}.
Both the $t-J$ and the \hm\ 
can be extended to include additional degeneracies \cite{schlo}.
Another interesting extension consists in introducing longer-range hopping
\cite{gebhard} or spin exchange. We shall not say much on these variants here.

Most of the methods discussed below for extracting the \LL\ parameters from
one of these models will work, with minor modifications, also for the others
with similar structure. When a model is not solvable by Bethe-Ansatz,
numerical diagonalization can provide similar information.
We therefore limit our discussion as much as
possible to the quite generic case of the Hubbard model and only briefly
discuss changes occurring when passing to other systems. In the following
section, we list some important elements of the Bethe-Ansatz which are
helpful for understanding the mapping onto the \LL .

\subsection{Bethe Ansatz}
\label{bethans}
The 1D Hubbard model has been solved exactly via Bethe Ansatz
by Lieb and Wu \cite{liebwu} (for pedagogical reviews on the Bethe Ansatz,
see Sutherland \cite{suth}, Korepin et al.~\cite{korerev},
Izyumov and Skryabin \cite{izsk}
or Nozi\`{e}res \cite{nobe}),
and the ground state energy and some thermodynamic quantities
can be obtained \cite{shiba}--\cite{ultz}. Also  the excitation spectrum
of some collective modes has been computed quite early \cite{ovchi,coll}.
The basic physical picture emerging from these initial studies is as 
follows. For $U>0$, the system is metallic whenever the band is not
half-filled ($N_{electrons} \neq N_{sites}= L/a$ with lattice constant 
$a$): the chemical potential
for adding a particle to $N$- and $N-1$-particle systems are equal.
Exactly at half-filling, one finds a difference ($\sim \sqrt{U} \exp(-1/U)$
for $U \rightarrow 0$ and $\sim U$ for $U \rightarrow \infty$ 
\cite{liebwu,ovchi}) between these
two quantities indicating that the system has turned into a Mott
insulator for any $U>0$. Finite $U>0$ obviously prohibits double 
occupancy of sites, all sites
are now (singly) occupied and no low-energy charge excitations possible.
The lower Hubbard band is completely filled, and the upper Hubbard band
is empty in the ground state.
The spins are coupled through an effective antiferromagnetic exchange integral
$J = 4 t^2/U$, and their dynamics reduces to a Heisenberg model. 

At half-filling, the $U<0$-sector is related to the
$U>0$ one by a particle-hole transformation 
$c_{i,s} \rightarrow (-1)^i c^{\dag}_{i,s}$ for a \em single \rm spin direction
only, say $s = \uparrow$,
exchanging the role of charge and spin degrees of freedom, and the charge
gap discussed above turns into a spin gap: occupying 
sites with two electrons with antiparallel spins
is favoured. These pairs are mobile and the charge excitations 
massless. There are no
singular features in the Bethe Ansatz for $U<0$  as a function of band-filling
implying that the picture applies to the whole $U<0$-sector
\cite{ultz}.

These results can be obtained qualitatively, and for $U/t \ll 1$ also 
quantitatively, with the \rg\ methods described in the preceding section.
The coupling constants are $g_{i\perp} = Ua, \; i=1, \ldots, 4$ ($g_{3\perp}$
only occurs for half-filled bands), and the Fermi velocity is $v_F = 
2 t a \sin (k_Fa)$. 

Bethe's Ansatz provides a solution for all interaction
strengths and band-fillings \cite{liebwu}. We sketch the principal ideas,
following Nozi\`{e}res \cite{nobe}. The Bethe Ansatz relies on the following
facts. (i) Due to energy and momentum conservation, in 1D
a two-particle collision classically and quantum-mechanically conserves both
momenta individually. The particles then only can be exchanged or phase-shifted,
and the two-particle wave-function asymptotically ($| x_1 - x_2 | \rightarrow
\infty$) obeys
\begin{equation}
\Psi(x_1,x_2) = a e^{i(k_1 x_1 + k_2 x_2)} + b e^{i( k_1 x_2 + k_2 x_1)} \;\;\;.
\end{equation}
The Bethe Ansatz postulates this behaviour for all distances between the
particles.
(ii) A three-particle collision does not conserve individual momenta
\em except if \rm the scattering matrix factorizes. This factorization
implies another conservation law. For $N$ particles, one then
has $N$ conservation laws, expressed by $\{ k'_i \} = \{ k_i \}$. 
(iii) The Hilbert space of the Hamiltonian separates in $N!$ quadrants
each characterized by a permutation $P$ of the $N$ particles, ordered in
one quadrant as $ 1 \leq x_{1} \leq x_{2} \leq \ldots x_{N} \leq L $. 
The $N$-particle wave-function there becomes
\begin{equation}
\Psi ( x_1 , \ldots , x_N ) = \sum_P A[P] e^{i k_{P_i} x_i} \;\;\;.
\end{equation}
Fermi or Bose statistics determines its continuation into the other sectors.
(iv) The amplitude $A[P]$ is determined by the conditions of continuity
of $\Psi$ as $x_i \rightarrow x_{i +1}$ and periodic boundary conditions
$\Psi (x_1, \ldots, x_N ) = \Psi (x_2, \ldots, x_N, x_1 + L)$. The problem
is the computation of $A[P]$. (v) Introducing spin, suppose we have
$N$ electrons, $M$ of which have spin $\downarrow$, on a 
lattice with $L$ sites $x_i$. One must then ensure that the factorization
of the $S$-matrix is not perturbed by the spin indices (Yang-Baxter
conditions). There is then a second permutation $Q$ for the spin labels, and 
the wave function where the $M$ down-spins
occupy the sites $x_1 \ldots x_M$ and the $N-M$ up-spins the sites
$x_{M+1} \ldots x_N$ is denoted by $\Psi(x_1, \ldots , x_M, x_{M+1}, \ldots,
x_N)$. The Bethe Ansatz
postulates that in each quadrant characterized by $Q$, 
i.e. $ 1 \leq x_{Q_1} \leq x_{Q_2} \leq \ldots x_{Q_N} \leq L $, 
the wave function is given by \cite{liebwu}
\begin{equation}
\label{betwv}
\Psi(x_1, \ldots , x_M, x_{M+1}, \ldots, x_N) = \sum_{P} A[Q,P] \exp 
\left( i \sum_{j=1}^N k_{Pj} x_{Qj} \right) \;\;\;.
\end{equation}
The $N$ numbers $k_i$ are determined from the coupled Lieb-Wu equations
($ u = U / 4 t$)
\begin{eqnarray}
\label{liwu1}
2 \pi I_j & = & L k_j - 2 \sum_{\beta = 1}^{M} \arctan \left(
\frac{\sin k_j - \Lambda_{\beta}}{u} \right) \;\;\;, \\
\label{liwu2}
2 \pi J_{\alpha} & = & 2 \sum_{j=1}^{N} \arctan \left( 
\frac{\Lambda_{\alpha} - \sin k_j }{u} \right)
- 2 \sum_{\beta =1}^M \arctan \left( 
\frac{\Lambda_{\alpha} - \Lambda_{\beta}}{2 u} \right) \;\;\;, \\
\label{liwubound}
I_j & = & \left\{ \begin{array}{l} \rm integer \\
				   \rm half-odd-integer  \end{array} \right.
\; \rm if \; M = \left\{ \begin{array}{l} \rm even \\
				   \rm odd  \end{array} \right. \;\;\;, \\
J_{\alpha} & = & \left\{ \begin{array}{l} \rm integer \\
				   \rm half-odd-integer  \end{array} \right.
\; \rm if \; N-M = \left\{ \begin{array}{l} \rm odd \\
				   \rm even  \end{array} \right. 
\;\;\;. \nonumber
\end{eqnarray}
The total energy and momentum of the system are then
\begin{equation}
\label{bethen}
E = - 2 t \sum_{i=1}^N \cos k_i\;\;\;, \hspace{0.5cm} P = \sum_{i=1}^N k_i
\;\;\;.
\end{equation}

Eqs. (\ref{betwv}) -- (\ref{bethen}) give the exact energy and wavefunction
of the 1D Hubbard model. The quantum numbers $k_i$ are the momenta 
of the particles characterizing the orbital
degrees of freedom. Unlike for free particles, they are not equally spaced
but shifted by the presence of the other particles.
The $\Lambda_{\alpha}$ are called rapidities and describe the 
spin state. On the other hand, the integers or half-odd-integers $I_i$ 
and $J_{\alpha}$ are equally spaced. The ground state is obtained by
occupying the levels with minimal $|I_i|$ and $|J_{\alpha}|$. Therefore
the distribution of 
$q_i = 2 \pi I_i / L$ and $p_{\alpha} = 2 \pi J_{\alpha} / L$ is given 
by a Fermi distribution $\Theta(k_{F \uparrow} + k_{F\downarrow} - q_i)$
and $\Theta(k_{F\downarrow} - p_{\alpha})$, respectively. In the absence
of a magnetic field, the ground state has $k_{F\uparrow} + k_{F\downarrow}
= 2k_F$ and $k_{F\downarrow} = k_F$, so that the $q_i$ have a doubled
Fermi wavevector while the $p_{\alpha}$ have the normal $k_F$. 

This splitting of the Fermi surface into two can be clarified further by
studying the elementary excitations. Two of them are obtained by making a hole
either in the $I_i$- or in the $J_{\alpha}$-distribution. In the first
case, one obtains a charged, spinless holon, in the second case a 
neutral spin-$1/2$ spinon. Both holon and spinon live in the lower 
Hubbard band. There are other solitonic excitations involving
doubly occupied sites which therefore build up the upper Hubbard band
\cite{halhub}. In general, holons and spinons are not independent, and
the Lieb-Wu equations (\ref{liwu1}), (\ref{liwu2})
imply that they interact. Introducing a real hole will affect both channels. 
Moreover, the representation of physical electrons and holes in terms of
holons and spinons is not known to date.

The interaction of holons and spinons complicates the calculation of
their dispersion. Simple results are obtained only for weak or strong 
$U$. Then, the holons obey
\begin{equation}
\label{eholon}
\varepsilon^{(h)}(q) = \left\{ \begin{array}{lll} 
4 t \cos (qa/2) - 2 t \cos(k_Fa) & {\rm for} & u \ll 1 \\
2 t \cos (qa) & {\rm for} & u \gg 1 
\end{array} \right. \;\;\; .
\end{equation}
Their Fermi surface is at $k_F^{(h)} = 2k_F, \; \varepsilon_F^{(h)} = - \mu$.
The spinon have dispersion
\begin{equation}
\label{espinon}
\varepsilon^{(s)}(q) = \left\{ \begin{array}{lll} 
2 t [ \cos (qa) - cos (k_Fa) ] & {\rm for} & u \ll 1 \\
(\pi/2) J_{eff} \cos (qa/n) & {\rm for} & u \gg 1 
\end{array} \right. \;\;\; .
\end{equation}
The effective exchange integral is
\begin{equation}
J_{eff} = \frac{4 t^2}{U} \left( n - \frac{\sin(2 \pi n)}{2 \pi}
\right) \;\;\;.
\end{equation}
The dispersion is only defined for $q \leq k_F^{(s)} = k_F$, and the 
energy becomes zero at $k_F$. The first feature translates the reduced
Brillouin zone of the compressed Heisenberg chain, and the second one
implies that spinon-antispinon pairs can be created spontaneously.

The wavefunction
(\ref{betwv}) is not of much practical value due to its enormous complexity:
in fact, there are about $N!^2$ expansion coefficients $A[Q,P]$! In the 
calculation of the wavefunction, there is no gain with respect to brute-force 
exact diagonalization. A calculation of correlation functions, and especially
of their asymptotic behaviour, based on the
Bethe Ansatz is therefore elusive.

Important
simplifications occur in the limit $U \rightarrow \infty$ \cite{ogata}.
The members on the left-hand sides of Eqs. (\ref{liwu1}) and (\ref{liwu2})
are of order $O(U^0)$. 
For the equalities to hold, the $\Lambda$ on the right-hand
sides must be proportional to $u$: $\Lambda_{\alpha} = 2 u \lambda_{\alpha}$
making the $\sin k_j$-terms negligible. This simplifies the Lieb-Wu equations
to
\begin{eqnarray}
\label{liwus1}
2 \pi I_j & = & L k_j + 2 \sum_{\beta=1}^M \arctan \left( 2 \lambda_{\beta}
\right) \;\;\;, \\
\label{liwus2}
2 \pi J_{\alpha} & = & 2 N \arctan \left( 2 \lambda_{\alpha} \right)
- 2 \sum_{\beta =1}^M \arctan \left( \lambda_{\alpha} - \lambda_{\beta} \right)
\;\;\;. 
\end{eqnarray}
The equations for $k_j$ and $\lambda_{\alpha}$ now decouple and can be
solved successively. Concomitant with this decoupling is a decoupling
of the wave function (for the quadrant $Q$) 
\begin{equation}
\label{wvsim}
\Psi(x_1, \ldots, x_M, x_{M+1}, \ldots, x_N) = 
(-1)^Q \det \left[ e^{i k_i x_{Qj}}
\right] \Phi( y_1, \ldots, y_M) 
\end{equation}
into a charge and a spin part. det[...] is a Slater determinant involving
only the particle positions irrespective of their spin, i.e. describing
free spinless fermions. $\Phi(y_1, \ldots, y_M)$ is the Bethe Ansatz
wave function
of a Heisenberg chain \cite{heis} of the $N$ spins, characterized through
the positions of the $M$ down-spins, on a \em compressed \rm lattice of
just $N$ sites. This decoupling of the wave function means a complete
charge-spin separation over all energy scales in the $U \rightarrow 
\infty$-Hubbard model and is correct to $O(1/u)$.

The wave function (\ref{wvsim}) can be evaluated numerically for
much bigger systems than (\ref{betwv}) and, combined with 
either finite-size scaling or further analytical
work, allows to discuss the asymptotic low-energy properties and the
critical exponents of correlation 
functions. Applications will be discussed in the following section.
We also note that there is a systematic large-$U$ expansion for 
the distribution functions of the momenta $k$ and rapidities $\Lambda$
\cite{caba}.

\subsection{Low-energy properties of one-dimensional lattice models}
\label{seclat}

In the first part of this section, we shall discuss various successful methods
to derive the correlation exponents of interacting electrons in 1D lattices,
taking the Hubbard model as an example. 
At the end, we briefly summarize the physical picture and then outline
the changes occurring when going to the variants introduced in Section
\ref{modls}.

Early studies of correlation functions heavily relied on numerical 
simulation of Hubbard and
extended Hubbard models. Hirsch and Scalapino used quantum Monte Carlo
techniques to directly study the density and spin density correlation functions
at various band-fillings and interaction strengths 
on lattices of 20 -- 40 sites at temperatures down to about $t/15$
\cite{hirsca}. 
Of course,
both the finite temperatures and the accuracy of the simulations did not
allow a determination of the correlation exponents and thus of $K_{\rho}$,
but one main point of concern was the doubling of the wave vector of
divergence in the charge density response from \tkf\ to \fkf\ as interactions
are increased and/or $V$ is turned on. This could be rephrased in terms of
the present language as under what conditions $K_{\rho} + 1 < (>) \;
4 K_{\rho} $ i.e. $K_{\rho} < (>) \; 1/3$ which marks the value where 
\tkf - and \fkf 
-responses are equally divergent. Quite generally, it was found that 
increasing $U$ \em decreases \rm the \tkf -CDW-correlations but somewhat
increases the \fkf -CDW as $U \rightarrow \infty$. The decrease at \tkf\
was less, however, if charge density fluctuations were measured on the
bonds between lattice sites (bond order wave, BOW)
rather than on the lattice sites themselves.
The \tkf -SDW correlations were enhanced by $U$.
This is quite easy to understand physically: $U$ generates antiferromagnetic
spin exchange but suppresses on-site charge fluctuations while intersite
fluctuations remain unaffected, at least at lowest order in $U$. 
On the other hand, for $U \rightarrow \infty$, the electrons behave as spinless
fermions with a Peierls divergence vector of $2 k_F^{(s=0)} = 4 k_F^{(s=1/2)}$.
Adding a nearest-neighbour repulsion $V$ strongly favours the \fkf -CDW,
especially on-site and in quarter-filled bands, also enhances the 
SDW and further suppresses the \tkf -CDW: the energies due to
both $U$ and $V$ are
minimized when the particles occupy every second site, i.e. forming a 
\fkf -CDW. 

The suppression of the \tkf -CDW by $U$ and $V$
is interesting in view of the general
expression for the density correlations in a Luttinger liquid, Eqs. 
(\ref{rcdw}) and (\ref{rq})
which imply that both exponents of divergence at \tkf\ and \fkf\ \em increase
\rm with decreasing $K_{\rho}$ and thus with increasing 
$U$ and $V$. The suppression of the \tkf -CDW then must be due to the 
influence of $U$ on its prefactor which must decrease
as $U$ increases. Hirsch and Scalapino demonstrate this by showing that,
at low enough temperature and various $U$, 
both SDW and CDW diverge with the same exponent but that the scale where
the asymptotic behaviour is observed, is vastly different and, in fact, very
low for the CDWs \cite{hirsca}. A related suppression of \tkf -CDW
correlations due to the prefactor (with a concomitant enhancement of SDW
and BOW) can also demonstrated for a half-filled band 
in \rg\ \cite{pref}. 

More extended
results on the influence of band-filling and interaction range on the
competition of \tkf - and \fkf -CDWs have been produced by Mazumdar, Dixit,
and Bloch \cite{maz}. They also propose a qualitative but systematic picture
predicting the appearance of \tkf - or \fkf -CDWs, in terms of the contribution
to the ground state wave function of certain extreme symmetry broken
configurations and the barriers to resonance between them. Specifically,
for the quarter-filled band, finite $V$ is necessary to promote a \fkf -CDW
but an eventual second-neighbour repulsion $V_2$ must be small: $V_2 < V/2$.
For $1/2 < n < 2/3$, however, a new kind of defect-CDW 
with periodicity $\pi /a$ is found possible and competes with the \fkf\ one,
depending on the precise values of the interaction constants. Long-range
interactions are necessary to stabilize a \fkf -CDW for $n > 2/3$,
and the generic CDW will be at \tkf . 
The competition between \tkf - and
\fkf -CDWs will reappear below in terms of the Luttinger parameter $K_{\rho}$
being smaller or larger than $1/3$. For the special case of $n=1/2$, 
electron-phonon coupling has also been included recently \cite{ung}.
Also, a more systematic theory for
a \LL\ floating on top of a commensurate CDW, e.g. with $q_{CDW} = \pi/a$,
will be given at the end of this section \cite{gomez}.

Subsequent work rather turned attention to single-particle properties
and to the question of (non)-Fermi-liquid behaviour in the 1D Hubbard model.
Several quantum Monte Carlo studies indicated a finite jump in the
momentum distribution function $n(k)$ at $k_F$, to be compared with the
Luttinger prediction (\ref{nkut}) \cite{sortor}-\cite{imada}. 
Notice, however, that Equation (\ref{nkut}) applies to an infinite system.
Finite size effects give $n(k)$ a finite jump at $k_F$ 
whose scaling is governed by the exponent $\alpha$ \cite{brech,bourmc}
\begin{equation}
\Delta n(k_F) = n(k_F - \pi / L) - n(k_F + \pi /L) \sim L^{-\alpha} \;\;\;,
\end{equation}
subsequently identified in improved simulations \cite{soreur}. The absence
of any significant rounding expected from the power-law behaviour in $n(k)$,
up to about 200 lattice sites 
indicates, however, that the asymptotic Luttinger regime in the 1D Hubbard
model is confined to a tiny momentum slice around the Fermi surface,
whose smallness, in fact, remains surprising. (With reference to the 
different scales in different quantities, identified 
by Hirsch and Scalapino \cite{hirsca} 
and discussed above, this does not necessarily
imply that Luttinger liquid correlations in all other quantities are confined
to such small momentum/energy scales.)

Sorella et al. also studied the divergences of the density and spin density
correlation functions of the 1D Hubbard model and could identify the
different exponents from finite-size scaling \cite{soreur}. 
In particular, they were able
to verify the scaling relations between $\alpha_{CDW}$, $\alpha_{SDW}$, and
$\alpha_{4k_F}$, Eqs. (\ref{rcdw}) -- (\ref{asdw}), implied by their dependence
upon $K_{\rho}$ alone ($K_{\sigma} = 1$ for SU(2)-invariance), and they
found an upper limit $\alpha = 1/8$ as $U \rightarrow \infty$, implying
$K_{\rho} \geq 1/2$. 

These exponents can be determined exactly from the $U \rightarrow \infty$
Bethe wave function \cite{ogata}.
We study the momentum distribution function 
\begin{equation}
\label{nog}
n(k) = \langle c^{\dag}_{ks} c_{ks} \rangle = \frac{1}{L} \sum_{j,l} 
\langle c^{\dag}_{js}
c_{ls} \rangle e^{i k (j - l)a} \;\;\;.
\end{equation}
At this stage, the real-space representation of the Bethe wave function
can be used. In order to transfer an electron from site $l$ to $j$, we
have to take out of the Slater determinant one spinless fermion at $l$
and reinsert it at $j$. At the same time, the spin configuration is changed:
we must take out of the Heisenberg chain 
the spin at $l'$, corresponding to the electron
at $l$ ($l \neq l'$ because of the compressed lattice), and
insert it again at $j'$ corresponding to the new electron site
$j$ which can also be viewed as permuting neighbouring spins successively
between $l'$ and $j'$. 
Then, one has to sum over all configurations of the spinless fermions,
as implied by the average $\langle ... \rangle$ 
in (\ref{nog}) taken over the wavefunction
(\ref{wvsim}). The permutation of two neighbouring spins is mediated by
the operator $2 {\bf S}_i \cdot {\bf S}_{i+1} +1/2$ so that the evaluation
of the spin contribution to $n(k)$ requires calculation a correlation 
function of $j-l$ of these operators for each configuration of spinless
fermions. The charge contribution, on the other hand, is just the product
of two Slater determinants.

Ogata and Shiba solved these functions numerically and obtained a
function $n(k)$ characterized by a jump at $k_F$ and, surprisingly at
that time, another weak singularity at $3k_F$, shown in Fig.~4.3. 
Both of these facts were
somewhat surprising because the spinless fermions' $n(k)$ jumps at
$2k_F$ which thus appeared a natural candidate for Fermi surface of the
$U \rightarrow \infty$-\hm . However, the electrons involved in $n(k)$
both contain a charge and a spin component, and the spins feed back into
the charges by the kernel on the right-hand side of (\ref{liwus1}).
The (re-)appearance of $k_F$ and $3k_F$
is due to oscillations with wavevector $\pm 2k_F$ in the spin contribution
to $n(k)$. A careful finite-size-scaling analysis showed that the apparent
jump at $k_F$ would fade away as $L \rightarrow \infty$ and that the 
variation of $n(k)$ with system size was compatible with the Luttinger
liquid power law (\ref{nkut}) with an exponent $\alpha \approx 0.14$. The
$3k_F$-singularity was shown to be due an excitation of $k_F$-fermions
to $3k_F$ accompanied by creation of electron-hole pairs with -\tkf ,
and is also required by the picture of Section 4.2.
They also studied the spin-spin correlation function at $q = 2k_F$. 
From the singularity observed as a function of $q$ they inferred a
decay in real space as $R_{SDW}(x) \sim \cos(2k_F x) x^{-1.44}$ while
their results for the Heisenberg model were consistent with the
known form $\sim \cos(\pi x) \ln^{1/2}(x) / x$ \cite{giam,heiscor}.

An analytical evaluation of these quantities is also possible 
\cite{parola}-\cite{parsor}. Parola and Sorella started from an 
evaluation of the spin-spin correlation function \cite{parola}
\begin{equation}
\label{spsp}
< {\bf S}_r \cdot {\bf S}_0 > = \sum_{j=2}^{r+1} P_{SF}^r(j)
S_H(j-1) \;\;\;,
\end{equation}
where $S_H(j)$ is the (known) spin correlation function of the Heisenberg model
and $P_{SF}^r(j)$ is the probability
of finding $j$ (spinless) particles between $0$ and $r$ with one at $0$ and
one at $r$. The evaluation of this latter quantity is difficult but at least
asymptotically possible, and one finds
\begin{equation}
\label{sps}
< {\bf S}_r \cdot {\bf S}_0 > \approx \cos(2k_F r) \frac{\ln^{1/2}r}{r^{3/2}}
\;\;\;,
\end{equation}
which is consistent with the \LL\ function (\ref{asdw}) 
provided $K_{\rho} = 1/2$.
This implies an exponent $\alpha =1/8$ for the momentum distribution function
$n(k)$ of the $ U = \infty$-Hubbard model at $k_F$, in quite good agreement
with the numerical data of Ogata and Shiba. Parola and Sorella
could recalculate analytically
the momentum distribution, following the procedure by Ogata and Shiba and,
fixing two open parameters so as to reproduce the behaviour at $k_F$, were
able to identify the exponent at $3k_F$ as $\alpha_{3k_F} = 9/8$ \cite{sorpar},
a value also found by others \cite{peso,aren}.

Anderson and Ren compute the correlation exponents 
of the Hubbard model in the $U \rightarrow \infty$-limit 
in a more physical way \cite{aren}.
They observe that the Ogata-Shiba wave function implies
charge-spin separation only for those excitations which take place solely
in one channel. If we consider correlation functions of excitations affecting
both channels, phase shifts will arise in the distribution of the momenta
$\{k_i\}$ due to changes in the rapidity-distribution. This is due to the
kernel on the right-hand side of Eq.~(\ref{liwus1}), and to the
parity effects in the distributions of the quantum numbers $I_i$ and  
$J_{\alpha}$ in the Lieb-Wu equations, cf. (\ref{liwubound}). 
As an example, the Green function for the holon at
+\tkf\ is $G^{(h)} (x,t)
\sim e^{2ik_Fx} / (x - v_{\rho} t)$ and involves only charge degrees of 
freedom: if one adds an $I_i$ to the system, the rapidities do not change.
Removing a spinon at $-k_F$, i.e. a $J_{\alpha}$, there is a phase shift
of $\delta_{\pm 2k_F} = \pi/2$ of all holon momenta in the same direction. 
The spinon Green function, which for the Heisenberg model is 
$e^{i k_F x} / \sqrt{x - v_{\sigma} t}$,
will therefore also contain a contribution from the phase
shift of the holons which reduces the overlap with the ``unshifted'' ground
state wave function. This introduces a factor $(x \pm v_{\rho} t)^{-(\delta /
2 \pi)^2}$ for the phase shifts on each side of the momentum distribution. 
The \tkf -spin-spin correlation function consists of a right-moving 
particle (spinon) and a left-moving hole (antispinon). Each of them
shifts the $k_i$-distribution in the same direction, so that the
total phase shift is $\delta_{\pm 2k_F} = 
\pi$ on each side. Putting everything together, we have
\begin{eqnarray}
\langle {\bf S}(xt) \cdot {\bf S}(00) \rangle & = &
\frac{\cos (2 k_F x)}{(x - v_{\sigma} t)^{1/2} (x + v_{\sigma} t)^{1/2} 
(x - v_{\rho} t)^{1/4} (x + v_{\rho} t)^{1/4} } \nonumber \\
& = & x^{-3/2} \cos ( 2 k_F x)  \;\;\; {\rm for} \;\;\; t=0 \;\;\;.
\end{eqnarray}
The \fkf -CDW only involves a right-moving holon and a left-moving antiholon,
and therefore is decoupled from the spins 
\begin{equation}
\langle O_{4k_F}(xt) O^{\dag}_{4k_F}(00) \rangle = \frac{ \cos (4k_F x)}{
(x - v_{\rho} t) (x + v_{\rho}t)} \;\;\;.
\end{equation}
Other interesting examples are provided by the $k_F$- and $3k_F$-pieces
of the single-particle Green function. Here, one has to take out (add)
both a holon and a spinon. One can take out the holon at \tkf\ and the
spinon at $-k_F$. The removal of the spinon shifts the holon momenta
by $\pi/2$ in the positive direction, and this phase shift cancels a
quarter of the $2 \pi$-shift caused by the holon removal at \tkf :
$\delta_{2k_F} = 3 \pi / 2$ while $\delta_{-2k_F} = \pi / 2$. 
This process determines the $k_F$-component. One can, however, also
take out the spinon at $+k_F$, and then the ensuing phase shift adds to
that of the holon removal $\delta_{2k_F} = 5 \pi / 2$. 
This determines the $3k_F$-Green function.
We have 
\begin{equation}
G_{k_F (3k_F)}(xt) = \frac{\exp [ i (2k_F \pm k_F) x]}{ 
(x - v_{\sigma} t)^{1/2} (x - v_{\rho} t)^{(1 \mp 1/4)^2} 
(x + v_{\rho} t)^{1/16} } \;\;\;.
\end{equation}
All these correlation functions agree with the \LL\ expressions taken 
at $K_{\rho} = 1/2$. 

The removal or addition of spinons and holons can also be interpreted
in terms of the charge and current excitations of a \LL\ in the
charge and spin channels. One can consider the $3k_F$-component of the
Green function as being due to an additional current excitation with momentum
\tkf\ with respect to the $k_F$-piece. Anderson and Ren also prove that
the correlation exponents of the Green function,
which can be derived from the kernels of the
Lieb-Wu equations (\ref{liwu1}) and (\ref{liwu2}) \cite{halphlet},
are precisely the Fermi-surface phase shifts due
to the insertion of an additional particle \cite{aren}.

The applicability of these methods is, however, quite restricted: (i) there are
many models which cannot be solved exactly, and (ii) even if a Bethe
Ansatz solution is available, manageable simplifications generally only occur
in special limits such as $U \rightarrow \infty$ for the Hubbard model.
On the other hand, the notion of a \LL\ is based on the low-energy properties
of a many-body problem, and a priori, a complete solution is not required.
Following Haldane \cite{Haldprl} and Section \ref{secbos}, a \LL\
can be identified and its characteristic parameters determined by using
only the low-energy spectrum of the lattice Hamiltonian. These can
be found reliably either from an exact solution or by numerical diagonalization.

One general method is due to Efetov and Larkin \cite{efe} and Haldane
\cite{Haldane,Haldprl} where it was formulated for a spinless fermion system,
and then extended by Schulz 
\cite{ijmp,schulz} for models of $S=1/2$-electrons. 
Here, one formulates an effective Luttinger Hamiltonian for the low-energy
physics and then identifies its parameters $K_{\nu}, v_{\nu}$ from the
properties of the exact solution. 
Central to this approach is the use of the relations between
the correlation exponents $K_{\nu}$ and the renormalized velocities
\cite{Haldprl}
$v_{N\nu} = v_{\nu}/ K_{\nu}, ;\; v_{J\nu} = v_{\nu} K_{\nu}$, 
Eq.~(\ref{speccoup}).
To identify the \LL\, one must,
in principle, determine the three velocities $v_{\nu}, v_{N\nu}$,
and $v_{J\nu}$ per degree of freedom and check that they satisfy 
(\ref{speccoup}).
$K_{\nu}$ is then obtained automatically. In practice, this programme is
rarely carried out to this point (with the notable exception of 
\cite{halphlet}). Rather, one assumes (\ref{speccoup}) to hold and determines
both $v_{\nu}$ and $v_{N\nu}$ which are sufficient to yield the remaining
parameters $K_{\nu}$. 
$v_{N\rho}= v_{\rho}/K_{\rho}$ 
is related to the compressibility $\kappa$ by Eq.~(\ref{susc})
${\kappa}^{-1} = L^{-1} \partial^2 E / \partial n^2 = {\pi} v_{N\rho} /2$,
and the change of the ground state energy with particle density $n$ 
can be readily determined by Bethe Ansatz or numerical methods. 
A similar relation (\ref{susc})
for $v_{N\sigma} = v_{\sigma}/K_{\sigma}$ to the magnetic
susceptibility can be explored in the spin sector. If the system is 
spin-rotation invariant, $K_{\sigma} = 1$ and $v_{\sigma}$ is found. 
In the charge sector, $v_{\rho}$ must be determined
independently from the low-energy spectrum of charge excitations.
To identify which type of excitations in the Bethe Ansatz is relevant,
one can realize, as does Schulz \cite{ijmp}, from the boson 
representation (\ref{cdw4}) that the \fkf -CDW operator involves
only charge degrees of freedom and then argue that power-law decay of this
correlation function must originate from
gapless excitations at that wavevector. These 
``particle-hole'' excitations have been known since a long time \cite{coll},
and their velocity is
\begin{equation}
v_{\rho} = \lim_{p \rightarrow 0} 
\frac{\varepsilon(k_0,p_0)}{p(k_0,p_0)} \;\;\;.
\end{equation}
Operationally, in the Bethe Ansatz wave function, 
take one particle with pseudo-momentum $k_0$ out of the filled (charge)
pseudo-Fermi sea and put it
into one of the empty states above at pseudo-momentum $p_0$. 
Find the energy $\varepsilon(k_0,p_0)$ and (physical) 
momentum $p(k_0,p_0)$ associated with this excitation; then take
the limit $p \rightarrow 0$. This gives $v_{\rho} $,
and $K_{\rho}$ is then determined, too. Ultimately, the 
correlation exponent $K_{\rho}$ is fully determined by thermodynamic 
properties \cite{Haldane,efe,schulz,Haldprl}.

This procedure does not suffer from the limitations of
perturbative \rg\ and allows an exact calculation of the
Luttinger liquid parameters. Being related to properties of the eigenvalue
spectrum, it
can easily be adapted to numerical exact diagonalization studies.
One concern might be finite size effects because the numerical solutions
are confined to small systems. They are, however, not critical here since
the energies of the low-lying 
states usually converge rather quickly to the infinite system
limit. 

One can also apply conformal field theory methods to determine the correlation
exponents of the Hubbard model. 
Conformal field theory, as we have sketched it in Section \ref{cft}, 
requires a Lorentz-invariant
system \cite{confft}. This is the case for spinless fermions (or models
of the same universality class) with only \em one \rm branch of gapless
excitations, and has consequently been applied to such problems with
great success \cite{mirozab}. 
The Hubbard model and all other models in the universality
class of a spin-1/2 Luttinger liquid are not Lorentz-invariant because
the spin and charge velocities $ v_{\sigma} \neq v_{\rho}$ play the roles of two
different velocities of light. Each channel $\nu$ taken by itself is conformally
invariant, however, described by a Virasoro algebra
with central charge $c_{\nu} =1$. The complete theory is then described
by a semidirect product of these Virasoro algebras, and the scaling dimensions
of operators now depend, instead of a single coupling constant, on an 
$N \times N$-matrix of coupling constants, the ``dressed charge matrix'', 
for an $N$-component system \cite{ncomp}. 

Frahm and Korepin have applied these methods to the 1D Hubbard model in order
to deduce the long-distance asymptotics of its correlation functions \cite{fk}. 
The idea is the following: the elements of the dressed charge matrix 
\begin{equation}
\label{dcm}
Z \equiv \left( 
\begin{array}{cc}
Z_{cc} & Z_{cs} \\
Z_{sc} & Z_{ss} 
\end{array}
\right) = \left( 
\begin{array}{cc}
\xi_{cc}(k_0) & \xi_{cs}(\Lambda_0) \\
\xi_{sc}(k_0) & \xi_{ss}(\Lambda_0)  
\end{array}
\right) 
\end{equation}
(and, of course, the velocities of the gapless excitations $v_{\nu}$) can
be evaluated from the Bethe Ansatz.
Its entries $\xi$ obey equations derived from and similar to the Lieb-Wu
equations (\ref{liwu1}), (\ref{liwu2}), with the 
limit $L \rightarrow \infty$ taken. 
$k_0$ and $\Lambda_0$ are the cutoffs in the distribution functions of 
the momenta and rapidities. The entries of the dressed charge matrix are 
related to thermodynamic
quantities of the model in much the same way as the effective coupling 
constant of spinless fermions is.
For example, Frahm and Korepin find \cite{fk} 
\begin{equation}
\label{dressch}
\xi_{cc}^2(k_0) =  \pi v_{\rho} n^2 \kappa
\end{equation}
where $\kappa$ is the compressibility and $n$ the charge density.
Comparing (\ref{dressch}) to (\ref{susc}), we see that $\xi_{cc}^2(k_0)$
is essentially $K_{\rho}$, up to a factor $n^2$. 
This formula has been derived
independently by Kawakami and Yang \cite{kawaya}, who use earlier 
Bethe Ansatz evaluations of
on the thermodynamic properties in order to get $\xi_{cc}$ as a function
of the system parameters. 
These authors, however, neglect the 
off-diagonal elements of the dressed charge matrix.

In analogy to Section \ref{cft}, conformal invariance then determines
the scaling dimensions of (primary and descendant) operators.
The role of the coupling constant $g$ of the Gaussian model is now 
played by the matrix $Z$. $\Delta N_{c(s)}$ and $D_{c(s)}$, 
later grouped into vectors ${\bf \Delta N}$ and ${\bf D}$, 
count the charges ($c$) 
and spins ($s$) added by the field $\phi$
to the Bethe Ansatz distribution and are 
(up to linear combination) the changes in the 
charge and current excitations of the \lm\ \cite{pesomag}.
The ground state has ${\bf \Delta N} = {\bf D} = {\bf 0}$. 
Allowed values of ${\bf D}$ depend on 
${\bf \Delta N}$. For example, for a fermion operator 
$c^{\dag}_{\pm k_F, \uparrow}$, ${ \bf \Delta N} = (1,0)$, ${\bf D} =  
(\pm 1/2, \mp 1/2)$, for  
$c^{\dag}_{\pm k_F, \downarrow}$, ${ \bf \Delta N} = (1,1)$, ${\bf D} =  
(0, \pm 1/2)$, and for the density operator $\rho$, one has ${\bf \Delta N} 
= {\bf 0}$. Numbers $N_{c(s)}^{\pm}$ characterize descendent
fields $\phi$ at level $\sum N_{c(s)}^{\pm}$. 
Primary fields have $N_{c,s}^{\pm}=0$, and finite
values describe secondary fields. 
The correlation functions of the primary and descendent fields 
$\phi_{\Delta^{\pm}}$ with scaling dimensions $\Delta^{\pm}$ 
are given by conformal field theory as
\begin{equation}
\label{cofcon}
\langle \phi_{\Delta^{\pm}}(x,t) \phi_{\Delta^{\pm}}(0,0) \rangle = 
\frac{ \exp[ -2 i D_c P_{F,\uparrow} x ] 
\exp[-2i (D_c + D_s) P_{F,\downarrow} x]}{ (x-iv_{\rho}t)^{2 \Delta_c^+}
(x+ iv_{\rho}t)^{2 \Delta_c^-} (x-iv_{\sigma}t)^{2 \Delta_s^+} 
(x+iv_{\sigma}t)^{2 \Delta_s^-} } 
\end{equation}
which is a direct generalization of (\ref{crophys}) to a two-component
system. 

As in Section \ref{cft}, the central charge and the scaling dimensions
can be obtained from the finite size corrections to the energy of the
ground state and the energies and momenta of low-lying excited states
of the system. For $c=1$ in both the charge and spin channel, (\ref{freen})
generalizes to
\begin{equation}
\label{gsen}
E_0(L) - L \varepsilon_0 = - \frac{\pi}{6 L} (v_{\rho} + v_{\sigma})
+ {\cal O} \left( \frac{1}{L} \right)\;\;\;,
\end{equation}
where the symbol ${\cal O}(1/L)$ stands for terms decaying faster than $1/L$.
$E_0$ is the ground state energy at size $L$, and $\varepsilon_0$ is the energy
density in the infinite system. 
Eq.~(\ref{gsen}) must be verified by the solution. Eqs.~(\ref{finen}) and
(\ref{finmom}) for
energy and momentum of the low-lying excitations are given by
\begin{eqnarray}
\label{ensca}
E({\bf \Delta N}, {\bf D}) - E_0 & = & \frac{2 \pi}{L} \left[
v_{\rho} (\Delta_c^+ + \Delta_c^-) + v_{\sigma} (\Delta_{s}^+ + \Delta_s^- )
\right] + {\cal O} \left( \frac{1}{L} \right)\;\;\;, \\
P ({\bf \Delta N}, {\bf D} ) - P_0 & =& \frac{2 \pi}{L} \left(\Delta_c^+ -
\Delta_c^- + \Delta_s^+ -\Delta_s^- \right) + 2 D_c P_{F \uparrow}
+ 2 ( D_c + D_s ) P_{F\downarrow} \;. \nonumber
\end{eqnarray}
On the other hand, these quantities can be computed from the Bethe Ansatz
(or numerically) as
\begin{equation}
\label{eex}
E( {\bf \Delta N}, {\bf D} ) - E_0   =  \frac{2\pi}{L} \left\{
\frac{1}{4} {\bf \Delta N}^T \cdot (Z^{-1})^T \cdot 
\left({\rm diag} [v_{\rho},v_{\sigma}] \right)
\cdot Z^{-1} \cdot {\bf \Delta N} + \right.  
\end{equation}
\begin{displaymath}
+  \left. {\bf D}^T \cdot Z \cdot \left({\rm diag}[v_{\rho},v_{\sigma}] 
\right) \cdot Z^T \cdot {\bf D} + v_{\rho} (N^+_c + N^-_c) +
v_{\sigma} (N_s^+ + N_s^-) \right\} + 
{\cal O} \left( \frac{1}{L} \right) \;\;\;, 
\end{displaymath}
\begin{displaymath}
P ({\bf \Delta N}, {\bf D}) - P_0  =  \frac{2 \pi}{L} \left\{
{\bf \Delta N}^T \cdot {\bf D} + N_c^+ - N_c^- + N_s^+ - N_s^- \right\}
+ 2 D_c P_{F,\uparrow} + 2 (D_c + D_s) P_{F,\downarrow} \;\;\;. 
\end{displaymath}
Comparing Eqs. (\ref{eex}) to (\ref{ensca})
one deduces the scaling dimensions
\begin{eqnarray}
2 \Delta_c^{\pm} ({\bf \Delta N}, {\bf D}) & = & \left( Z_{cc} D_c +
Z_{sc} D_s \pm \frac{Z_{ss} \Delta N_c - Z_{cs} \Delta N_s}{
2 \det Z } \right)^2 + 2 N_c^{\pm} \;\;\;, \nonumber \\
\label{dims}
2 \Delta_s^{\pm} ({\bf \Delta N}, {\bf D}) & = & \left( Z_{cs} D_c +
Z_{ss} D_s \pm \frac{Z_{cc} \Delta N_s - Z_{sc} \Delta N_c}{
2 \det Z } \right)^2 + 2 N_s^{\pm} \;\;\;.
\end{eqnarray}

In order to obtain the correlation functions of the Hubbard model, 
on would have to expand the physical operators $O$ in terms of the conformal
fields. This is usually not possible. On the other hand, the quantum numbers
of the intermediate states can be determined from the representation of
the operators $O$ in terms of the electron creation and annihilation operators.
We have given examples above. Then, the complete asymptotic behaviour
of the correlation functions can be given in terms of the scaling dimensions
and thus in terms of the dressed charge matrix $Z$ whose entries depend on
$U$ and the bandfilling $n$. As an example, the first few terms of the
complete density-density
correlation function are given by Frahm and Korepin \cite{fk} and, in 
the absence of a magnetic field, reduce to Eqs.~(\ref{rroh})--(\ref{r4kf})
when the appropriate $K_{\rho}$ is inserted there.
In this way, the conformal fields contributing to the density-density
correlations are not explicitly identified. The knowledge of their scaling
dimensions is sufficient to determine their contribution to the correlation
function. Penc and Solyom have finally deduced explicit 
Tomonaga-Luttinger coupling constants
$g_i$ from the dressed charge matrix and the scaling dimensions of the
Hubbard model \cite{pesomag}. 

While the asymptotic correlation exponents agree with the approach by
Schulz \cite{ijmp,schulz} and Kawakami and Yang \cite{kawaya}, 
there are some subtle differences. In general, in the absence
of magnetic fields, $Z$ is not diagonal as naively
expected for charge-spin separation. 
However, one of the matrix elements $Z_{cs} =0$
and $Z_{sc} = Z_{cc} / 2$ which gives critical exponents identical to
those for a charge-spin separating system as assumed by Schulz.
One can include an external magnetic field 
\cite{fk2}. Then, there is no longer a simple relation between the elements
of $Z$, the exponents now differ from those derived under the assumption
of charge-spin separation, and charge and spin are strongly coupled. 
On the other hand, the dressed charge matrix is probably not a good
quantity to ``measure'' charge-spin separation, because it does not
change in any essential way in the limit $U \rightarrow \infty$ where
we know \cite{ogata} that the product form of the Bethe wavefunction 
implies complete charge-spin separation.

The dependence of $K_{\rho}$ on $U$ and $n$ is shown in Fig.~4.4,
and that of the velocities $v_{\nu}$ in Fig.~4.5 \cite{schulz}.
For small $U$, the variation of $K_{\rho}$ with $U$ is consistent
with the perturbative result $K_{\rho} \approx 1 - U / \pi v_F$,
and the slope varies with bandfilling due to the $n$-dependence
of the Fermi velocity $v_F = 2 t \sin (\pi n /2)$. At larger $U$,
$K_{\rho}$ deviates from a straight line and $K_{\rho} \rightarrow 1/2$
for $U \rightarrow \infty$ for all $n$. $K_{\rho} = 1/2$ is also the
limit for $n \rightarrow 0$ for any $U>0$ which is quite obvious due
to the $n$-dependence of $v_F$. Also $K_{\rho} \rightarrow 1/2$ for
$n \rightarrow 1, (U>0)$, cf. below. The velocities $v_{\nu} \rightarrow v_F$
for $U \rightarrow 0$ as expected, and as $U \rightarrow \infty$,
$v_{\rho} = 2t \sin(\pi n)$ and $v_{\sigma} = (2 \pi t^2 / U) [1 - 
\sin(2 \pi n)/(2\pi n)]$. While $v_{\rho} \propto n$ for all $U$ and small 
$n$, $v_{\sigma} \propto n^2$ for $U>0$ and $\propto n$ for $U=0$. 

These parameters can then be inserted into the results obtained
in Section \ref{secprop} to obtain the
correlation functions of the Hubbard model as a function of $U$ and $n$. 
In particular, for $U \rightarrow \infty$, one obtains $\alpha \rightarrow
1/8$, $\alpha_{CDW,SDW} \rightarrow 1/2$ and $\alpha_{4k_F} \rightarrow 0$.
The properties
of the charge degrees of freedom in this limit can be straightforwardly
understood in
terms of spinless fermions, in agreement with the
factorization of the Bethe wave function. E.g. the \fkf -part of
the density-density correlations is simply the \tkf -CDW of free spinless 
fermions with a doubled Fermi wavevector. The large-$U$ limit of $v_{\rho}$
is simply the Fermi velocity of free spinless fermions with a hopping
integral $t$. Also close to half-filling even at finite $U$, 
a spinless fermion picture
applies: here one best thinks in terms of a few holes doped into the
insulating half-filled band, and the repulsion $U$ is accounted for
by treating them as spinless fermions. When there are very few of them,
their mutual interaction will be negligible. This explains the value
$K_{\rho} = 1/2$ found for all $U$ as $n \rightarrow 1$. 
The spinless fermion picture also implies that the prefactor of the
\tkf -part of the density-density correlation function must vanish as
$U \rightarrow \infty$. 
More care has to be taken for spin or single-particle correlations.
The ground state of the Hubbard model can be viewed as containing a number
of holons appropriate to the doping level but no spinons. In this way,
it becomes clear that the characteristic wave vector for the SDW oscillations
\tkf\ shifts with doping due to the introduction of holes
although in a local picture, there are no configurations of
parallel neighbouring spins \cite{ijmp}. The motion of holons
disrupts the spin correlations
and therefore leads to a more rapid decay of the spin-spin correlations
than in the half-filled band or a Heisenberg antiferromagnet. Introducing
a hole (or an electron) creates, however, a holon at $\pm 2k_F$ 
and a spinon at $\pm k_F$,
and therefore the single-particle Green function oscillates with wavevectors
$k_F$, $3k_F$, etc. 

The low-energy spectral function of the Hubbard model, obtained by inserting
$\alpha = 1/8$ for the limit $U \rightarrow \infty$ into the \lm , is shown 
in Fig.~3.6. It is clearly dominated by the spectral weight between 
$v_{\sigma}$ and $v_{\rho}$, and the weight above/below $\pm v_{\rho}$
is quite negligible. Comparison to functions of models with either charge-spin
separation or anomalous dimensions only, suggests that charge-spin separation
is the dominant non-Fermi-liquid feature in the 1D \hm\ 
\cite{spec2,spec1,ms}. Even for infinite repulsion, the anomalous correlations
are quite \em weak. \rm Physically, this implies that the power-laws in
the correlations are most sensitive to the range of the interaction, taken
finite in the Luttinger but zero in  the \hm , while the influence of 
short-range interactions is strong on charge-spin separation. This
allows to rationalize the small momentum range, where \LL\ behaviour is
seen in $n(k)$ in Figure 4.3. Unfortunately, no signature of charge-spin
separation has been detected in a Monte-Carlo simulation of the spectral
function directly of the Hubbard model \cite{preuss}. This could be due
to finite system size and/or temperature, but certainly needs further study.

Correlation functions of the $t-J$-model behave in a similar manner
and also identify it as a \LL\ 
\cite{kawatj,tj}. Specifically, at the supersymmetric point $J/t=2$, where
the model is solvable by Bethe Ansatz, conformal field theory allows
to derive the dependence of $K_{\rho}$
on band-filling \cite{kawatj}, in a similar manner as for the Hubbard model. 
It obeys to the same limits as for the
$U > 0$ Hubbard model $1 \geq K_{\rho} \geq 1/2 $ but tends towards 
the free value for the nearly empty band, while in this limit the Hubbard
model behaves as if $U \rightarrow \infty$. On going away from the 
supersymmetric point, the model is no longer solvable, and one has to turn
to numerical diagonalization on small clusters to obtain the correlation
exponents \cite{tj}. Again, one uses Eq. (\ref{susc}) to obtain $v_{\rho}
/ K_{\rho}$ and separately studies the spectrum of the charge excitations.
While $K_{\rho}$ continues to obey to the lower bound $K_{\rho} \geq 1/2$
(the equality holding for empty bands at $J< 2t$, half-filled bands at
$ J < 3.5 t$ and at $J=0$ for any filling), $K_{\rho} > 1$ now occurs for
larger values of $J$. Eqs.~(\ref{ass}) and (\ref{alts}) imply a 
region of dominant superconducting fluctuations.
According to the general scaling
arguments above, logarithmic corrections would favour
the triplet type if the spins are massless, while opening of a spin gap
would make singlet superconductivity dominate. Evidence for a spin gap
has been produced by using variational wavefunctions, a procedure to
be discussed below \cite{helme4}. 
Imada and Hatsugai
also measured spin correlation functions in their Monte Carlo simulations
\cite{imada}. While for small $J/t$, their
results are quite close to those of the Hubbard model found by
Hirsch and Scalapino \cite{hirsca},
the spin correlations become commensurate, i.e. peaked
at $q = \pi/a$ rather than at \tkf\, as $J$ increases. In this regime,
the holes in the $t-J$-model probably act as mobile defects in a
short-range-ordered antiferromagnetic background.
Finally, in the
large-$J$ region ($> 2 \ldots 3.5 t$ depending on $n$), phase separation
occurs: here the attraction due to the interaction terms in Eq. (\ref{hamtj})
is optimized at the expense of the kinetic energy.
The point $J=2t$, $n =0$
is possibly singular and $K_{\rho}$ there may depend on the order of the limits.

The phase diagram and the Luttinger liquid correlations of the 
$t-J$-model have also been established from variational wave functions
\cite{helme4,helme1,helme2}. This result is particularly noteworthy because
these functions can be generalized into higher dimensions where exact
solutions generally are not possible and numerical studies are severely
limited by finite size effects (Section \ref{twochains}). Recall that, on
a technical level, a major problem in treating the $t-J$-model with analytical
methods, is the implementation of the constraint of excluded double occupancy.
This constraint is implemented, however, in a variational wave function
due to Gutzwiller \cite{gutz}
\begin{equation}
\label{guwf}
| \Psi_G \rangle = \prod_i (1 - n_{i \uparrow} n_{i \downarrow} ) | FS 
\rangle \hspace{3mm} {\rm with} \hspace{3mm} | FS \rangle = \prod_{| k |
 < k_F}
c^{\dag}_{k \uparrow} c^{\dag}_{k \downarrow} | 0 \rangle \;\;\;,
\end{equation}
where $| FS \rangle $ is the filled Fermi sea and $| 0 \rangle$ 
the vacuum.
This wave function yields rather good energies but the correlation hole
between two particles
it contains is too short. The momentum distribution has a sharp jump at $k_F$
but the spin correlations (at $n=1$) are pretty close to the exact ones 
for a Heisenberg chain \cite{gevo}. To find a way to increase the range
of correlations, notice the following. $| \Psi_G \rangle$ 
provides an exact solution to spin chains with an exchange integral
falling off as $J \propto r^{-2}$ \cite{halshas}. In the course of this
solution, it has been shown that $| \Psi_G \rangle$ can be rewritten as
\begin{equation}
| \Psi_G \rangle = \sum \det( e^{i q_i r_j} ) \det (e^{i p_i r_j})
\prod_j S^-( r_j) |FM \rangle  \propto \prod_{i < j} \sin^2 
\left[ \frac{\pi}{L} \left( r_i - r_j \right) \right] \;\;,
\end{equation}
where $| FM \rangle$
denotes the fully (up)- spin-polarized ferromagnetic chain and $j$ labels
the sites of the \em overturned \rm spins. The size of the correlation
hole can now be increased simply by increasing the power of the sines
(Jastrow factor):
\begin{equation}
\label{jastrow}
| \psi_{\nu} \rangle = \prod_{{\rm all} \; i<j} \left| \sin \left[ 
\frac{\pi}{L} (r_i - r_j) \right] \right|^{\nu} \left| \psi_G \right\rangle
\;\;\;,
\end{equation}
where the notation under the product sign emphaised that in this product,
the positions enter irrespective of the particles' spin direction.
($| \Psi_{\nu} \rangle $ is also related to the
quantum Hall effect as will be seen in Section \ref{fqh}).
This can be seen quite
explicitly from $| \psi_{\nu} |^2 = \prod_{i<j} \exp ( - V_{ij})$
with $V_{ij} \propto - \nu \ln | z_i - z_j |$ and 
$z_i = \exp (2 \pi i r_i / L)$, which represents the partition function
of hard core objects with a logarithmic interaction 
\cite{helme2}. 
It therefore can serve
as a natural starting point for a variational treatment of the $t-J$-model.
Correlation functions show power-law behaviour compatible with the
\LL\ form, Section \ref{secprop}, whose exponents 
now depend on the optimal value
of $\nu$ which is obtained from variational Monte Carlo simulations. 
One can establish an explicit relation 
\begin{equation}
\label{knuvar}
K_{\rho} = \frac{1}{2 \nu + 1}
\end{equation}
to the Luttinger exponent $K_{\rho}$, and the
numerical data are in good agreement. (\ref{knuvar}) can be derived either
by finding a solvable model whose ground state is given by
$| \psi_{\nu} \rangle$. $K_{\rho}$ can then be extracted from the spectrum of
low-lying states exactly as for the Hubbard model above \cite{kyvar}.
Another possibility is to computed explicitly the momentum distribution
and then identify the exponent to Eq.~(\ref{gree}) \cite{helme5}. Finally,
by applying increasing powers of the Hamiltonian to $| \Psi_{\nu} \rangle$,
one can obtain increasingly accurate approximations to the exact ground
state (provided that it is not orthogonal to 
the trial state $| \Psi_{\nu} \rangle$), 
and this method has allowed to uncover
evidence for the formation of a spin gap in a region of very low carrier
density and large $J/t$ close to the phase separation instability. 
In this limit, the system is a singlet superconductor. The resulting
phase diagram is given in Figure 4.6, where ``attractive Luttinger''
stands for dominant TS and ``repulsive Luttinger'' implies dominant SDW
correlations in a \LL . The $t-J$-model can be generalized to include
a $J \propto r^{-2}$ exchange, and its solutions are quite close to the 
Gutzwiller-Jastrow form discussed above.

One can also formulate Hubbard- and $t-J$-type models with long-range
\em hopping \rm which, in the limit of half-filled band, reduce to the
Haldane-Shastry spin chain \cite{gebhard}. These models are exactly
solvable but the solution in general is not Jastrow form.
Away from half-filling, they have Luttinger liquid low-energy
physics. One important element of these models is chirality, i.e.
the hopping term must be constructed in such a way that the electronic
dispersion contains only a single linear branch, corresponding to
right-(or left-) moving particles alone. In this situation, the
only allowed effective interaction of the electrons is of $g_4$-type, 
Eq.~(\ref{h4}),
while $g_1 = g_2 = g_3 =0$. This still allows for charge-spin separation
because $g_{4\perp} \neq 0$ but the renormalized effective coupling constant
$K_{\rho} = 1$, the value for free fermions. This is an interesting situation
because all $q$- or $\omega$-dependent correlation functions will be 
indistinguishable from a Fermi liquid, still there are no quasi-particle
excitations. From the discussion in Sec.~\ref{secprop} one would conclude, e.g.,
that the momentum distribution of such a model is a step function 
$n(k) = \Theta(k_F - k)$ but the spectral function $\rho(q,\omega)$
is purely incoherent with spectral weight between $v_{\sigma} q$ and
$v_{\rho} q$ and square-root singularities at these frequencies. 

At half-filling, the model exhibits a metal-insulator transition for
$U \geq 2 \pi t$, as borne out by a jump in the chemical potential
at $n =1$. How is this possible if Umklapp processes are forbidden?
As $U \rightarrow 2 \pi t$ from below, the charge velocity diverges,
corresponding to a divergence of $g_4$! At the same time, the compressibility
goes to zero and the Drude weight of the conductivity also diverges.
This is due to the dispersion exhibiting a jump discontinuity at the
Brillouin zone boundary \cite{gebhard}. 
This is pretty opposite to the standard scenario, where relevant Umklapp
processes generate the Mott-Hubbard transition, the charge
velocity vanishes, and the Drude weight has a finite jump 
(Sec.~\ref{motttr}). 
Here, the compressibility
diverges and the Drude weight vanishes. The properties
of the model in the charge sector therefore are peculiar and strongly
affected by the pathological dispersion. The spin fluctuations, on the
other hand, are more normal with a strong
peak in $\chi(T)$ at $T \sim J = 4 t^2 /U$, 
and the instantaneous spin-spin correlation
function has a logarithmic divergence at $q=\pi$, corresponding to 
$1/r$-decay in space, as for the half-filled Hubbard model. 

A further example for application of these methods, especially Eq. 
(\ref{susc}), is provided by the extended Hubbard model, Eq. (\ref{hehm}).
At half-filling, this model has been studied by many methods both
analytical and numerical \cite{ehm,ehmhalf}. This model possesses a rich
phase diagram but, due to the importance of
Umklapp scattering here, the system is insulating when the interactions are
repulsive \cite{ehm,ehmhalf}. Correlation functions 
and the identification of the phases as \LL s (eventually only in a single
channel) have been studied using \rg\ \cite{ehm}. 
The physics away from half-filling is also interesting. 
For $U \rightarrow \infty$, we recover spinless fermions, and the model can 
be mapped onto an anisotropic
Heisenberg chain which again can be solved by Bethe Ansatz, so that
the correlation exponents can be deduced \cite{lusg,luso}. For finite
$U$, Mila, Zotos,
and Penc \cite{milzot,milpen} used 
numerical diagonalization combined with Eq. (\ref{susc}) to evaluate the phase
diagram and correlation functions of the quarter-filled band. The phase
diagram together with lines
of constant $K_{\rho}$ in the positive $U,V$-region 
is given in Figure 4.7. 
On a technical level, this study is noteworthy because it is one of the
few instances where the \em three \rm
velocities $v_{N\rho}, v_{\rho}$, and $v_{J\rho}$
have been determined explicitly and their consistency with the \LL\
relations has been verified. For $v_{J\rho}$, one can use the relation
of the Drude weight of the conductivity (precisely $2v_{J\rho}$) to the
dependence of the ground state energy on an external flux \cite{flux}
\begin{equation}
\label{vflux}
v_{J\rho} = \frac{\pi}{2L} \left. \frac{\partial^2 E_0 (\phi)}{\partial \phi^2} 
\right|_{\phi = 0} \;\;\;.
\end{equation}
At weak coupling, the system is a Luttinger liquid, and its correlations
are described by a parameter $K_{\rho}$, indicated in Fig.~4.7 as dashed
lines. This parameter can now become smaller
than $1/2$, the $U \rightarrow \infty$-limit of the Hubbard model. Due  to
the relevance of Umklapp scattering terms which
have a scaling dimension of $2 - 8 K_{\rho}$, there is
a lower limit $K_{\rho} = 1/4$ in the metallic phase. Beyond, the system
goes insulating, and $K_{\rho}$ discontinuously jumps to zero. 
The dominant correlations at weak-coupling are
SDW but \tkf -CDWs are only logarithmically weaker. For $K_{\rho} < 1/3$,
however, the divergence of the \fkf -CDW correlations becomes stronger than
the SDW one, indicating gradual charge localization on alternating sites. 
In the insulating phase, this charge modulation is long-range ordered,
and the system can be viewed as a Wigner crystal. Fourth-order processes
in $t$ still give an antiferromagnetic exchange interaction between 
occupied sites, and a \tkf -SDW consequently is superposed on the \fkf -CDW.
As one goes to larger $V$ (unphysical if one thinks in terms of 
electron-electron interaction alone but conceivable if on-site
phonons are included), 
the effective correlations get weaker,
and $K_{\rho}$ can become even larger than unity. If the system is still
a Luttinger liquid in this range, it would be dominated by 
triplet superconducting correlations before giving way to phase separation.
(The caveat is important, since a
conclusive study of the properties of the spectrum of the Hamiltonian 
on which the derivation of the Luttinger parameters is based, was not
possible \cite{milzot,milpen}, and Mila et al. conjecture about
a two-fluid picture where a \LL\ would coexist with a liquid
of local singlet pairs.) A more complete diagram,
including attractive interactions is also available \cite{milpen}. 
Here, one can find \tkf -CDWs, SS when a spin gap opens, 
another TS when there is no spin gap, and a phase separation
regime when all interactions are attractive, as in the half-filled band
(commensurability is unimportant at this point). Most of these results
agree with a similar study by Sano and Ono \cite{sano} who, however, find
evidence for a spin gap in the large-$V$--small-$U$ region, and who therefore
would favour an SS phase preceding phase separation. Moreover, these authors
extend these results to a third-filled band ($n = 2/3$) where similar
results obtain \em except for strong repulsion. \rm Here, no evidence for a
transition to an insulating phase is found. While the general absence of 
such a transition is somewhat surprising, the Umklapp operators which could
mediate such a transition are less relevant than at quarter-filling, and
the transition may have escaped detection because
is expected to occur at stronger coupling. Notice that for a third-filled
band, $2k_F = 4k_F = \pi / 3 a \; ({\rm mod} \; 2 \pi / a)$, and that the
Umklapp operator couples charges and spins (Section \ref{motttr}). 
One thus predicts
the opening of a charge gap to be accompanied by opening of a spin gap.
The competition of SDW and TS at positive $U$ and negative $V$,
a situation that might be generated in two-band models, has been studied
in detail by Kuroki et al. who, however, find somewhat poor agreement
with \rg\ predictions even at weak coupling \cite{kuroki}. 

At general bandfilling $n$, the mapping onto a Heisenberg chain produces
a finite magnetic field \cite{lusg,luso}. Still, one can derive Luttinger
parameters, and $K_{\rho}$ now can become as small as $1/8$ \cite{Haldprl},
allowing $\alpha$ to increase up to about $1.5$. This confirms the
earlier statement that the correlation exponents are strongly sensitive 
to the interaction range. A finite range is required to produce really
strong correlations from strong interactions. This is exemplified 
most dramatically by considering a long-range Coulomb potential 
$V(r) \approx 1/r$ \cite{schuwi}. In a continuum system,
the charge fluctuations no longer
have a linear spectrum at low $q$ but rather go as $\omega_{\rho}(q)
\approx \sqrt{q^2 \ln q}$, so that formally $v_{\rho} = 0$. The spin 
fluctuations behave normal. The logarithmic low-energy spectrum also
gives a peculiar dependence to the correlation functions involving
density operators. The density-density correlations decay as
\begin{equation}
\label{rhowig}
\langle \rho(x) \rho(0) \rangle = A_1 \cos (2k_F x) 
\frac{\exp(-c \sqrt{\ln x})}{x} + A_2 \cos (4k_F x) \exp (-4 c \sqrt{\ln x})
\;\;\;,
\end{equation}
i.e. slower than any power of $x$ in its \fkf -component. Comparing to
(\ref{rcdw}) and (\ref{r4kf}), one formally would have $K_{\rho} = 0$. 
The Green function behaves as $G(x) \approx \exp ( ik_F x - c \ln^{-3/2} x)$
and decays faster than any power of $x$ (formally $\alpha \rightarrow \infty$).
The system is at the edge of Wigner crystallization. It retains some
marginal \LL\ character because the quantum version of the Mermin-Wagner 
theorem \cite{mermwa,quant} forbids a 
real phase transition into a long-range ordered 1D crystal.

On the lattice, a
systematic investigation of the effects of band-filling on the structure
of the ground state and thus the dominant CDW instability can be performed
in the strong coupling limit under quite general conditions of convexity
for a long-range density-density interaction
\cite{Hubtcnq,wigcrys}. In particular, from the minimization of the electronic
interaction energy (i.e. in the atomic limit)
one finds a series of generalized Wigner lattices as the band-filling $n$ is
varied. For $n = 1/m$, one has every $m$-th site singly occupied with
$m-1$ empty sites in between. For $n = 2/(2m+1)$, the singly occupied
sites are separated alternatingly by $m$ and $m-1$ empty sites. Other 
configurations can be constructed with a simple algorithm 
\cite{Hubtcnq,wigcrys}. 
As one dopes the system away from these rational band-fillings, one introduces
solitons with fractional charge $q = \pm \rho e$ into the ground states. 
These particles are mobile if the hopping integral $t$ is finite, and
they will experience an effective interaction. They will form again
a Luttinger liquid, and their effective velocities $v_{\nu}$ and correlation
exponent $K_{\nu}$ can be calculated as a function of the 
filling factor (the analysis is practical only for infinite interactions)
\cite{gomez}.

\section{Electron-phonon interaction and impurity scattering}
\label{phonie}

How stable is the \LL\ with respect to electron-phonon coupling? 

There are several models describing different aspects of this 
interaction. Coupling of electrons to acoustic phonons is modelled
by the Hamiltonian
\begin{equation}
\label{ssh}
H_{\rm SSH} = - \sum_{i,s} \left( t_0 - \alpha_{SSH} [ u_{i+1} - u_i ] \right)
\left( c^{\dag}_{i+1,s} c_{i,s} + {\rm H.c.} \right) + 
\sum_i \left( \frac{P_i^2}{2M} + \frac{K}{2} \left[ u_{i+1} - u_{i} \right]^2
\right) \;\;\;.
\end{equation}
The electron-phonon coupling arises from the first-order modulation of the
hopping integral by the relative displacements $u_{i+1} - u_i$ 
of two neighbouring sites. $K$ is the spring constant, $P_i$ the momentum
operator, and $M$ the ion mass. Electron-electron interactions can be added
if required. This model has been proposed to describe the essential physics
of conducting polymers, and polyacetylene in particular, by Su, Schrieffer,
and Heeger (SSH) and most often has been studied close to
half-filling \cite{ssh,sshrev}.

Electrons may also be coupled to intramolecular vibrations (optical phonons)
which modulate the energy levels $\varepsilon_i$ of the lattice sites
\begin{equation}
\label{holst}
H_{\rm Hol} = - t_0 \sum_{i,s} \left( c^{\dag}_{i+1,s} c_{i,s} + {\rm H.c.}
\right) + \sum_i \left( \frac{P_i^2}{2M_r} + \frac{f}{2} Q_i^2 \right) +
g \sum_i Q_i n_i \;\;\;.
\end{equation}
This model is due to Holstein \cite{holpap} and has played a central role
in the polaron problem \cite{pol}. Here, the phonons are dispersionless, 
the spring constant is called $f$, and $g$ is the first order coupling of
a molecular energy level to a vibrational coordinate $Q$ with an associated
reduced mass $M_r$. 
Another model somewhat intermediate between the SSH and Holstein models,
where the electrons couple to the librational motion of rings in a polymer
chain, has been introduced and discussed recently \cite{ge}. It combines
dispersionless phonons with coupling to the hopping integral $t$. 
Phonon dispersion and different structures of the coupling terms lead
to important differences in the physics of these models. Being not central
to the subject of this article, we shall not detail them here but rather
emphasize the common features of phonon-coupled Luttinger liquids.

Mean-field theory, the crudest form of an adiabatic approximation, does not
lead to \LL\ behaviour \cite{Peierls}. Rather, a gap opens and CDW long-range
order obtains. Compatibility with the Mermin-Wagner theorem \cite{mermwa,quant}
for incommensurate systems is reestablished by the Goldstone mode, 
the sliding of the CDW, but this only gives a gapless charge density 
excitation. The spins remain gapped, and the generic physics of such models
is discussed further in Section \ref{spingaps}. 
For genuine \LL\ behaviour, one must therefore 
go beyond phonon mean-field theory and/or 
include electron-electron interaction. 
Due to the importance of electron-phonon backscattering in 1D \cite{Peierls},
we give a brief discussion
using \rg , directly extending Section \ref{scatt} to include electron-phonon
interaction. 

The generic electron-phonon coupling Hamiltonian has the boson representation
\cite{phonons}
\begin{eqnarray}
\label{hphon}
H & = & H_1^{e-p} + H_2^{e-p} \;\;\;, \\
\label{h1phon}
H_1^{e-p} & = & \frac{\gamma_1 }{\pi \alpha} \int \! dx \left\{ \exp \left[
\sqrt{2} i \Phi_{\rho} (x) \right] \cos \left[ \sqrt{2} \Phi_{\sigma}(x)
\right] \varphi_{2k_F} (x) + {\rm H.c.} \right\} \;\;\;, \\
\label{h2phon}
H_2^{e-p} & = & \frac{\gamma_2}{\sqrt{L}} \int \! dx 
\left[ \rho_+(x) + \rho_-(x) \right] \varphi_0 (x) \;\;\;.
\end{eqnarray}
$H_{1,2}^{e-p}$ describe electron-phonon backward and forward
scattering. $\varphi_{2k_F}(x)$
is the \tkf -component of the displacement field $Q_i$ or $u_{i+1} - u_i$,
scaled by $\sqrt{M}$, and $\varphi_0(x)$ the $q \approx 0$-component. 
$\varphi_0(x)$ is a real field, but $\varphi_{2k_F}(x)$ is complex because
the scattering processes with $\pm 2k_F \neq \pi/a$ are physically different.
The coupling constants are
\begin{equation}
\gamma_1^{\rm (SSH)} =  4 i \alpha_{SSH} \sin (k_F a) \;, \;\;\; \gamma_2^{\rm
(SSH)} = 0\; ; \;\;\; \gamma_1^{\rm (Hol)} = \gamma_2^{\rm (Hol)} = g \;\;\;.
\end{equation}
The vanishing of $\gamma_2$ for acoustic phonons only holds at $q=0$ 
and is a consequence of the linear dispersion in the centre of the Brillouin 
zone. The influence of the finite-$q$ contribution is of order 
$(v_s/v_{\rho})^2$ where $v_s$ is the sound velocity of the phonons, and 
can be neglected in most realistic situations \cite{phofor}. This may perhaps
not be permitted close to a Mott transition where $v_{\rho} \ll v_F$ and
may become comparable to $v_s$ \cite{marlo}.

Electron-phonon forward scattering in a \lm\ can be diagonalized exactly
\cite{phofor}. Equivalent results are obtained by including it into the 
\rg . First, we integrate out the phonons to generate effective, \em retarded 
\rm electron-electron interactions (in imaginary time formalism)
\begin{eqnarray}
H_1^{\rm eff} (\tau - \tau')
& = & 2 \left( \frac{| \gamma_1 |}{2 \pi \alpha} \right)^2
\int \! dx \int \! dx' D_0(x-x,\tau - \tau') \exp \left( \sqrt{2} i
\left[ \Phi_{\rho}(x,\tau) - \Phi_{\rho}(x',\tau') \right] \right) 
\nonumber \\
\label{hphoneff1}
& & \times \sum_r \cos \left( \sqrt{2 } \left[ \Phi_{\sigma}
(x,\tau) +r \Phi_{\sigma}(x',\tau') \right] \right) + {\rm H.c.} \;\;\;, 
\end{eqnarray}  
\begin{equation}
\label{hphoneff2}
H_2^{\rm eff} (\tau - \tau')  =  \gamma_2^2 \int \! dx \int \! dx'
\sum_{r,r'} \rho_r(x,\tau) D_0(x-x',\tau - \tau') \rho_{r'}(x',\tau') 
\end{equation}
with the bare phonon propagator
\begin{equation}
D_0 (x, \tau) = \frac{1}{2 \omega_{ph}} \delta(x) \exp \left( - 
\omega_{ph} | \tau | \right) \;\;\;.
\end{equation}
When deriving scaling equations for the effective coupling constants, 
$\alpha$ is interpreted as a cutoff which is also extended to the 
$\tau$-direction (Section \ref{scatt}). In order not to loose the 
short-time contributions which are important at high phonon frequencies
$\omega_{ph}$, one must integrate the effective retarded interactions
between $\tau - \tau' = 0$ and $\alpha / v_F$, giving effective \em 
instantaneous \rm interactions, plus (\ref{hphoneff1}) and (\ref{hphoneff2})
with a cutoff $\alpha/v_F$ in $\tau$ \cite{scpol}. The instantaneous
interactions are added to the electronic terms, and the retarded pieces
are included into the \rg . Following Section \ref{scatt}, we obtain
the following set of scaling equations \cite{phonons,scpol}
\begin{eqnarray}
\frac{d K_{\rho}^{-1}}{d \ell } & = & \frac{1}{2} \frac{v_{\rho}}{v_{\sigma}}
\left( Y_1^{(ph)} - Y_2^{(ph)} \right) {\cal D}(\ell) \;\;\;, \nonumber \\
\frac{d K_{\sigma}^{-1}}{d \ell } & = & \frac{1}{2}
\left( Y_{\sigma}^2 + Y_1^{(ph)} {\cal D}(\ell)  \right) \;\;\;, \nonumber \\
\label{rgphon}
\frac{ d Y_{\sigma} }{d \ell} & =& Y_{\sigma} \left( 2 - 2 K_{\sigma} \right)
- Y_1^{(ph)} {\cal D}(\ell) \;\;\;, \\
\frac{ d Y_1^{(ph)} }{d \ell} & = & Y_1^{(ph)} \left( 3 - K_{\rho} - K_{\sigma}
- Y_{\sigma} \right) \;\;\;, \nonumber \\
\frac{d Y_2^{(ph)}}{d \ell} & = & Y_2^{(ph)} \;\;\;, \nonumber \\
\frac{d v_{\nu}}{d \ell} & = & \ \frac{1}{2} v_{\nu} K_{\nu} 
\left( Y_1^{(ph)} - Y_2^{(ph)} \right) {\cal D} (\ell) \;\;\;. \nonumber
\end{eqnarray}
The abbreviations are
\begin{equation}
Y_{\sigma} = \frac{g_{1\perp}}{\pi v_{\sigma}} \;, \;\;\; Y_1^{(ph)} = 
\frac{ | \gamma_1 |^2}{\pi v_{\sigma} \omega_{ph}^2 } \;, \;\;
{\cal D} = \frac{ \alpha_0 \omega_{ph}}{ v_{\sigma}} 
\exp \left( - \frac{ \alpha(\ell) \omega_{ph} }{v_{\sigma}} \right) \;\;\;.
\end{equation}
$Y_2^{(ph)} = Y_1^{(ph)}$ for dispersionless modes, and zero 
(for the present purposes) for acoustic modes.

A few important points are immediately apparent. (i) The phonon frequency
$\omega_{ph}$, through ${\cal D}(\ell)$, is the decisive quantity controlling
the interplay of repulsive electron-electron and attractive electron-phonon
interactions. At a scale $\ell_{ph} = \ln (E_F/\omega_{ph})$, all retardation
effects are scaled out, and the model behaves (and eventually continues to
renormalize) as effectively instantaneous. 
The influence of electron-phonon interaction is the stronger
the lower $\omega_{ph}$. (ii) The charge degrees of freedom remain gapless
for any nonvanishing phonon frequency. While $K_{\rho}$ rather strongly
decreases for
acoustic phonons, its sense of renormalization for dispersionless modes 
depends on the relative importance of forward and backward scattering. 
The intial $K_{\rho} (\ell =0)$ contains a contribution from the short-time
part of $D_0$, as discussed above. (iii) A gap may open in the spin 
fluctuations, and in fact does so for low enough phonon frequency and / or
high enough electron-phonon coupling. The system then is no longer a \LL\
and its physics will be discussed in more detail in Section \ref{spingaps}. 
Here SS or CDW correlations dominate, depending on $K_{\rho}$ 
\cite{scpol,kzl}. 
(iv) In the opposite limit of high phonon frequency and/or weak coupling,
there is a \LL\ regime. Depending on the renormalized $K_{\rho}$, 
we have SDW or TS correlations. The properties at low energies, below
$\omega_{ph}$, are then given by the \LL\ correlations (Section \ref{secprop})
with the fixed-point values $K_{\rho}^{\star}$ and $v_{\nu}^{\star}$. 
The scaling equations
respecting spin-rotation invariance, $K_{\sigma}^{\star} = 1$ is guaranteed
for gapless spin fluctuations. At energies above $\omega_{ph}$, there
will be deviations from the \LL\ properties. An example, the Holstein
contribution to the optical conductivity, involving phonon emission,
will be discussed in the next section \cite{hol}. Corrections to the
spectral functions of a model with forward scattering only
\cite{phofor,marlo}, have also
been evaluated \cite{mepho}. (v) The velocities $v_{\nu}$ of
the charge and spin fluctuations are renormalized by electron-phonon interaction.
Consequently, this interaction has a pronounced 
influence on the thermodynamic properties
such as specific heat, compressibility and susceptibility \cite{hol}. 
It is analogous to the enhancement of the effective mass, or the density of
states at the Fermi level, familiar from higher-dimensional systems. In 
contrast to higher dimensions, the electron-phonon interaction couples charge
and spin fluctuations and therefore strongly renormalizes
the magnetic properties of the \LL .

Renormalization group is also  very useful to study the influence of 
impurity scattering on the low-energy properties of \LL s \cite{thierry,apel}.
The forward 
($ q \approx 0$) and backward 
($q \approx 2k_F$) electron-impurity scattering components
can be represented by two Gaussian fields $\eta(x)$ and $\xi(x)$
with white noise correlations $
P_{\xi} = \exp [ - D_{\xi}^{-1} \int |\xi(x)|^2 dx] $ and a similar expression
for $P_{\eta}$ \cite{abri}.
$D_{\eta(\xi)} = v_F / \tau_{\eta(\xi)}$ and $\tau$ is the scattering time.
The interaction Hamiltonian is 
\begin{eqnarray}
\label{hforw}
H_f & = & \sum_{rs} \int \! dx \: \eta(x) \Psi_{rs}^{\dagger}(x) \Psi_{rs} (x)
= - \frac{\sqrt{2}}{\pi} \int \! dx \: \eta(x) \frac{\partial 
\Phi_{\rho}(x)}{\partial x} \\
H_b & = & \sum_s \int \! dx \left[ \xi(x) \Psi^{\dagger}_{+s}(x) \Psi_{-s}(x)
+ \xi^{\star}(x) \Psi^{\dagger}_{-s}(x) \Psi_{+s}(x) \right] \nonumber \\
\label{hback}
& = & \frac{1}{\pi \alpha} \int \! dx \left\{ \xi(x) e^{i [\sqrt{2} 
\Phi_{\rho}(x) + 2k_F x ]} \cos [ \sqrt{2} \Phi_{\sigma}(x) ] + {\rm H.c.} 
\right\} \;\;\;.
\end{eqnarray}
This Hamiltonian is of the same structure as the electron-phonon interaction
(\ref{h1phon}) and (\ref{h2phon}) except that $\eta(x)$ and $\xi(x)$ are
static fields while $\varphi_0(x)$ and $\varphi_{2k_F}(x)$ posses dynamics. 
The \rg\ treatment therefore is parallel to the phonon problem up to
two differences: (i) the ``phonon frequency'' $\omega_{ph} = 0$ here to reflect
the static nature of the impurity fields; (ii) for the same reason, 
forward scattering can be completely eliminated by simply shifting 
\begin{equation}
\label{phishift}
\Phi_{\rho}(x)
\rightarrow \tilde{\Phi}_{\rho}(x) = \Phi_{\rho}(x) - 
\frac{\sqrt{2} K_{\rho}}{v_{\rho}} \int^x \! dz \: \eta(z)
\end{equation}
and completing the square.
More importantly, if one uses the replica trick to treat backscattering,
the resulting action only contains \em differences \rm of $\Phi_{\rho}$-fields
so that they are not affected by the shift (\ref{phishift}). 
Also unaffected by this shift are the $\Pi_{\rho}(x)$- and 
$\Theta_{\rho}$-fields because they are generated from $\Phi_{\rho}$ by
time derivatives. This immediately implies that both the conductivity
and the pairing fluctuations (SS and TS) are unaffected by electron-impurity
forward scattering. The charge and spin density wave correlation functions,
on the other hand will decay exponentially with distance
\begin{equation}
\label{cdfor}
R_{CDW,SDW}(x,t) = e^{- D_{\eta} (\frac{K_{\rho}}{v_{\rho}})^2 | x |}
R_{CDW,SDW}(x,t) \mid_{\eta \equiv 0} \;\;\;.
\end{equation}
Decay with time is not affected. 

Also the influence of electron-impurity backscattering $\xi(x)$ is dramatic.
The \rg\ equations (\ref{rgphon}) can be taken over directly \cite{thierry} 
with the phonon ${\cal D}$ dropped, 
$Y_2^{(ph)} = 0$, and $Y_1^{(ph)}$ is replaced by a new
\begin{equation}
\label{abbre}
{ \cal D} = \frac{2 D_{\xi} \alpha }{\pi v_{\sigma}^2 } 
\left( \frac{v_{\sigma}}{v_{\rho}} \right)^{K_{\rho}} \;\;\;.
\end{equation}
The scaling dimension of the impurity backscattering operator ${\cal D}$
is $3 - K_{\rho} - K_{\sigma}^{\star}$ and determines its (ir)relevance 
in the limit ${\cal D}, \; Y_{\sigma} \rightarrow 0$ where mutual
renormalization effects can be neglected \cite{apel}. $\cal{D}$ goes
relevant except 
for $K_{\rho} > 2$ for a SU(2)-invariant system ($K_{\sigma}^{\star}
=1$) and for $K_{\rho} > 3$ for a spin-gapped system ($K_{\sigma}^{\star}
=0$): disorder is always relevant except deep in the superconducting region,
and more so for a triplet superconductor than for a singlet one. 
The fact that even weak disorder becomes relevant for weak superconducting 
correlations ($1 < K_{\rho} < 2$) shows that Anderson's
theorem \cite{andthe} fails in 1D.
A \LL\ only occurs when the coupling constants of the
pure system were such that it is strongly TS ($K_{\rho}^{\star} \geq 
2$ required), and TS correlations then continue to dominate.
Disorder can also be irrelevant when a spin gap opens ($Y_{\sigma} \rightarrow
-\infty$) with SS correlations strongest, but very large $K_{\rho}^{\star}
\geq 3$ is called for here. 
In all other cases, both $\cal{D}$ and $-Y_{\sigma}$ flow to infinity.
Disorder is relevant, and localization occurs. Moreover, one always
has a spin gap, and the physics then is best described as a CDW
($K_{\rho}^{\star} = 0$) pinned by impurities \cite{flr} (charge density
glass). For strong enough repulsion between the electrons, one would
however expect localization in the presence of antiferromagnetic 
correlations. Such a random antiferromagnet 
is, in fact, conjectured by Giamarchi and Schulz
\cite{thierry} and their failure to find it identified as an artefact
of the development leading to the \rg\ equations.
The main features of the RG equations
can also be rationalized by realizing that the impurity backscattering operator
linearly couples to the CDW-operator (\ref{cdw}) 
while the other types of fluctuations
(SDW, SS, TS) are only influenced in higher order. 
It is therefore clear,
that the charge density glass phase descending from CDWs 
strongly extends in the phase diagram.

The influence of impurities and electron-phonon scattering on transport
is a subject of the following section.

\section{Transport in Luttinger liquids}
\label{lltrans}

\subsection{Electron-electron scattering}
In the presence of band curvature and in lattice models, the Hamiltonian 
does not commute with the current which 
no longer is proportional to the momentum, and nonvanishing
conductivity at finite frequencies is possible. 
There are several processes contributing. The most obvious ones are
Umklapp processes whereby $n$ electrons are transferred from one side
of the Fermi surface to the other, carrying with them momentum of the
order $\pm 2 n k_F$. These processes are possible at low energy only 
in commensurate
systems where the Fermi wave vector has a rational relation to a reciprocal
lattice vector $G$: $k_F = (m/n) G$. Away from these commensurate band-fillings
they involve states separated from the Fermi level by a finite energy gap
$\Delta_{m/n}$ 
and will therefore contribute to the conductivity only at frequencies
or temperatures above $\Delta_{m/n}$ \cite{giam1,giamil}. These issues will
be discussed further in Section \ref{motttr}.
Another contribution
comes from band curvature. In the presence of interactions, band curvature
will also renormalize the current operator \cite{Haldane,giamil}. 

Band curvature in an incommensurate system 
adds a term (\ref{hnl}) to the Luttinger Hamiltonian (\ref{hlutt}). 
The current operator is given by Eq.~(\ref{currop}), and its commutator
with the Hamiltonian for small $q$ reduces to
\begin{equation}
\label{curcom}
\lim_{q \rightarrow 0} \left[ H, j(q) 
\right] = \frac{\lambda}{48 m^2 v_F} + \ldots
\end{equation}
Then, using Eq. (\ref{conint}), one obtains at $T=0$
\begin{equation}
\label{sigminc}
\sigma(\omega>0) = \frac{1}{8 \pi} \left( \frac{\lambda}{12 m^2 v_F} 
\right)^2 \frac{K_{\rho}-K_{\rho}^{-1}}{4 v_{\rho}^3} \omega^3 + \ldots
\end{equation}
i.e. a universal (interaction-independent) $\omega^3$-law \cite{giamil}. 
This result is
essentially perturbative in the band-curvature. Both for 
$K_{\rho} \rightarrow 1$ (noninteracting electrons) and $\lambda/m^2 
\rightarrow 0$
[free electrons with $k^2/2m$-dispersion, i.e. Galilei invariance $(\lambda 
\rightarrow 0)]$, or Lorentz invariance $(m \rightarrow \infty)$] 
the finite frequency contribution
disappears, as it must according to Chapter \ref{chaplm}. A direct
calculation of the temperature dependence of the dc-conductivity is not
possible, but Giamarchi and Millis \cite{giamil} give arguments for a
divergence faster than any power of $1/T$ as $T \rightarrow 0$. 

In these calculations, there is no mechanism for dissipation.
Ogata and Anderson argue that special boundary conditions must be used
in order to allow for dissipative effects \cite{ogpwa}. Further neglecting
vertex corrections, they find that the dc-resistivity and conductivity vary as
\begin{equation}
\rho(T) = \frac{1}{\sigma(T)} \sim T^{1-2\alpha} 
\end{equation}
where $\alpha$ is the 
single-particle exponent. For $\alpha \ll 1$ as we have in the 1D \hm\ 
the resistivity varies nearly linearly with temperature. This behaviour
should be closer to real systems than the Hamiltonian-based calculations
outlined before. The frequency dependence then is determined by
a relaxation rate linear in $\omega$: 
\begin{equation}
\label{ogeq}
\sigma(\omega)
\sim \frac{\omega^{2\alpha}}{i \omega + \omega \tan \pi \alpha} \;\;\;. 
\end{equation}
That the optical
conductivity in real materials could essentially
probe the density of states and thus depend on
powers of $\alpha$ had also been conjectured earlier \cite{bas}.

Finally, some information can also be obtained by other methods. Carmelo
and Horsch \cite{carhor} calculate the weight of the $\delta(\omega)$-peak
in the conductivity of the 1D Hubbard model directly from the Bethe-Ansatz
wave function and provide an interpretation from the Landau-Luttinger-liquid
point-of-view (Section \ref{lalu}). 

Obviously, while electron-electron scattering is one conduction-limiting 
mechanism, experiments often probe other influences: scattering off
impurities and phonons. Much work has been done on the impurity problem,
less on electron-phonon scattering, and we start with the former. 

\subsection{Electron-impurity scattering}
We shall
proceed in two steps: (i) a single (or double) impurity and (ii) a system
containing a finite concentration of impurities. 

If we consider a single impurity in a \LL , it is convenient to compute
the conductance $G$ rather than the conductivity $\sigma$ of the system.
The conductance is defined on a sample of finite dimensions by $G = 1/R$
in terms of the resistance, and related macroscopically to the conductivity
$\sigma = 1/\rho$ by $G = \sigma A / L$ where $A$ is the cross section
($=1$ in our 1D problems) and $L$ the length of the system. 
Microscopically, $G$ can be computed via a Kubo formula \cite{kafi}
\begin{equation}
\label{gcon}
G = \lim_{\omega \rightarrow 0} \frac{1}{\hbar L \omega}
\int_0^L \! dx \int \! d\tau e^{i \omega \tau} \langle T_{\tau} \:
J(x\tau) J(00) \rangle \;\;\;.
\end{equation}
For reference, one can evaluate this expression for the impurity-free \LL\
and finds
\begin{equation}
\label{gres}
G = n K_{\rho} \frac{e^2}{h} \;\;\;,
\end{equation}
where $h = 2 \pi \hbar$ and $n$ is the number of channels ($n=2$ for spin-1/2
electrons). This result has been found earlier by Apel and Rice \cite{apri}.
Notice that in contrast to $\sigma$, $G$ is renormalized by
the electron-electron interactions. At first sight, this is not surprising
since the finite length breaks the translational invariance of the system
which is the basis of the independence of $\sigma$ of electronic interactions.
On the other hand, $G$ can also be defined for $L \rightarrow \infty$
and then gives a results different from the Drude weight in the conductivity.
It thus appears that the limit $L \rightarrow \infty$ and the process of
turning on the electron-electron interaction do not commute.

We now include a single or double impurity, mainly following Kane and Fisher
\cite{kafi}. Equivalent results have been obtained by Furusaki and Nagaosa
\cite{funa}. 
For simplicity, we restrict ourselves to spinless fermions
($n=1, K_{\rho} \rightarrow K$). There are two complementary starting points:
a weak impurity where perturbation theory in the impurity potential 
works, and a strong impurity which can be viewed as two weakly connected
semi-infinite \LL s. In the first case, the Hamiltonian for an impurity at
$x = 0$ is
\begin{equation}
\label{hamimp}
\delta H = \int \! dx \: V(x) \: \Psi^{\dagger}(x) \Psi(x)
\end{equation}
with $V(x)$ strongly peaked around $x=0$. The action for its
dominant contribution, backscattering of $m$ electrons 
across the Fermi surface, i.e. transferring momentum $\pm 2 m k_F$, is
\begin{equation}
\label{acimp}
\delta S \approx \sum_{m = - \infty}^{\infty} \frac{v_m}{2} \int \! d\tau
e^{i 2 m \Phi(x=0,\tau)} \;\;\;.
\end{equation}
$v_m$
is the Fourier transform of $V(x)$ at $2mk_F$. Here, we have included the
higher harmonics from $\Psi(x)$ which occur in the \LL\ 
(Section \ref{nonldisp}). 
The weak link, on the other hand, can be modelled by a hopping Hamiltonian
\begin{equation}
\label{hop}
\delta H \approx -t \left[ \Psi_l^{\dagger}(x=0) \Psi_r(x=0) + {\rm H.c.}
\right] \;\;\;.
\end{equation}
Now one traces over the degrees of freedom
away from the impurity ($x \neq 0$) to obtain an effective action for
$x=0$ only. This can then be used to compute the conductance through
the impurity. To this end, one derives renormalization group equations
for $v_m$ or $t_m$ 
\begin{equation}
\label{rg1}
\frac{ d v_m}{d \ell}  = \left(1 - m^2 K \right) v_m \;\;, \hspace{1cm}
\frac{ d t_m}{d  \ell} = \left(1 - \frac{m^2}{K} \right) t_m \;\;.
\end{equation}
For repulsive interactions $K < 1$, 
the most relevant backscattering term $v_1$ increases under scaling, i.e. an
initially weak impurity behaves effectively as a strong one. This qualitative
conclusion is supported by the
strong-coupling limit where $t_1$ (and all higher $t_m$)
is irrelevant, i.e. the
two \LL s are effectively isolated. The impurity thus produces total reflection
for repulsive interactions. On the other hand, for attractive interactions, 
$K > 1$, all $v_m$ are irrelevant and at least $t_1$ is relevant, i.e.
the impurity allows for total transmission. At the fixed point,
in the former case the resulting conductance is $G=0$, in the latter one has 
the ideal \LL\ conductance $G = K e^2 / h$.

When temperature $T$, frequency $\omega$, or voltage $V$ are finite, they 
provide an effective cutoff to the \rg\ 
flow and produce power-law corrections to
the fixed point conductances. One finds [with $\Omega = \max(\omega,T,V)$
and $n=1$ for spinless fermions]
\begin{equation}
\label{concorr}
G(\Omega) = \frac{e^2}{h} \left[ K - \sum_{m=1}^{\infty} a_{m\Omega}
\mid v_m \mid^2 \Omega^{2(m^2 K -1)} \right] \;\;\;.
\end{equation}
The expansion coefficients $a_{m\Omega}$ are nonuniversal but their ratios
are universal. Power-laws similar to the second terms on the right-hand side
may be derived for transport through a weak link \cite{kafi,funa}. 

Including spin degrees of freedom, the physics becomes much richer. 
The Hamiltonian for scattering off an impurity now becomes
\begin{eqnarray}
\delta H & = & \sum_s \int \! dx \: V(x) \: \Psi_s^{\dagger} (x) \Psi_s(x)
\nonumber \\
\label{acimps}
\delta S & = & \sum_{m_{\rho},m_{\sigma}}
v_{m_{\rho},m_{\sigma}} \int \! d\tau \cos \left(\sqrt{2} m_{\rho} \Phi_{\rho}
\right) \cos \left(\sqrt{2} m_{\sigma} \Phi_{\sigma} \right) 
\end{eqnarray}
and for a weak link connecting two semi-infinite spin-1/2 \LL s
\begin{eqnarray}
\delta H & = & -t \sum_s \left[ \Psi_{rs}^{\dagger}(0) \Psi_{ls}(0)
+ {\rm H.c.} \right] \nonumber \\
\label{achops}
\delta S & = & \sum_{m_{\rho},m_{\sigma}} t_{m_{\rho},m_{\sigma}} \int \! 
d \tau \cos \left( \sqrt{2} m_{\rho} \Theta_{\rho}\right)
\cos \left( \sqrt{2} m_{\sigma} \Theta_{\sigma} \right) \;, \hspace{0.15cm}
m_{\rho} = m_{\sigma} \; {\rm mod} \: 2 \;.
\end{eqnarray}
Notice how the impurity couples charge and spin degrees
of freedom. 
In addition to the (charge) conductance $G \equiv G_{\rho}$,
Eq.~(\ref{gcon}), a spin conductance $G_{\sigma} = 2 K_{\sigma} e^2 / h$
can be defined.
Then several phases are possible in principle, depending on the interactions
$K_{\nu}$: (i) the impurity can be irrelevant for charge and spin so that
one recovers the perfect conductor $G_{\nu} = 2 K_{\nu} e^2 /h$; (ii) the
impurity can be relevant in one channel only, i.e. one has a charge
conductor and spin insulator ($G_{\rho} = 2 K_{\rho} e^2 /h$ and
$G_{\sigma} = 0$) or vice versa; (iii) the impurity is relevant and one
has a perfect insulator $G_{\nu} = 0$. Treating the impurity potential
or the weak link by renormalization group, one finds that the most relevant
term is $2k_F$-backscattering of an electron on the impurity. 
In the limit of small impurity potential, the renormalization equation for
$v_{m_{\rho},m_{\sigma}}$ is
\begin{equation}
\label{rgimp}
\frac{d v_{m_{\rho},m_{\sigma}}}{ d \ell} = \left( 1 - 
m_{\rho}^2 \frac{K_{\rho}}{2} - m_{\sigma}^2 
\frac{K_{\sigma}}{2} \right) v_{m_{\rho},m_{\sigma}} \;\;\;.
\end{equation}
In the spin-symmetric case ($K_{\sigma}=1$), the lowest term
with $m_{\rho} = m_{\sigma} = 1$ is
relevant [case (iii)] for repulsive interactions ($K_{\rho} <1$), marginal
for free electrons ($K_{\rho} =1$) and irrelevant [case (i)] for attractive
interactions ($K_{\rho} > 1$), as in the spinless case. Also corrections
to the fixed point conductances can be evaluated, and one finds expressions
very similar to Eq. (\ref{concorr}) from the spinless case.
Case (ii) obtains when, for some reason, a potential component with $m_{\rho}
> 1$ much larger than $v_{1,1}$ and has to be incorporated first. Its
principal effect is to fix the phase of $\Phi_{\rho}$ at the impurity to
a preferred value. With respect to this phase-quenched situation, 
$v_{1,1}$ is irrelevant if $K_{\rho} < 2$ and $K_{\sigma} > 2$ in general,
and if $K_{\sigma} > 1/2$ for symmetric potentials. All indices $\rho
\leftrightarrow \sigma$ if a potential component with $m_{\sigma} > 1 $
is large, and the criterion
of symmetry of the scattering potential is replaced by spin symmetry of
the barrier. There are also several
fixed points at intermediate couplings where explicit calculations are
possible \cite{kafi}. 

For finite impurity strength, consistency requires to include the irrelevant
electronic interactions into the \rg\ scheme. This applies to $g_{1\perp}$ 
in the spin-1/2 case. Matveev et al.
\cite{matv} treat a Fermi gas plus perturbative interactions. 
$g_{1\perp}$ renormalizes $K_{\sigma}$ which
enters the conductance exponent. Translated into a conductance, one
obtains logarithmic corrections to the power-laws characterizing the
(electronic) fixed-point properties. Matveev et al. also claim that
the temperature dependence of the conductance changes to nonmonotonic
for $g_1$ more repulsive than a critical strength. 
Electron-electron backscattering
can be quenched by applying a magnetic field.
This gives an interesting
crossover behaviour to the conductance. At energy scales larger than 
$ 2 \mu_B B$ ($\mu_B$ is the Bohr magneton), backscattering is present
and the system behaves as a \LL . Below $2 \mu_B B$, the external
field blocks the backscattering contribution, and the scaling behaviour
of spinless electrons applies. One then has to match both sets of
equations at $\ell_B = \ln (2 \mu_B B )$. Matveev et al. also argue that
backscattering can be restored by applying a finite bias $V$, and predict
a cusp-singularity in the differential conductance at $V = 2 \mu_B B /e$.

The analysis can be generalized to a situation with two impurities
creating an island between two semi-infinite \LL s  \cite{kafi,furuna}. 
By fine-tuning a parameter, e.g. the energy of the (noninteracting and
spinless) electrons incident on the barrier, one finds resonances with perfect 
transmission at certain energies although they have a finite width even at 
$T=0$. As one turns on repulsive interactions, interesting changes take place.
One still needs to tune the energy of the incident electrons in order to
make the $2k_F$-backscattering matrix element $v_1$ vanish. 
(i) Then, however, the resonances become infinitely sharp as 
$T \rightarrow 0$ --
the interactions suppress all off-resonance conductance. (ii) The conductance
exactly at resonance depends on the strength of the electron-electron
repulsion and, eventually, on the impurity strength at higher multiples of
\tkf . In particular, for $1 > K > 1/2$ (and also for $1/2 > K > 1/4$ provided
that $v_2$ is small enough), one recovers the full \LL\ conductance $K e^2 / h$
at resonance with zero conductance off resonance. Only for $K < 1/4$, or
$K < 1/2$ and $v_2$ large enough, is zero conductance obtained. Of course,
for attractive interactions, the barriers become irrelevant, there
are no resonances and one recovers the full \LL\ conductance without 
fine-tuning. 

Including spin, there are important differences, as can be seen easily
in the limit of very strong barriers \cite{kafi}. 
The charge on the island now is
discrete; if it is odd, there will be a spin degeneracy as for a local
magnetic moment. This is reminiscent of the Kondo effect, where a
magnetic impurity ($s=1/2$ in the simplest case) is embedded in a Fermi
sea of electrons. Resonant transmission through the island is again
possible upon fine-tuning one (or several) parameters.

Generically, there are two types of resonances which can be achieved
tuning one parameter only. The Kondo resonance is the generalization
of the spinless  fermion resonance discussed above. Suppose that we
have tuned the \tkf -backscattering term (\ref{acimps}) to zero. Transmission
will then depend on whether the next-to-leading terms ($m_{\rho}$ or 
$m_{\sigma} = 2$) are relevant or not. For $K_{\sigma} = 1$, $v_{1,2}$ is
harmless, and $v_{2,1}$ blows up only if $K_{\rho} < 1/2$.
This implies that, for $1 > K_{\rho} > 1/2$, both spin and charge are
perfectly transmitted on resonance (although a single barrier would be
totally reflecting), but for $K_{\rho} < 1/2$, charge
is totally reflected while spin is transmitted on resonance. Off resonance,
there is no conductance, as for spinless fermions. Of course, for attractive
interactions ($K_{\rho} > 1$) the barriers are irrelevant altogether. 
Allowing for $K_{\sigma} \neq 1$, 
one can also find a phase which transmits charge and reflects spin. 

Another resonance is possible when both $v_{1,1}$ and $v_{2,1}$
become relevant, i.e. $K_{\rho} < 1/2$. For a symmetric potential and
$v_{1,1}$ only, one obtains charge and spin insulating barriers. For
$v_{2,1}$ only, the barriers are charge insulating and spin conducting.
As a function of $v_{1,1}/v_{2,1}$,
one will have a charge resonance with finite conductance
in between two charge insulating phases. In a case with broken spin-rotation
invariance, this intermediate fixed point is accessible perturbatively, and
its properties can be computed in some detail \cite{kafi}. Its main interest
lies in its finite charge conductance, because the generic Kondo resonance
in this regime has $G_{\rho} = 0$ and is thus difficult to observe. 

More relevant experimentally is the shape of the resonance (an 
$I-V$-characteristic, for example) 
as a control parameter $\delta$ (e.g. the gate voltage on
the island) is tuned through the resonance. Perfect resonance is achieved
when the renormalized potential $v^{\star} = 0$. Off resonance, the
conductance will be determined by the growth of $v$ as it flows away
from the fixed point $v^{\star} = 0$. According to Eq. (\ref{rgimp}), 
$v_{1,1}$ grows with an exponent 
$\lambda = 1-K_{\rho}/2-K_{\sigma}/2$ (resp. $1-K$ 
for spinless fermions). Close to the critical point, there will be a
vanishing frequency scale $\Omega \sim \delta^{1/\lambda}$. Here, one then
expects the conductance depend in a universal way on the ratio $\Omega/T$
\begin{equation}
\label{scalfun}
G(T,\delta) \sim \tilde{G}(c \delta / T^{\lambda})
\end{equation}
where $\tilde{G}$ is a universal scaling function. For small argument,
one can expand $\tilde{G}$ to second order about the fixed-point value 
$G^{\star}$. For large arguments, i.e. far from the critical point, one
can match onto the conductance at finite temperature in the 
single-strong-impurity limit. In this way, one finds  
\begin{equation}
\label{ginf}
\tilde{G}(X) \sim X^{-2/K_{\rho}} 
\end{equation}
(drop the index $\rho$ for fermions without spin). Only for a noninteracting
system is the line shape Lorentzian. If interactions are present ($K_{\rho}
\neq 1$), the tails of the resonance line will be suppressed (repulsion)
or enhanced (attraction). For spinless fermions, these scaling arguments
can be backed up by an exact nonperturbative calculation for a special
value of the coupling constant $K=1/2$. Moreover, for another value $K = 1/3$,
relevant for the \LL\ description of the fractional quantum Hall effect
at $\nu = 1/3$ (where $\nu$ is the Landau level filling factor)
\cite{wen} (Section \ref{fqh}), quantum
Monte Carlo simulations give excellent agreement with the scaling prediction
(\ref{ginf}) and, in addition, provide the complete scaling function
$\tilde{G}(X)$ for all $X$ \cite{moon}.

A particularly detailed discussion of the line shapes is given by Furusaki
and Nagaosa either for the tail region of a strong resonance or 
in the limit of strong barriers (weak link) 
\cite{furuna}. They showed that, only for strong electron-electron interaction
(no matter what its sign), both width and height of the peak
vary monotonically with temperature. Nonmonotonic behaviour
in one of these quantities is observed for weaker interactions $1/2 < K < 2$:
for repulsive interactions, the peak \em height \rm
passes through a minimum between
its high-temperature value and the low-$T$ fixed point conductance $K e^2 /h$,
while for moderate attraction, the peak \em width \rm passes through a minimum.
The crossover in both cases is determined by the ratio of temperature to
the island quantization energy $\delta \epsilon \sim v_F / R$. 

These results apply to short-range i.e. well-screened interactions. This
is presumably relevant for quasi-1D organic metals but a doubtful hypothesis
for semiconductor quantum wires where the electron density both in the
wire and in its environment is low. Then, the long-range nature of the
Coulomb interactions has to be taken into account \cite{schuwi}. 
There are two essential modifications \cite{michele}: 
(i) the conductance of even the pure
\LL\ is length ($L$) dependent as
\begin{equation}
\label{gwi}
G(L) = \frac{3 e^2 \nu}{h \sqrt{R_{\perp}/L}} \;\;\;,
\end{equation}
where $R_{\perp}$ is the transverse extension of the quantum wire, and
$\nu = (2/3) \sqrt{(1+g_1) \pi / 4 \zeta}$ (with $g_1$ the 
$2k_F$-component of the Coulomb potential, $\zeta = e^2 / \kappa v_F$,
and $\kappa$ a dielectric constant simulating the environment). The 
length-dependence of $G$ apparently indicates a vanishing of the Drude weight 
in the conductivity of the infinite system. This is interpreted as being
due to
the vanishing compressibility of this system with long-range interactions,
Eqs.~(\ref{susc}) and (\ref{druwei}) 
with $K_{\rho} \rightarrow 0$ [Section \ref{sec1d}, after Eq.~(\ref{rhowig})]. 
(ii) The conductance through an impurity vanishes faster than any power
of $T$ or $V$ (replace $T \rightarrow eV$ below)
\begin{equation}
\label{glr}
G(T) \sim \exp \left[ -\nu \ln^{3/2} (T_0/T)\right]  \;\;\;.
\end{equation}
This is very much reminiscent of threshold behaviour. 

Essentially in disagreement with this work, 
an earlier calculation \cite{fks} finds
a power-law variation of the resistivity with temperature 
by treating the scattering with a \em finite \rm density of impurities 
in Born approximation. Using the same method, Ogata and Fukuyama later
studied the crossover taking place as a function of system size and
temperature, between regimes of quantized conductance 
implying infinite dc-conductivity, and finite conductivity
\cite{ogafuk}.
They show that, for small $L$, one better
thinks in terms of a conductance $G = 2 K_{\rho} e^2 / h$ (including
spin) while beyond a given system size (determined by temperature and
the elastic mean free path), the conductance crosses over to a
$1/L$-behaviour which implies finite dc-conductivity. While details
of their prediction may depend on computational procedures, the
existence of such a crossover seems quite plausible both for 
short and long range interactions.

We now pass on to the problem of many randomly positioned impurities. Here,
the interference of the scattered electrons becomes important and can
lead to localization \cite{leera}. In a noninteracting
1D electron gas in the presence of disorder, all states will be localized
\cite{mottwo}. This need no longer be so if electron-electron
interaction is turned on, and we have given a general discussion of their
mutual renormalization in Section \ref{phonie}. Here we sketch their 
influence on transport.

One important feature is already apparent when comparing the \rg\ flow
of a single impurity [(\ref{rgimp}) with $m_{\rho} = m_{\sigma} = 1$]
and of many impurities (Section \ref{phonie})
in a \LL . For the noninteracting system, the single impurity is marginal,
i.e. does not change its conductance, while a finite concentration of
impurities is strongly relevant and leads to localization (the interacting
system follows the same logic). The difference is due to quantum interference,
i.e. an electron multiply scattered off impurities interferes with its time
reversed shadow. This leads to an effective backward scattering of the electron
and enhanced localization. The process is absent for a single impurity. 

The \rg\ equations allow to determine, in some cases, 
the localization length $L_{loc}$
and temperature $T_{loc} = v_F/L_{loc}$ beyond/below which localization 
takes place. For small disorder ${\cal D} \rightarrow 0$, 
and close to the critical surface where the disorder is marginal, one
finds respectively
\begin{equation}
\label{lloc}
L_{loc} \sim {\cal D}^{-1/(3-K_{\rho} -K_{\sigma}^{\star})} \;\;, 
\hspace{0.5cm}
L_{loc} \sim \exp \left( \frac{K_{\rho}-2}{{\cal D} - (K_{\rho}-2) Y_{\sigma}}
\right) \;\;\;.
\end{equation}
In the spin-gap phases, these equations change [put $K_{\sigma}^{\star} = 0$
and replace $\exp ( \ldots)$ by $\exp\{1/\sqrt{9{\cal D} - (K_{\rho}-3)^2}\}$ 
in (\ref{lloc})]. 

To determine the temperature-dependent conductivity $\sigma(T)$, we observe that
thermal fluctuations will break coherence at a length scale $\xi_T = v_F / T$,
and renormalization will stop there. The conductivity can then be calculated
in Born approximation. In the delocalized phase, one obtains
$\sigma \sim T^{-1-\gamma} $
with $\gamma = K_{\rho}^{\star} - 2$ in the TS region and $\gamma = 
K_{\rho}^{\star} - 3$ in the SS region, and $K_{\rho}^{\star}$ is
the \rg\ fixed point value. In the localized region,
$T_{loc}$ sets a crossover scale: for $T>T_{loc}$, conductivity first
increases with decreasing temperature, and only for $T< T_{loc}$ the
quantum interference leading to localization and a decreasing $\sigma(T)$
sets in. Above (below) $T_{loc}$ ($L_{loc}$), quantum interference is 
unimportant and one has a diffusive regime (absent for noninteracting 1D 
electrons). Furusaki and Nagaosa have refined this picture by pointing
out that there is another temperature scale $T_{dis} = v_F / k_B R > T_{loc}$
where $R$ is the mean impurity distance \cite{funa}. For $T_{dis} > T >
T_{loc}$, there is no localization (quantum interference) and the impurities
behave as isolated. Electron-electron interactions are present however,
and the conductivity of the system is governed by the scattering off 
individual impurities as considered in the beginning of this Section.
$T_{dis}$ necessarily exceeds $T_{loc}$ because localization can take
place only on length scales beyond the mean impurity distance.

\subsection{Electron-phonon scattering}
One may finally inquire about the influence of electron-phonon scattering
on the conductivity of a  \LL . The \rg\ equations for the electron-phonon
problem (\ref{rgphon}) are strongly controlled by $\omega_{ph} / E(\ell)$.
For $\ell > \ln (E_F/\omega_{ph})$, they reduce to
a purely electronic problem with renormalized starting parameters. 
For an incommensurate system where there are no non-Luttinger
interactions left in the charge channel, once all retardation effects
have been renormalized away one expects to find a Drude peak
with a renormalized weight $2 v_{\rho}^{\star} K_{\rho}^{\star}$ where
$v_{\rho}$ and $K_{\rho}$ are obtained from (\ref{rgphon}).  
These parameters generically decrease 
under renormalization so that the conductivity is lowered by 
electron-phonon scattering. An exception occurs only in the 
high-phonon-frequency
regime of models with significant forward scattering which do have 
dominant superconducting fluctuations \cite{scpol}. At temperatures
above the phonon frequency, the renormalization stop is determined by
$T$, and one can take over the conductivity results from the impurity
scattering problem. 

Interesting effects occur in the optical conductivity $\sigma(\omega)$.
In the presence of phonons, a new absorption process (Holstein absorption)
is allowed: upon absorbing a photon ($\omega$ finite, $q \approx 0$), 
one creates a particle-hole pair [$\omega_{p-h} = \omega - \omega_{ph}(q)$] 
and a phonon [$\omega_{ph}(q)$] whose essential task is
to take up the momentum imparted by the particle-hole pair. Such a process
is possible, of course, only for $\omega > \omega_{ph}$, and the 
additional optical conductivity generated, has been computed in
second order for a 
\tkf -phonon as
\cite{hol}
\begin{equation}
\label{condhol}
\sigma_{hol} (\omega) \sim \Theta(\omega - \omega_{2k_F}) \mid
\omega - \omega_{2k_F} \mid^{1 - \alpha_{CDW}} \;\;\;,
\end{equation}
where $\alpha_{CDW}$ is the CDW correlation function exponent (\ref{rcdw}).
In higher orders, one has to take account of the lattice softening induced
by the electron-phonon coupling, and there will be Holstein conductivity 
for all $\omega > 0$ varying again as a power-law for small $\omega$.
Small momentum scattering, on
the other hand, in inefficient in generating additional conductivity in
the absence of band curvature. Physically, this is so because the 
particle and the hole generated travel with the same group velocity and
therefore will recombine with probability one. In the presence of 
band-curvature, there is a finite forward contribution to the Holstein
conductivity.

\section{The notion of a Landau-Luttinger liquid}
\label{lalu}
We have seen in Section \ref{bethans} that the Bethe Ansatz solution of
the Hubbard model (\ref{betwv})
is generated from the distributions of two quantum numbers $\{I_i\}$
and $\{J_{\alpha}\}$ via the Lieb-Wu equations (\ref{liwu1}) and 
(\ref{liwu2}). 
In the ground state, $\{I_i\}$ and $\{J_{\alpha}\}$ occupy consecutive
integer or half-odd-integer values once and only once, so that the
distribution functions become
\begin{equation}
\label{laludis}
M_c^0(q) = \Theta(2k_F - |q|) \;\;, \;\;\; 
N_{\downarrow}^0(p)  = \Theta(k_{F \downarrow} - |p|) 
\end{equation}
in the limit $L \rightarrow \infty$. The single occupancy of $q$- and
$p$-states suggests that (pseudo)-
particles associated with these quantum numbers behave as fermions. 
Carmelo and collaborators have used this fact to construct a formalism
which allows an interpretation of the Bethe-Ansatz solution in terms
of a generalized Fermi liquid of charge- and spin-pseudo-particles,
and proposed the name ``Landau-Luttinger liquid'' to integrable quantum
systems exhibiting this structure \cite{carhor}, \cite{car1}-\cite{carend}.
The low-energy physics is fully controlled by departures of the
distribution functions from their ground state forms (\ref{laludis}).

Consider a Hubbard model off half-filling in a magnetic field. The 
$SO(4)$-symmetry is therefore broken down to $U(1) \times U(1)$. In this
case, all low-energy excitations of the \hm\ are given by real roots
$\{k_i\}$, $\{ \Lambda_{\alpha}\}$ of the Lieb-Wu equations (\ref{liwu1})
and (\ref{liwu2}), and are functionals of the distributions $M_c(q)$ and
$N_{\downarrow}(p)$. Quantities like the energy $E$ or the momentum 
$P$ of a state, Eq.~(\ref{bethen}), or magnetization and particle number,
are therefore functionals of $M_c(q)$ and $N_{\downarrow}(p)$, too. 
All low-energy states only have \em small \rm deviations $\delta_c(q)$,
$\delta_{\downarrow}(p)$ from their ground state distributions
\begin{equation}
M_c(q) = M_c^0(q) + \delta_c(q) \;\;, \hspace{0.8cm} 
N_{\downarrow}(p) = N_{\downarrow}^0(p) + \delta_{\downarrow}(p) \;\;.
\end{equation}
The smallness of $\delta_c(q)$ and $\delta_{\downarrow}(p)$ allows an
expansion of the energy in powers of these deviations $E = E_0 + E_1
+ E_2 + \ldots$. Here, $E_0$ is the ground state energy, and
\begin{eqnarray}
E_1 & = & \frac{L}{2 \pi} \left\{ 
\int_{-\pi}^{\pi} \! dq \:  \delta_c(q)\: \varepsilon_c (q) + 
\int_{-k_{F \uparrow}}^{k_{F \uparrow}} \! dp \: \delta_{\downarrow} (p) \:
\varepsilon_{\downarrow}(p) \right\} \;\;\;, \\
E_2 & = & \frac{L}{(2 \pi)^2} \left\{ \int_{\pi}^{\pi} \! dq 
\int_{-\pi}^{\pi} \! dq' \delta_c(q) \frac{f_{cc}(q,q')}{2} \delta_c (q')
\right. \nonumber \\
& & + \int_{-k_{F \uparrow}}^{k_{F \uparrow}} \! dp 
\int_{-k_{F \uparrow}}^{k_{F \uparrow}} \! dp' \delta_{\downarrow} (p)
\frac{f_{ss}(p,p')}{2} \delta_{\downarrow} (p') \nonumber \\
& & \left. + \int_{-\pi}^{\pi} \! dq 
\int_{-k_{F \uparrow}}^{k_{F \uparrow}} \! dp \delta_c(q) f_{cs}(q,p)
\delta_{\downarrow} (p) \right\}
\end{eqnarray}
in precise analogy to the Fermi liquid [Eq. (2.1)]! 
The quantities $\varepsilon_c(q)$
and $\varepsilon_s(p)$ are the renormalized pseudo-particle energies,
and the functions $f_{cc}, \; f_{ss},$ and $f_{cs}$ describe the
pseudo-particle interactions. The momentum, particle number, and magnetization
are linear in the deviations and therefore independent of the 
pseudo-particle interactions. Also the low-lying excitations involve
only a single pseudo-particle, and the interaction term is of order $1/L$
with respect to the kinetic energy and unimportant. On the other hand,
the asymptotic decay of correlation functions is controlled by finite
densities of pseudo-particles with low energy, and therefore determined
by the pseudo-particle interactions $f$. These interactions
can be related to both the elements of the dressed charge matrix $Z$
(\ref{dcm}) and to the scattering phase shifts at the Fermi surface
\cite{aren}. The influence of the interaction $U$ and the external fields
$H$ and $\mu$ on the low-energy properties essentially is through the
pseudo-particle interactions $f$. 

Though extremely similar in structure to the Fermi liquid, these 
Landau-Luttinger liquids differ in some important ways. 
Unlike the Fermi liquid, 
the pseudo-particles describe collective charge and spin modes of the
physical system (Section \ref{bethans}), and a construction of the
physical electrons in terms of these pseudo-particles has not been
achieved yet. Single-particle excitations constructed
out of one holon and one spinon do not map onto free electrons 
as $U \rightarrow 0$. Finally, the pseudo-particle excitations here
refer to exact eigenstates of the Hubbard Hamiltonian whereas the Fermi
liquid quasi-particles are made from a superposition of such eigenstates
and therefore decay with time.

Similarities and differences to the Fermi liquid can be gauged quite
accurately from a study of two-particle excitations \cite{carhor,car3}. 
The dynamical charge- or spin-susceptibility 
($\nu = \rho, \; \sigma$) has the spectral decomposition
\begin{equation}
\chi^{(\nu)}( \bk, \omega ) = - \sum_j \left| \left\langle
j \left| \nu( \bk) \right| 0 \right\rangle \right|^2
\frac{2 \omega_{j0} }{\omega_{j0}^2 - (\omega + i0)^2} \;\;\;,
\end{equation}
where $|j \rangle $ is an eigenstate with energy $\omega_{j0}$ relative to the
ground state $|0 \rangle$. Carmelo and Horsch observe that even in a
Fermi liquid, in the limit $\bk \rightarrow 0$, only matrix elements
involving single-pair excitations connect to the ground state
\cite{carhor,car3}. In this limit, the pair excitations become real
one-electron--one-hole excitations, and the corresponding matrix element
with the ground state reduces to unity. This fact can be taken as
a two-particle criterion for Fermi liquids. Unlike the quasi-particle
residue, these matrix elements (there are four of them, taking $\rho$ and
$\sigma$ between states with holon or spinon excitations and the ground
state) do not vanish in the generalized Landau-Luttinger liquids,
and are determined by the pseudo-particle interactions $f$. 
The matrix formed by these elements regularly tends towards
the unit matrix as $U \rightarrow 0$, indicating that the long-wavelength
two-particle properties of the Landau-Luttinger liquid smoothly 
evolve out of those of the free Fermi gas as the interactions are turned
on. Adiabatic continuity therefore holds in the long-wavelength
two-particle excitations.
From a study of the charge and spin currents, one can determine
both the charge and spin Drude weights of the conductivities, but also
the charge and spin of the pseudo-particles themselves. All except the
spinon charge (zero) depend on $U$ and the external fields and are
not fixed to canonical values.

From the Fermi liquid character of the pseudo-particle excitations, we
expect that we can find a framework similar to Chapter \ref{chaplm}
for their low-energy description, and this is indeed the case \cite{carend}.
In particular, one can formulate operator descriptions of these 
pseudo-particles, separately in the charge and spin sectors, which obey
to a fermionic algebra. The low-energy structure can also be analyzed
with conformal field theory, where one finds the typical
tower structure of charge, current and sound excitations both for charges
and spin. The particle-hole excitations are described by a
$U(1)$-Kac-Moody algebra with central charge $c=1$, and the 
Virasoro generators can be constructed explicitly from the currents.
Of course, one can then construct an effective boson description of the
fermionic pseudo-particles.

However, one always has to remember that the
pseudo-particles are not perturbatively related to physical particles,
and their quantum numbers are not the quantum numbers of real excitations.
While they clarify the structure of the theory to a considerable extent,
they still do not allow for a straightforward computation of the
physical correlation functions. 

The single-particle properties of Luttinger liquids are distinctively
different from Fermi liquids. Their two-particle properties are
distinct at larger wavevectors, but in the centre of the Brillouin zone,
they are very similar. In this sense, they can be considered as almost
Fermi liquids. The notion of a Landau-Luttinger liquid is one formal way
of making these connections explicit.

\chapter{Alternatives to the Luttinger liquid: 
spin gaps, the Mott transition, and phase separation}
\label{mottch}

The \LL\ is one possible low-energy state of 1D fermions, realized when there
is no gap in the excitation spectrum. There are several other possibilities:
states with a gap in the spin excitations or/and in the charge excitations
(the Mott insulator) or phase separation. 
(A more detailed review of the Mott transition in 1D has been written
by Schulz \cite{heinzmott}.) In many models, 
there is a duality between the spin and charge degrees of freedom, and 
consequently the methods to describe them are closely related. 

\section{Spin gaps} 
\label{spingaps}
Spin gaps occur in spin-rotationally invariant models
when the electron-electron backscattering is effectively attractive. We
have seen an example for \rg\ scaling in this situation in Section 
\ref{scatt}, Eqs.~(\ref{kot}), when $g_1 < 0$.
Scaling was towards strong coupling indicating that the \LL\ fixed point
was unstable but naturally, \rg\ alone cannot tell us
much about the physics in this situation. The generic model for this problem
is given by the Hamiltonian (\ref{hlutt}) plus (\ref{hperp}) with $g_{1\perp}
< 0$, and an easy solution was provided 
by Luther and Emery \cite{lutem}. For explicit 
spin-rotation invariance, we add a process $g_{1\|}$ to the Hamiltonian,
whose effect is to renormalize the $g_{2\nu} \rightarrow g_{2\nu} -
g_{1\|} / 2$. After the canonical
transformation (\ref{canon}), $H_{\sigma}$, Eq.~(\ref{hdiag}) is diagonal
with a renormalized velocity $v_{\sigma}$ (\ref{vnu}), and using (\ref{phastf}),
$H_{1\perp}$ becomes
\begin{equation}
\label{hlem}
H_{1\perp} = \frac{2 g_{1\perp}}{(2 \pi \alpha)^2} \int \! dx
\cos \left[ \sqrt{8 K_{\sigma}} \Phi_{\sigma}(x) \right] \;\;\;.
\end{equation}
The essential observation now is that for $K_{\sigma} = 1/2$, 
i.e. $ g_{1\|} - 2 g_{2\sigma} = - 6 \pi v_F / 5$, a sizable 
attractive interaction for a spin-rotation invariant model ($g_{2\sigma} = 0$),
(\ref{hlem}) can be written as a bilinear in spinless fermions (\ref{slf})
\begin{equation}
H_{1\perp}  =  \frac{g_{1\perp}}{2 \pi \alpha} 
\int \! dx \left[ \Psi_+^{\dag}(x) \Psi_-(x)
e^{2 i k_F x} + {\rm H.c.} \right] 
\end{equation}
in an external potential $ (g_{1\perp}/\pi \alpha )\cos (2  k_F x)$. 
The kinetic energy $(2 \pi v_{\sigma} / L) \sum_{rp} \sigma_r(p) $
$ \sigma_r(-p)$ can also be written in fermion representation, and the
total Hamiltonian
\begin{equation}
\label{hamfe}
H' = v_{\sigma} \sum_{r,k} r k  c^{\dag}_{r,k} c_{r,k} +
\frac{g_{1\perp}}{2 \pi \alpha} \sum_{k} \left( c^{\dag}_{+,k} c_{-,k-2k_F}
+ {\rm H.c.} \right)
\end{equation}
can be diagonalized by a Bogoliubov transformation. In (\ref{hamfe}), 
$k \approx rk_F$. From the sine-Gordon form of the boson representation 
(\ref{hlem}), 
it is apparent that the $\Psi_r^{\dag}(x)$ create solitons rather than
electrons.
The eigenvalue spectrum is
\begin{equation}
E_{r,\pm}(k) = v_{\sigma} k_F \pm \sqrt{(k - rk_F)^2 + \Delta_{\sigma}^2}
\;\;\;, \hspace{1cm} \Delta_{\sigma} = \frac{g_{1\perp}}{2 \pi \alpha} \;\;\;,
\end{equation}
where $\Delta_{\sigma}$ is the spin gap at the Fermi level.

When $K_{\sigma} \neq 1/2$, the problem can no longer be solved exactly.
Renormalization group arguments, however, support the existence of a
spin gap 
\begin{equation}
\label{spingap}
\Delta_{\sigma} \sim \frac{v_F}{\alpha} \exp \left( \frac{\pi v_F}{ g_1 }
\right)
\end{equation}
for all negative values of $g_{1\|} = g_{1\perp}$ and, with a different
functional dependence, for all $g_{1\perp} < 0 , \; g_{1\|} < | g_{1\perp} |$
\cite{chui}. This conclusion is reached
by scaling the model onto the Luther-Emery
line $K_{\sigma}(\ell_{LE}) = 1/2$ and relating the gap to the Luther-Emery gap 
$\Delta_{LE}$ by
the length of the scaling trajectory $\Delta_{\sigma} = \Delta_{LE} 
\exp(- \ell_{LE})$. The gap may also be obtained from a homogeneity requirement
of the partition function \cite{elp} or by using the exact solution of the
sine-Gordon model \cite{lusg}.

The Hamiltonian (3.49) plus (5.1) is recognized as the quantum-sine-Gordon
model which is equivalent to the massive Thirring model \cite{mandel,colem}.
Both models are related to the spin-$1/2$ Heisenberg chain \cite{lusg,luso},
and it is not surprising that they can be solved by Bethe Ansatz resp. the
quantum-inverse-scattering method \cite{korerev,izsk,bergtha,fowzo,skly}.
Haldane constructed a renormalized Bethe Ansatz solution and could determine
some correlation functions of these models \cite{hasigo}.

This gap has dramatic consequences for the physical properties. 
It implies long-range order in the $\Phi_{\sigma}$-field. This is best
seen by going back to the boson representation of $H_{\perp}$ which 
has the form of a quantum-sine-Gordon Hamiltonian. For $g_{1\perp}$ negative
and scaling to strong coupling, the energy will be minimized by 
$\langle \cos (\sqrt{8} \Phi_{\sigma} \rangle = 1$, i.e. 
$\sqrt{8} \Phi_{\sigma}
= 0 \: {\rm mod} \: 2 \pi$. The $\Theta_{\sigma}$-field gets disordered, and
correlation functions containing exponentials of $\Theta_{\sigma}$ will decay
exponentially with a correlation length $\xi_{\sigma} = v_{\sigma} / 
\Delta_{\sigma}$. 
This cuts off the divergences in the SDW and
TS correlation functions, while SS and CDW continue to diverge. Their
exponents can be obtained by setting formally $K_{\sigma} = 0$ in
(\ref{rcdw}) and (\ref{ass}). 
(Due to the breaking of the $SU(2)$ spin-symmetry to $U(1)$ by our abelian
bosonization scheme, the cutting off of the divergence is obvious only for the 
$S_z = \pm 1$-components of TS and the $x$- and $y$-component of SDW. 
The representation of the TS$_0$ and SDW$_z$ 
(\ref{sdwz}) rather suggests a cancellation of two individually divergent 
terms with
ordered $\Phi_{\sigma}$-fields, which is also found in
a \rg\ calculation of the correlation functions \cite{pref}. 
The conclusion of non-diverging TS and SDW correlations in all components
is firm, however, and required by spin-rotation invariance.)

The negative-$U$ \hm\ falls into this universality class, 
and the spin gap can also be calculated
from the Bethe Ansatz \cite{ultz}. A physical mechanism for the generation
of attractive interactions is electron-phonon interaction, and
a \rg\ treatment of this problem has been presented in Section \ref{phonie}
\cite{phonons}. In an incommensurate 
system where repulsively interacting electrons are
coupled to phonons of finite frequency, both the electron-phonon coupling and
the phonon frequency determine if the system scales towards a \LL\ fixed
point or into a strong-coupling region with a spin gap. The first alternative
has been discussed in Section \ref{phonie}. If the phonon frequency 
$\omega_{ph}$ is small and the coupling constant $\gamma_1$ big enough,
the system will pass beyond the \LL\ fixed point $g_{1\perp}^{\star} = 0,
\; K_{\sigma}^{\star} = 1$ before all retardation effects are scaled out,
and flow into the spin gap region. 
The scaling out of the retardation effects
then provides another factor $\omega_{ph} / E_F$ on the right-hand side
of (\ref{spingap}), and $g_1 \rightarrow g_1(\ell = \ln[E_F/\omega_{ph}])$
in the exponent in (\ref{spingap}).  
Usually, CDW correlations are dominant, except for
a Holstein-type electron-phonon coupling at sufficiently high phonon frequency,
where superconductivity is found \cite{scpol}. 
Here, the electron-phonon system is in the universality class of the 
Luther-Emery model. If the phonon frequency is low enough, it may be more
appropriate to start out from the Peierls mean-field limit \cite{Peierls}, 
and correct it by quantum fluctuations and interactions \cite{phonons}. 
In any case, the formation of a 3D CDW is preceded by the opening of a 1D 
spin gap on the chains. 

Examples for opening of spin gaps when the spin-$SU(2)$ is broken, are given
by Giamarchi and Schulz \cite{giaso}.

\section{The Mott transition}
\label{motttr}
In half-filled bands, a Mott metal-insulator transition may occur as a
results of commensurability, manifest in Umklapp scattering (\ref{h3perp})
becoming relevant. The problem is completely analogous to the spin-gap 
situation discussed above. In the special case of the Hubbard model at 
$n = 1$,
there is a duality transformation relating positive to negative $U$
\begin{equation}
c_{i\uparrow}  \rightarrow (-1)^i c_{i\uparrow}^{\dag} \;\;\;,
\hspace{1cm} c_{i\downarrow} \rightarrow c_{i\downarrow} \;\;\;,
\end{equation}
i.e. a particle-hole transformation on one spin species only. In the boson
representation, $U \rightarrow -U$ simply leads to an exchange of the
roles of charge and spin fluctuations. All results of the preceding section
then carry over, and the spin gap becomes the Mott-Hubbard charge gap, 
separating the upper and lower Hubbard (sub)bands. 

This picture can be extended to include the effects of doping (in the spin-gap
problem, this corresponds to the introduction of a magnetic field 
which, however, is required to be unrealistically strong to have visible 
effects). We take the doping level $\delta = 1 - n$ and, due
to charge conjugation symmetry at $n=1$, do not 
distinguish electron from hole doping. 
Due to the
commensurability pinning, it is expected that $k_F$ does not respond
immediately to doping, and that one will rather create charged defects
in a commensurate SDW background. 
Therefore the gap structure is expected to 
persist for some finite doping range, but
the chemical potential will move above (below) the gap to accommodate the
additional charge carriers. 

A more refined formulation of the model is necessary, however, to obtain
a detailed picture of the Mott transition \cite{giam1}. 
We consider the Hamiltonian
of the charge degrees of (\ref{hlutt}) plus (\ref{h3perp}) and apply 
the canonical transformation (\ref{canon}). Unlike Section \ref{hamdia},
we do not require the non-diagonal terms $(2 g_{2\rho}/L) \sum_p
\rho_+(p) \rho_-(p)$ to vanish but just transform so that $K_{\rho} = 1/2$.
In general then, a finite $g_2$-type interaction, not diagonal in the bosons,
will remain. The Hamiltonian then becomes \cite{luso}
\begin{eqnarray}
\label{hsub0}
H & = & H_0 + H_1 \;\;\;, \\
H_0 & = & \sum_k \left[ \left(vk + \mu \right)
: c_{+,k}^{\dag} c_{+,k}: - \left( vk - \mu \right) : c_{-,k}^{\dag} c_{-,k}:
\right] + \frac{g_{3\perp}}{2 \pi \alpha} \int dx \left[ c^{\dag}_{+,k}
c_{-,k} + {\rm H.c.} \right] \;\;\;, \nonumber \\
H_1 & = & \frac{\pi v_{\rho} \sinh (2 \theta)}{L} \sum_p
\left( 2 \rho_+(p) \rho_-(- p) - f_1 \sum_r : \rho_r (p) \rho_r (-p) : \right)
\;\;\;, \nonumber \\
v & = & v_{\rho} \left( \cosh (2 \theta) + f_1 \sinh ( 2\theta) \right)
\;\;\;, \hspace{1cm} \exp ( -2 \theta) = 2 K_{\rho} \;\;\;. \nonumber
\end{eqnarray}
The fermions $c_{rk}$ are spinless, and the boson operators $\rho_r(p)$
refer to these spinless fermions. 
$H_0$ is of the Luther-Emery form and can be diagonalized. For the
half-filled system, the interactions simply renormalize the gap as 
in the preceding section. 
The $f_1$ term 
has been introduced by Schulz \cite{schuga} and does not affect the gap.
$f_1$ is arbitrary and can
be fixed as convenient. 

At half-filling, the chemical potential is in the centre of the gap, and
the lower (upper) band is completely filled (empty). Doping will
shift it into the upper ($ \delta < 0$) or lower ($\delta > 0$) subband
generating a finite occupation of negative or positive carriers there.
For very low energies ($ \ll \Delta_{\rho}$),
the physics is determined by the partially occupied band only.
The band structure can then be linearized again
around the new Fermi level $k_c = \pi |\delta|$, and one keeps only
interaction processes at the new Fermi surface. $f_1$ can now be fixed
so as to cancel all $g_4$-type terms arising, and the new subband
Hamiltonian is \cite{giam1}
\begin{eqnarray}
H & = & \sum_k v_c k \left( a_{+,k}^{\dag} a_{+,k}  - a_{-,k}^{\dag} 
a_{-,k} \right) + 2 \pi v_{\rho} \sinh (2 \theta) f(k_c)
\sum_p \rho^{(a)}_+ (p) \rho^{(a)}_- (-p) \;\;\;, \nonumber \\
v_c & =& \frac{\partial E}{\partial k} = \frac{v^2 k_c}{
\sqrt{(v k_c)^2 + (\Delta_{\rho}/2)^2}} \;\;\;, \\
f_1 & = & \frac{1}{\sqrt{1 + (2 v k_c / \Delta_{\rho})^2 }} \;\;\;, 
\hspace{1cm} f(k_c) = \frac{v_c^2}{v^2} \;\;\;. \nonumber
\end{eqnarray}
The fermions $a_{rk}$ now refer to the partially occupied subband, 
and the $\rho^{(a)}_r(p)$ are constructed from these fermions. This
spinless Hamiltonian can now be diagonalized as in Section \ref{secslf},
and the exponent $K$ governing the decay of correlation functions is
given by
\begin{equation}
K = \frac{1}{2} \left[ 1 - \frac{4 v_{\rho} k_c}{\Delta_{\rho}}
\sinh ( 2 \theta) \right] \;\;\;.
\end{equation}
Notice the following: (i) As one goes towards the half-filled band 
$\delta \rightarrow 0$, the Fermi velocity in the partially occupied
band vanishes ($v_c \sim k_c/\Delta_{\rho} \rightarrow 0$) which could
be interpreted as a diverging effective mass $ m \sim 1/\delta$
\cite{gube} as $v_c = k_c / m$ in the vicinity of the Fermi surface.
However, caution is required with this argument, because when only a
few holes are left, the Fermi sea and velocity become ill-defined.
The vanishing of $v_c$ indicates that one has reached the bottom of the
upper subband, and there one recovers the parabolic dispersion of free
particles with a finite effective mass.
(ii) The interactions between the spinless fermions 
vanish $\propto \delta^2$ on account of
the $f(k_c)$-factor, i.e. one is always in the weak-coupling limit
close to the half-filled band. This ultimately justifies the separation
of the Hamiltonian as in Eqs.~(\ref{hsub0}). (iii)
$K \rightarrow 1/2$ as $\delta \rightarrow 0$, which is consistent
with the behaviour of the Hubbard model, Section \ref{sec1d}. 
We do however not make use of any specific feature of this model, so
that these results are valid for any \LL\ close to half-filling
\cite{giam1}. 
(iv) Densely packed, strongly coupled spin-1/2 fermions map onto dilute,
weakly coupled spinless fermions (holons, solitons). This kind of
mapping can be fruitfully applied to many other problems \cite{emcuo}. (v)
Of particular interest is the Drude weight $D$ of the 
conductivity which, as was pointed out by Kohn \cite{flux}, can be taken
as an order parameter for the metal-insulator transition. The spinless
fermion current is proportional to the one of the spin-carrying 
fermions, and the
Drude weight is therefore $D \propto v_c K $, vanishing linearly as 
$\delta \rightarrow 0$ with a slope $ \sim 1/\Delta_{\rho}$. 
(vi) From the mapping onto the 2D commensurate-incommensurate transition,
it follows that the Mott-Hubbard transition is in the 
universality class of the Pokrovsky-Talapov transition
\cite{schuga,gube,pokt}. This is 
the case quite generally for the doping behaviour of models
which, exactly at commensurability, display a Kosterlitz-Thouless transition
as a function of the coupling constant, Eq.~(\ref{kot}), and therefore
applies universally to the metal-insulator transition at \em even \rm
commensurability ratios in 1D. 

At higher energies, the other (completely occupied) band contributes 
to the properties. To treat this case, Gul\'{a}csi and Bedell have 
proposed a bosonization scheme which decomposes the physical fermion
into four new particles, a right- and left-moving fermion for each band
\cite{gube}.
The Hamiltonian then takes the form of a Luttinger model for the
partially occupied band and of a sine-Gordon model for the gapped band. 
One can then calculate various correlation functions and, concerning
e.g. the momentum distribution function $n(k)$, finds a sum of
a Luttinger function (\ref{nkut}), weighted by the doping level $\delta$,
and a term linear in $k$ characteristic for gapped systems
\cite{brech}, rather independent of doping. 

A mapping of strongly coupled spinning fermions onto weakly interacting
spinless holons can also be operated in the Bethe-Ansatz formalism for
the \hm\ \cite{stami}. The Bethe-Ansatz equations can be reformulated
in terms of the charge excitations only and, for small doping, can be
mapped onto weakly coupled holons. The Bethe Ansatz allows the introduction
of a magnetic flux through the Hubbard ring, and the Drude weight can
then be obtained from the second derivative of the ground state energy,
Eq.~(\ref{vflux}). Also, the total optical spectral weight
\begin{equation}
\label{optsum}
\pi N_{\rm tot} \equiv \int_0^{\infty} {\rm Re} [\sigma(\omega)] d \omega
= \frac{\pi}{2L} \langle T \rangle = \frac{\pi}{2L} \left( E_0 - U
\frac{\partial E_0}{\partial U} \right)
\end{equation}
can be computed quite easily, so that the optical properties can be
discussed in some detail.

At half-filling, on a ring of circumference $L$, the Drude weight varies
exponentially $D(L) \sim \exp[ -L / \xi(U) ]$ as $L \rightarrow \infty$,
defining a coherence length $\xi(U)$. $\xi(U) \sim 1/\Delta_{\rho}$
for $U \rightarrow 0$ but vanishes only logarithmically $\xi \sim 1/
\ln(U)$ for $U \rightarrow \infty$. $\xi$ also determines the exponential
decay of the Green function at $n = 1$, and comparing to the sine-Gordon
form of the Hubbard Hamiltonian, is identified as the typical length
of the solitons introduced upon doping. The divergence of $\xi$ as 
$U \rightarrow 0$ suggests that $U=0, \; n=1$ is a quantum critical point,
and power-counting gives $D$ the
scaling dimension zero. Consequently, the singular part of $D$ should
be a dimensionless scaling function
\begin{equation}
D^{\rm sing}(n, L, U ) = Y_{\pm} (\xi \delta, \xi /L) \;\;\;,
\end{equation}
where the index $\pm$ refers to $U >(<) 0$. $Y_{\pm}$ can be determined
both analytically and numerically in various limits. One remarkable
result is that $D$ has a universal jump of $2/\pi$ as $U$ goes from $0^+$
to $0^-$  at $n = 1$ and $L = \infty$ -- the system is insulating at
positive $U$ and metallic at $U<0$. Moreover, at small doping, $Y_+(\xi 
\delta,0) \sim \xi \delta$ as $ \xi \delta \rightarrow 0$. 
The Drude weight grows linearly with
doping, as we have already seen above. It saturates for $ \xi \delta \sim 1$.
On the other hand, the sum rule (\ref{optsum}) is rather independent of 
$\delta$, $L$ and $U$ in the critical region. At $\delta =0$, all the 
spectral weight is in the upper Hubbard band. As one dopes the system,
spectral weight is simply transferred from the upper to the lower Hubbard
band where it goes into the Drude peak, a result which also holds in
higher dimensions \cite{eskes}. When $\xi \delta \sim 1$, most of the
spectral weight resides in the Drude peak, and the gap structure has
been destroyed.

The charge velocity vanishes as $\delta \rightarrow 0$ (the Bethe Ansatz
explicitly gives a finite effective mass to the holons). Close to the metal
insulator transition, the charge entropy then is much higher than the
spin entropy and will dominate the thermopower. This can then be
evaluated from the spinless fermions \cite{thermo}, 
and one finds a hole-like thermopower for $n < 1$ and an electron-like
sign for $n > 1$ with coefficients varying as $1/\delta^2$
\cite{ijmp,stami}. On the other hand, 
the Fermi surface is given by the number of 
electrons in agreement with Luttinger's theorem \cite{lt}.

Kolomeisky used \rg\ to provide a general framework of the Mott transition
in a 1D metal of spinless fermions in an external potential with periodicity 
$k_Fa = (p/q) \pi$ ($p, \; q$ integers) \cite{kolo1,kolo2}. 
There is forward and backward single-particle scattering from this
potential, and the electron-electron interaction is of spinless Luttinger
form ({hslf}). Forward scattering, though, can be eliminated by the same
argument as in (\ref{phishift}) so long as one is interested only in
the conductance which can be used as an order parameter for the transition. 
The backscattering terms are then mapped onto a
model for the commensurate -- incommensurate transition on 2D surfaces,
which occurs at a critical $K_c = 2 / q^2$ \cite{cic}. 
This mapping again identifies the universality classes of the Mott transition. 
Unlike the Hubbard model, a finite critical interaction strength $K_{c} < 1$ is 
necessary for the transition to occur. We can approach it two different
ways: (i) one can decrease $K \rightarrow K_c$ at fixed band-filling
$k_Fa = (p/q) \pi$ (corresponding to a soliton density $n_s =
q k_F / \pi - pa = 0$ in
the 2D problem) (Kosterlitz-Thouless universality class \cite{kothou}); 
(ii) one can vary the bandfilling $k_Fa \rightarrow
(p/q) \pi$ ($n_s \rightarrow 0$) at fixed $K < K_c$ (Pokrovsky-Talapov
university class \cite{pokt}). In both cases, 
there is a universal jump of $K$ and thus of the conductance at
the transition. The insulating phase of course has $G \equiv 0$, and
in the metallic phase
\begin{eqnarray}
\label{nojump}
G & = & \frac{e^2}{h} \left( K + \frac{| n_s |^{2(q^2K-2)} }{q^2}
\right)
\;\;\;, \hspace{0.5cm} (K>K_c) \;\;\;, \\
\label{jumpc}
G & = & \frac{e^2}{h} \left( \frac{2}{q^2}  -\frac{ \rm const.}{q^2
\ln |n_s |} \right) \;\;\;, \hspace{0.5cm} (K = K_c) \;\;\;, \\
\label{jumpd}
G & \rightarrow & \frac{e^2}{h} \frac{1}{q^2} \;\;\;,
\hspace{0.5cm} (K < K_c, \; n_s \rightarrow 0) \;\;\;.
\end{eqnarray}
Eq. (\ref{nojump}) gives the conductance in case (i) above the transition,
and here as well as in (\ref{jumpc}), we have indicated the corrections
predicted for slightly doping ($n_s \ll k_Fa$)
away from the ideal commensurability. Of course,
$K \geq K_c$ includes the renormalization by the irrelevant scattering
processes from
the lattice (in total analogy to the logarithmic corrections found in
Section \ref{scatt}).
These are responsible for the nonanalytic correction terms. 

The universal jumps of the conductance can also be rationalized quite
easily in a language closer to the main development of this article.
One would describe commensurability effects in the presence of
electron-electron interaction by introducing suitable Umklapp operators
$H_{3}^{(q)} \sim \int \! dx \cos [ 2 q \Phi (x) ]$,
transferring $q$ electrons across the
Fermi surface, generalizing Section \ref{scatt}. 
Their scaling dimension will depend on $q$ as $2 - q^2 K$ and therefore
imply a critical value of $K_c(q) = 2/q^2$ for each $q$. The universal 
jump of the
conductance is then immediately obtained by inserting $K_c(q)$ into
(\ref{gres}). Little work has been done on the problem in the presence of
spin. At $q$-even commensurabilities, the above analysis
is extended straightforwardly: the $2 q k_F$-transfer Umklapp operators
become $H_{3}^{(S=0,q)} \sim \int \! dx \cos [ \sqrt{2}q \Phi_{\rho}(x)]$,
where in the superscript we have indicated the fact that the $q$ particles 
transfer a total spin $S=0$. These operators are more relevant than those
transferring finite spin and have scaling dimensions $2 - q^2 K_{\rho}/2$.
The critical $K_{\rho}$ scales as $K_{\rho,c} = 4/q^2$. The half-filled
Hubbard model and the half- and quarter-filled extended Hubbard models
\cite{ehm,ehmhalf,milzot} obey to this relation. An example of an Umklapp 
operator transferring finite spin is given by 
\begin{equation}
\label{parum}
H_{3\|} = \frac{2 g_{3\|}}{(2 \pi \alpha)^2} \int \! dx \cos \left[
\sqrt{8} \Phi_{\rho}(x) \right] \cos \left[ \sqrt{8} \Phi_{\sigma} (x)
\right] \;\;\; 
\end{equation}
which couples charge and spin. It is important in the half-filled extended
Hubbard model where the CDW-SDW transition is 
continuous at weak coupling but becomes first order beyond a tricritical
point at about $U = 2V \approx 4 \dots 5 t$. At the origin is 
$H_{3\|}$ with a scaling dimension $(2 - 2K_{\rho} - 2K_{\sigma})$ 
\cite{ehm,ehmhalf}. For $q$-odd commensurabilities, half-odd-integer spin
is necessarily transferred in Umklapp scattering. The Umklapp operators
therefore  must be of the form (\ref{parum}) although $\Phi_{\rho}$ and
$\Phi_{\sigma}$ have prefactors different
from $\sqrt{8}$. Here, charges and spins
are strongly coupled, and the Mott transition is accompanied by the opening
of a spin gap. The physics of such an odd-$q$ Mott insulator has not been
explored yet.

Transport in commensurate systems is very interesting because
Umklapp processes provide an important relaxation mechanism for charge carriers.
The problem here is that the conductivity does not have a regular
perturbation expansion in the Umklapp operators $g_3$, Eq.~(\ref{h3perp}).
A way out is provided by the memory function formalism \cite{goewoe}. 
Assuming that the system is a normal metal with a finite dc-conductivity
(a strong assumption which needs justification), one can rewrite 
Eq.~(\ref{kubcon}) as
\begin{equation}
\label{mem}
\sigma(\omega) = \frac{2 i v_{\rho} K_{\rho} }{\pi} \frac{1}{\omega + 
M(\omega)} \;\;\;,
\end{equation}
and a perturbative calculation of the memory function $M(\omega)$ is
well-defined. It involves the commutator $[ H, j]$ introduced before
which, for the case of Umklapp processes in a half-filled band, 
Eq.~(\ref{h3perp}), reads
\begin{equation}
\label{comum}
[j(xt),H] = \frac{8 g_{3\perp}}{(2 \pi \alpha)^2} i v_{\rho} K_{\rho}
\sin \left[ \sqrt{8K_{\rho}} \Phi_{\rho}(xt) + \delta x \right] \;\;\;,
\end{equation}
where $\delta =4 k_F - 2 \pi / a$ measures an eventual doping level with
respect to the  half-filled band. 

If $g_{3\perp} \ll 1$, one obtains $\sigma(\omega) \sim 1/\omega$ for
$K_{\rho} > 1$ and $\sigma(\omega) \sim \omega^{3-4K_{\rho}}$ for $K_{\rho} <
1$ if $\omega \gg T$. In the opposite case $T \gg \omega$, one has 
$\sigma(0,T) \sim T^{3-4K_{\rho}}$. Of course, these results are valid only
at sufficiently high $T$ and $\omega$ because at smaller scales, there
will be a charge gap and one expects an activated conductivity. In case
that $g_{3\perp}$ is finite, one can extend these results by performing
a renormalization group calculation such as the one in Section \ref{scatt}
which will be stopped by the finite temperature at a scale $\ell_T 
= \ln(E_F/T)$.
Then, additional temperature dependence in the conductivity
will be generated by inserting
the renormalized values of $g_{3\perp}(\ell_T)$ and $K_{\rho}
(\ell_T)$ into the memory functions. 

Surprising results obtain at lower frequencies where one expects the
influence of the charge gap. Here, one can use the 
Luther and Emery solution \cite{lutem} to diagonalize the 
charge part
of the Hamiltonian in terms of new spinless fermions, and then express
the current in the new fermions \cite{giam1}. The
dc-conductivity comes out infinite
at any finite temperature. The explanation is quite obvious:
the only scattering mechanism for our charge carriers were the Umklapp processes
which, however, have been diagonalized exactly by the Luther-Emery 
transformation. Thermally excited carriers have no dissipation mechanism left.

Away from half-filling, there are two regimes. If $T, \omega \gg v_{\rho} 
\delta$, the energy scales are too high for the Umklapp processes to be
quenched by doping, and one basically recovers the half-filled band results. 
If $T$ falls below $v_{\rho} \delta$, the latter quantity will act as a 
cutoff to renormalization and freeze out the Umklapp scattering.
One will then have a crossover to presumably exponential increase with
$1/T$ of the conductivity characteristic for the incommensurate system. On
the other hand, at zero temperature, as one approaches the half-filled band,
the weight of the Drude $\delta(\omega)$-part vanishes linearly with doping
and with a slope that depends on the charge gap \cite{giam1}, for the \hm\
in agreement with the Bethe Ansatz results \cite{stami}.

Similar results for $\sigma(\omega)$
can be obtained by using Eq. (\ref{conint}). On the other hand, 
the predictions for $\sigma(T)$ agree with the memory function
approach \cite{giam2} only in certain cases \cite{giamil}. 
Giamarchi and Millis suggest that the use of
memory functions is particularly dangerous in cases of infinite dc-conductivity
where the underlying conservation laws may not be incorporated correctly into
this method \cite{giamil}. 
In fact, it seems that there is an obvious contradiction between
the \em assumed \rm finite dc-conductivity of the memory function method
and the infinite conductivity in the Luttinger liquid.

The $q \approx 0$-component of the
two-particle charge-charge spectral function [${\rm Im} R_{\rho}(q,\omega)$
with $R_{\rho}$ from Eq. (\ref{rroh})] can be studied in detail
close to the Mott transition because the charge operator here
has a simple representation in terms of the spinless fermions
$\rho(xt) = \Psi_+^{\dag}(xt) \Psi_+(xt) + \Psi_-^{\dag}(xt) \Psi_-(xt)$
\cite{mori}. At finite doping $\delta$, one finds a two-peak structure:
a low-energy peak with linear dispersion arises from the long-wavelength
density fluctuations within the partially occupied subband. It looses
weight $\propto \delta \rightarrow 0$ as the Mott transition is approached.
This peak is modelled accurately by the effective \LL\ description of Chapter 
\ref{chapll}. However, a second peak at higher energies ($\sim
\Delta_{\rho}$) is quite pronounced 
over a significant doping range. It represents
density fluctuations between the upper and lower Hubbard subbands.
In the Mott insulating phase ($\delta =0$), 
it is the only signal present. The region
in $q$ and $\omega$ in which a simple \LL\ description is valid, therefore
is very small close to the Mott transition and may contain excitations
carrying very little spectral weight \cite{mori}.

Shankar has considered the effect of impurities in
commensurate models \cite{shankar}.
For a half-filled spinless fermion system with nearest-neighbour repulsion
$V \sum_i n_i n_{i+1}$, which has a Mott transition into a CDW state 
at $V = 2t$, he finds that the Mott gap is destroyed by even a small
amount of disorder. This is supported by numerical density matrix
\rg\ \cite{pang}. 
On the other hand, this work points towards persisting
differences between impurities in systems with and without a Mott gap:
if one dopes the system with an additional charge, it is strongly localized
when the interactions are strong enough to open a charge gap while localization
is quite weak in the absence of the gap, for identical disorder configurations.
This situation corresponds to a Mott transition as a function of band-filling
at fixed interaction strength, and Kolomeisky has argued that the elementary
excitations remain solitons, as in the pure Mott system, which, however,
become localized in the presence of arbitrarily small disorder \cite{kolo3}. 
In contrast, for a half-filled (S=1/2) Hubbard model, Shankar predicts the 
Mott-Hubbard charge gap $\Delta_{\rho}$ to survive so long as
the variations of the random potential are bounded to $|\xi (x)|
\ll \Delta_{\rho}$ in agreement with general arguments \cite{kolo3}
and real-space renormalization group
\cite{ma}. The difference may again be due to the impurities coupling 
linearly to the CDW fluctuations which build up the order parameter in 
the spinless model but which are suppressed in the half-filled \hm .

In a different line of work, Horsch and
Stephan \cite{horste} consider the conductivity of a single hole doped
into a half-filled $t-J$- and large-$U$ Hubbard model. They find that, in
addition to the Drude peak, there is conductivity at finite frequencies
varying as $\omega^{-1/2}$ and $\omega^{3/2}$, respectively. These analytical
results are supported by numerical diagonalization studies on rings as 
big as 19 sites. Currently, it is not understood why $\sigma(\omega >0)$
in both models differs qualitatively and also differs from the Luttinger
liquid prediction $\omega^3$. On the other hand, one could speculate
that a crossover to the $\omega^3$ behaviour could occur as a 
finite \em concentration \rm of holes is doped into the Mott insulator and a
Fermi surface forms.

The Ogata-Shiba wavefunction Eq.~(\ref{wvsim}) \cite{ogata} 
also allows to find the spectral properties of one hole
doped into a half-filled $U = \infty$-Hubbard model, the prototypical
Mott-Hubbard insulator. This problem had been examined long time ago
by Brinkman and Rice \cite{briri} who assumed N\'{e}el order for the
spin configuration. They had found a complete localization of the hole 
becoming totally incoherent. Expressed in terms of the spectral function
\begin{equation}
\label{spbrrii}
A(k,\omega) = - \frac{1}{\pi} {\rm Im} G(k,\omega) = 
\frac{1}{2 \sqrt{\omega^2 -4 }} \;\;\;. 
\end{equation}
There is no quasi-particle pole, and $A$ is independent of $k$. 
This result, however, is mainly due to the assumed static antiferromagnetic
N\'{e}el order. In the $U \rightarrow \infty$-Hubbard model, the magnetic
ground state is far from N\'{e}el, and taking the real Heisenberg
ground state, Sorella and Parola get quite different spectral functions
\cite{sorpar}. Generically, a three-peak structure is found which can
be understood qualitatively as a convolution of a holon and spinon 
Green function [$ \sim 1/\{\omega - \varepsilon^{(h)}(k)\} $ resp.
$\sim 1/\sqrt{\omega - \varepsilon^{(s)}(k) }$ \cite{aren}, where 
$\varepsilon^{(h,s)}(k)$ has been defined in Eqs.~(\ref{eholon}) and
(\ref{espinon})]. 
At special wavevectors, two singularities may
coalesce giving the peaks observed. 
Unfortunately, qualitatively different results were published
slightly later by the same authors \cite{parsor} where
only a two-peak behaviour is found. While the latter is in agreement
with the spectral function of a \lm\ with charge-spin separation
only and no anomalous dimension ($\alpha =0$ --
there is no partner for the
holes to $g_2$-interact) \cite{spec1,fogedby},
the reason for the discrepancy between both results is not clear.
The dynamics of a single hole in the 1D $t-J$-model has also been studied
by Horsch and coworkers \cite{horsch} both by numerical diagonalization,
and analytically
within the subspace generated by applying the hopping operator
to the state obtained by annihilating a fermion with $(k,s)$ in the 1D
N\'{e}el state. The two
methods agree in their essential features. The inclusion of quantum fluctuations
in the spin background changes the results in several essential ways
with respect to the dispersionless, incoherent Brinkman-Rice continuum.
(i) It generates
interesting dispersion in the spectra. While the lowest eigenstate disperses
on a scale $J$, the first moment of the spectral function 
$\int_{-\infty}^{\infty} \! d\omega \omega A(k,\omega)$ disperses on a scale
$t$. (ii) To the extent one can gauge from the finite lattice data, there
seem to be three peaks, and the above dispersion behaviour suggests that
spectral weight is mainly transferred, as a function of $k$ between peaks
which have different dispersion. This is not unlike the three-peak structure
initially found by Parola and Sorella for the Hubbard model \cite{sorpar} 
but different from later work by the
same authors \cite{parsor}. (iii) The lower edge of the spectrum
disperses little and remains sharp 
close to $\omega \approx - 2t$, but the high-energy
edge of the Brinkman-Rice continuum gets washed out  into a tail of states.
Comparing the work of the different groups it appears that
the dynamics of a single hole doped into a Mott insulator
is not fully clarified.

\section{Phase separation}
\label{phassep}
For strongly attractive interactions $g_{2\rho} + g_{4\rho} = - \pi v_F$,
$K_{\rho}$ becomes infinite, indicating an instability of 
the Luttinger liquid \cite{matl}. 
The physical interpretation of this transition 
becomes obvious by going back to Eqs.~(\ref{speccoup}) and 
(\ref{susc}) which shows that a divergence in $K_{\rho}$ implies
\begin{equation}
\kappa = \frac{\partial n}{\partial \mu} \rightarrow \infty \;\;\;,
\hspace{1cm} v_{N\rho} \rightarrow 0 \;\;\;.
\end{equation}
Both facts indicate that a particle can be added to the system without
cost in energy; the electron clump into droplets i.e. one has phase
separation. This transition takes place e.g. in the extended Hubbard
model with nearest neighbour attraction \cite{ehm,ehmhalf,milzot,milpen,kuroki}
or in the $t-J$-model with sufficiently large $J$ \cite{tj,helme4,helme2}.

\chapter{Extensions of the Luttinger Liquid}
\label{ext}

This chapter discusses three extensions of the \LL\ picture. The first
two extensions are intimately related: models with two or more bands,
and models with several coupled chains. In the third part, we outline
the important extension to chiral \LL s arising in the fractional 
quantum Hall effect, where a \LL\ with central charge $c \neq 1$ is found.

There are very strong similarities between models with several
bands and models with several chains. The former in fact often model the
bandstructure  of materials with several chains per unit cell. 
Coupling N chains by a hopping matrix element produces N bands. They
also can originate from applying a strong magnetic field to a chain. 
We somehow artificially separate this topic into two
parts on multi-component and multi-chain models mainly because the 
physical questions asked in both parts are rather different. 

To see the similarities more closely, consider first a two-band model
\begin{equation}
\label{htwob}
H = \sum_{k,s,\alpha} \epsilon_{\alpha}(k) c^{\dag}_{ks\alpha} c_{ks\alpha}
+ H_{int} \;\;\;.
\end{equation}
$\alpha$ is the band index, and $\epsilon_{\alpha}(k)$ is the dispersion.
The Fermi momenta $k_{F\alpha}$ and velocities $v_{F\alpha}$ may be different
in general, and the $v_{F\alpha}$ may be positive (electron) or negative 
(hole bands at the centre of the Brillouin zone) as shown in Fig.~6.1.
$H_{int}$ is the interaction Hamiltonian
which contains all the processes $g_i$ discussed before, both within every
band $\alpha$ and between the different bands.
For electrons in a magnetic field $H \| \hat{z}$, 
the single-particle Hamiltonian is
\begin{equation}
\label{hmag}
H_0 = \sum_{ks} \epsilon (k) c^{\dag}_{ks} c_{ks} 
+ h \sum_{k} \left( c^{\dag}_{k \uparrow} c_{k \uparrow} - 
c^{\dag}_{k \downarrow} c_{k \downarrow } \right) \;\;\; , \;\;\; 
h = g \mu_B H /2 \;\;\;.
\end{equation}
With $s \rightarrow \alpha$, this reduces to a spinless variant of (\ref{htwob})
with $\epsilon_{\alpha} (k) = \epsilon (k) \pm h$. Coupling two chains $i=1,2$
with a single-particle tunneling matrix element $\tperp$ gives
\begin{equation}
\label{twoch}
H_0 = \sum_{ksi} \epsilon (k) c^{\dag}_{ksi} c_{ksi}  - \tperp 
\sum_{ks} \left( c^{\dag}_{ks1} c_{ks2} + {\rm H.c.} \right) \;\;\;.
\end{equation}
Here the bonding and antibonding bands disperse with $\epsilon_{0,\pi}(k) = 
\epsilon (k) \mp 2 \tperp$ and are labeled by their
transverse momenta $0$ and $\pi$. The spinless version of (\ref{twoch})
also describes electrons in a transverse magnetic field $H \| \hat{x}$,
and the dispersions $\epsilon_{0,\pi}(k)$ can then also be obtained by
rotating the field around the $y$-axis align it with $\hat{z}$. 
Under this rotation, the interactions transform nontrivially.
The transformation does not affect the ``isocharge''. In the ``isospin''
channel, the $g_1$-processes (and consequently also $g_{2\sigma}$) transform as
\cite{ners}
\begin{eqnarray}
H_1 & = & - g_{1\|} \int \! dx : \sigma_+(x) \sigma_-(x) : 
+ g_{1\perp} \sum_s \int \! dx \: \Psi_{+,s}^{\dag}(x)
\Psi^{\dag}_{-,-s}(x) \Psi_{+,-s}(x) \Psi_{-,s}(x) \nonumber \\
& \rightarrow & - g_{1\perp} \int \! dx : \sigma_+(x) \sigma_-(x) : 
+ \frac{g_{1\|} + g_{1\perp}}{2} \int \! dx \: \Psi_{+,s}^{\dag}(x)
\Psi^{\dag}_{-,-s}(x) \Psi_{+,-s}(x) \Psi_{-,s}(x) \nonumber \\
\label{gftraf}
& & - \frac{g_{1\|} - g_{1\perp}}{2} \int \! dx \: \Psi_{+,s}^{\dag}(x)
\Psi^{\dag}_{-,s}(x) \Psi_{-,-s}(x) \Psi_{+,-s}(x) \;\;\;. 
\end{eqnarray}
The last process does not conserve the total spin of the scattering partners.
In  a single-band model, it can arise from spin-orbit scattering, and work
on this topic is relevant here \cite{giaso}. The coupling constant
is commonly denoted by $g_f$ and describes interband backscattering and does
not conserve the number of particles on a given branch of a band. More 
interactions of this kind can arise in systems where the starting model
has internal degrees of freedom. Of course, in all these multicomponent 
problems, the standard fluctuation operators from Section (\ref{secprop}) 
can be extended to
include inter-component fluctuations, giving a flavour of the richness
of the physics that can be described.

\section{Multi-component models} 
\label{multi}
We first describe a multi-component \tlm , and then some of the 
instabilities occurring in more complicated models when scaling
does not go towards a Tomonaga-Luttinger fixed point in all channels.

The Hamiltonian for \tlm\ with $N$ components (colours, labelled by $\lambda$,
including spin and chirality index) 
is \cite{pesomag}
\begin{eqnarray}
H & = & \sum_{\lambda = 1}^N H_{\lambda} + \sum_{\lambda,\lambda' = 1}^N
H_{\lambda \lambda'} \;\;\;, \nonumber \\
\label{hncomp}
H_{\lambda} & = & v_{\lambda} \sum_k (k - k_{F \lambda} ) c^{\dag}_{k \lambda}
c_{ k \lambda} \;\;\;, \\
H_{\lambda \lambda'} & = & \frac{1}{2L} 
\sum_p g_{\lambda \lambda' } (p)
\rho_{\lambda}(p) \rho_{\lambda' } (-p) \;\;\;. \nonumber 
\end{eqnarray}
We assume that the Fermi velocities and momenta 
are pairwise ($v_{\lambda}, - v_{\lambda}$), ($k_{F \lambda}, - k_{F \lambda}$),
corresponding to a symmetric dispersion, and that the coupling constants
satisfy $g_{\lambda \lambda'} = g_{\lambda' \lambda}$. The 
standard \tlm\ (\ref{hlutt})
is obtained for $N=4$ and the two pairs of Fermi velocities and
momenta equal. The density operators commute as
\begin{equation}
\label{ncompcom}
\left[ \rho_{\lambda}(p), \rho_{\lambda'}(-p) \right]  = 
- \frac{pL}{2 \pi} \delta_{\lambda , \lambda'} 
\delta_{p, p'} \: {\rm sign} \: v_{\lambda} \;\;\;.
\end{equation}
One can now repeat the arguments of Section \ref{hamdia} to show that (i) the 
single-particle Hamiltonian $H_{\lambda}$ can be written as a boson
bilinear, and (ii) that the equivalence to the boson description is
complete only upon including charge ($N_{\lambda}$) excitations 
with respect to the vacuum. The Hamiltonian
then becomes 
\begin{eqnarray}
\label{ncompbos}
H & = & \frac{2 \pi}{L} \sum_{\lambda, \lambda',  p > 0} 
A_{\lambda \lambda'} (p) \rho_{\lambda}(p) \rho_{\lambda'}(-p)
+ \frac{\pi}{L} \sum_{\lambda \lambda'} 
A_{\lambda \lambda'} (p=0) N_{\lambda} N_{\lambda'}
\;\;\; , \\
\label{amat}
A_{\lambda \lambda'} (p) & = & | v_{\lambda} | \delta_{\lambda
\lambda'} + \frac{g_{\lambda \lambda'}(p)}{2\pi} \;\;\;.
\end{eqnarray}
This Hamiltonian being a bilinear form in the bosons, it can be diagonalized
by an $N$-component generalization of a Bogoliubov transformation 
\cite{pesomag}
\begin{eqnarray}
\label{nhdiag}
H & = & \frac{2 \pi}{L} \sum_{j, p>0} | v_j | \rho_j ( p \: {\rm sign}v_j )
\: \rho_j ( - p \: {\rm sign}v_j ) + \frac{\pi}{L} \sum_j | v_j |
\sum_{\lambda \lambda'} N_{\lambda} \alpha^{(j)}_{\lambda \lambda'}
N_{\lambda'} \;\;\;, \\
P & = &  \sum_{\lambda} k_{F \lambda} N_{\lambda} + \frac{\pi}{L}
\sum_{j \lambda \lambda'} {\rm sign} (v_j )  
N_{\lambda} \alpha^{(j)}_{\lambda \lambda'}
N_{\lambda'} + \\
& & +
\frac{2 \pi}{L} \sum_{j, p>0} {\rm sign} (v_j ) \: \rho_j ( p \: {\rm sign}v_j )
\: \rho_j ( - p \: {\rm sign}v_j ) \;\;\;. \nonumber
\end{eqnarray}
$P$ is the momentum operator. 
The index $j$ denotes the new operators and parameters. The renormalized sound
velocities $v_j$ and the matrix $\alpha^{(j)}_{\lambda \lambda'}$
are obtained as the solution of the eigenvalue problem
\begin{eqnarray}
{\cal A} \cdot {\cal B} | w^{(j)} \rangle & = & v_j | w^{(j)} \rangle 
\;\;, \\
{\cal A} & = & \left( A_{\lambda \lambda'} \right) \;\;,
\;\;\;\; {\cal B} = \left( \delta_{\lambda \lambda'} 
{\rm sign} v_{\lambda} \right) \;\;, \;\;\;\;
\alpha^{(j)}_{\lambda \lambda'} = w_{\lambda}^{(j)} 
w_{\lambda'}^{(j)} \;\;. \nonumber
\end{eqnarray}
Correlation functions 
\begin{displaymath}
G_{\lambda_1 \ldots \lambda_m} (\bx'_1, \ldots, \bx'_m ;
\bx_1, \ldots, \bx_m ) = 
\end{displaymath}
\begin{equation}
\label{ncorr}
(-i)^m \langle { T} \Psi_{\lambda_1} (\bx'_1)
\ldots \Psi_{\lambda_m} (\bx'_m) \Psi_{\lambda_m}^{\dag} (\bx_m)
\ldots \Psi_{\lambda_1}^{\dag}(\bx_1) \rangle
\end{equation}
can then be evaluated either by generalization of the bosonization formula
(\ref{bos}) to $N$ components, or by combining the Ward identities 
associated with (\ref{hncomp}) with equation-of-motion methods. [In 
(\ref{ncorr}), \bx\ stands for the space-time point $(x,t)$ and ${ T}$
is the time-ordering operator.] One finds
\begin{equation}
\label{ncompres}
G_{\lambda_1 \ldots \lambda_m} (\bx'_1, \ldots, \bx'_m ;
\bx_1, \ldots, \bx_m ) 
 =  \prod_l 
G_{\lambda_l}
(\bx'_l , \bx_l)
\prod_{l' < l} 
\frac{ f_{\lambda_l \lambda_{l'}} (\bx_l - 
{\bx'}_{l'}) f_{\lambda_l \lambda_{l'}} ({\bx'}_l - 
\bx_{l'}) 
}{ 
f_{\lambda_l \lambda_{l'}} (\bx_l - 
\bx_{l'}) f_{\lambda_l \lambda_{l'}} ({\bx'}_l - 
{\bx'}_{l'}) } 
\;, 
\end{equation}
\begin{eqnarray}
G_{\lambda}(\bx', \bx) & = & G_{\lambda}^{(0)} (\bx' , \bx) \:
f_{\lambda \lambda} (\bx - \bx') \;\;\;, \nonumber \\
f_{\lambda \lambda'}(x,t) & =& \Lambda^{- 2 \alpha'} 
\left[ x - v_{\lambda} t + \frac{i}{\Lambda} {\rm sign} (v_{\lambda}t)
\right]^{\delta_{\lambda \lambda'} } \prod_j 
\left[ x - v_j t + \frac{i}{\Lambda} {\rm sign} (v_j t) \right]^{- 
\alpha^{(j)}_{\lambda \lambda'} } \;\;\;. \nonumber
\end{eqnarray}
Here $G^{(0)}(\bx)$ is the noninteracting Green function, $\Lambda$ is
the momentum transfer cutoff familiar from Section \ref{lumod}, and
$\alpha' = \sum_j \alpha^{(j)}_{\lambda \lambda'}$ with the sum
going only over those $j$ where $v_j v_{\lambda} < 0$. Asymptotically,
this gives
\begin{equation}
\label{cofmult}
G(\{ \bx \}, \{ \Delta N_{\lambda} \} ) \sim e^{- i x \sum_{\lambda}
\Delta N_{\lambda} k_{F\lambda}} \prod_j (x - v_j t)^{-2 \Delta_j}
\end{equation}
with 
\begin{equation}
\label{tlmdim}
\Delta_j = \sum_{\lambda \lambda'} N_{\lambda} 
\alpha^{(j)}_{\lambda \lambda'} N_{\lambda'} \;\;\;.
\end{equation} 
$\Delta N_{\lambda}$
is the number of fermions of colour $\lambda$ propagating from \bf 0 \rm 
to \bx , weighted by sign$v_{\lambda}$, i.e. the charge excitation introduced
by the operator whose correlations are to be computed. This is in agreement 
with the conformal field theory prediction, if the $\Delta_j$ are interpreted
as scaling dimensions. That this is indeed justified is seen by
evaluating energies and momenta in a state with a definite number of charge and 
particle-hole excitations ($N_{\lambda}, n_j = Lq/2\pi$ respectively)
\begin{eqnarray}
E(N_{\lambda}, n_j) - E_0 & = & \frac{2 \pi}{L} \sum_j | v_j |
\left( \Delta_j + n_j \right) + \ldots \;\;\;, \\
P(N_{\lambda}, n_j) & = & \sum_{\lambda} N_{\lambda} k_{F\lambda}
+ \frac{2 \pi}{L} \sum_j \left( \Delta_j + n_j \right) {\rm sign} v_j \;\;\;, 
\end{eqnarray}
i.e. one obtains 
the typical tower structure of conformal field theories. 
Comparing with 
Eqs. (\ref{finen}), (\ref{finmom}), $\Delta_j$ is identified as the scaling
dimension of a primary operator, and we thus have generalized these
expressions to an $N$-component system \cite{pesomag}.

One example where these expressions can be fruitfully applied, is the
Hubbard model in a magnetic field which scales towards an $N=4$-\tlm . 
Of course, one could take the 
perturbative renormalization approach \cite{mont}. More accurate
results, valid at any coupling, are obtained, however, from the
the Bethe-Ansatz solution in a finite magnetic field. There are two 
practical possibilities: either use conformal field theory directly
to obtain the correlation exponents \cite{fk2} or perform a mapping
on the $N$-color \tlm\ by identifying the low-energy spectral properties
between both models \cite{pesomag}. We briefly comment on the second
method. 

The similarity
to the ${\bf \Delta N}$ and ${\bf D}$ in the conformal field theory treatment
of the \hm\ in Section \ref{seclat} should be apparent, as well as the
similarity between Eqs.~(\ref{cofmult}) and (\ref{cofcon}).
The requirement that the correlation exponents of the \tlm\ and the \hm\
be equal, implies for the scaling dimensions
\begin{equation}
\Delta_j = \Delta_{c(s)}^{\pm} \;\;\;,
\end{equation}
where the latter quantity is evaluated in the Hubbard model and related
by (\ref{dims}) to the elements of the dressed charge matrix. 
Also the ${\cal A}$-matrix, and therefore the coupling constants 
$g_{\lambda \lambda'}$ of the \tlm , can be found. This basically
involves constructing an expansion of the scaling dimensions (\ref{tlmdim})
in terms of the excitations in the Bethe Ansatz wavefunction
of the Hubbard model. Of course, the coupling constants $g_i$ are found
linear in $U$ at small $U$. As $U \rightarrow \infty$, they saturate,
as implied by the saturation in $K_{\rho}$. More interesting is the
finding that, at fixed $U$, the $g_i$ have a nonanalytic variation with $h$
as $h \rightarrow 0$, translating into a nonanalytic $h$-dependence of
$K_{\rho}$ \cite{fk2}, and meaning that a magnetic field $h$ can never
be regarded as a small perturbation. 

Models involving interactions other than forward scattering alone often
do not scale towards the $N$-colour \LL\ fixed point (\ref{hncomp})
and most often have been studied for two bands ($A,B$)
with or without spin degrees of freedom. 
The most important interaction not contained in single-band models
is interband backscattering, the last term in (\ref{gftraf}) with 
coupling constant $g_f$. In Eq. (\ref{gftraf}), the spin index labels the
two bands $s=A,B$. (There are more interband backscattering terms;
momentum conservation suppresses all of them but $g_f$ when the
two Fermi momenta are not nearly equal.) For spinning fermions, we can
have $g_{f\|} \neq g_{f\perp}$. For half-filled bands (here $k_{F,A} + k_{F,B}$
or $2k_{F,A(B)} = 2 \pi / a$), one must further add 
the corresponding Umklapp scattering process $g_3$. 
It is also useful to think in terms
of charge fluctuation interactions (scattering processes changing the
number of particles in a band like $g_f$ does) and exchange interactions
(interband processes conserving the charge but 
changing the spin in a band). 

These models have been studied most often with the same methods 
used for the single-band problems:
\rg\ starting from a bosonic or fermionic
description and eventually strong-coupling field theory. 
We neglect from our subsequent discussion all non-Luttinger intraband
interactions whose effects have been discussed in the preceding chapters.
The $2$-colour \LL\ (\ref{hncomp}) is a hyperplane of critical fixed points
$g_3 = g_f = 0$ \cite{mutem}. It is stable for $2 g_{2AB} < - | g_{2AA} + 
g_{2BB} | $, and then is attractive for [$g_f = 0$ and $(2 g_{2AB} +
g_{2AA} + g_{2BB}) < c | g_3 |$] or [$g_3 = 0$ and $(2 g_{2AB} -
g_{2AA} - g_{2BB} ) < - c | g_f |$], where $c$ is a constant related
to the difference in the Fermi velocities. 
In the symmetric case ($v_{F,A} = v_{F,B}, \; g_{2AA} = g_{2BB}$) 
the two bands decouple, and one can compute correlation functions 
by the standard methods presented earlier \cite{giaso}. In the absence
of intraband 
backscattering, the \rg\ equations have the Kosterlitz-Thouless structure
(\ref{kot}), and there are both massive and massless phases. If $g_{2AB} - g_{2AA}
< 0 $ and $| g_{2AB} - g_{2AA} | < | g_f | $, scaling will go
to weak coupling, and there will be a massless two-component \LL\ with 
dominant CDW correlations. The combination $g_{2AB} + g_{2AA}$ controls
if they are of inter- or intraband type (corresponding to SDW$_z$ and CDW
of a s=1/2-single-band model, respectively). 
If one of the preceding inequalities
is violated, scaling will go  to strong coupling and, depending on 
$g_{2AB} + g_{2AA}$, one will have either intraband superconducting pairing,
or one of two new types of interband CDWs (corresponding to SDW$_{x,y}$
in a spin-1/2 single-band model). Their structure can be seen more clearly
if one imagines the two bands arising from two spinless fermion chains
coupled by $\tperp$ \cite{ners}. There, they correspond to (i) a CDW with
a charge density modulation on the bonds of the chains and (ii) to 
a configuration where currents circulate around the plaquettes of the ladder
in an alternating pattern. This can be viewed as an orbital antiferromagnet
and is directly related to the staggered flux phases discussed some time
ago in the {\htc -}\-problem \cite{flupha}.

In the asymmetric model, the bands do not decouple. At weak coupling, this
is quite apparent from the \rg\ equations. At strong coupling, the Hamiltonian
can be decoupled into two sine-Gordon models involving phase fields which
are linear combinations of those describing the bands, with
one condition on the coupling constants \cite{mutem}, and
the excitation spectra and correlation exponents can be determined. 
One consequence of the coupling between the bands is that the correlation
exponents for the different intraband and the interband fluctuations may
all be different. For example, when $g_3 = 0 $, $g_f$ may become 
relevant and open a gap in one of the
sine-Gordon models, the other remaining massless. Depending on the precise
value of the interactions, either an intraband SS or an interband CDW
have divergent fluctuations. It is interesting then that the conditions
for divergent SS can be realized from purely repulsive interactions, 
something impossible in the single-band case \cite{mutem}. 

Up to now, we have discussed only the physics of the charge fluctuations.
The interband exchange processes are interesting, too, and require an 
extension of the previous models by spin degrees of freedom
\cite{pesot,vaza}. Due to the proliferation of the coupling constants, 
a general discussion of such a model is a formidable task, and will not
be attempted here. An interesting limit is $v_A \ll v_B$, i.e. a band
of light electrons (B) coupled to heavy electrons (A). On a technical
level, a small velocity $v_A$ in (\ref{htwob}) is generated by hybridizing
a dispersionless A-band with the B-electrons [take the two-chain Hamiltonian
(\ref{twoch}) and put $t_{\|} = 0$ for one spin direction only]. 
For $v_A \ll v_B$, the charge fluctuations between the bands may scale
out of the problem. In that case, exchange between the bands is the only
remaining coupling, and the physics then becomes very similar to the 
single-impurity Kondo
problem \cite{pesot,vaza}. In particular, for positive exchange coupling
constants, scaling goes to weak coupling, in analogy to the ferromagnetic
Kondo impurity. For negative exchange constants, on the other hand, scaling
is to strong coupling as in the antiferromagnetic Kondo problem. 
The carriers in the two bands will bind into interband singlets, and an
interband spin gap will open. The dominant response functions are then
interband CDW and SS, depending on the remaining marginal intraband couplings.
In this way, one can model a 1D Kondo lattice.

Pursuing the analogy of this two-band model to the Kondo problem and
identifying the heavy carriers with spins, complete Kondo screening can
occur because there are always sufficient light electrons to screen
out the spins ($k_{FA} < k_{FB}$ in Fig.~6.1) \cite{nozkon}.
Caron and Bourbonnais have studied directly a 1D Kondo lattice with
$n_s$ impurity spins and $2n_c$ carriers \cite{caboko}. They verify by
second order \rg\ Nozi\`{e}res' criterion \cite{nozkon}, stating that
complete Kondo screening only occurs when there are sufficient carriers
$2n_c \geq n_s$. If this criterion is violated, the spins become RKKY-coupled
by the electrons and form a \tkf -SDW, at least for weak exchange integrals.
For stronger exchange, a Kondo regime may be reestablished.

Finally, we add for completeness that Emery has solved the 1D version 
\cite{emcuo} of the two-band model which was proposed in 2D for the
$CuO$-\htcs\ \cite{cuo}. Here, the band splitting is produced by a
term in the Hamiltonian $(\epsilon/2) \sum_{n,s} (-1)^n c^{\dag}_{ns} c_{ns}$ 
modelling the energy difference between the copper and oxygen orbitals.
In contrast to the models discussed above, the two bands do not 
overlap in energy, and the physics taking place upon doping is qualitatively
similar to that arising from a half-filled one-band model with a charge gap,
which was discussed in Section \ref{motttr}.

\section{Crossover to higher dimensions}
\label{twochains}
On a macroscopic level, 1D systems are well known for their opposition to
long-range order. At finite temperature, the entropy associated with the
defects in an ordered phase more than outweighs the cost in energy for
their creation \cite{lndlif}, and thermal fluctuations thus destroy 
ordered phases. At $T=0$, the influence of quantum fluctuations is more subtle. 
From the equivalence of 1D quantum systems to 2D classical statistical 
mechanics, the Mermin-Wagner theorem \cite{mermwa} suggests that phases
where a continuous symmetry would be broken, could not possess long-range 
order even
at zero temperature. That quantum fluctuations indeed destroy long-range
order associated with continuous broken symmetries was demonstrated by
Takada \cite{quant}.
Finite temperature phase transitions therefore must be a consequence of
3D coupling between the chains. Two mechanisms can couple electrons on 
different chains: finite-range Coulomb interactions and interchain tunneling.

The electron dynamics must also be affected by 3D tunneling. Assume that
we have a \lm\ on-chain dispersion $v_F (rk - k_F^{1D})$ and a hopping
matrix element $\tperp$ between neighbouring chains. The 3D dispersion
of the electrons then becomes
\begin{equation}
\varepsilon^{3D}_r ( \bk ) = v_F (r k - k_F^{1D}) - 
2 \tperp \cos (k_{\perp b} b) - 2 \tperp \cos (k_{\perp c}c) \;\;\;,
\end{equation}
where $b$ and $c$ are the transverse lattice constants (the longitudinal one
is denoted $a$). We find for the
Fermi surface
\begin{equation}
\bk_{F,r}^{3D} =  \left( \begin{array}{c}
r \left[ k_F^{1D} + \frac{2 \tperp}{v_F} \cos (k_{\perp b} b) +
\frac{2 \tperp}{v_F} \cos (k_{\perp c} c) \right] \\
\bkp
\end{array}\right) \;\;\;.
\end{equation}
If $\tperp$ is of the order of $t_{\|} \sim v_F / a$, the Fermi surface
will be closed and we better start from an anisotropic 3D system.
For smaller $\tperp$, however, the Fermi surface consists of two warped
sheets, Fig.~6.2, and retains some 1D character. The issue now is
the transverse coherence of the electronic motion. Naively, we expect
that at high temperatures $T > \tperp$, 
where there is ``thermal blurring'' of the Fermi
surface of the order $T$, the electrons are unable to sense the warping 
and behave essentially 1D. 
At $T < \tperp$, the warping and thus the 3D
aspects of the Fermi surface can be probed, and transverse coherence
would emerge. We shall see below that the actual picture is significantly
more complicated. 

The reasons for these complications reside in the electron-electron
interaction which may possibly confine electrons on their chains. 
In 1D, here are two dramatic differences from the
free particle picture implictly assumed in the preceding arguments:
the vanishing quasi-particle residue at the Fermi energy, and charge-spin
separation. $\tperp$ transfers particles or at least quasi-particles. 
Its efficiency may therefore be severly reduced if there are only collective
excitations. While $z(k = \tperp/ v_F) \neq 0$ in general, we still expect that
a reduced quasi-particle residue
$z(\tperp) \ll 1 $ will delay the establishment of 3D coherence in the
single-electron dynamics \cite{firsov,bourbonnais,prigofi}. 
Charge-spin separation could also confine particles
on their chains because a holon and spinon must tunnel together -- yet in
general they are separated \cite{pwaconf}.

The theory we have outlined in the preceding paragraphs relies heavily
on the 1D nature of our models. The key point were the 1D conservation
laws of charge and spin currents, i.e. of charge and spin on each branch 
of the dispersion separately, Eq. (\ref{conssep}) which no longer hold
at $\tperp \neq 0$.
Is the theory of the preceding chapters therefore 
limited to strictly one dimension, or can it be extended beyond? 
What is its relevance for highly anisotropic quasi-1D problems? 
Can it provide a framework to describe the physics of real quasi-1D materials
such as the organic (super-)conductors? 

To answer this questions, we consider an array of coupled chains to mimic
a 2D or 3D situation. The Hamiltonian then becomes 
\begin{eqnarray}
\label{hamcoup}
H & = & H_{\|} + H_{\perp}^{(t)} + H_{\perp}^{(g)}  \;\;\;, \\
\label{hpar}
H_{\|} & = & \sum_n H^{\|}_n \;\;\;, \\
\label{hperpt}
H_{\perp}^{(t)} & = & - t_{\perp} 
\sum_{<m,n>,r,s} \int \! dx \Psi^{\dag}_{m,r,s}(x) \Psi_{n,r,s}(x)
+ {\rm H.c.} \\
& = & -2 t_{\perp} \sum_{k\approx rk_F,{\bf k}_{\perp}, r, s} 
\left[ \cos ( k_{\perp b}b )  + \cos (k_{\perp c} c ) \right]
c^{\dag}_{k,{\bf k_{\perp}}, r, s} c_{k,{\bf k_{\perp}}, r, s} 
\;\;\;, \nonumber \\
\label{hperpv}
H_{\perp}^{(g)} & = & \frac{2}{L} \sum_{k, {\bf k_{\perp}}} 
g_2^{(\perp)}(k, {\bf k_{\perp}}) 
\rho_+(\kkp ) \rho_-( -k , -k_{\perp} ) + \\
& + & \frac{1}{L} \sum_{ \kkp , r} g_4^{(\perp)} (\kkp ) \rho_r (\kkp ) 
\rho_r (-k, -k_{\perp} ) +
\nonumber \\
\label{h1perpv}
& + & \sum_{<m,n>,s,s'} g_{1,m,n}^{(\perp)} \int \! dx \Psi_{m,+,s}^{\dag}(x) 
\Psi_{n,-,s'}^{\dag}(x) \Psi_{n,+,s'}(x) \Psi_{m,-,s}(x) \;\;\; .
\end{eqnarray}
Here, $H^{\|}_n$ is the one-chain Luttinger Hamiltonian (\ref{hlutt}) for chain 
$n$ where the density operators acquire an additional chain ($n$) or 
transverse momentum (${\bf k_{\perp}}$) label.
$t_{\perp}$ is the transverse hopping integral, $m,n$ in the sums
denotes the chains, 
and $r$ and $s$ are the branch and spin index,
respectively. We have allowed for different lattice constants in all three
directions. $g_2^{(\perp)}$ and $g_4^{(\perp)}$ 
are the interchain forward scattering constants
of $g_2$- and $g_4$-type, respectively, which, in the case of a Coulomb 
interaction, must be equal. 
$g_1^{(\perp)}$ measures
the strength of the interchain backscattering. If this term is included
into the Hamiltonian, consistency would require that one include also
the intrachain backscattering Hamiltonian (\ref{hperp}) into $H^{\|}_n$,
in order to treat both interactions on an equal footing. 

We first discuss the interchain Coulomb interaction ($t_{\perp} =0$). 
Quite generally, transverse Coulomb coupling screens the effective
on-chain interactions or, if negative initially, makes them more negative
\cite{gukle,klr,gord} -- a tendency reminiscent of Little's old suggestion
in favour of quasi-1D materials as candidates for high-temperature 
superconductivity \cite{little}. However, a more detailed investigation
leads to rather different conclusions.
Interchain forward 
scattering ($g_2^{(\perp)}, g_4^{(\perp)}$) 
respects the conservation of total charge and spin on each
branch of the dispersion on each chain separately. Moreover, $g_2^{(\perp)}$ 
is a marginal
operator and only couples the charge fluctuations; 
therefore it will influence the dimensions of operators, and the
exponents of those correlation functions which are
sensitive to charge fluctuations. The Hamiltonian can be diagonalized
exactly, and one obtains renormalized values of $K_{\rho}({\bf k_{\perp}})$
depending now on the perpendicular wavevector ${\bf k_{\perp}}$ 
\cite{gukle,klr,gord}. 
Exponents of on-chain correlation functions then contain
integrals over ${\bf k_{\perp}}$ involving $K_{\rho}({\bf k_{\perp}})$ 
or functions thereof while the spin parts can be taken over unchanged
from the single-chain problem. 
Physically, the system remains a \LL\ but, concerning
the competition between SS and CDWs, SS is favoured at the expense of CDWs,
not unlike Little's suggestion \cite{little}.
The situation is completely different if interchain backscattering 
($g_1^{(\perp)}$) is
allowed. Firstly, as on  a single chain, it violates the separate conservation
of total spin on each dispersion branch. But, unlike the intrachain 
backscattering, (\ref{h1perpv}) also violates separate charge conservation.
This indicates that, if this interaction process can become relevant, gaps
may open in the charge and spin fluctuations with the concomitant possibility
of long-range CDW order. This is indeed what happens, at least in the regime
of attractive on-chain backscattering, but also for repulsive backscattering
if the complete intra- and inter-chain potential has $\delta$-function shape
\cite{gukle,klr,gord}. Depending on the sign of $g_1^{(\perp)}$, 
long-range CDW order
can be stabilized with a wavevector $Q_< = (2k_F,0,0)$ (i.e. a CDW in phase
on neighbouring chains) for $g_1^{(\perp)} < 0$, 
and $Q_> = (2k_F, \pi/b , \pi/c)$
(i.e. the CDWs on neighbouring chains are out of phase) for $g_1^{(\perp)}>0$. 
Physically, these results are quite easy to understand 
for dominant on-chain-CDW fluctuations because the system
can gain Coulomb energy from the charge modulations on the neighbouring chains
with $Q_<$ resp.~$Q_>$. 

Up to this point, we have allowed general coupling contants $g_i$ both
for the intra- and interchain interactions. Of course, the physical Coulomb
potential is $V({\bf q}) = 4 \pi e^2 / {\bf q}^2$, and finite forward 
scattering constants only arise as a consquence of screening ($g_1$ can
be considered as constant but may depend on details of the wavefunctions
of particular materials). The problem of screening of the divergence of
the Coulomb potential can be solved on an array of chains 
\cite{schucou,baris}: the interaction of two electrons on a given chain
can be screened by the electrons on the others. 

The Hamiltonian
\begin{equation}
\label{hamcoul}
H_C = \frac{e^2}{2} \sum_{n,n'} \int \! dz dz' \frac{\rho_n(z) 
\rho_{n'}(z')}{ \sqrt{a^2[ (n_x - n'_x)^2 + (n_y - n'_y)^2] + (z-z')}}
\end{equation}
[where $e$ is the electron charge, $\rho_n(z)$ the \em total \rm charge
density operator (\ref{rhox}) at position $z$ on chain $n$, and $a$ the 
transverse lattice constant] can be mapped onto the form (\ref{hnu})
with the couplings
\begin{equation}
\label{gq}
g_{2 \rho}({\bf q}) = g_{4 \rho}({\bf q}) = \frac{4 \pi e^2}{ a^2 
(\varepsilon_{\|} q_z^2 + \varepsilon_{\perp} q_{\perp}^2 ) 
} \;\;\;.
\end{equation}
(In principle, there is an infinite sum over transverse reciprocal lattice 
vectors, but the matrix elements then will depend again on details of
wavefunctions and are not universal \cite{schucou}). $\varepsilon_{\|,\perp}$
are background dielectric constants. The $g_{i\sigma}$ remain unaffected.
One now can diagonalize the Hamiltonian for each ${\bf q_{\perp}}$ leading
to ${\bf q_{\perp}}$-dependent velocities and coupling constants.
The energy of the charge excitations is 
\begin{equation}
\label{omegrho}
\omega_{\rho}({\bf q}) = \sqrt{v_F^2 q_z^2 + \frac{\omega_{pl}^2 q_z^2}{
\varepsilon_{\|} q_z^2 + \varepsilon_{\perp} {\bf q_{\perp}}^2}}
\;\;\;, \;\; \omega_{pl}^2 = \frac{8 e^2 v_F}{a^2} = \frac{4 \pi e^2 n}{m}
\;\;.
\end{equation}
For any finite ${\bf q_{\perp}}$, $\omega_{\rho}(q_z) \propto | q_z|$
in the limit $q_z \rightarrow 0$, 
allowing to define a renormalized charge velocity 
$v_{\rho}({\bf q_{\perp}})$. For $| {\bf q} | \rightarrow 0$, one obtains
the plasma frequency of the anisotropic system ($\varepsilon_{\|} =
\varepsilon_{\perp} = \varepsilon$ for simplicity)
\begin{equation}
\label{pls}
\omega_{\rho}(\Theta) = \frac{\omega_{pl}}{\sqrt{\varepsilon}} |\cos \Theta|
\;\;\;.
\end{equation}
$\Theta$ is the angle between ${\bf q}$ and the $z$-axis. 
A renormalized coupling constant $K_{\rho} ({\bf q_{\perp}})$ can be
defined from the diagonalization of the Hamiltonian. Its usefulness 
for determining the asymptotic decay of 
correlation functions is, however, 
not as immediate as for the single-chain problem. In fact, when calculating
the correlation functions, one obtains ${\bf q_{\perp}}$-dependent 
expressions for their decay exponents which have to be integrated over
${\bf q_{\perp}}$ \em at the end. \rm
It is \em not \rm allowed to use an integrated
$\int d^2 q_{\perp} K_{\rho}({\bf q_{\perp}})$ in the standard Luttinger
expressions. As a consequence, the simple scaling relations between the
exponents of the various correlation functions (single-particle
Green function, SDW, CDW, SS, \ldots) break down! Each function has its
own, independent exponent. 

The screening effect strongly depends on the density of carriers and
on the anisotropy of the lattice. Denser chain packing means better screened
interaction. In the limit of vanishing packing density, one 
crosses over to the single-chain case \cite{schuwi} with a vanishing
plasma frequency, a correction $\propto \ln |q_z|$ to the Fermi velocity,
and a formally vanishing $K_{\rho}$-exponent, as discussed at the end of
Section \ref{seclat}.

Interchain single-particle tunneling $\tperp$ 
can lead to further new physics. $\tperp$ can generate transverse coherence
in the electron dynamics, i.e. a crossover from essentially 1D 
to effectively 3D behaviour. It is not clear at this time, if the effectively
higher-dimensional behaviour is necessarily of Fermi-liquid type or not.
Interchain tunneling also can generate transverse pair tunneling
(either of particle-particle or particle-hole type) which propagate
the dominant on-chain correlations in the transverse directions, and eventually 
a finite-temperature phase transition into a symmetry-broken ground state
occurs. In both cases, the 1D \LL\ is unstable. 

We now consider the Hamiltonian (\ref{hamcoup}) with $H_{\perp}^{(g)} \equiv 0$
(\ref{hperpv}). 
The essential qualitative 
physics can be seen from a scaling argument due to Schulz 
\cite{ijmp} and Wen \cite{wenperp}. Consider the free energy of a system
with small $\tperp$ at finite temperature and investigate the relevance of 
different terms generated by an expansion in $\tperp$. The free energy of 
the strictly 1D system is $F^{(0)} \propto T^2$. In second order in $\tperp$,
we obtain a correction
\begin{equation}
\label{df}
\delta F^{(2)} \approx \tperp^2 \int \! dx d \tau \: G_{rs}^2 (x,\tau) \;\;\;,
\end{equation}
where $G_{rs}(x,\tau)$ is the single-chain Green function at imaginary
time $\tau$. This function behaves as
\begin{eqnarray}
\label{greent}
G_{rs} ( x_1 - x_2 , \tau_1 - \tau_2 ) & \approx & \mid 1 - 2 \mid^{-1-\alpha}
\;\;\;, \\
{\rm with } \;\;\; \mid 1 - 2 \mid & \equiv & \frac{v_F}{2 \pi T}
\sqrt{\cosh \left[ 2 \pi T \frac{x_1 - x_2 }{v_F} \right] -
\cos \left[ 2 \pi T ( \tau_1 - \tau_2 ) \right] } \;\;\; . \nonumber
\end{eqnarray}
$\alpha$ is the single-particle exponent from Section (\ref{secprop}).
The correction to the free energy then scales as
\begin{equation}
\label{dftwo}
\delta F^{(2)} \propto \tperp^2 T^{2\alpha} \;\;\;.
\end{equation}
If $\alpha < 1 $ (i.e. $3 - \sqrt{8} < K_{\rho} < 3 + \sqrt{8}$), a case
encountered in many models (Section \ref{sec1d}), 
this terms will become more important
than $F^{(0)}$ at sufficiently low temperature no matter how small $\tperp$,
indicating that $\tperp$ then is a relevant perturbation. If $\tperp$ is the
most relevant perturbation, we expect the system to show a single-particle
1D--3D crossover at some temperature $T_X^1$. Only if $\alpha > 1$ interchain
single-particle tunneling will be irrelevant. But then, look at the next
order in the expansion of the free energy, corresponding to interchain
pair tunneling
\begin{eqnarray}
\delta F^{(4)} & \approx & \tperp^4 \int \! d1\: d2 \: d3 \: d4 
\left[ \frac{\mid 1-3 \mid \mid 2-4 \mid}{\mid 1-2 \mid \mid 3-4 \mid} 
\right]^{(K_{\rho} - 1/K_{\rho})/2} \left[ \mid 1-4 \mid \mid 2-3 \mid 
\right]^{-2 - 2\alpha} \nonumber \\
& \approx & \tperp^4 \max \left( 
T^{4\alpha}, T^{2K_{\rho}}, T^{2/K_{\rho}} \right) \;\;\;.
\end{eqnarray}
The first term corresponds to two uncorrelated single-particle events and
is the square o $\delta F^{(2)}$. The second term is generated
from coherent tunneling of a particle-hole pair and is more important 
than $F^{(0)}$ whenever $K_{\rho} < 1$, i.e. for repulsive interactions.
Coherent pair tunneling generates the third term which dominates the 
zero-order term
for attractive interactions. The particle-particle and particle-hole pair
terms are more important than the single-particle terms for $1/K_{\rho}$
resp. $K_{\rho} < (2/\sqrt{3} - 1)$ i.e. rather strong interactions. If this
happens, one expects to find a two-particle 1D--3D crossover at a temperature
$T_X^2 > T_X^1$, and the system very likely will undergo a symmetry-breaking
phase transition to a ground state corresponding to the most dominant
intrachain fluctuation. For spinless fermions, single-particle tunneling
is relevant for $\alpha < 1/2$ i.e. $K$ resp. $1/K < 2 + \sqrt{3}$, but 
two-particle tunneling is stronger than single-particle tunneling already
for $K$ resp. $1/K > 1 + \sqrt{2}$ \cite{kusma}.

This \rg\ argument is good for infinitesimal transverse 
coupling only. It will certainly fail for bigger $\tperp$: one expects the
warping of the Fermi surface and deviations from perfect nesting
to cut off the Peierls divergence. Below the temperature where this cutoff
happens, superconductivity in general will be the only 
possible instability remaining.
A description
of the system for finite $\tperp$ is, however, a difficult task. We only
briefly review the major achievements and refer the reader to more
extended treatments \cite{firsov,bourbonnais,prigofi} for further details. 

Early work starts from the exact solution of the 1D models and adds $\tperp$
as a perturbation \cite{prigofi,gukle,braya}. In this way, one generates
an effective transfer of a pair of particles. There are no real interchain
single-particle transitions, and the pair motion takes place through virtual
events. The perturbation theory only becomes well defined if there is either
a gap in the charge or spin fluctuations, or if the Luttinger model 
interactions are sufficiently strong [basically, the integral in Eq.
(\ref{df}) must converge]. This is a good assumption for attractive
backscattering where we have a spin gap, but not for the generic repulsive
\LL . Doing then mean-field theory in $\tperp$ with spin gap, one finds
a transition to a SS or CDW phase at a finite critical temperature
\cite{gukle}. When only one type of fluctuation is divergent on a single
chain, tunneling will stabilize it into an ordered phase. When both are
divergent, tunneling will favour SS. 

The question of a single-particle crossover from
1D to 3D behaviour was addressed by Prigodin and Firsov \cite{prigofi}. 
When the 1D interactions are weak, the system will behave as a 1D \LL\
at higher energies. Naively, one would expect a crossover to 3D behaviour
at an energy of the order of the transverse bandwidth $\tperp$. $\tperp$ is,
however, renormalized by the 1D intrachain correlations, and the 1D--3D
crossover will only take place at a temperature $T_X^1 \approx \tperp' / \pi$
determined by the renormalized $\tperp'$ which can be significantly lower
than $\tperp$. $\tperp'$ is determined 
self-consistently from the requirement $\tperp' = \tperp z(\tperp')$
where $z(\tperp')$ is the quasi-particle residue at the energy scale
$\tperp'$. The renormalization of interchain tunneling $\tperp \rightarrow
\tperp'$ indicates a tendency of the electrons towards confinement on the
chains induced by their on-chain correlations.
Below the crossover temperature, the interference between the
Peierls and Cooper channels is destroyed. The transition temperatures
to a symmetry-broken
ground state and the competition of various types of order
then can be determined from standard summation of ladder
diagrams. In the 1D high-energy regime, the perturbative treatment
does not allow for generation of interchain pair tunneling. This, again,
could only take place in the case of strong interactions or presence of
a spin or charge gap, a problem that also is present in later work
by Brazovski\u{i} and Yakovenko \cite{braya}. 
Interestingly, however, in such a gapped regime,
Prigodin and Firsov obtain a maximum of the
superconducting $T_c$ as a function of $\tperp$ for $\tperp \sim \Delta$,
where $\Delta$ is the 1D spin or charge gap \cite{prigofi}. $T_c$ at maximum
is significantly higher than in the 3D limit $\tperp \rightarrow t_{\|}$. 
The reasons for this become apparent upon realizing that the maximal $T_c$
just occurs at the point where the Peierls state breaks down: here one
has an optimal combination of 1D effects (phonon softening at \tkf\ and
high density of states at $E_F$) with the 3D tunneling necessary to establish
superconductivity. 

Bourbonnais and Caron proposed  a \rg\ scheme both for the on-chain interactions
$g_i$ and the interchain tunneling $\tperp$ with respect to the free Fermi
gas which 
generates interchain pair tunneling from
single particle tunneling even at weak coupling \cite{bourbonnais,sc}.
The basic mechanism is shown in Fig.~6.3, where we display an expansion
of the vertex corrections in terms of the $g_i$ and $\tperp$. In particular,
the last diagram corresponds to  the coherent hopping of a pair to neighbouring
chains. This scheme produces all kinds of pair tunneling processes
because the general diagrammatic structure of 
Fig.~6.3 applies to all combinations of propagation directions and spins.
The \rg\ transformations under a scaling  
of the bandwidth cutoff from $E_0$ to $E_0(\ell)$ generate terms
of the type
\begin{equation}
H_{\rm pair} = \frac{1}{4} \sum_{\mu,i,j,q,\omega_n} V_{\mu}(\ell)
O_{\mu,i}^{\dag} (q, \omega_n)
O_{\mu,j} (q, \omega_n) \;\;\;,
\end{equation}
where finite pair tunneling matrix elements $V_{\mu}$ arise from $\tperp$ 
through
\begin{eqnarray}
\label{rgv}
\frac{d V_{\mu}({\bf k_{\perp}})}{d \ell} & = & f_{\mu}(\ell) \left[ 
\cos ( k_{\perp b} b) + \cos ( k_{\perp c} c) \right] + 
V_{\mu} ({\bf k_{\perp}}, \ell) \frac{ d \ln 
\bar{X}_{\mu}(\ell)}{ d \ell} - \frac{\left[ V_{\mu} ({\bf 
k_{\perp}}, \ell) \right]^2 }{2 \pi v_F} \\
\label{rgf}
f_{\mu} & = & \pm  2 \pi v_F \left( \frac{\tperp '}{E_0(\ell)} \right)^2
g_{\mu}^2(\ell) \;\;\;.
\end{eqnarray}
Here, the index $\mu$ = CDW,SDW,SS,TS
denotes the different kinds of fluctuations, the operators $O_{\mu,i}$
describe these fluctuations on chain $i$ as in Eqs.~(\ref{cdw}) or
(\ref{sdwx})--(\ref{sdwz}), and the
$g_{\mu}$ denote effective combinations of the coupling constants $g_i$
relevant for the respective operators. In (\ref{rgf}), the plus-sign applies
for $\mu = $CDW,SDW and the minus-sign for $\mu = $SS,TS. 
$\bar{X}_{\mu}(\ell)$ is  an auxiliary pair correlation function for 
fluctuations of type $\mu$. The last term is an RPA-like interchain ladder
contribution, the second term is the pair vertex correction whose
strong-coupling limit basically
has been treated in the earlier work, and the
first term generates finite $V_{\mu}$ from the initial value $V_{\mu} = 0$. 

The fluctuation $\mu$ has a tendency to long-range order at that wavevector
${\bf k_{\perp}}$ for which $V_{\mu} ({\bf k_{\perp}}) $ is negative and
extremal. One now integrates the \rg\ equations (approximately)   and 
finds the effective pair-tunneling amplitudes in all situations of interest,
both for weak and strong coupling, and on high and low energy scales.
Interestingly, the solutions allow for a regime where, for weak on-chain
interactions, scaling of $V_{\mu}$ goes to strong coupling at energies
above $T_X^1$. This corresponds to the growth of critical fluctuations
of type $\mu$ gaining 3D coherence. There can thus be a two-particle
1D -- 3D crossover temperature $T_X^2 > T_X^1$ where interchain pair
tunneling becomes coherent despite essentially 1D single-particle dynamics.
This complements, at weak coupling, earlier work in the strong-coupling
or gapped regime \cite{braya}. 
Moreover, since the interchain interactions strongly depend on temperature,
they can change from repulsive at high temperature to attractive at low
temperature. In this case, one will find antiferromagnetic fluctuations
coexisting with singlet superconductivity. At lower temperatures $T < 
T_X^1$, in the absence of perfect nesting, a transition to superconductivity
occurs. Here, interchain Cooper pairs form, and the superconducting gap
has a line of zeros on the Fermi surface.
By combining the quasi-1D \rg\ with approaches like RPA and parquet summation,
a variety of useful results on critical temperatures and
response functions in the presence of interchain tunneling can be computed
in all relevant regimes \cite{bourbonnais}.

By 1991, it was believed that this series of work provided quite detailed
a picture for the crossover from one into three dimensions both concerning
the electron dynamics and the establishment of long-range order of some
symmetry-broken phase. Then, Anderson pointed out that both our simple
\rg\ argument at the beginning of this section as well as the series of
detailed calculations reviewed thereafter, are irrelevant because they
neglect charge-spin separation \cite{pwaconf}. This is supposed to be
a particularly serious flaw because charge-spin separation is a consequence
of the restricted phase space in 1D and a nonperturbative effect. 
According to Anderson, there is a well-defined order in which to turn
on interactions and $\tperp$: interactions first, and then $\tperp$. In this
way, electrons on the chains would first separate into holons and spinons
and inactivate $\tperp$, because it requires both of them, i.e. an
electron, to tunnel. Consequently, the electrons would be confined to a
1D chain. Only pair tunneling of holons and spinons would be allowed, and
therefore, in the language of the preceding paragraphs, one would necessarily
have a two-particle 1D--3D crossover. A single-particle crossover,
presumably to a Fermi-liquid state, would be forbidden. 
In addition, anomalous fermion dimensions could, of course, strengthen the
intrachain confinement \cite{csa}. Anderson also
pointed out that two chains are enough to study confinement which introduces
considerable simplification in the actual calculations.

Anderson's suggestion has spun off a flurry of activity on two-chain Hubbard
and \lm s. Many of these do not follow Anderson's prescription and diagonalize
the bandstructure first and then turn on the interactions. Moreover, 
charge-spin separation is often neglected, again.
In this way, one arrives at the effective two-band models discussed before.

Charge-spin separation is present in a one-branch \LL\ where the only allowed
interaction is $g_{4\perp} \neq g_{4\|}$. The question ``confinement or not?''
can be studied here on a minimal model. 
Fabrizio and Parola have produced an exact solution of
such a two-chain one-branch \LL\ following precisely Anderson's prescription
in that they first produce charge-spin separation and then turn on $\tperp$
\cite{fab,fabprb}. 
The Hamiltonian is (\ref{hperpt}) with (\ref{hfree}) and (\ref{h4}) for each
chain. Keeping only the right-moving particles and dropping the 
corresponding index $r=+$ on the operators,
\begin{eqnarray}
\label{twochob}
H & = & \frac{2 \pi}{L} v_{\rho} \sum_{p>0} \left[ \rho_1(p) \rho_1 (-p)
+ \rho_2(p) \rho_2(-p) \right] \\
& + & \frac{2 \pi}{L} v_{\sigma} \sum_{p>0} \left[ \sigma_1(p) \sigma_1 (-p)
+ \sigma_2(p) \sigma_2(-p) \right] 
-  \tperp \sum_{k,s}\left[ c^{\dag}_{ks1} c_{ks2} + {\rm H.c.} \right] 
\;\;\;. \nonumber
\end{eqnarray}
The index 1,2 labels the chains. The on-chain part of the Hamiltonian has
already been diagonalized and exhibits charge-spin separation $v_{\rho} \neq
v_{\sigma}$. We now can use the bosonization identity (\ref{bos}) for
the interchain part $H_{\perp}$ in order to get its representation in terms
of the on-chain phase fields $\Phi_{\nu}$ and $\Theta_{\nu}$, Eqs.~(\ref{phi}) 
and (\ref{theta}). Introducing symmetric and antisymmetric
combinations of the charge and spin density operators $\nu_{1,2}(p)$ of the
different chains, the resulting boson Hamiltonian can be refermionized simply
by inverting the bosonization identity (\ref{bos}). Then going through
a series of unitary transformations, one finally obtains a Hamiltonian
bilinear in fermions with four dispersion branches 
\begin{equation}
\label{perpex}
\begin{array}{lll}
\epsilon_1(q) = v_{\rho} q & & 
\epsilon_3(q) = \frac{(v_{\rho} + v_{\sigma})q}{2}
+ \sqrt{\left[ \frac{(v_{\rho}- v_{\sigma})q}{2} \right]^2 + 4 \tperp^2} \\
\epsilon_2(q) = v_{\sigma} q & &\epsilon_4(q) = 
\frac{(v_{\rho} + v_{\sigma})q}{2}
- \sqrt{\left[ \frac{(v_{\rho}- v_{\sigma})q}{2} \right]^2 + 4 \tperp^2} \;\;\;.
\end{array}
\end{equation}
A particle-hole transformation having been performed in the course of the
calculation, these expressions are only defined for $q>0$. 

Two excitation branches 1,2 retain the original Luttinger dispersion: 
they originate from the chain-symmetric linear combinations of the density
operators which remains unaffected by $\tperp$. The antisymmetric combinations
are shifted by $\tperp$. At $q = 0$, they are split from the Fermi level
by $\pm 2 \tperp$ and for $q \rightarrow \infty$, they approach the two
Luttinger branches: $\epsilon_{3,4} \rightarrow v_{\nu} q$. The ground state
of the system is thus obtained by occupying the branch 4 up to $Q = 2 \tperp
/ \sqrt{v_{\rho} v_{\sigma}}$, i.e. up to $\epsilon_4 (Q) = 0$, the chemical
potential. Information on the confinement of the carriers on individual chains
can be obtained from several quantities. The ground state energy change due
to $\tperp$ is 
determined by the occupation of the branch 4
\begin{equation}
\frac{\Delta E}{L}  = - \frac{1}{2\pi} 
\frac{4 \tperp^2 }{v_{\rho} - v_{\sigma}}
\log \left( \frac{v_{\sigma}}{v_{\rho}} \right) \;\;\;.
\end{equation}
Being of order $\tperp^2$, it indicates that there is a finite ground state
expectation value of $H_{\perp}^{(t)}$ 
(e.g.~$\Delta E \propto \tperp^2$ is obtained
for free particles where $\tperp$ shifts the bands). The occupation number
difference between the bonding and anti-bonding bands (labelled by their
$k_{\perp}$-values $0$ and $\pi$) is
\begin{equation}
\label{nshift}
\frac{\langle N_0 - N_{\pi} \rangle }{L} = \frac{4 \tperp}{2 \pi }
\frac{1}{v_{\rho} - v_{\sigma}} 
\log \left( \frac{v_{\sigma}}{v_{\rho}} \right) \;\;\;,
\end{equation}
also indicating a shift between the bonding and anti-bonding bands $\propto 
\tperp$. 
The difference in occupation numbers corresponds to a shift of the Fermi 
wavevector between the bonding and antibonding bands of $2\Delta k_F$ with
\begin{equation}
\label{deltakf}
\Delta k_F = \frac{\tperp}{v_{\rho} - v_{\sigma}} \log \left( 
\frac{v_{\rho}}{v_{\sigma}} \right) \;\;\;.
\end{equation}
All three quantities show that interchain tunneling does shift the
bonding with respect to the antibonding band, and that there is no confinement
of electrons on individual chains \em despite \rm charge-spin separation.
On the other hand, the nature of the spectrum 
indicates that charge-spin separation is a phenomenon robust
against interchain coupling so that it could conceivably survive 
under certain circumstances in more
than one dimension, and with it some \LL\ physics.

This is made more clear in the single- and many-particle dynamics.
The Green function for particles with transverse momentum $0$ or $\pi$
can be calculated from a cumulant expansion and becomes
\begin{equation}
\label{tcgreen}
\langle \Psi_{0,\pi} (x,t) \Psi_{0,\pi}^{\dag} (0,0) \rangle \sim
e^{i(k_F \pm \Delta k_F)x} (x - v_{\rho} t)^{-3/8} (x - v_{\sigma}t)^{-3/8}
(x-v_r t)^{-1/4} \;\;\;,
\end{equation}
where $v_r = 2 v_{\rho} v_{\sigma} / (v_{\rho} + v_{\sigma})$ and the
$+(-)$-signs go with $k_{\perp}=0,\pi$, respectively. The corresponding
spectral function still is purely incoherent -- there is no quasi-particle-like
feature -- but now, it exhibits
a three-peak structure at wavevectors small with respect to $\tperp/
(v_{\rho} - v_{\sigma})$,
compared to two peaks for
the isolated chains. In the new excitation dispersing with 
$v_r$, charge and spin strongly interact
through $\tperp$. At large wavevectors, on the contrary, $\tperp$ seems
to be inefficient, and the spectral function reduces to that of two 
uncoupled chains \cite{spec2,spec1,fogedby}.
The long-wavelength charge and spin density correlation functions are also
changed. The on-chain spectral function for the charge fluctuations
contains a pole contribution from the Luttinger branch which has the form
of isolated chains. In addition, there are incoherent pieces close to
$v_{\rho} q \pm 2 \tperp$
indicating branch cuts in
the correlation function, which originate from the coupling between
charge and spin fluctuations generated by $\tperp$ as well as the
nonlinear dispersion of $\epsilon_{3,4}(q)$. As $q$ is increased, spectral
weight is transferred from the incoherent features into the central pole.

For the special problem of the two-chain--one-branch \LL , the same results
are found by proceeding in the opposite sense: first diagonalize the
band structure and then turn on the interactions \cite{fink}. Here, one
bosonizes the bonding and antibonding fermions, but now the antiparallel-spin
interactions lead to a Hamiltonian which is highly nonlinear in the boson
operators [of the structure of the backscattering Hamiltonian $H_{1\perp}$
in Eq. (\ref{hperp}) but with right- or left-moving fields only], 
so that an exact solution no longer is feasible.
Still, the Hamiltonian separates into four pieces corresponding to the
four excitations found in Eq. (\ref{perpex}). These spectra can be determined
from thermodynamics and using special symmetries of the model, 
and agree with (\ref{perpex}). The Green function then obtains as
Eq. (\ref{tcgreen}). 

Although obtained from an exceedingly simple Hamiltonian, these results
are extremely important. They tell us (i) that even in the case where
the interchain tunneling is turned on \em after \rm the establishment
of charge-spin separation on the chains, there is no confinement of electrons
by this special kind of 1D interactions,
and their interchain dynamics can become coherent; (ii) the results do
not depend on the order of turning on charge-spin separation and interactions.
Since charge-spin separation was the only important feature left out in
the work reviewed at the beginning of this section, we get additional confidence
in the relevance of its conclusions.

One can now include the $g_2$-forward scattering into a two-chain 
model. This couples right-and left-moving electrons and gives rise to the
anomalous dimensions on a single chain parameterized by $K_{\rho}$. On a
more qualitative level, one can study confinement by looking at the
stability of the 1D momentum distribution function $n(k)$, Eq. (\ref{nkut}),
with respect to $\tperp$. Turning on $\tperp$ at the end, one finds
\cite{metz}
\begin{equation}
\delta n( \bk)  = \left\{ 
\begin{array}{ll} \tperp \cos k_{\perp} \left[ C + D ( k_{\|} - k_F
)^{2 \alpha - 1} \right] \hspace{0.5cm} & {\rm for} \hspace{0.5cm} \alpha < 1 \\
\tperp \cos k_{\perp} \left[ C + D (k_{\|} - k_F
) \right] & {\rm for} \hspace{0.5cm} \alpha \geq 1 \;\;\;.
\end{array}
\right.
\end{equation}
There is a singular correction to the momentum distribution in the 
neighbourhood of $k_F$ for small $\alpha$ while it vanishes for large $\alpha$.
This essentially reproduces the conclusions of our 
scaling argument on the level $\delta F^{(2)}$
from the beginning of this section, Eq. (\ref{dftwo}). 
Of course, in the large-$\alpha$
limit, we will find relevant pair-tunneling processes which are not covered
by this argument. 

A more detailed analysis is possible if one accepts first diagonalizing the
band structure -- but from the behaviour of the two-chain--one-branch model
above, we expect this procedure to be safe. In a \rg\ analysis 
of Luttinger models ($g_2, g_4 \neq 0$) coupled by $\tperp$, new 
relevant interactions are generated by the RG \cite{fink}. Specifically,
an interaction term
\begin{equation}
\label{hlam}
H_{\lambda} = \frac{\lambda}{4L} \sum_{\mid p \mid \ll k_F} 
\left[ \rho_{+,0}(p) -  \rho_{+,\pi}(p) \right] 
\left[ \rho_{-,0}(p) -  \rho_{-,\pi}(-p) \right] 
\end{equation}
is generated and goes relevant, and its coupling constant $\lambda$ increasing
towards large negative values under renormalization -- independent of the
sign of the on-chain $g_2$! It is driven by interband
forward scattering with antiparallel spins involving opposite branches
(i.e. of interband-$g_2$-type but formally of structure similar to 
the usual backscattering Hamiltonian) which does not conserve the total
spin on each of the four excitation branches.
In Eq. (\ref{hlam}), $\rho_{r,k_{\perp}}(p)$
denotes the right- or left-moving ($r=+,-$) 
density fluctuations obtained from the bonding or antibonding ($k_{\perp} =
0, \pi$) fermions with parallel momentum $p$. Due to the shift between
the bands brought about by $\tperp$, the differences in $[\ldots]$ do not
equal the on-chain densities. One can now bosonize the strong-coupling
fixed point and find that a spin gap opens in the system. At this level,
it is not completely clear if this favours SS or CDW correlations,
but in a 3D array of pairs of chains, the Peierls
divergence will be cut off by $\tperp$ and superconductivity comes up
\cite{fink}. Here, the Cooper pairs can be formed predominantly
either on or between the chains, depending on the sign of $g_2$. For repulsive
interactions, one would find pairing between the chains, in qualitative 
agreement with Bourbonnais and Caron \cite{bourbonnais}.

One can go one step further and consider the Hubbard model on two-chains --
finally, this is the problem we are most interested in! This requires
the inclusion of all kind of backscattering processes, and one has to treat
a problem with fifteen independent coupling constants (excluding commensurate
situations where additional Umklapps come up) \cite{fabprb,fab3}. 
The phase diagram is given in Fig.~6.4.
There is a trivial \LL\ (LL1) at large $\tperp$ where the upper
band is empty and only the bonding band is filled. When the two bands only 
slightly overlap, there is another \LL\ (LL2) whose existence
is related to a big difference between the bonding and antibonding
Fermi velocities. However, there is a finite value of $\tperp^{\rm eff}$
and no confinement. Decreasing $\tperp$, one enters strong-coupling phases.
The phase III at large $U$ and small $\tperp$ does have confinement
($\tperp^{\rm eff} = 0$) and strong pair hopping between the chains. 
It is interesting that an interband pair susceptibility, pairing particles
of the bonding with those of the antibonding band, has the most divergent
fluctuations, but on-chain SDWs diverge, too. At smaller $U$, one enters
another phase I where the dominant pairing fluctuations involve pairs
from the bonding or from the antibonding bands. An interchain SDW diverges, too,
though less strongly. Intercalated between those phases may be a third
phase II with conventional on-chain Cooper pairing.

The correlations on two Hubbard chains can also be studied with numerical
calculations \cite{noack}. 
A density-matrix \rg\
study finds a spin gap at half-filling with exponentially decaying 
spin-spin correlations. The correlation
length is a few lattice constants. Singlet pairing correlations also decay 
exponentially, and their correlation length is even shorter which may
indicate a liquid of disordered singlets. If one dopes the systems
with holes, the spin gap is quite robust and persists down to at least
$n \sim 3/4$ i.e. a doping level of 25\%. On the other hand, the doping
greatly favours the pairing correlations (both on and between the chains
i.e. in a d-wave like pattern) which change from exponential
to power-law and therefore will be dominant at long distance. 
However, they decay as $1/r^2$, like free fermions which is definitely
weaker than the divergences predicted from \rg . In addition, no sign
of the subdominant SDW divergences predicted by \rg\ is reported from
the numerical calculations. 
Another Quantum Monte Carlo study where the reduced density matrix of
the superconducting correlations was computed, does find evidence for
enhanced superconducting correlations with respect to the uncorrelated
system at $n = 3/4$ \cite{asai}. There is some structure in this enhancement
when $\tperp$ is varied at fixed $U$ which has been associated with 
the different phases LL2, SC1, and SC2 in Fig.~6.4.  When
$\tperp$ becomes so large that only the bonding band is occupied (LL1),
no enhancement can be detected. On the other hand, a finite size analysis
of the correlations assuming the \LL\ power laws (\ref{rcdw}) and (\ref{ass})
suggests that the exponent $K_{\rho} < 1$ which would imply dominant density
wave and only leave space for subdominant SS correlations \cite{asai}.
This conclusion does not agree with the \rg\ work \cite{fabprb,fab3}.

Much information has also been gathered on arrays of coupled $t-J$-chains
following a suggestion that they could be used as a model for certain
cuprate compounds \cite{gopa}. Two or more chains described by the standard 
$t-J$-Hamiltonian (\ref{hamtj}) 
are coupled by transverse hopping ($\tperp$) and  
and exchange ($J_{\perp}$) integrals. In an 
undoped system of two (or an even number of)
coupled Heisenberg chains \cite{twoheis}, if 
$J_{\perp} \gg J$, of course singlet pairs will form across the rungs of the
ladder, and  the excitations
will have a spin gap. This singlet-triplet gap survives not only down
to the isotropic point $J_{\perp} = J_{\|}$ but there is evidence that it
does so for any finite $J_{\perp}$ \cite{twoheis}. On the other hand, for
an odd number of chains, no such dimer state is possible, and it turns
out that its excitations are gapless like in the single-chain model.
Introducing holes into two chains
will lower but not destroy the spin gap \cite{kvesh,tsune}. One can now
combine this fact with a bosonization analysis to inquire what type 
of correlations will govern the physics of this two-chain system. 
In a spin-singlet state, the effective exponents for the charge 
degrees of freedom become
\begin{equation}
\frac{1}{K_{\rho}^{\pm}} =  \sqrt{1 + \frac{J}{\pi t} \langle {\bf S}_{ij}
\cdot {\bf S}_{i+1j} \rangle \pm \frac{J_{\perp}}{\pi t} \langle {\bf S}_{i1}
\cdot {\bf S}_{i2} \rangle } \;\;\;,
\end{equation}
where $\pm$ stands for the chain-symmetric (antisymmetric) combination of
the phase fields $\Phi_{\nu}$ and $\Theta_{\nu}$ used to bosonize the
single chain, ${\bf S}_{ij}$ is the spin operator at site $i$ of chain
$j=1,2$, and the expectation values do not depend on the
site index $i$ \cite{kvesh}. In general, one has $K_{\rho}^+ > 1$ corresponding
to attractive interactions in the bonding channel which are generated by
the preferred singlet 
($\langle {\bf S}_{i1} \cdot {\bf S}_{i2} \rangle < 0$) correlations
across the rung of the ladder. This analysis cannot determine if $K_{\rho}^- >
\; {\rm or} \; < 1$, which would correspond to modified d-wave SS correlations
or a special CDW phase where an alternating 
flux $\Phi = 2k_F$ is enclosed in a plaquette \cite{kvesh}. Such an ``orbital
antiferromagnet'' had been discovered earlier in a study of a two-chain
model of spinless fermions with nearest-neighbour interaction \cite{ners}. 
Numerical calculations seem to prefer the d-wave SS correlations \cite{tsune}.
The orbital antiferromagnet is a two-chain version of the flux states discussed
for the 2D \htcs ; these flux phases have been discussed also for anisotropic
2D systems as we consider them here \cite{vome}. They model systems with
both open and closed orbitals in the neighbourhood of the Fermi surface, and
in the anisotropic limit, instabilities reminiscent of the 1D systems are
found.

There is also a detailed picture of the excitations created upon hole-doping 
a $t-J$-ladder \cite{tsune}. The lowest excitation
at half-filling is a spin-triplet above the gap. Doped holes (with
concentration $\delta$) will pair
so that the spins can take advantage of the singlet binding across the rungs.
One obvious magnetic excitation is the triplet, again, which can propagate
in the ladder with an exchange integral $J/2$ while the hole pair moves
with $1/(J_{\perp} - 4/J_{\perp})$. The number of such possible
triplets goes as $(1-\delta)$. But there is another possibility: one can
form quasi-particles, singly occupied rungs, carrying charge and spin-1/2, 
as a bound holon-spinon pair. These quasi-particles will move with a 
hopping element $t/2$ and their number scales with $\delta$. They have
triplet spin correlations. In a wide range of $J_{\perp}$, the creation
of such quasi-particles is energetically favourable, and the spin
gap is then reduced from the singlet-triplet gap of the half-filled model. 
Consequently, the dynamics and thermodynamics of the spin excitations
is dominated by different energy scales as the doping level is varied:
increasing doping will bring up a new low-energy scale associated with
the quasi-particle excitations \cite{tsune}. The quasi-particles also
show up in the single-particle spectral function although there are sizable
incoherent contributions. There is a coherent peak dispersing towards
the Fermi energy as $k \rightarrow k_F$,
until the spin gap is reached. In this limit, the particle peak at $\omega >0$
has acquired a strong shadow component at $\omega < 0$, as in a superconductor
\cite{tsune}. On the other hand, this strong shadow component is the direct
continuation of the spectral weight at $\omega <0$ found for the \LL\
in Section \ref{secprop}, Fig.~3.6, 
to a situation where a fully developed spin gap exists.

A generalization of this picture for four coupled $t-J$-chains is now
available \cite{poil}. Also, the interplay of superconductivity and
phase separation has been studied in the regime of large $J/t$ \cite{riera}.
Here, a strong possibility for phase separation between the chains is
found, and this is precisely the range where strong signals of 
superconductivity are detected. Notice, however, that these $J$ values 
are out of the range which can be derived from large-$U$ Hubbard models,
although more general models do allow them \cite{pafo}.

Very recently, the partition and Green functions have been derived for
Luttinger liquids on an arbitrary number of chains coupled by interchain
hopping, including charge-spin separation \cite{boies}. This analysis
essentially confirms the earlier results \cite{bourbonnais} where this
feature had been neglected. Provided the interactions are not so strong
that a two-particle crossover would occur before the single-particle
crossover, this work provides us with Green functions containing explicitly
a quasi-particle residue indicative of a Fermi liquid ground state in
this case, plus correction terms containing the remnants of the 1D \LL .
Others give explicit spectral functions for the case of selfconsistently
screened Coulomb interactions \cite{schucou,baris} including interchain
hopping \cite{kopie} although some approximations made may  
overestimate the anomalous 1D component of the spectra.

An important virtue of the variational wavefunctions used for mapping
out the \LL\ correlations of the 1D $t-J$-model in Section \ref{sec1d} is
the possibility to generalize them to 2D \cite{vale}. The \LL\ state with 
nontrivial $K_{\rho}$
is stabilized by gains in kinetic energy with respect to the Gutzwiller
wave function describing a Fermi liquid. It possess the anomalous
dimensions of a \LL\ but no charge-spin separation. Moreover, as in the
1D case, one can apply the power method to obtain increasingly accurate
approximations to the true ground state which conserve the typical 
power laws \cite{chenle}. 
There have been claims of charge-spin separation in the 
2D $t-J$-model based on high-temperature expansion \cite{putti} but
this interpretation of the data has been opposed by others 
\cite{chenotr}.

An approach rather different from the work above has been taken by
Castellani et al.~who consider fermions with short-range interactions
in continuous dimensions $1 \leq D \leq 2$ \cite{newmetz}. Here, the
Fermi surface is closed and isotropic in the $D$-dimensional reciprocal space
while the approaches coupling 1D \LL s with $\tperp \ll t_{\|}$ all imply open
warped Fermi surfaces which conserve a strong 1D character.
They find
that while the 1D conservation laws for total charge and spin on a
branch of the dispersion, Eqs.~(\ref{cont1}) and (\ref{cont2}),
are no longer satisfied exactly, similar laws for
charge and spin associated with directions radially outward from the
$D$-dimensional Fermi surface are still obeyed asymptotically. This allows
the formulation of corresponding asymptotic Ward identities which strongly
constrain the low-energy physics close to the Fermi surface. In particular,
a Fermi liquid fixed point is found for all dimensions $D > 1$. However,
for dimensions $D \leq 2$, there are dramatic corrections to quasi-particle
behaviour away from the fixed point. They are strongly reminiscent of
the behaviour of 1D \LL s but now in radial direction. 
Such a ``tomographic \LL '' behaviour at finite energy could completely
mask the Fermi liquid fixed point physics, and provide a realization of
Anderson's suggestion \cite{ar}. Singular interactions, as proposed
by Anderson, could then conceivably stabilize such physics also at the
fixed point. 

We finally mention that there is a variety of work in 2D producing
evidence for non-Fermi-liquid and possibly \LL\ low-energy physics using
peculiar, often long-range, interaction Hamiltonians \cite{ll2d}. 
Others attempt to describe the Fermi liquid with methods borrowed from
the 1D systems reviewed here, such as bosonization and \rg\ \cite{bosfl}.
In some sense, one goes the way opposite to the one we took in Section 
\ref{altmet} where we applied standard techniques of  Fermi liquid theory
in 1D. Further development of these methods will hopefully sharpen our
understanding of scenarios for a possible breakdown of Fermi liquid theory
in higher dimensions, and for similarities and differences to the 1D case
reviewed here.

\section{Edge states in the  quantum Hall effect}
\label{fqh}
When a 2D electron gas which can be created in the inversion layer of a
metal-oxide-semiconductor or a semiconductor heterostructure, 
is exposed to a strong magnetic field, it is observed that the Hall conductance
is quantized in units of the elementary conductance $\sigma_{xy} = \nu e^2/h$ 
\cite{qhebook}. The initial observation was that $\nu$ is integer 
\cite{klitzing}
but subsequently fractional $\nu$ were discovered, too \cite{tsui}.
In both cases, the quantization is due to the existence of a mobility gap
at the Fermi level in the bulk of the sample, 
although its microscopic origin is different:
in the integer effect, disorder leads to localized states while in 
the fractional effect, correlations condense the particles into a new 
collective state whose excitations are gapped. Due to the mobility gap in the 
bulk, transport must take place on the edge of the sample -- a 1D manifold. 

The basic model for the integer effect was put forward by Halperin 
\cite{halp} and developed further by B\"{u}ttiker \cite{buetti} and others
\cite{wenrev}.
In Fig.~6.5 an annulus with inner radius
$r_1$ and outer radius $r_2$ is considered and the disorder is supposed to
be confined to the bulk of the annulus. 
The edges are shifted upward in energy 
because of the boundary condition of vanishing
wavefunction at the sample boundaries. Although all bulk states at $E_F$ are 
localized (if there are any), at the fields where $\sigma_{xy}$ shows a 
plateau, low-energy excitations are possible at the edges. 
The excitations living on the edges are ordinary electrons. 
They form a 1D chiral Fermi liquid -- Fermi liquid now understood in the
sense of the higher-dimensional systems. In the \lm\ (\ref{hlutt}),
there is only one of the two dispersion branches:
due to the
orbital coupling, all electrons move in the same direction, i.e. have a definite
chirality. This leaves $g_4$ as the only possible interaction. However,
due to Zeeman coupling, the electrons are fully spin-polarized, and
$g_{4\perp}$, in principle able to generate charge-spin separation, 
is quenched. Remains 
$g_{4\|}$ which only renormalizes the Fermi velocity and does not destroy
the quasi-particle pole in the Green function.

Gapless edge excitations also exist in the fractional quantum Hall effect,
and Wen has clarified their nature and dynamics in considerable detail
\cite{wen}. We attempt to follow his proceeding here, because his way
from very general principles (essentially only gauge invariance, locality
of the theory, and incompressibility of the ground state)
to a detailed operator description of
the low-energy properties is in some sense opposite to the bulk of our
earlier presentation, and highly instructive. Related work has been
performed by Stone \cite{stone}.

We first fix the general form of the action of the edge excitations from
the requirement of gauge invariance.
Assume that a system displays the quantum Hall effect with $\sigma_{xy}
= \nu e^2 / h$ in an magnetic vector potential $\bar{A}_{\mu}$. We do
not know the detailed Hamiltonian, but due to the gap in the quasi-particle
excitations, we know that the electrons can be integrated out safely, 
resulting in an effective Lagrangian
\begin{eqnarray}
\label{cslag}
{\cal L}_{\rm eff} [\delta A_{\mu}]
& = & \frac{\nu e^2}{4 \pi} \delta A_{\mu} \partial_{\lambda}
\delta A_{\kappa} \epsilon^{\mu \lambda \kappa} + \frac{1}{4 g_1^2} 
(\delta F_{01} )^2 - \frac{1}{4 g_2^2} (\delta F_{12} )^2 + \ldots \;\;\;, \\
\delta A_{\mu} & = & A_{\mu} - \bar{A}_{\mu} \;\;\;, \hspace{1cm}
\delta F_{\mu \lambda} = \partial_{\mu} \delta A_{\lambda} 
+ \partial_{\lambda} \delta
A_{\mu} \;\;\;, \hspace{1cm} \mu = 0,1,2 \;\;\;. \nonumber
\end{eqnarray}
$\delta F_{\mu \lambda}$ is the strength of the magnetic field. 
The first term on the right-hand side is called Chern-Simons term, and its
coefficient is given by the Hall conductance. The detailed properties
of ${\cal L}_{\rm eff}$ are unimportant in what follows. On a compactified
space, say a torus, the action $S_{\rm bulk} = \int \! d^3 x \: 
{\cal L}_{\rm eff}
[\delta A_{\mu}]$ is invariant under gauge transformations $A_{\mu}
\rightarrow A_{\mu} + \partial_{\mu} f(x)$. This is not so on a bounded
space where
\begin{equation}
\label{gaugebulk}
S_{\rm bulk}[A_{\mu} + \partial_{\mu} f(x)] = S_{\rm bulk} [ A_{\mu} ]
+ \frac{\nu e^2}{4 \pi}\int \! dx_0 \: d\sigma f(x) \delta F_{\sigma 0}
(x) \;\;\;.
\end{equation}
The variable $ \sigma $  parameterizes the boundary, and $x_0=t$. 
Gauge invariance is violated by a boundary term generated by the Chern-Simons
term. A gauge invariant action can now be obtained by adding a boundary
action $S_{\rm edge}$, 
from which the properties of the boundary excitations can be constructed. 
$S_{\rm edge}$ alone must not be gauge invariant and must
transform as (\ref{gaugebulk}) but with a minus-sign
in front of the Chern-Simons contribution, and is given by
\begin{equation}
\label{edgeac}
S_{\rm edge} = \frac{1}{2} \int \! dt \: d\sigma \: dt' \: 
d\sigma' \delta A_{\alpha}
(t,\sigma) R_{(j)}^{\alpha \beta} (t-t',\sigma - \sigma') \delta A_{\beta} (t',
\sigma') \;\;, \hspace{0.5cm} (\alpha, \beta = 0, \sigma) \;\;,
\end{equation}
where $R_{(j)}^{\alpha \beta}$ is the time-ordered 
current-current correlation function
(containing, in the notation of Chapter \ref{chaplm}, $R_{\rho\rho}$,
$R_{jj}$ and $R_{\rho j}$). It has the properties
\begin{equation}
\label{curprop}
- k_{\alpha} R_{(j)}^{\alpha \beta} = \frac{\nu e^2}{4 \pi} 
\epsilon^{\alpha \beta} k_{\alpha}
\;\;\;, \hspace{1cm} R_{(j)}^{\alpha \beta}(k_{\alpha}) = 
R_{(j)}^{\alpha \beta}(-k_{\alpha})
= \left[ R_{(j)}^{\alpha \beta}  (- k_{\alpha}) \right]^{\star} \;\;\;,
\end{equation}
where $k_{\alpha} = (\omega, k)$ is a reciprocal space vector. 
We recognize the first equation as a Ward identity, up to the different
right-hand side identical to (\ref{wi1}). The second equation implements
the required symmetries (even for the symmetric and odd for the nonsymmetric)
and the reality of the correlation functions. 

One can imagine knowing the Hamiltonian for the edge excitations and their
coupling to the gauge field $A_{\alpha}$. By integrating out the
edge excitations, one would then obtain the action (\ref{edgeac}). We
proceed inversely and try to get information on the dynamics of the excitations
from the effective action. Assume that all edge excitations are gapped,
and that the theory is local.
Then $R_{(j)}$ is smooth near $\omega =0$ and $k=0$. But a smooth function
cannot satisfy (\ref{curprop}). There must thus be gapless excitations
(labelled by $i$). 
If their dispersion is linear $\omega(k) = v_i k$, and if $R_{(j)}$ has pole
structure for the gapless excitations (the gapped ones can only contribute
polynomials), its singular part must be of the form
\begin{eqnarray}
\label{urcor}
R_{(j),ij}^{\alpha \beta \; \rm sing}(\omega , k) & = 
& \frac{\delta_{ij} \eta_i
S^{\alpha \beta}(\omega , k) }{2 \pi (\omega - v_i k)} \;\;\;, \\
S^{00} = k & , & S^{0 \sigma}=S^{\sigma 0} = (\omega + v_i k)/2
\;\;, \;\;\;\; S^{11} = v_i \omega \;\;\;. \nonumber
\end{eqnarray}
$\eta_i = {\rm sign} (v_i) q_i^2 $, where 
$q_i$ is the charge of the excitations.
There must be an operator $j_i^{\alpha}(k)$ which generates a state with
energy $\omega(k) = v_i k $ from the vacuum. One can then deduce
first the vacuum expectation values of the commutators of $j^{\pm}_i(k)
= [j_i^0(k) \pm j_i^{\sigma}(k)/v_i]/2$ 
from the correlation functions, and then, under some very general
further assumptions, the algebra of the operators themselves
\begin{equation}
\label{algebra}
[ j_i^+ (k), j_j^+(k') ] = | \eta_i | k \delta_{i,j} \delta_{k,-k'} 
\;\;\;, \hspace{1cm} [ H, j_i^+ (k) ] = c k j_i^+ \;\;\;.
\end{equation}
For sign$(v_i)$\-sign$(k) < 0$, the operator $j_i^+(k)$ acts as an annihilation
operator. For sign$(v_i)$sign$(k) >0$, it generates the harmonic spectrum
of $H$. The operator $j^-_i(k)$ acts as a null operator. 

We recognize the algebra of the $j_i^+(k)$ (\ref{algebra}) 
as a $U(1)$-Kac-Moody algebra. Unlike  Eq.~(\ref{kacmood}),
however, the prefactor of the right-hand side is $|\eta_i|$ rather than
unity. This suggests that our edge excitations on each branch $i$ are described
by a $U(1)$-Kac-Moody algebra with a central charge $c = | \eta_i | = q_i^2 $. 
Following our discussion in Section \ref{cft}, we 
then can construct both an effective action as a bilinear in
boson currents $j_i^{\pm}$ and in terms of chiral fermions $\Psi_i$
coupled to a gauge field,
and the latter reads
\begin{equation}
\label{sedge}
S_{\rm edge} = i \sum_i \int \! dt d\sigma \Psi_i^{\dag} \left[ (\partial_t
+ i q_i \delta A_0) + v_i ( \partial_{\sigma} + i q_i \delta A_{\sigma} )
\right] \Psi_i
\end{equation}
with charges $q_i$ satisfying the sum rule
\begin{equation}
\label{rfr}
\sum_i \eta_i = \sum_i {\rm sign}(v_i) q_i^2 = \nu e^2 \;\;\;.
\end{equation}
For the integer effect ($\nu$ integer), the charges are integer 
i.e. the fermions describe real electrons, but for the
fractional quantum Hall effect, they are irrational in general. 
If there is a single branch, we have $ q_i = \sqrt{\nu} e$.
The fermions rather refer to solitons than real electrons.
So long as all edge excitations move in the same direction, there is no
possibility to open a mass gap. Only when velocities have different signs
(an issue we touch upon below) can a gap be opened, as a
consequence of backward and Umklapp scattering. In the
same way, it should be apparent from the discussion of impurity scattering
in Section \ref{lltrans} that the edge excitations travelling in the
same direction, are not sensitive
to scattering by impurities. 

For practical calculations, of course, the boson representation is more
convenient, and the Hamiltonian corresponding to the boson action is
\begin{equation}
\label{chill1}
H = \sum_i \sum_k \frac{\pi v_i }{q_i^2} \rho_i(k) \rho_i(-k) \;\;\;.
\end{equation}
There may be also interactions between different excitations, i.e.
\begin{equation}
\label{chill2}
H \rightarrow H + \delta H \;\;\;, \hspace{1cm} \delta H = \sum_{ij}
\sum_k g_{ij} \rho_i (k) \rho_j(-k) \;\;\;.
\end{equation}
This has the structure of the Luttinger Hamiltonian (\ref{hlutt}) with $g_2$ and
$g_4$ processes, depending on the sign of the velocities of the branches,
or of its multicomponent generalization (\ref{hncomp}). 
In general, the properties will be different, however, because $c \neq 1$.
(\ref{chill2}) can be diagonalized by a Bogoliubov transformation, leading
to new operators $\tilde{\rho}_i$, renormalized velocities $\tilde{v}_i$,
and to renormalized charges $\tilde{q}_i = \sum_j U_{ij} q_j$ where $(U)_{ij}$
is the transformation matrix. It is these renormalized charges and velocities
which are experimentally measurable in edge magnetoplasmon excitations.
Fractional charges can also be found in the integer effect
when several branches of excitations are present. 
Wen showed that the renormalized fractional charges can be measured 
experimentally as the strength of the 
peaks in the absorption spectrum of a rotating electric  field. Also the
width of the resonances is related through the charges to the edge resistance.

Up to now, we have only considered particle-hole excitations out of the
Fermi sea on the edges, i.e. we have restricted to the charge-zero sector
of the theory. We now consider the charge excitations on the edges, to
give a more systematic basis to the notion of irrational charges,
and to construct a relation between the physical fermions $\Psi$
and the bosons living on the edges. We assume a single excitation branch for the
moment with $q = \sqrt{\nu}$. There is thus a definite chirality.
The fermions act as charge-raising operators 
\begin{equation}
[ \Psi, Q ] = e \Psi \;\;\;, \hspace{1cm} Q = \int \! d \sigma j^0 (\sigma)
\;\;\;.
\end{equation}
The Hilbert space is therefore composed of sectors with charge $Q$, and 
within these sectors, $j^+$ creates particle-hole excitations. Within
each sector $Q$, it generates 
the Fock space of the harmonic oscillators defined
by the $U(1)$-Kac-Moody algebra (\ref{algebra}). One can now introduce
a bosonic field $\varphi(x)$ via the current and charge
\begin{equation}
j^{\alpha}(t,\sigma) = \frac{\sqrt{\nu}}{2 \pi} \epsilon^{\alpha \beta} 
\partial_{\beta} \varphi(t,\sigma) \;\;\;, \hspace{1cm} Q = \sqrt{\nu} 
N_{{\rm sign} \: v} \;\;\;.
\end{equation}
The charge-raising operators of a chiral boson theory are in general,
Eq.~(\ref{bosfera})
\begin{equation}
\label{chiralbos}
\Psi(x) = : e^{ i \gamma \varphi(x)} : \;\;\;.
\end{equation}
From the requirement that they anticommute, we derive
$ \gamma^2 = 2n+1$, an odd integer. Moreover, $\Psi$ must carry a unit
charge. From $[ Q, \Psi]$ we find that the charge of $\Psi$ is $\gamma
\sqrt{\nu}$ which must equal unity. This imposes a restriction on the
filling fractions $\nu$ which can be described by a single branch
of excitations $\nu = 1 / (2n+1)$. All other filling fractions must
possess more than one edge excitation -- a conclusion 
also verified in numerical calculations \cite{john}. 
But even in the single-branch situation, a physical electron carrying unit
charge, added to
the edge, is fragmented into solitonic excitations of charge $\sqrt{\nu}$. 
However, 
the charges of excited states are integers, not multiples of $\sqrt{\nu}$.

In principle, one can now calculate all the correlation functions on the
edge. Due to the fractional charges, or equivalently to the central charge
being different from unity, anomalous powers arise even in the one-branch
situation, contrary to the standard \LL\ of Chapter \ref{chaplm}.
This can be seen quite easily from (\ref{chiralbos}) where the fractional
charge $\nu = K$ plays the same role as the stiffness constant played in
Chapter \ref{chaplm}. The mapping is provided by comparing to 
Eq.~(\ref{currop}). 
As an example, the single-particle Green function is
\begin{equation}
\label{chiralgf}
\langle \Psi(t,\sigma) \Psi^{\dag}(0,0) \rangle \sim \left(
\frac{i}{t+\sigma} \right)^{1/\nu} \;\;\;.
\end{equation}
Rewriting this in the form of a standard \LL , one concludes that
the chiral single-particle exponent is
\begin{equation}
\label{chiralpha}
\alpha_{\rm chiral} = \frac{1}{\nu} - 1 = 2n
\end{equation}
for a filling fraction $\nu = 1/(2n+1)$. The scaling relation between
$\alpha$ and $K$ in a chiral \LL\ is different from the non-chiral system. 
The remarkable fact here is that
the exponents become \em universal \rm and are fully determined by the
filling of the Landau levels. This means that one can
precisely predict the exponents of the correlation
functions probed by specific experiments. Moreover, $\nu$ being a topological
invariant, the exponents are expected to be robust against perturbations. 
Wen has proposed the label ``chiral \LL '' for the low-energy physics
of the quantum Hall edge states.

An interesting generalization, which could also be relevant for situations
with several edges with opposite velocities, is provided by considering
the edge states on a cylinder threaded by the magnetic flux. There are now
two edges with excitations moving in opposite directions, and the \LL\
is no longer chiral. Of course, it can be built on the chiral \LL\ of a
disk, and the structure of the theory is quite similar to the previous one. 
There are two important differences, however: (i) charge excitations can
be transferred between the edges, which complicates somewhat the quantization 
rules relating $\gamma$ in (\ref{chiralbos}) to the filling fraction $\nu$;
(ii) if the edges are not too far from each other, tunneling both of
electrons and of Laughlin quasi-particles, the fractionally charged
objects introduced earlier, may take 
place between them. Electron tunneling can be described by the operator
\begin{equation}
\label{qhetunnel}
H_{\rm tunnel} = g \int \! dx \cos \left[ \frac{\varphi(x)}{\sqrt{\nu}}
\right] \;\;\;,
\end{equation}
which is isomorphic in form to our the backscattering operator (\ref{hperp}). 
The scaling dimension of this operator is $2 - 1/\nu$, and the operator is
irrelevant for $\nu < 1/2$ while it is relevant for $\nu > 1/2$. This would
imply that in the integer effect ($\nu =1$), a gap could open on the edges
when they are brought close enough together. On the other hand, for the
fractional effect $\nu = 1/(2n+1) \leq 1/3$, the electron 
tunneling operator is irrelevant. Tunneling of Laughlin quasi-particles
is possible because the two edges are connected by the quantum Hall fluid
(and not vacuum), and described by an operator (\ref{qhetunnel}) where,
however, the factor $1/\sqrt{\nu}$ is replaced by $\sqrt{\nu}$ \cite{wenrev}. 
This operator is always relevant. An important problem, however, is that
of a local constriction on a Quantum Hall cylinder: here the inter-edge
tunneling
only takes place at $x=0$, and the dimension of this operator is then
$1- \nu$. It is marginal in the integer effect and relevant for all 
fractional single-edge situations. The problem can also be formulated
in terms of scattering off impurities in a non-chiral \LL\ 
\cite{kafi}, and out
conclusion agrees with the analysis of Section \ref{lltrans}.

The dynamical excitations on the edges have a very rich structure which
reflects the topological order in the quantum Hall effect. Wen
\cite{wen,wenrev} and others \cite{stone} 
have developed here a chiral \LL\ theory as a framework for a 
detailed description of their properties. It is similar to the \LL\
discussed in the preceding chapters but differs (at least) in one 
important way: the charges of the edge excitations are fractional 
(or irrational), and the universality classes of the chiral and 
standard \LL s are different: the central charge describing the edge 
excitations is different from unity.

\chapter{The normal state of quasi-one-dimensional metals -- a Luttinger 
liquid?}
\label{materials}

A wide variety of materials with low-dimensional structural and
electronic properties is available now, and experimentalists have used
them to search for \LL\ correlations. 
In the main body of this chapter,
we shall concentrate on the quasi-1D organic conductors such as \ttf\,
and superconductors like \tmx\ or the related 
(nonsuperconducting) series \tmt. 
We summarize evidence
that electronic correlations are important in these materials, that
in their normal state, they are sufficiently anisotropic so that 1D
models of interacting electrons are indeed relevant for their description,
and that experiments can be interpreted consistently within the
theoretical framework set up in this article. At the end, we briefly
touch upon semiconductor heterostructures, a new and rapidly growing
branch in the field of correlated 1D electrons.

\section{Organic conductors and superconductors}
\label{eliac}

Tetrathiafulvalene-tetracyanoquinodimethane (\ttf ) crystalizes in
a herringbone pattern of two segregated stacks of $TTF$ and
$TCNQ$ molecules, respectively. 
A charge transfer from the TTF to
the TCNQ molecules of 0.57 electron/molecule produces partially
filled electron- and hole-like bands, i.e. the material behaves as a
two-chain conductor. 

Between 54K and 38K, \ttf ~undergoes a series of
phase transitions into a CDW state, accompanied by a periodic lattice
modulation due to electron-phonon coupling. Although the traditional
picture due to Peierls would consider electron-phonon
interaction as the driving force of such a CDW transition
\cite{Peierls,js}, the actual 
situation is more complicated, probably radically different. 

In fact, there is ample evidence for important repulsive interactions
on both chains. In X-ray experiments, diffuse X-ray scattering is 
not only detected at \tkf\ but surprisingly also at \fkf ,
even up to very high temperature \cite{pouget,kago}.
While \tkf -fluctuations are expected as precursors 
of the Peierls transition \cite{js}, 
the existence of \fkf\ fluctuations can only be explained 
assuming sizable Coulomb correlations \cite{emery}, as can be seen
from comparing the exponents of the \tkf - and \fkf -CDW correlation
functions (\ref{rcdw}) and (\ref{r4kf}). Notice
from Eqs.~(\ref{rgphon}) that coupling to lattice phonons ($Y_2^{(ph)}=0$)
reduces $K_{\rho}$. When electronic correlations are dominant, phonons
can enhance them further, and they certainly outweigh the logarithmic
``advantage'' of the SDWs against the \tkf -CDWs. 
The Pauli susceptibility is significantly enhanced over the free electron value
\cite{klotz,taka} and the finite frequency 
optical conductivity is much larger than the dc-con\-duc\-ti\-vi\-ty \cite{bas}.
Further evidence for Coulomb correlations
stems from the analysis of systematic variations of
physical properties through entire families \cite{torrance}
of closely related compounds such 
as $(NMP)_x (Phen)_{1-x} (TCNQ)$ \cite{maz,pougnmp} or the
$1:2-TCNQ$ salts \cite{skov}.

There is thus an 
alternative, well supported view that the CDWs 
in $TTF-TCNQ$ are, in fact the consequence
of electronic correlations rather than of electron-phonon interaction;
the latter then would just probe the electronic fluctuations without
feeding back on them in any significant manner, and
couple the preformed charge density modulation
into the lattice and make it visible to X-rays. This is different
from other CDW systems to be discussed in Section \ref{cdwover} below.

Closely related are the single-chain conductors \tmx\ and
\tmt\ (``Bechgaard salts''), 
where $TMTSF$ stands for the molecule 
tetramethyl-te\-tra\-se\-le\-na\-ful\-va\-le\-ne.
A sketch of this molecule and of the stacking pattern, immediately
suggestive of one-dimensionality, is shown in Figure 7.1.
\tmp\ undergoes a metal--insulator transition 
into a SDW state at ambient pressure at $T_{SDW} = 12K$. 
Evidence for SDWs
is provided by peculiar NMR relaxation behaviour \cite{jersch} but
most convincingly by observation of the nonlinear conductivity associated
with a sliding SDW \cite{Silvia}. 
Under 12 kbar pressure, however, superconductivity can be stabilized 
at $T_c = 0.9K$
\cite{jerome}. Other members of the family show superconductivity, too
\cite{parkin}.
By substituting the four
selenium atoms by sulfur, one obtains \tmt. Generically, these sulfur-based
systems undergo a 
charge localization 
transition (``Wigner crystallization'') around 
$200K$ (seen e.g.~as a minimum in the resistivity)
and reduce to an effective spin chain below. Finally, around
20 K, one observes spin-Peierls transitions into a spin-singlet state,
accompanied by a lattice deformation \cite{jersci}. However, by applying
pressure, a behaviour more akin to the SDW state of \tmp\ can be obtained.

The importance of Coulomb interactions in the \tmx ~and \tmt\ is more 
readily apparent than in \ttf . We discussed various \LL\ correlation functions
in Chapter \ref{chapll}, and SDW correlations require repulsive interactions. 
In the \tmt-series, the Coulomb 
repulsion is even stronger than in \tmx : the charge localization transition
around 200K can be interpreted as a transition into a \fkf -CDW
\cite{emery,Hubtcnq,maz}. 
Other properties indicate strong Coulomb interaction, too. 
The Pauli susceptibility of the conduction electrons is enhanced considerably
($\times 3 - 5$) with respect to a simple band picture
\cite{forro}. Moreover, it
is temperature dependent and, at low temperature, close
to what is expected for a 1D antiferromagnet \cite{jersci}. Optical properties
are also 
unconventional. Apparently it is observed that 
$\sigma(\omega)$ in the infrared is higher than the dc-conductivity which
is not compatible with free electrons and suggests rather localized charges.
In these infrared measurements one also observes 
totally symmetric molecular vibrations
\cite{pecile,bozio}.
These vibrations are IR-forbidden for free electrons but can be activated by 
local charge modulations such as a CDW \cite{Rice}. In the \tmx\ and \tmt\,
such a CDW would be at \fkf\, and the observation of the activated vibrations
in the IR then suggests a considerable degree of charge localization.
Strong electronic
repulsion generates antiferromagnetic spin fluctuations which 
can be, and have been, observed in NMR \cite{NMR,wzie}. 
There is 
also a detailed theory for the analysis
of these experiments which lends further support to a strongly correlated
picture for the electrons in the Bechgaard salts \cite{clamag,claude}.

The basic building blocks of the materials discussed in the preceding
section are large planar molecules with $\pi$-orbitals directed out
of the molecular plane. 
In general, the lattice parameters, the lattice dynamics, and the
elastic constants are anisotropic though not
strongly so, and are best viewed as three-dimensional  \cite{jersch}.
The electronic 
properties, however, are strongly anisotropic. 
Essentially, the overlap between wave functions on neighbouring
molecules, and matrix elements of a Hamiltonian between 
them, both depend exponentially on the intermolecular separation:
minor variations in the structure are then dramatically amplified in the
electron dynamics. The sensitive dependence on the intermolecular distances
is further amplified by the strong directionality of many of the molecular
orbitals involved. 
However, relatively short
distances between $Se$ or $S$ atoms in the \tmx ~and \tmt ~give nonnegligible
interstack contacts and some effectively 2D character to the 
\tmx\ salts \cite{ducasse}. 

The central question is therefore if the Fermi surface of these organic
conductors is closer to the parallel sheets characterizing 1D or to
the cylinder obtained for layered 2D materials, i.e. open vs. closed
orbits, and if, for given external parameters ($T,P$), the electrons
hop coherently or not from one chain to its neighbour.
There seems to be general agreement \cite{jersci} 
that the picture of two sheets, warped by the finite
hopping integrals perpendicular to the chains, is most appropriate.
Values often used are $t_{\|} \equiv t_a = 150 \ldots 200 meV$,
$t_{\perp} \equiv t_b = 20 \ldots 30 meV$, and $t_c < t_b /10$ 
for \tmx~\cite{jersci}
although next-nearest-neighbour transfer integrals also seem to play
some role \cite{ducasse}. Apparently \ttf~is significantly more
anisotropic and the Fermi surface has been suggested to consist
out of two parallel, diffuse sheets. Representative transfer integrals are
$t_{\|} = 110 meV$ for the $TCNQ$ chain, $t_{\|} = 50 meV$ for $TTF$
\cite{jersch}, and $t_{\perp}$ as low as
$5 meV$ \cite{sossc}. 

Transport measurements probe
the anisotropy and coherence of the electron dynamics. 
In optical absorption, a pronounced plasma edge in the $1eV$-range is
observed for electric fields polarized parallel to the chain axis
both in \ttf ~and \tmx, indicating band formation along
the chains, and longitudinal bandwidths of the order of $1eV$ have
been derived within simple tight-binding models.
In \tmp, a transverse plasma edge at about $0.1eV$, 
implying coherent perpendicular
transport, is observed only  at $T=25K$ \cite{jersch}. In \ttf,
to the best of the author's knowledge, the establishment of a transverse
plasma edge has never been observed. These measurements support
the picture of a very anisotropic band and rule out the possibility of a closed
Fermi surface in these materials.

Depending 
on the magnitude of $\tau_{\|} t_{\perp} / \hbar$, the transverse
transport can be either coherent or diffusive. Here $\tau_{\|}$ is the 
on-chain collision  time after which the electron wave function looses its 
coherence, i.e.~a measure of cleanliness,  and $t_{\perp} < t_{\|}$
is assumed. 
The transverse escape rate
 $1/\tau_{\perp} \approx 2 \pi
\tau_{\|} t_{\perp}^2 / \hbar$ can be measured by NMR \cite{soda}.
The long-wavelength spin fluctuations on the chains are certainly diffusive
and then described by a 1D random walk which gives a $1/\sqrt{\omega_e}$
power spectrum to the NMR relaxation rate $1/ T_1 T$ and strong field-dependent
deviations from the Korringa law through the electronic Larmor frequency
$\omega_e$. At low-fields, the spin dynamics becomes 3D and $1/ T_1 T$ 
is field independent. A crossover is observed for $\omega_e \tau_{\perp}
\approx 1$ in \ttf\ \cite{soda} and \tmx\ \cite{stein}.
$\tau_{\|}$ can be estimated e.g.~from conductivity.
Putting things together, one concludes that, in \ttf, the perpendicular
transport is always diffusive while in \tmp, it is diffusive at higher
temperature but transverse coherence is established at lower temperatures.
One also obtains the above values of $t_{\perp}$. This suggests that
theories discussing the establishment of transverse coherence in the 
single- and two-particle dynamics can be critically tested here.

The information available points towards a picture of strong Coulomb interaction
and pronounced one-dimensionality. What is the experimental evidence in
favour of \LL\ behaviour in these ``metals''? 
The instabilities observed at low temperatures are
certainly suggestive of 1D physics. Of course, they do not tell us much
about the normal-state properties; still the observation of high-temperature
\fkf -CDW fluctuations  in \ttf\ \cite{pouget,kago} places constraints on
the effective Luttinger parameters: $K_{\rho} < 1/2$ for their divergence,
and $K_{\rho} < 1/3$ for them being stronger than the \tkf -CDWs. 
These low values of $K_{\rho}$ indicate that a simple Hubbard model is
certainly inappropriate for the description of this compound, and that
longer range interactions cannot be neglected.
The optical spectrum of \ttf\ is very unusual (Figure 7.2). There is a
pronounced pseudogap present already at 85 K, i.e. far above the Peierls
temperature, which deepens as the temperature is lowered into the CDW phase. 
It is tempting to connect this with the pseudogap observed in the density
of states \cite{bas}, and in fact, $\sigma(\omega)$ is not incompatible with the
expression (\ref{ogeq}) derived by Ogata and Anderson \cite{ogpwa}, 
if the large $\alpha$-values implied by the \fkf -fluctuations are inserted. 

In the \tmx -materials, NMR has uncovered anomalous correlations 
\cite{NMR,wzie} (Figure 7.3). The spin-lattice relaxation rate $T_1^{-1}$
strongly deviates from the Korringa law 
$T_1^{-1} \propto T$ at lower temperatures.
This is believed to result from the temperature dependent SDW-correlations. 
In fact, $T_1^{-1}$ contains two contributions, from the long-wavelength
and the \tkf -spin-fluctuations: $T_1^{-1} \propto T + T^{1-\alpha_{SDW}}
= T + T^{K_{\rho}}$. Here, we have a quantity which can give direct
information on $K_{\rho}$! Below 
a certain temperature $T_0$, the SDW contribution
$T^{K_{\rho}}$ dominates over the long-wavelength part $T$. From the latest
work \cite{wzie}, one deduces $K_{\rho} \sim 0.15$. This is a surprisingly
big value, and again indicates effective strong and long-range interactions. 
One problem here, in contrast to \ttf\, is that the \fkf -fluctuations 
predicted in this limit, are not observed in X-rays (but they are neither
in the \tmt -series where \fkf -localization is established quite firmly). 

Similar values of $K_{\rho}$ are suggested by the photoemission experiments
shown in Figure 7.4. This experiment measures the occupied ($\omega < 0$)
part of the single-particle density of states $N(\omega)$ (\ref{nomega}).
There is no spectral weight at the Fermi surface, and it rises smoothly
as one goes to lower energies. The origin of the peak at $-1 eV$ is not
clear -- this scale lies at the lower edge of the valence band, or even
below. Close to the Fermi energy, the smooth variation is compatible 
with the \LL\ form $| \omega |^{\alpha}$ if one assumes an exponent
$\alpha > 1$. The NMR-$K_{\rho} = 0.15$ implies $\alpha \sim 1.25$, and
this value describes the data quite well. 

Summarizing, there are experimental indications in favour of \LL\ correlations
in these quasi-1D organic conductors which, if taken serious, would place
them in the limit of strong, long-range effective interactions. The experiments
described probe, however, only the anomalous operator-dimensions. There
has been no successful attempt to measure charge-spin separation. 
From Section \ref{secdyn}, this would
require angle-resolved photoemission or inelastic neutron scattering.
However, the relevance of a \LL\ description, even within a 1D framework,
has been questioned \cite{milseoul}. In fact, the \tmx -materials
are slightly dimerized, and therefore the Fermi level lies in an effectively
half-filled subband, where even weak interactions can lead to charge 
localization and an associated charge gap, cf. Chapter \ref{mottch}.
The unusual phenomena observed would, in this view, either be
extrinsic or mainly reflect the consequences of a Mott transition. At the 
time of writing, the controversy over the appropriate model for describing
the organic conductors is pretty open.

\section{Inorganic charge density wave materials}
\label{cdwover}
Many inorganic crystals, such as e.~g.~$K_{0.3}MoO_3$, $(TaSe_4)_2I$,
$NbSe_3$ etc.~undergo, at temperatures between $50 K$ and $250 K$, 
a Peierls transition
into a CDW ground state, giving rise to fascinating nonlinear transport
phenomena \cite{CDW}. Apart $NbSe_3$ which apparently is quite
3D and where pieces of a Fermi surface persist even into the CDW state,
the materials, generically, are rather 1D and do not show indications of 
strong electronic correlations.

The driving mechanism for CDW formation is the electron-phonon interaction
\cite{Peierls}, and accordingly, this interaction is supposed to be the
dominant one in these materials. Considerable success in describing
their normal state properties has been achieved based on the model of
a fluctuating Peierls insulator \cite{lra}. The basic idea here is
that 1D fluctuations will lower the actually observed
critical temperature $T_P$ below
the 1D mean-field temperature $T_P^{MF}$ by as much as a factor of $4-5$.
In the fluctuation regime, there is a pseudo-gap in the electronic
density of states which develops
into a real gap only at $T_P$ and which accounts for the unusual thermodynamic
and transport properties above $T_P$. There is no pseudo-gap 
left beyond $T_P^{MF}$.

The fluctuating Peierls insulator model is a single-particle
picture, i.e. charge and spin degrees of freedom behave symmetrically.
This is observed in some materials, such as $(Ta Se4)_2I$, but not in
others, e.g. $K_{0.3}MoO_3$, where
the dc-conductivity is totally unaffected by the conjectured pseudo-gap
while the susceptibility shows temperature dependence reminiscent of thermal
activation. Moreover, the transition in conductivity is extremely
sharp -- but gradual in susceptibility \cite{bluebr}. The Lee-Rice-Anderson
picture seems to imply that
above $T_P^{MF}$ the system reduces to a normal metal. The corresponding
temperature-independent Pauli susceptibility above $T_P^{MF}$ has, however, 
not been observed in any of the 1D CDW materials. 

Photoemission experiments question the validity of the picture of a
fluctuating Peierls insulator. As in the organic conductors,
the single-particle density of states shows no significant weight at
the Fermi energy in the blue bronze $K_{0.3} Mo O_3$, and appreciable
spectral weight is only found a sizable fraction of an $eV$ below \cite{ne}.
Angle-resolved studies on related materials either show a fading away of
the signal as the Fermi surface is approached \cite{hwu}, or weight
dispersing some finite energy below $E_F$ \cite{smith}. Clearly, there
have been speculations about 1D correlations at the origin of the mysterious
behaviour.

Experimental evidence points against simple \LL\ behaviour in these CDW
materials. On the other hand, our \rg\ analysis (\ref{rgphon}) in Section
\ref{phonie} shows that, in such a 1D scenario, a spin gap $\Delta_{\sigma}$
must open as
a precursor of CDW formation while the charge fluctuations remain massless
in strictly 1D. Such a system is in the Luther-Emery universality class
\cite{lutem}. The Pauli susceptibility rising with increasing temperature,
found in all CDW materials,
can be interpreted as evidence for such a spin gap. The conductivity of 
the blue bronze $K_{0.3} Mo O_3$ also provides evidence for massless
charge fluctuations \cite{bluebr}. One can compute, at least the diagonal
part of the single-particle density of states for this model and finds
\cite{spec2}
\begin{equation}
N(\omega) \approx \Theta( |\omega| - \Delta_{\sigma} )
(|\omega| - \Delta_{\sigma} )^{2 \gamma_{\rho}} \;\;\;,
\end{equation}
where $\gamma_{\rho}$ has been defined in (\ref{gammanu}). 
Such a behaviour is consistent
with the photoemission properties of the blue bronzes. We note, however,
that (i) optical experiments \cite{degior} are in excellent agreement with the 
fluctuating-Peierls-insulator model when one goes beyond the 
Lee-Rice-Anderson treatment and includes the thermally excited motion
of the lattice \cite{mcken}, and their precise
relation to the photoemission experiments is not understood to date;
(ii) there is no evidence for charge-spin separation in CDW-materials
other than $K_{0.3} Mo O_3$ which therefore are \em not \rm in the 
Luther-Emery universality class. 

Similar unusual photoemission behaviour: absence of spectral weight at
the Fermi surface and a smooth rise below, has also been reported for
another inorganic 1D material, $Ba V S_3$ \cite{fujimori}.

\section{Semiconductor heterostructures}
Semiconductor heterostructures may open a wide new field for the study
of 1D interacting electrons, in regimes usually inaccessible to organic
crystals. There are two principal directions: (i) quantum wires and (ii)
quantum Hall effect edge states. 

Traditionally, a 2D electron gas is induced 
at interfaces in such structures
by applying a gate voltage across the structure.
Very recent progress has allowed the fabrication of narrow channels in the
heterostructures where the electrons can be confined \cite{hetero} 
into a quantum wire. The electronic system can be made
truly one-dimensional by appropriate design of the structure. 
Periodic conductance oscillations are observed
as a function of the carrier density, and it has been
speculated that they could be either due to the formation of a charge
density wave or of a Wigner crystal \cite{hetero}. Wigner crystal formation,
i.e. the formation of \fkf -CDWs has been discussed in Chapter
\ref{chapll} \cite{Hubtcnq,schuwi,gomez,wigcrys}. Furthermore,
the 1D channel can be constricted, 
and we can verify the predictions for transport through an impurity,
discussed in Section \ref{lltrans}.

Such a constriction can also be built on samples showing the quantum Hall
effect. From Section \ref{fqh}, we know that the edge states are described
as chiral \LL s. For the $\nu = 1/3$ quantum Hall state where we have
a single edge, the Luttinger stiffness constant is $K = \nu = 1/3$. 
The exponents of all correlation functions are fully determined by the
filling-factor $\nu$ of the Landau levels! Eq.~(\ref{concorr}) shows that
the corrections to the \LL\ conductance diverge as $T \rightarrow 0$. 
An equation equivalent for the opposite case, 
tunneling across the constriction, predicts the tunneling
conductance across the constriction to vary as $G(T) \sim T^4$ \cite{kafi}. 
Such an experiment has been performed \cite{webb}, and the result
shown in Figure 7.5 is in accurate agreement with the theoretical prediction.
For comparison, a temperature independent conductance is expected for
a Fermi liquid ($K=1$), and is indeed found in the integer $\nu = 1$-state
\cite{webb}. The scaling with applied point-contact voltage is similar.
Again, for the $\nu = 1/3$-state, a strong variation with voltage is
found while the variation for $\nu = 1$ is much weaker \cite{webb}. 
It would be very interesting to see if such constrictions can also
be used to separate charge and spin of the electrons on the two sides,
as suggested in Section \ref{lltrans}. 

\chapter{Summary}

We have gone a long way from the simplest 1D model Hamiltonian, the \lm ,
to exotic correlations in complicated materials, often uncovered only under
extreme conditions. It is certainly useful to briefly summarize the 
essential steps and achievements. 

Fermi liquid theory based on a quasi-particle picture as in higher dimensions, 
does not work in 1D because of two new features with respect to 3D: 
a logarithmic divergence in the particle-hole bubble, due to the perfect
nesting of the 1D Fermi surface, and because of charge-spin separation in
1D. Both effects are connected to the fact that momentum transfer cannot
be neglected in the scattering processes in 1D. 

Quasi-particles are unstable in 1D, and the elementary excitations are
bosonic collective charge and spin fluctuations. 
The \lm\ incorporates these essential features and can therefore be taken
as a basis for the description of gapless 1D quantum systems. This model
has been solved by several methods: bosonization, equation of motion and 
Green's function techniques, conformal field theory. All of them are 
related to the symmetries and conservation laws of the 1D Fermi surface,
but incorporate them in a different manner. While the use of Ward identities
in the Green's function method emphasizes strong similarities to the Fermi
liquid, bosonization rather displays the differences. In particular,
we used bosonization to compute correlation functions. 
They decay as non-universal power-laws, and the scaling relations between
their exponents are parameterized by a single effective coupling constant
$K_{\nu}$ per degree of freedom. In addition, there is a renormalized
velocity of each collective mode. It renormalizes the thermodynamic and
transport properties, but its most spectacular consequence, 
charge-spin separation, is only 
visible in dynamical correlations at large wave-vector, such as the 
single-particle spectral function close to $k_F$ or the charge and
spin structure factors at \tkf .

The effective coupling constant is defined from the velocities associated
with three different low-energy excitations: particle-hole excitations,
charge ($\pm k_F$-symmetric addition of particles) and current
($k_F \leftrightarrow -k_F$-transfer of particle) excitations, and thus
from the eigenvalue spectrum alone. This structure persists in 
a low-energy subspace of more
complicated models containing nonlinear dispersion, (irrelevant) 
large-momentum transfer scattering, coupling to external degrees of freedom,
etc., and carries to the notion of a ``\LL ''. It implies that 
the low-energy properties of these models are described by a renormalized
\lm , provided their excitations are gapless. The \LL\ is the universality
class of these gapless 1D quantum systems. By a controlled mapping of
1D models ranging from the Hubbard model to electron-phonon systems,
an asymptotically exact solution of the 1D many-body problem is achieved. 
We have discussed several procedures, some applicable only where an
exact solution by Bethe Ansatz is possible, others generally applicable 
and therefore also suited for non-soluble models. We often used
\rg\ which is not as powerful as methods based on an exact eigenvalue
spectrum because it is based on perturbative developments and thus limited
to weak coupling. While failing quantitatively at stronger coupling, most
of its predictions are qualitatively valid beyond the weak-coupling range.
In particular, it allows to derive logarithmic corrections to correlation
functions which lift the unphysical degeneracies implied by their exponents. 
Moreover, it is flexible enough to allow a treatment of problems beyond
the reach of both exact and numerical solutions such as electron-phonon
coupling and scattering off impurities. It is therefore essential for
a determination of transport properties, to which we devoted much space.
Of course, \LL\ behaviour is also found in multi-band and multi-component
models, and most methods generalize straightforwardly to these problems.

Not all 1D fermion systems are \LL s. When backward or Umklapp scattering
operators become relevant, as they do for attractive interactions or 
commensurate band-fillings, respectively, a gap opens in the spin or
in the charge channel. Passing back and forth between strong- and 
effective weak-coupling models, a detailed picture of their properties
can be constructed. We have done this in particular for the Mott transition
in commensurate, repulsively interacting systems, and emphasized phase
diagrams, critical interaction strengths, and the scaling behaviour
of transport properties and correlations. 

The solution of the Luttinger, Hubbard and other models relies in an
essential way on the strong conservation laws provided by the small
phase space of 1D. 
An important problem therefore is the stability of the \LL\ with respect to
transverse coupling. Coupling by interchain Coulomb interaction often
gives only quantitative modifications of the 1D behaviour, except for
backscattering which can stabilize charge density wave correlations into
a long-range ordered phase. Interchain tunneling on the other hand can
lead to transverse coherence either in the single- or in the two-particle
dynamics, depending on the on-chain correlations, 
and stabilize a variety of phases. If a single-particle crossover occurs
first, as the temperature is lowered, the Peierls-Cooper interference
in destroyed, and a low-temperature phase transition may take place in
one channel alone. Despite intense research, the normal-state properties
above such a transition, are not fully clear to date. There seems to
be agreement that charge-spin separation is an essential feature in this
situation, but there is disagreement on the extent to which it confines
the electron dynamics onto a single chain. A two-particle crossover
may occur before the single-particle one, and the on-chain correlations
are then propagated by transverse particle-particle or particle-hole 
pair hopping, leading again to low-temperature phase transitions. Here,
the normal state is of 1D Luttinger type. An important problem is the
Hubbard model on two or more coupled chains, and at least the two-chain
variant provides evidence for possible superconducting correlations 
at repulsive interactions. 

Many experiments have provided evidence for \LL\ correlations in low-dimensional
materials, although there is often some controversy about their precise
interpretation. Quasi-1D organic conductors and superconductors, for example,
show (\ttf ) diffuse X-ray scattering at \tkf\ and \fkf\ and strongly
depressed low-frequency optical conductivity, and others (\tmx ) 
power-law deviations from the Korringa law in NMR, and vanishing spectral
weight at the Fermi surface in photoemission.  These experiments
point towards really strong, and in particular long-range, interactions.
Instrumental to this conclusion are bounds on the Luttinger coupling constants
$K_{\nu}$ derived from mapping various lattice models onto the \lm .
Suggesting single-particle exponents $\alpha$ in excess of unity, in fact, there
are no known lattice models which naturally would provide values so big. 
This certainly is a major problem for future research and for the modelling
of these materials. 

Similarly surprising is the absence of spectral weight at $E_F$ in inorganic
charge density wave systems although these are believed to be dominated
by electron-phonon coupling. From our discussion of phonon-coupled \LL s,
we have suggested that they fall into the Luther-Emery universality class,
and that spin gap formation is quite generally a precursor of charge density
wave formation. Studies of spectral functions for these models are certainly
called for. 

Finally, edge state transport in the fractional quantum Hall effect is an
exciting new area of low-dimensional physics. We have discussed how these
gapless edges can be modelled as chiral \LL s. While they are similar
to those discussed before, having a central charge $c \neq 1$, they fall into
a different universality class. They also have power-law correlations, but
their charges in general are irrational. Still, much of what has been
said about correlation functions and transport for the normal \LL\ carries
over to the chiral variant. The remarkable feature here is that, at least
for the single-edge situations, the
renormalized coupling constant $K$ is fully determined by the Landau level
filling fraction $K = \nu$, and therefore all correlation exponents (i)
are known in advance, and (ii) can be tuned accurately by varying $\nu$.
A recent experiment on tunneling through a barrier on the edges at $\nu =
1/3$ is in agreement with the chiral Luttinger prediction and 
apparently provides a
first evidence for the relevance of this picture.

\section*{Acknowledgements}
I should like to thank the following colleagues for stimulating interaction,
helpful suggestions, criticism, and essential support, often over many years:
Jim Allen,
Natan Andrei, 
Claude Bourbonnais,
Sergei Brazovski\u{i},
Helmut B\"{u}ttner,
David Campbell,
Laurent Caron,
Michele Fabrizio,
Florian Gebhard (especially for many constructive comments on the present
article), 
Thierry Giamarchi,
Daniel Malterre,
Thierry Martin,
Eugene Mele,
Philippe Nozi\`{e}res,
J\"{u}rgen Parisi,
Jean-Paul Pouget,
Dierk Rainer,
Mario Rasetti and the Institute for Scientific Interchange in Torino,
Heinz Schulz,
Markus Schwoerer,
Andr\'{e}-Marie Tremblay,
Joe Wheatley.
The responsibility for flaws in this article is, however, entirely mine.
My research is supported by Deutsche Forschungsgemeinschaft
through SFB 279--B4.


\newpage
\section*{List of Figures}

{\bf Figure 2.1:}
The Peierls instability: the presence of an interaction component
with wave vector $q \approx 2k_F$ in a 1D electron gas (left) hybridizes
the two Fermi points $\pm k_F$ and, in a mean-field description, opens
a gap at the Fermi level (middle). Responsible is the 
logarithmic \tkf\ divergence
in the particle-hole bubble (right).\\

\noindent
{\bf Figure 2.2:}
Diagrams contributing to the self-energy to second order.\\

\noindent
{\bf Figure 2.3:}
The Bethe-Salpeter equation. $\Gamma$ is the complete particle-hole
interaction, $I$ the irreducible one. Solid lines are Green's functions.\\

\noindent
{\bf Figure 3.1:}
Particle-hole excitations in 1D (left). The spectrum
(right) has no low-frequency excitations with $0 \leq \mid q \mid \leq 2k_F$
unlike in higher dimensions where these states are filled in.\\

\noindent
{\bf Figure 3.2:}
The Luttinger model. Dispersion (left) and forward scattering
processes (right). Solid lines denote electrons propagating with $k_F$
and dashed lines those propagating with $-k_F$.\\

\noindent
{\bf Figure 3.3:}
Backward ($g_{1\perp}$) and Umklapp ($g_{3\perp}$) scattering
not included in the Luttinger model. Scattering particles have antiparallel
spin here.\\

\noindent
{\bf Figure 3.4:} 
Dyson equation for the density correlation function $R_{\rho\rho}$.\\

\noindent
{\bf Figure 3.5:}
Dyson equation for the polarization. $\lambda^{\rho}$ represents the bare
vertex.\\

\noindent
{\bf Figure 3.6:} 
Luttinger model spectral function $\rho_+(q,\omega)$ 
for $q \geq 0$ and $\alpha=0.125$. The $\omega < 0$-part has been 
multiplied by $10$ for clarity. \\

\noindent
{\bf Figure 3.7:}
Dynamical charge and spin structure factors of the \lm . $S(q,\omega)$
and $S_4(q,\omega)$ are the \tkf - and \fkf -CDW structure factors, and
$\chi(q,\omega)$ is the magnetic structure factor close to \tkf . \\

\noindent
{\bf Figure 4.1:}
Linearized renormalization group 
flow of $g_{1\perp}$ and $K_{\sigma}$ for the backscattering
Hamiltonian. The line $g_{1\perp}^{\star} = 0, \; K_{\sigma}^{\star} \geq 1
$ is the \LL\ fixed line. \\

\noindent
{\bf Figure 4.2:}
Phase diagram of the one-dimensional Fermi liquid off half-filling.
The system is a \LL\ at $g_{1\perp} \geq 0$ where $K_{\sigma}^{\star} = 1$. 
Fluctuations indicated in parenthesis have the same exponents as the
dominant ones but are logarithmically weaker. At $g_{1\perp} < 0$, there
is a spin gap, and formally $K_{\sigma}^{\star} = 0$. Here, fluctuations
appearing in parenthesis diverge with a smaller power-law exponent than
the dominant ones. \\

\noindent 
{\bf Figure 4.3:}
Momentum distribution $n(k)$ of the quarter-filled Hubbard model in the limit
$U/t \rightarrow \infty$ as calculated from the Bethe Ansatz equations
(\ref{liwus1}) and (\ref{liwus2}). (Anti)periodic boundary conditions were
used for $4n+2$-($4n$-)site lattices, respectively.
From ref. \cite{ogata}, Fig.~3. \\

\noindent
{\bf Figure 4.4}
The correlation exponent $K_{\rho}$ of the Hubbard model 
as a function of bandfilling $n$ for
different values of $U$ ($U/t = 1,2,4,8,16$ from top to bottom). 
From ref. \cite{ijmp}, Fig.~3. \\

\noindent
{\bf Figure 4.5:}
The charge and spin velocities $v_{\rho}$ (full line) and $v_{\sigma}$
(dash-dotted lines) of the Hubbard model as a function of bandfilling for
different $U/t$. $U/t = 1,2,4,8,16$ from top to bottom for $v_{\sigma}$
and from bottom to top for $v_{\rho}$ in the left part of the Figure.
From ref. \cite{ijmp}, Fig.~1. $u_{\rho,\sigma}$ are denoted 
$v_{\rho,\sigma}$ in our text. \\

\noindent
{\bf Figure 4.6:} 
Phase diagram of the $t-J$-model determined from variational wavefunctions.
``Repulsive Luttinger'' stands for a \LL\ with dominant SDW correlations,
and ``attractive Luttinger'' for one with dominant TS. 
The spin-gap phase has dominant SS. The dashed line corresponds to
$K_{\rho} = 1$. From ref. \cite{helme4}, Fig.~1. \\

\noindent
{\bf Figure 4.7:}
Phase diagram of the extended Hubbard model at quarter-filling. I is
an insulating \fkf -CDW state, M the metallic, repulsive \LL\, and SC
an attractive \LL\ with superconducting correlations. The dashed lines
are lines of constant $K_{\rho}$. 
From \cite{milzot}, Fig.~1. \\

\noindent
{\bf Figure 6.1:}
Dispersion relations of two hybridized bands and their linearized
approximations. In (a) two particle-like bands hybridize. In (b) 
a particle-like band hybridizes with a hole-like band leading to
different signs of the Fermi velocities on the same side of the
dispersion.
From Ref. \cite{pesot}, Figs. 1+2. (a) contains (a) from both Fig.~1
and Fig.~2; same applies to (b). \\

\noindent
{\bf Figure 6.2:}
Cut at $k_y = 0$ through the Brillouin zone of a system of weakly coupled
chains. The shaded area indicates occupied electron states.
From Ref.~\cite{jersch}, Fig. 1.7. Their $k_F^0$ is denoted by $k_F^{1D}$
in our text. \\

\noindent
{\bf Figure 6.3:} 
Generation of transverse hopping corrections to the propagation of 
correlated particle-particle or particle-hole pairs. The thick (thin)
lines refer to 3D (1D) propagators in the high-energy shell eliminated
by the \rg\ transformation. Time arrows and spin indices can be put 
depending on the (CDW, SDW, SS, TS)-operator under consideration, and
the full square then denotes its effective combination of coupling constants.
From Ref.~\cite{bourbonnais}, Fig.~9. \\

\noindent
{\bf Figure 6.4:}
Phase diagram for the quarter-filled two-chain \hm\ from \rg . The different
phases are explained in the text. 
From Ref.~\cite{fabprb}, Fig.~9.  $\rho$ in the inset is denoted by $n$
in our text.\\

\noindent
{\bf Figure 6.5:} Electronic structure for the integer quantum Hall effect
on an annulus of radii $r_1$ and $r_2$. The shaded areas indicate localized
states.
From Ref.~\cite{halp}, Fig.~3.\\

\noindent
{\bf Figure 7.1:} Structure of the molecule tetramethyl-tetraselenafulvalene
and schematic stacking pattern of the \tmx -crystals. \\

\noindent
{\bf Figure 7.2:}
Real part of the optical conductivity of \ttf\ at 85 K
along the chain axis for two different samples (trial 1 and 2). Notice
the suppression of conductivity with respect to a Drude model at low
frequency. From Ref.~\cite{bas}, Fig.~6. \\

\noindent
{\bf Figure 7.3:} 
$^{77}Se$-NMR spin-lattice relaxation rate $T_1^{-1}$ as a function of temperature
in $(TMTSF)_2 ClO_4$. The different symbols denote different fields. 
Shown in the inset is the theoretical profile for $T_1^{-1}$ for two
values of the exponent $\alpha_{SDW}$. $T_X^1$ is the single-particle
crossover temperature, and $E_0$ marks the temperature where the
\tkf -SDW becomes stronger than the $q \approx 0$-fluctuations.
From Ref.~\cite{NMR}, Fig.~1 (main figure) and 5(a) (inset). $\eta$ is
$\alpha_{SDW}$ in our text and $T_X$ is $T_X^1$. \\

\noindent
{\bf Figure 7.4:}
HeII photoemission spectrum of \tmp\ at $T=50 K$. The HeI-spectrum 
in the insert has a better statistics and
clearly shows that there is no spectral weight near the
Fermi surface. 
From Ref.~\cite{photem}, Figure 1.\\

\noindent
{\bf Figure 7.5:}
Tunneling conductance through a constriction in a $\nu=1/3$ quantum Hall
state as a function of temperature. Different curves refer to different
voltages on the point contact forming the constriction. The \LL\
prediction is $G(T) \sim T^4$. 
From the preprint of Ref.~\cite{webb} and Physics Today (p.23).\\


\begin{thebibliography}{100}
\bibitem{landau} L. D. Landau, Sov. Phys. JETP {\bf 3}, 920 (1957);
	{\bf 5}, 101 (1957); {\bf 8}, 70 (1959).
\bibitem{Noz} 
	P. Nozi\`{e}res, \em Interacting Fermi Systems, \rm
	W. A. Benjamin Inc, New York (1964).
\bibitem{Haldane} 
	F. D. M. Haldane, J. Phys. C {\bf 14}, 2585 (1981).
\bibitem{lt} J. M. Luttinger, Phys. Rev. {\bf 119}, 1153 (1960).
\bibitem{Peierls} R. Peierls, \em Quantum Theory of Solids, \rm
	Oxford University Press, London (1955).
\bibitem{bcs} J. Bardeen, L. N. Copper, and J. R. Schrieffer, Phys. Rev.
	{\bf 108}, 1175 (1957).
\bibitem{lut} J. M. Luttinger, J. Math. Phys. {\bf 4}, 1154 (1963).
\bibitem{tom} S. Tomonaga, Prog. Theor. Phys. {\bf 5}, 544 (1950).
\bibitem{matl}
	D. C. Mattis and E. H. Lieb, J. Math. Phys. {\bf 6}, 304 (1963).
\bibitem{efe} K. B. Efetov and A. I. Larkin, Sov. Phys.  JETP {\bf 42},
	390 (1976).
\bibitem{ar} 
	P. W. Anderson and Y. R. Ren, Proceedings of the Los Alamos
	Conference on High-$T_c$-Superconductivity, Addison 
	Wesley Publ. Comp., 1990,
	p. 3; P. W. Anderson, Phys. Rev. Lett. {\bf 64}, 1839
	and {\bf 65}, 2306 (1990).
\bibitem{fl2d} J. R. Engelbrecht and M. Randeria, Phys. Rev. Lett. {\bf 65},
	1032 (1990); D. Coffey and K. S. Bedell, ibid. {\bf 71}, 1043 (1993).
\bibitem{mfl} C. M. Varma, P. B. Littlewood, S. Schmitt-Rink, E. 
	Abrahams, and A. E. Ruckenstein, Phys. Rev. Lett. {\bf 63}, 
	1996 (1989); N. Mitani and S. Kurihara, 
	Physica C {\bf 192}, 230 (1992).
\bibitem{mflimp} I. Perakis, C. M. Varma, and A. E. Ruckenstein,
	Phys. Rev. Lett. {\bf 70}, 3467 (1993); G.-M. Zhang and L. Yu, 
	ibid. {\bf 72}, 2474 
	(1994); C. Sire, C. M. Varma, A. E. Ruckenstein, and T. Giamarchi,
	ibid. p. 2478. See also G. M. Eliashberg, JETP Lett. {\bf 46}, S81
	(1988); B. R. Alascio and C. R. Proetto, Sol. St. Comm. {\bf 75},
	217 (1990).
\bibitem{jersch} 
	D.~J\'{e}r\^{o}me and H.~J.~Schulz, Adv. Phys. {\bf 31}, 299 (1982).
\bibitem{icsm} 
	Proceedings of recent conferences on Synthetic Metals, 
	e.~g.~Synth. Met. {\bf 27} (1988) -- {\bf 29}
	(1989); {\bf 41} -- {\bf 43} (1991); {\bf 55} -- {\bf 57} (1993);
	{\bf 69} -- {\bf 71} (1995).
\bibitem{kondo} P. Nozi\`{e}res and A. Blandin, J. Phys. (Paris) {\bf 41},
	193 (1980); N. Andrei and C. Destri, Phys. Rev. Lett. {\bf 52}, 
	364 (1984); A. W. W. Ludwig and I. Affleck, Phys. Rev. Lett. 
	{\bf 67}, 3160 (1991).
\bibitem{solyom} J. S\'{o}lyom, Adv. Phys. {\bf 28}, 201 (1979).
\bibitem{emrev} V. J. Emery, in \em Highly Conducting One-Dimensional
	Solids, \rm ed.~by J. T. Devreese, R. E. Evrard, and V. E. van Doren,
	Plenum Press, New York (1979).
\bibitem{nonabbos} I. Affleck, in: \em Fields, Strings, and Critical
	Phenomena, \rm ed. by E. Br\'{e}zin and J. Zinn-Justin,
	Elsevier Science Publishers B. V., Amsterdam, 1989.
\bibitem{firsov} Yu. A. Firsov, V. N. Prigodin, and Chr. Seidel, 
	Phys. Rep. {\bf 126}, 245 (1985).
\bibitem{bourbonnais} C. Bourbonnais and L. G. Caron, Int. J. Mod. Phys.
	B {\bf 5}, 1033 (1991).
\bibitem{jersci} D.~J\'{e}r\^{o}me, Science {\bf 252}, 1509 (1991).
\bibitem{et} J.~M.~Williams, A. J. Schultz, U. Geiser, K. D. Carlson,
	A. M. Kini, H. H. Wang, W.-K. Kwok, M.-H. Whangbo, and
	J. E. Schirber, Science {\bf 252},
	1501 (1991).
\bibitem{magog} \em Low-Dimensional Conductors and Superconductors, \rm
	ed. by D. J\'{e}r\^{o}me and L. G. Caron, Plenum Press, New York,
	1987.
\bibitem{suth} B. Sutherland, in \em Exactly Solvable Problems in
	Condensed Matter and Relativistic Field Theory, \rm Lecture
	Notes in Physics {\bf 242}, 1 (1985). 
\bibitem{korerev} V. E. Korepin, N. M.
	Bogoliubov, and A. G. Izergin, \em Quantum Inverse Scattering
	Method and Correlation Functions, \rm Cambridge University Press,
	Cambridge, 1993.
\bibitem{izsk} Yu. A. Izyumov and Yu. N. Skryabin, \em Statistical
	Mechanics of Magnetically Ordered Systems, \rm Consultants
	Bureau, New York, 1988, chapter 5.
\bibitem{confft} A. A. Belavin, A. M. Polyakov, and A. B. Zamolodchikov,
	Nucl. Phys. B {\bf 241}, 333 (1984); \em Fields, Strings, and 
	Critical Phenomena, \rm ed. by E. Br\'{e}zin and J. Zinn-Justin,
	Elsevier Science Publishers B. V., Amsterdam, 1989; C. Itzykson
	and J.-M. Drouffe, \em Statistical Field Theory, \rm Cambridge
	Universtity Press, Cambridge, 1989, vol.~2;
	Y. Grandati, Ann. Phys. Fr. {\bf 17}, 159 (1992);
	A. W. W. Ludwig, Trieste Lectures 1992.
\bibitem{bychkov} Yu. A. Bychkov, L. P. Gorkov, and 
	I. E. Dzyaloshinski\u{i}, Sov. Phys. JETP {\bf 23}, 489 (1966).
\bibitem{heiden} R. Heidenreich, R. Seiler,
	and A. Uhlenbrock, J. Stat. Phys. {\bf 22}, 27 (1980).
\bibitem{lupe} A. Luther and I. Peschel, Phys. Rev. B {\bf 9}, 2911 (1974).
\bibitem{mattis} D. C. Mattis, J. Math. Phys. {\bf 15}, 609 (1974).
\bibitem{dzya} I. E. Dzyaloshinski\u{i} and A. I. Larkin, 
	Sov. Phys. JETP {\bf 38}, 202 (1974).
\bibitem{everts} H. U. Everts and H. Schulz, Sol. State Comm. {\bf 15}, 1413
	(1974).
\bibitem{apostol} M. Apostol, J. Phys. C {\bf 16}, 5937 (1983).
\bibitem{mechtha} This argument was suggested by Michele Fabrizio.
\bibitem{kronig} R. de L. Kronig, Physica {\bf 2}, 968 (1935).
\bibitem{witten} E. Witten, Commun. Math. Phys. {\bf 92}, 455 (1984).
\bibitem{mandel} S. Mandelstam, Phys. Rev. D {\bf 11}, 3026 (1975).
\bibitem{halhfl} F. D. M. Haldane, Phys. Rev. Lett. {\bf 47}, 1840 (1981).
\bibitem{ijmp} H. J. Schulz, Int. J. Mod. Phys. B {\bf 5}, 57 (1991).
\bibitem{flux} W. Kohn, Phys. Rev. {\bf 133}, A171 (1964); X. Zotos, 
	P. Prelovsek, and I. Sega, Phys. Rev. B {\bf 42}, 8445 (1990);
	B. S. Shastry and B. Sutherland, Phys. Rev. Lett. {\bf 65}, 243 (1990).
\bibitem{giam1} T. Giamarchi, Phys. Rev. B {\bf 44}, 2905 (1991).
\bibitem{giamil} T. Giamarchi and A. J. Millis, Phys. Rev. B {\bf 46}, 9325
	(1992).
\bibitem{metzner} W. Metzner and C. Di Castro, Phys. Rev. B {\bf 47}, 
	16107 (1993).
\bibitem{shankar} R. Shankar, Int. J. Mod. Phys. B {\bf 4}, 2371 (1990).
\bibitem{schulz} H. J. Schulz, Phys. Rev. Lett. {\bf 64}, 2831 (1990).
\bibitem{Haldprl} F. D. M. Haldane, Phys. Rev. Lett. {\bf 45}, 1358 (1980).
\bibitem{ehm}
	J. Voit, Phys. Rev. B {\bf 45}, 4027 (1992).
\bibitem{suzu} Y. Suzumura, Prog. Theor. Phys. {\bf 63}, 5 (1980).
\bibitem{schucou} H. J. Schulz, J. Phys. C {\bf 16}, 6769 (1983).
\bibitem{spec2} J. Voit, J. Phys. CM {\bf 5}, 8305 (1993).
\bibitem{schome} K. Sch\"{o}nhammer and V. Meden, Phys. Rev. B {\bf 47},
	16205 (1993) and (E) {\bf 48}, 11521 (1993).
\bibitem{theu} A. Theumann, J. Math. Phys. {\bf 8}, 2460 (1967).
\bibitem{gutsch} H. Gutfreund and M. Schick, Phys. Rev. {\bf 168}, 418 (1968).
\bibitem{brech}
	M. Brech, J. Voit, and H. B\"{u}ttner, Europhys. 
	Lett. {\bf 12}, 289 (1990).
\bibitem{emery} V. J. Emery, Phys. Rev. Lett. {\bf 37}, 107 (1976).
\bibitem{spec1} J. Voit, Phys. Rev. B {\bf 47}, 6740 (1993).
\bibitem{ms} V. Meden and K. Sch\"{o}nhammer, Phys. Rev. B {\bf 46}, 15753
	(1992).
\bibitem{fogedby} H. C. Fogedby, J. Phys. C {\bf 9}, 3757 (1976).
\bibitem{spec3} J. Voit, in the
	Proceedings of the NATO Advanced Research Workshop on
	\em The Physics and Mathematical Physics of the Hubbard Model, \rm 
	San Sebastian, October 3--8, 1993, edited by D. Baeriswyl, 
	D. K. Campbell, J. M. P. Carmelo, F. Guinea, 
	and E. Louis; Plenum Press, New York (1995).
\bibitem{schoensum} K. Sch\"{o}nhammer and V. Meden, Phys. Rev. B {\bf 48},
	11390 (1993).
\bibitem{seoul} J. Voit, Synth. Met. {\bf 70}, 1015 (1995). 
\bibitem{peso} K. Penc and J. S\'{o}lyom, Phys. Rev. B {\bf 44}, 12690 (1991).
\bibitem{metzcdc} C. Di Castro and W. Metzner, Phys. Rev. Lett. {\bf 67},
	3852 (1991); 
\bibitem{leechen} D. K. K. Lee and Y. Chen, J. Phys. A {\bf 21}, 4155 (1988).
\bibitem{schmeltz} D. Schmeltzer, Phys. Rev. B {\bf 43}, 8650 (1991).
\bibitem{mudry} C. Mudry and E. Fradkin, Phys. Rev. B {\bf 50}, 11409 (1994).
\bibitem{phastr} S.-K. Ma, \em Modern Theory of Critical Phenomena, \rm
	Benjamin/Cummings Publ. Comp., Reading, MA, 1976.
\bibitem{bloeaff} H. W. J. Bl\"{o}te,  J. L. Cardy, and M. P. Nightingale,
	Phys. Rev. Lett. {\bf 56}, 742 (1986); I. Affleck, ibid. p. 746.
\bibitem{friedan} D. Friedan, Z. Qiu, and S. Shenker, Phys. Rev. Lett.
	{\bf 52}, 1575 (1984).
\bibitem{cardy} J. Cardy, 
	J. Phys. A {\bf 17}, L385 (1984) and Nucl. Phys. B {\bf 270},
	[FS16], 186 (1986).
\bibitem{mirozab} A. D. Mironov and A. V. Zabrodin, Phys. Rev. Lett. {\bf 66},
	534 (1991).
\bibitem{halphlet} F. D. M. Haldane, Phys. Lett. {\bf 81A}, 153 (1981). 
\bibitem{chui} S.-T. Chui and P. A. Lee, Phys. Rev. Lett. {\bf 35}, 325 
	(1975).
\bibitem{kothou} J. M. Kosterlitz and D. J. Thouless, J. Phys. C {\bf 6},
	1181 (1973); J. M. Kosterlitz, J. Phys. C {\bf 7}, 1046 (1974).
\bibitem{pref} J. Voit, J. Phys. C {\bf 21}, L1141 (1988).
\bibitem{giam} T. Giamarchi and H. J. Schulz, Phys. Rev. B {\bf 39}, 4620 
	(1989).
\bibitem{black} J. L. Black and V. J. Emery, Phys. Rev. B {\bf 23}, 429
	(1981).
\bibitem{heis} H. A. Bethe, Z. Phys. {\bf 71}, 205 (1931).
\bibitem{hubbard} J. Hubbard, Proc. Roy. Soc. A {\bf 240}, 539 (1957);
	{\bf 243}, 336 (1958); {\bf 276}, 238 (163). 
\bibitem{liebwu} E. H. Lieb and F. Y. Wu, Phys. Rev. Lett. {\bf 20}, 1445
	(1968).
\bibitem{babla} P. A. Bares and G. Blatter, Phys. Rev. Lett. {\bf 64},
	2567 (1990).
\bibitem{suttj} B. Sutherland, Phys. Rev. B {\bf 12}, 3795 (1975).
\bibitem{schlotj} P. Schlottmann, Phys. Rev. B {\bf 36}, 5177 (1987).
\bibitem{bergtha} H. Bergknoff and H. B. Thacker, Phys. Rev. D {\bf 19},
	3666 (1979).
\bibitem{bosgas} E. Lieb and W. Liniger, Phys. Rev. {\bf 130}, 1616 (1963).
\bibitem{Hubtcnq} J. Hubbard, Phys. Rev. B {\bf 17}, 494 (1978).
\bibitem{schuwi} H. J. Schulz, Phys. Rev. Lett. {\bf 71}, 1864 (1993).
\bibitem{kssh} S. Kivelson, W.-P. Su, J. R. Schrieffer, and A. J. Heeger,
	Phys. Rev. Lett. {\bf 58}, 1899 (1987).
\bibitem{pai} A. Painelli and A. Girlando, in Ref. \cite{iacel}.
\bibitem{iacel} 
	\em Interacting Electrons in Reduced Dimensions, \rm ed. by
	D. Baeriswyl and D. K. Campbell, Plenum Press, New York (1989).
\bibitem{cmp} D. K. Campbell, J. T. Gammel, and E. Y. Loh Jr.
	Phys. Rev. B {\bf 42}, 475 (1990).
\bibitem{buz} F. Buzatu, Phys. Rev. B {\bf 49}, 10176 (1994).
\bibitem{hirsch} J. E. Hirsch, Physica C {\bf 158}, 326 (1989).
\bibitem{pafo} F. C. Zhang and T. M. Rice, Phys. Rev. B {\bf 37}, 3759 
	(1988); A. Fortunelli and A. Painelli, Sol. State Comm. {\bf 89}
	771 (1994).
\bibitem{schlo} K.-J.-B. Lee and P. Schlottmann, Phys. Rev. Lett. {\bf 63},
	2299 (1989); P. Schlottmann, Phys. Rev. B {\bf 43}, 3101 (1991).
\bibitem{gebhard} F. Gebhard and A. E. Ruckenstein, Phys. Rev. Lett.
	{\bf 68}, 244 (1992); F. Gebhard, A. Girndt, and A. E. Ruckenstein, 
	Phys. Rev. B {\bf 49}, 10926 (1994).
\bibitem{nobe} P. Nozi\`{e}res, Lecture Notes, Coll\`{e}ge de France, 
	Paris, 1991/92.
\bibitem{shiba} H. Shiba, Phys. Rev. B {\bf 6}, 930 (1972).
\bibitem{takahashi} M. Takahashi, Prog. Theor. Phys. {\bf 47}, 69 (1972).
\bibitem{ultz} T. B. Bahder and F. Woynarovich, Phys. Rev. B {\bf 33}, 2114
	(1986); K. Lee and P. Schlottmann, Phys. Rev. B {\bf 38}, 11566 (1988).
\bibitem{ovchi} A. A. Ovchinnikov, Sov. Phys. JETP {\bf 30}, 1160 (1970).
\bibitem{coll} C. F. Coll III, Phys. Rev. B {\bf 9}, 2150 (1974).
\bibitem{halhub} F. D. M. Haldane and Yuhai Tu, unpublished.
\bibitem{ogata} M. Ogata and H. Shiba, Phys. Rev. B {\bf 41}, 2326 (1990).
\bibitem{caba} J. Carmelo and D. Baeriswyl, Phys. Rev. 
	B {\bf 37}, 7541 (1988).
\bibitem{hirsca} J. E. Hirsch and D. J. Scalapino, Phys. Rev. B {\bf 27},
	7169 (1983) and {\bf 29}, 5554 (1984).
\bibitem{maz} S. Mazumdar and A. N. Bloch, Phys. Rev. Lett. {\bf 50}, 207,
	(1983); S. Mazumdar, S. N. Dixit, and A. N. Bloch, Phys. Rev. B
	{\bf 30}, 4842 (1984); S. Mazumdar and S. N. Dixit, ibid. {\bf 34},
	3683 (1986).
\bibitem{ung} K. C. Ung, S. Mazumdar, and D. Toussaint, Phys. Rev. Lett.
	{\bf 73}, 2603 (1994).
\bibitem{gomez} G. Gom\'{e}z-Santos, Phys. Rev. Lett. {\bf 70}, 3780 (1993).
\bibitem{sortor} S. Sorella and M. Parinello, in Ref.~\cite{iacel}.
\bibitem{sor} S. Sorella, E. Tosatti, S. Baroni, R. Car, and M. Parinello,
	Int. J. Mod. Phys. B {\bf 1}, 993 (1988).
\bibitem{bourmc} C. Bourbonnais, H. N\'{e}lisse, A. Reid, and A.-M. S. 
	Tremblay, Phys. Rev. B {\bf 40}, 2297 (1989).
\bibitem{imada} M. Imada and Y. Hatsugai, J. Phys. Soc. Jpn. {\bf 58},
	3752 (1989).
\bibitem{soreur} S. Sorella, A. Parola, M. Parinello, and E. Tosatti,
	Europhys. Lett. {\bf 12}, 721 (1990).
\bibitem{heiscor} T. A. Kaplan, P. Horsch, and J. Borysowicz, Phys. Rev.
	B {\bf 35}, 1877 (1987); R. R. P. Singh, M. E. Fisher, and R. Shankar, 
	Phys. Rev. B {\bf 39}, 2562 (1989).
\bibitem{parola} A. Parola and S. 
	Sorella, Phys. Rev. Lett. {\bf 64}, 1831 (1990).
\bibitem{sorpar} S. Sorella and A. Parola, J. Phys. CM {\bf 4}, 3589 (1992).
\bibitem{parsor} A. Parola and S. Sorella, Phys. Rev. B {\bf 45}, 13156
	(1992).
\bibitem{aren} Y. Ren and P. W. Anderson, Phys. Rev. B {\bf 48}, 16662 (1993).
\bibitem{ncomp} A. G. Izergin, V. E. Korepin, and N. Yu. Reshetikhin,
	J. Phys. A {\bf 22}, 2615 (1989); F. Woynarovich, ibid., p. 4243;
	H. Frahm and N.-C. Yu, ibid. {\bf 23}, 2115 (1990).
\bibitem{fk} H. Frahm and V. E. Korepin, Phys. Rev. B {\bf 42}, 10553
	(1990).  
\bibitem{kawaya} N. Kawakami and S.-K. Yang, Phys. Lett. A {\bf 148}, 359 
	(1990).
\bibitem{pesomag} K. Penc and J. S\'{o}lyom, Phys. Rev. B {\bf 47}, 6273 
	(1993).
\bibitem{fk2} H. Frahm and V. E. Korepin, Phys. Rev. B {\bf 43}, 5653 (1991).
\bibitem{preuss} R. Preuss, A. Muramatsu, W. von der Linden, P. Dieterich,
	F. F. Assaad, and W. Hanke, Phys. Rev. Lett. {\bf 73}, 732 (1994).
\bibitem{kawatj} N. Kawakami and S.-K. Yang, Phys. Rev. Lett. {\bf 65}, 2309,
	(1990) and J. Phys. CM {\bf 3}, 5983 (1991).
\bibitem{tj}
	M. Ogata, M. Luchini, S. Sorella, and 
	F. F. Assaad, Phys. Rev. Lett. {\bf 66}, 2388 (1991).
\bibitem{helme4} C. S. Hellberg and E. J. Mele, Phys. Rev. B {\bf 48}, 646
	(1993).
\bibitem{helme1} C. S. Hellberg and E. J. Mele, Phys. Rev. B {\bf 44}, 1360
	(1991). 
\bibitem{helme2} C. S. Hellberg and E. J. Mele, Phys. Rev. Lett. {\bf 67}, 2080
	(1991).
\bibitem{gutz} M. C. Gutzwiller, Phys. Rev. Lett. {\bf 10}, 159 (1963).
\bibitem{gevo} F. Gebhard and D. Vollhardt, Phys. Rev. Lett. {\bf 59}, 
	1472 (1987).
\bibitem{halshas} B. Sutherland, Phys. Rev. A {\bf 4}, 2019 (1971) and {\bf 5},
	1372 (1972); F. D. M. Haldane, Phys. Rev. Lett. {\bf 60}, 635 (1988);
	B. S. Shastry, ibid., p. 639.
\bibitem{kyvar} N. Kawakami and P. Horsch, Phys. Rev. Lett. {\bf 68}, 3110 
	(1992).
\bibitem{helme5} C. S. Hellberg and E. J. Mele, Phys. Rev. Lett. {\bf 68}, 3111
	(1992).
\bibitem{ehmhalf} B. Fourcade and G. Spronken, Phys. Rev. B {\bf 29}, 
	5089 and 5096, (1984); J. E. Hirsch, Phys. Rev. Lett. {\bf 53},
	2327 (1984); H. Q. Lin and J. E. Hirsch, Phys. Rev. {\bf 33}, 8155
	(1986); J. W. Cannon and E. Fradkin, Phys. Rev. B {\bf 41}, 9435
	(1990);
	J. W. Cannon, R. T. Scalettar, and E. Fradkin, Phys. Rev. B
	{\bf 44}, 5995 (1991).
\bibitem{lusg} A. Luther, Phys. Rev. B {\bf 14}, 2153 (1976).
\bibitem{luso} A. Luther, Phys. Rev. B {\bf 15}, 403 (1977).
\bibitem{milzot} F. Mila and X. Zotos, Europhys. Lett. {\bf 24}, 133 (1993).
\bibitem{milpen} K. Penc and F. Mila, Phys. Rev. B {\bf 49}, 9670 (1994).
\bibitem{sano} K. Sano and Y. Ono, J. Phys. Soc. Jpn. {\bf 63}, 1250 (1994).
\bibitem{kuroki} K. Kuroki, K. Kusakabe, and H. Aoki, Phys. Rev. B {\bf 50},
	575 (1994).
\bibitem{mermwa} N. D. Mermin and H. Wagner, Phys. Rev. Lett. {\bf 17},
	1133 (1966); P. C. Hohenberg, Phys. Rev. {\bf 158}, 383 (1967).
\bibitem{quant} S. Takada, Prog. Theor. Phys. {\bf 54}, 1039 (1975).
\bibitem{wigcrys} P. Bak and R. Bruinsma, Phys. Rev. Lett.
	{\bf 49}, 249 (1982); G. V. Uimin
	and V. L. Pokrovsky, J. Phys. (Paris) Lett. {\bf 44}, L865 (1983);
	L. A. Bol'shov, V. L. Pokrovsky, and G. V. Uimin, J. Stat. Phys.
	{\bf 38}, 191 (1985).
\bibitem{ssh} W.-P. Su, J. R. Schrieffer, and A. J. Heeger, Phys. Rev.
	Lett. {\bf 42}, 1698 (1979) and Phys. Rev. B {\bf 22}, 2099 (1980).
\bibitem{sshrev} A. J. Heeger, S. Kivelson, J. R. Schrieffer, and W.-P. Su,
	Rev. Mod. Phys. {\bf 60}, 781 (1988).
\bibitem{holpap} T. Holstein, Ann. Phys. {\bf 8}, 325 and 343 (1959). 
\bibitem{pol} D. Feinberg, S. Chiuchi, and F. de Pasquale, Int. J. Mod.
	Phys. B {\bf 4}, 1317 (1990). 
\bibitem{ge} J. M. Ginder and A. J. Epstein, Phys. Rev. B {\bf 41}, 10674
	(1990); D. Baranowski, H. B\"{u}ttner, and J. Voit, ibid. {\bf 45},
	10990 (1992) and {\bf 47}, 15472 (1993). 
\bibitem{phonons} J. Voit and H. J. Schulz, Phys. Rev. B {\bf 34},
	7429 (1986), {\bf 36}, 968 (1985), and {\bf 37}, 10068 (1988);
	L. G. Caron and C. Bourbonnais, Phys. Rev. B {\bf 29}, 4230 (1984);
	C. Bourbonnais and L. G. Caron, J. Phys. (Paris) 
	{\bf 50}, 2751 (1989).
\bibitem{phofor} S. Engelsberg and B. B. Varga, Phys. Rev. {\bf 136}, A1583
	(1964); J. Voit and H. J. Schulz, Molec. Cryst. Liq. 
	Cryst. {\bf 119}, 449 (1985). An error in the treament of acoustic
	phonons in this paper has been corrected by Y. Chen, D. K. K. Lee,
	and M. U. Luchini, Phys. Rev. B {\bf 38}, 8497 (1988). 
\bibitem{marlo} D. Loss and T. Martin, Phys. Rev. B {\bf 50}, 12160 (1994).
\bibitem{scpol} J. Voit, Phys. Rev. Lett. {\bf 64}, 323 (1990).
\bibitem{kzl} G. T. Zimanyi, S. A. Kivelson, and A. Luther, Phys. Rev. Lett.
	{\bf 60}, 2089 (1988).
\bibitem{hol} J. Voit, Synth. Met. {\bf 27}, A41 (1988).
\bibitem{mepho} V. Meden, K. Sch\"{o}nhammer, and O. Gunnarsson, Phys. Rev.
	B {\bf 50}, 11179 (1994).
\bibitem{thierry} T. Giamarchi and H. J. Schulz, Europhys. Lett. {\bf 3},
	1287 (1987), and Phys. Rev. B {\bf 37}, 325 (1988).
\bibitem{apel} S.-T. Chui and J. W. Bray, Phys. Rev. B {\bf 16}, 1329 (1977);
	W. Apel, J. Phys. C {\bf 15}, 1973 (1982); Y. Suzumura and H.
	Fukuyama, J. Phys. Soc. Jpn. {\bf 52}, 2870 (1983) and {\bf 53}, 3918
	(1984).
\bibitem{abri} A. A. Abrikosov and J. A. Ryzhkin, Adv. Phys. {\bf 27},
	147 (1978).
\bibitem{andthe} P. W. Anderson, J. Phys. Chem. Sol. {\bf 11}, 26 (1959).
\bibitem{flr} H. Fukuyama and P. A. Lee, Phys. Rev. B {\bf 17}, 535 (1978).
\bibitem{ogpwa} M. Ogata and P. W. Anderson, Phys. Rev. Lett. {\bf 70}, 3087,
	(1993).
\bibitem{bas} 
	H. Basista, D. A. Bonn, T. Timusk, J. Voit, D. J\'{e}r\^{o}me,
	and K. Bechgaard, Phys. Rev. B {\bf 42}, 4088 (1990).
\bibitem{carhor} J. M. P. Carmelo and P. Horsch, Phys. Rev. Lett. {\bf 68}, 
	871 (1992).
\bibitem{kafi} C. L. Kane and M. P. A. Fisher, Phys. Rev. Lett. {\bf 68}, 
	1220 (1992), Phys. Rev. B {\bf 46}, 7268 and 15233 (1992).
\bibitem{apri} W. Apel and T. M. Rice, Phys. Rev. B {\bf 26}, 7063 (1982).
\bibitem{funa} A. Furusaki and N. Nagaosa, Phys. Rev. B {\bf 47}, 4631 
	(1993).
\bibitem{matv} K. A. Matveev, D. Yue, and L. I. Glazman, Phys. Rev. Lett.
	{\bf 71}, 3351 (1993).
\bibitem{furuna} A. Furusaki and N. Nagaosa, Phys. Rev. B {\bf 47}, 3827
	(1993).
\bibitem{wen} X. G. Wen, Phys. Rev. Lett. {\bf 64}, 2206 (1990),
	Phys. Rev. B {\bf 41}, 12838 (1990) and {\bf 43},
	11025 (1991).
\bibitem{moon} K. Moon, H. Yi, C. L. Kane, S. M. Girvin, and M. P. A.
	Fisher, Phys. Rev. Lett. {\bf 71}, 4381 (1993).
\bibitem{michele} M. Fabrizio, A. O. Gogolin, and S. Scheidl, Phys. Rev.
	Lett. {\bf 72}, 2235 (1994).
\bibitem{fks} H. Fukuyama, H. Kohno, and R. Shirasaki, J. Phys. Soc. Jpn.
	{\bf 62}, 1109 (1993).
\bibitem{ogafuk} M. Ogata and H. Fukuyama, Phys. Rev. Lett. {\bf 73}, 468
	(1994).
\bibitem{leera} P. A. Lee and T. V. Ramakrishnan, Rev. Mod. Phys. {\bf 57},
	287 (1985).
\bibitem{mottwo} N. F. Mott and W. D. Twose, Adv. Phys. {\bf 10}, 107 (1961).
\bibitem{car1} J. Carmelo and A. A. Ovchinnikov, J. Phys. CM {\bf 3},
	757 (1991).
\bibitem{car2} J. Carmelo, P. Horsch, P. A. Bares, and A. A. Ovchinnikov,
	Phys. Rev. B {\bf 44}, 9967 (1991).
\bibitem{car3} J. M. P. Carmelo, P. Horsch, and A. A. Ovchinnikov, 
	Phys. Rev. B {\bf 45}, 7899 and {\bf 46}, 14728 (1992).
\bibitem{carend} J. M. P. Carmelo, A. H. Castro Neto, and D. K. Campbell,
	Phys. Rev. Lett. {\bf 73}, 926 and Phys. Rev. B {\bf 50},
	3667 and 3683 (1994).
\bibitem{heinzmott} H. J. Schulz, in \em Strongly Correlated Electronic
	Materials: The Los Alamos Symposium 1993, \rm ed. by K. S. Bedell
	et al., Addison-Wesley, Reading, 1994, p. 187.
\bibitem{lutem} A. Luther and V. J. Emery, Phys. Rev. Lett. {\bf 33}, 589
	(1974); P. A. Lee, Phys. Rev. Lett. {\bf 34}, 1247 (1973).
\bibitem{elp} V. J. Emery, A. Luther, and I. Peschel, Phys. Rev. B {\bf 13},
	1272 (1976).
\bibitem{colem} S. Coleman, Phys. Rev. D {\bf 11}, 2088 (1975).
\bibitem{fowzo} M. Fowler and X. Zotos, Phys. Rev. B {\bf 24}, 2634 (1981).
\bibitem{skly} E. K. Sklyanin, L. A. Takhtadzhyan, and L. D. Faddeev, Theor.
	Mat. Phys. {\bf 40}, 688 (1979).
\bibitem{hasigo} F. D. M. Haldane, J. Phys. A {\bf 15}, 507 (1982).
\bibitem{giaso} T. Giamarchi and H. J. Schulz, Phys. Rev. B {\bf 33},
	2066 (1986) and J. Phys. (Paris) {\bf 49}, 819 (1988).
\bibitem{schuga} H. J. Schulz, Phys. Rev. B {\bf 22}, 5274 (1980).
\bibitem{gube} M. Gul\'{a}csi and K. S. Bedell, Phys. Rev. Lett. {\bf 72},
	2765 (1994).
\bibitem{pokt} V. L. Pokrovsky and A. L. Talapov, Phys. Rev. Lett. 
	{\bf 42}, 65 (1979).
\bibitem{emcuo} V. J. Emery, Phys. Rev. Lett. {\bf 65}, 1076 (1990).
\bibitem{stami} C. A. Stafford and A. J. Millis, Phys. Rev. B {\bf 48},
	1409, (1993).
\bibitem{eskes} H. Eskes and A. Ole\'{s}, Phys. Rev. Lett. {\bf 73}, 1279
	(1994).
\bibitem{thermo} P. M. Chaikin, R. L. Greene, S. Etemad, and E. Engler,
	Phys. Rev. B {\bf 13}, 1627 (1976).
\bibitem{kolo1} E. B. Kolomeisky, Phys. Rev. B {\bf 47}, 6193 (1993).
\bibitem{kolo2} J. P. Straley and E. B. Kolomeisky, Phys. Rev. B {\bf 48},
	1378 (1993).
\bibitem{cic} B. Horovitz, T. Bohr, J. M. Kosterlitz, and H. J. Schulz,
	Phys. Rev. {\bf 28}, 6596 (1983).
\bibitem{goewoe} W. G\"{o}tze and P. W\"{o}lfle, Phys. Rev. B {\bf 6}, 1226
	(1972).
\bibitem{giam2} T. Giamarchi, Phys. Rev. B {\bf 46}, 342 (1992).
\bibitem{mori} M. Mori, H. Fukuyama, and M. Imada, J. Phys. Soc. Jpn. 
	{\bf 63}, 1639 (1994).
\bibitem{pang} H. Pang, S. Liang, and J. F. Annett, Phys. Rev. Lett. {\bf 71},
	4377 (1993).
\bibitem{kolo3} E. B. Kolomeisky, Phys. Rev. B {\bf 48}, 4998 (1993).
\bibitem{ma} M. Ma, Phys. Rev. B {\bf 26}, 5097 (1982).
\bibitem{horste} P. Horsch and W. Stephan, Phys. Rev. B {\bf 48}, 10 595, 
	(1993).
\bibitem{briri} W. F. Brinkman and T. M. Rice, Phys. Rev. B {\bf 2}, 1324
	(1970).
\bibitem{horsch} K. J. von Szczepanski, P. Horsch, W. Stephan, and 
	M. Ziegler, Phys. Rev. B {\bf 41}, 2017 (1990); M. Ziegler
	and P. Horsch, in \em Dynamics of Magnetic Fluctuations in 
	High-Temperature Superconductors, \rm ed. by G. Reiter and 
	P. Horsch, Plenum Press, New York, 1991, p.329.
\bibitem{pesot} K. Penc and J. S\'{o}lyom, Phys. Rev. B {\bf 41}, 704 (1990).
\bibitem{ners} A. A. Nersesyan, Phys. Lett. A {\bf 153}, 49 (1991).
\bibitem{mont} G. Montambaux, M. H\'{e}ritier, and P. Lederer, Phys. 
	Rev. B {\bf 33}, 7777 (1986).
\bibitem{mutem} K. A. Muttalib and V. J. Emery, Phys. Rev. Lett. {\bf 57},
	1370 (1986).
\bibitem{flupha} I. Affleck and J. B. Marston, Phys. Rev. B {\bf 37}, 3774
	(1988); A. B. Harris, T. C. Lubensky, and E. J. Mele, Phys. Rev.
	B {\bf 40}, 2631 (1989).
\bibitem{vaza} C. M. Varma and A. Zawadowski, Phys. Rev. B {\bf 32}, 7399 
	(1985).
\bibitem{nozkon} P. Nozi\`{e}res, Ann. Phys. Fr. {\bf 10}, 19 (1985).
\bibitem{caboko} L. G. Caron and C. Bourbonnais, Europhys. Lett. {\bf 11},
	473 (1990).
\bibitem{cuo} V. J. Emery, Phys. Rev. Lett. {\bf 58}, 2794 (1987);
	C. M. Varma, S. Schmitt-Rink, and E. Abrahams, Sol. State Comm.
	{\bf 62}, 681 (1987).
\bibitem{lndlif} L. D. Landau and E. M. Lifshitz, \em Statistical
	Physics, \rm Pergamon Press, London (1959), p.~482.
\bibitem{prigofi} V. N. Prigodin and Yu. A. Firsov, Sov. Phys. JETP {\bf 49},
	369 and 813 (1979).  
\bibitem{pwaconf} P.~W.~Anderson, Phys. Rev. Lett. {\bf 67},
	3844 (1991). 
\bibitem{gukle} R. A. Klemm and H. Gutfreund, Phys. Rev. B {\bf 14}, 1086 
	(1976).
\bibitem{klr} P. A. Lee, T. M. Rice, and R. A. Klemm, Phys. Rev. B {\bf 15},
	2984, (1977).
\bibitem{gord} L. P. Gor'kov and I. E. Dzyaloshinski\u{i}, Sov. Phys. JETP
	{\bf 40}, 198 (1975).
\bibitem{little} W. A. Little, Phys. Rev. {\bf 134}, A1416 (1964).
\bibitem{baris} S. Bari\u{s}i\'{c}, J. Phys. (Paris) {\bf 44}, 185 (1983);
	S. Botri\'{c} and S. Bari\u{s}i\'{c}, ibid. {\bf 45}, 185 (1984).
\bibitem{wenperp} X. G. Wen, Phys. Rev. B {\bf 42}, 6623 (1990).
\bibitem{kusma} F. V. Kusmartsev, A. Luther, and A. Nersesyan, JETP Lett.
	{\bf 55}, 724 (1992); V. Yakovenko, ibid. {\bf 56}, 510 (1992).
\bibitem{braya} S. Brazovski\u{i} and V. Yakovenko, J. Phys. (Paris) Lett.
	{\bf 46}, L-111 (1985); Sov. Phys. JETP {\bf 62}, 1340 (1985);
	J. Phys. (Paris) {\bf 47}, 175 (1986).
\bibitem{sc} C. Bourbonnais and L. G. Caron, Europhys. Lett. {\bf 5}, 209
	(1988).
\bibitem{csa} D. G. Clarke, S. P. Strong, and P. W. Anderson, Phys. Rev.
	Lett. {\bf 72}, 3218 (1994).
\bibitem{fab} M. Fabrizio and A. Parola, Phys. Rev. Lett. {\bf 70}, 
	226 (1993).
\bibitem{fabprb} M. Fabrizio, Phys. Rev. B {\bf 48}, 15838 (1993).
\bibitem{fink} A. M. Finkel'stein and A. I. Larkin, Phys. Rev. B {\bf 47},
	10461 (1993).
\bibitem{metz} C. Castellani, D. di Castro, and W. Metzner, 
	Phys. Rev. Lett. {\bf 69}, 1703 (1992).
\bibitem{fab3} M. Fabrizio, A. Parola, and E. Tosatti, Phys. Rev. B {\bf 46},
	3159 (1992).
\bibitem{noack} R. M. Noack, S. R. White, and D. J. Scalapino, Phys. Rev.
	Lett. {\bf 73}, 882 (1994).
\bibitem{asai} Y. Asai, Phys. Rev. B {\bf 50}, 6519 (1994).
\bibitem{gopa} T. M. Rice, S. Gopalan, and M. Sigrist, Europhys. Lett. {\bf 23},
	445 (1993).
\bibitem{twoheis} S. P. Strong and A. J. Millis, Phys. Rev. Lett. {\bf 69},
	2419 (1992); T. Barnes, E. Dagotto, J. Riera, and E. S. Swanson, 
	Phys. Rev. B {\bf 47}, 3196 (1993);
	S. Gopalan, T. M. Rice, and M. Sigrist, Phys. Rev. B {\bf 49},
	8901 (1994).
\bibitem{kvesh} D. V. Khveshchenko, Phys. Rev. B {\bf 50}, 386 (1994).
\bibitem{tsune} H. Tsunetsugu, M. Troyer, and T. M. Rice, Phys. Rev. B {\bf 49},
	16078 (1994).
\bibitem{vome} J. Voit and E. J. Mele, Synth. Met. {\bf 43}, 3911 (1991).
\bibitem{poil} D. Poilblanc, H. Tsunetsugu, and T. M. Rice, Phys. Rev. B 
	{\bf 50}, 6511 (1994).
\bibitem{riera} J. A. Riera, Phys. Rev. B {\bf 49}, 3629 (1994).
\bibitem{boies} D. Boies, C. Bourbonnais, and A.-M. S. Tremblay, Phys.
	Rev. Lett. {\bf 74}, 968 (1995).
\bibitem{kopie} P. Kopietz, V. Meden, and K. Sch\"{o}nhammer, Phys. Rev.
	Lett. {\bf 74}, 2997 (1995).
\bibitem{vale} R. Valent\'{i} and C. Gros, Phys. Rev. Lett. {\bf 68}, 2402
	(1992) and (E) {\bf 69}, 996 (1992); 
	C. Gros and R. Valent\'{i}, Mod. Phys. Lett. B {\bf 7},
	119 (1993).
\bibitem{chenle} Y. C. Chen and T. K. Lee, Z. Phys. B {\bf 95}, 5 (1994).
\bibitem{putti} W. O. Putikka, R. L. Glenister, R. R. P. Singh, and
	H. Tsunetsugu, Phys. Rev. Lett. {\bf 73}, 170 (1994).
\bibitem{chenotr} Y. C. Chen, A. Moreo, F. Ortolani, E. Dagotto, and
	T. K. Lee, Phys. Rev. B {\bf 50}, 655 (1994); C. Gros and 
	R. Valent\'{i}, ibid., p. 11313; T. Tohyama, P. Horsch,
	and S. Maekawa, Phys. Rev. Lett. {\bf 74}, 980 (1995).
\bibitem{newmetz} C. Castellani, C. Di Castro, and W. Metzner, Phys. Rev. Lett.
	{\bf 72}, 316 (1994).
\bibitem{ll2d} T. Holstein, R. E. Norton, and P. Pincus, Phys. Rev. B {\bf 8},
	2649 (1973); M. Yu. Reizer, Phys. Rev. B {\bf 39}, 1602 (1989) and
	{\bf 40} 11571 (1989); F. Guinea and G. Zimanyi, Phys. Rev. B {\bf 47},
	501 (1993); P. A. Bares and X.-G. Wen, Phys. Rev. B {\bf 48}, 8636
	(1993); D. V. Khveshchenko, R. Hlubina, and T. M. Rice,
	Phys. Rev. B {\bf 48}, 10766 (1993); R. Hlubina, Phys. Rev. B {\bf 50},
	8252 (1994).
\bibitem{bosfl} A. Luther, Phys. Rev. B. {\bf 19}, 320 (1979); 
	R. Shankar, Physica A {\bf 177}, 530 (1991) and 
	Rev. Mod. Phys. {\bf 66}, 129 (1994); A. Houghton and J. B. Marston, 
	Phys. Rev. B {\bf 48}, 7790 (1993); A. Houghton, H.-J. Kwon, and
	J. B. Marston, Phys. Rev. B {\bf 50}, 1351 (1994); A. H. Castro Neto
	and E. Fradkin, Phys. Rev. Lett. {\bf 72}, 1393 (1994) and Phys. Rev.
	B {\bf 49}, 10877 (1994).
\bibitem{qhebook} \em The Quantum Hall Effect, \rm ed. by R. E. Prange and
	S. M. Girvin, Springer Verlag, New York, 1987.
\bibitem{klitzing} K. von Klitzing, G. Dorda, and M. Pepper, Phys. Rev. Lett.
	{\bf 45}, 494 (1980).
\bibitem{tsui} D. C. Tsui, H. L. S\"{o}rmer, and A. Gossard, Phys. Rev. Lett.
	{\bf 48}, 1559 (1982).
\bibitem{halp} B. I. Halperin, Phys. Rev. B {\bf 25}, 2185 (1982).
\bibitem{buetti} M. B\"{u}ttiker, Phys. Rev. B {\bf 38}, 9375 (1988).
\bibitem{wenrev} For a review, see X.-G. Wen, Int. J. Mod. Phys. B {\bf 6},
	1711 (1992).
\bibitem{stone} M. Stone, Ann. Phys. (NY) {\bf 207}, 38 (1991) and Int. J.
	Mod. Phys. B {\bf 5}, 509 (1991).
\bibitem{john} M. D. Johnson and A. H. MacDonald, Phys. Rev. Lett. 
	{\bf 67}, 2060 (1991).
\bibitem{js} D.~J\'{e}r\^{o}me and H.~J.~Schulz, in \em Extended Linear
	Chain Compounds, \rm Vol.~2, edited by J.~S.~Miller, Plenum Press, New
	York (1982).
\bibitem{pouget} J. P. Pouget, S. K. Khanna, F. Denoyer, R. Com\`{e}s, A. F.
	Garito, and A. J. Heeger, Phys. Rev. Lett. {\bf 37}, 437 (1976).
\bibitem{kago} S. Kagoshima, T. Ishiguro, and H. Anzai, J. Phys. Soc.
	Japan {\bf 41}, 2061 (1976).
\bibitem{klotz} S. Klotz, J. S. Schilling, M. Weger, and K. Bechgaard,
	Phys. Rev. B {\bf 38}, 5878 (1988).
\bibitem{taka} T. Takahashi, D. J\'{e}r\^{o}me, F. Masin, J. M. Fabre, and
	L. Giral, J. Phys. C {\bf 17}, 3777 (1984).
\bibitem{torrance} J. B. Torrance, in Ref.~\cite{magog}.
\bibitem{pougnmp} J. P. Pouget, R. Com\`{e}s, A. J. Epstein, and J. S. Miller,
	Molec. Cryst. Liq. Cryst. {\bf 85}, 1593 (1982).
\bibitem{skov} J. Skov Pedersen and K. Carneiro, Rep. Prog. Phys. {\bf 50},
	995 (1987).
\bibitem{Silvia} S. Tomi\'{c}, J. R. Cooper, and K. Bechgaard, Phys. Rev. Lett.
	{\bf 62}, 462 (1989).
\bibitem{jerome} D. J\'{e}r\^{o}me, A. Mazaud, M. Ribault, and K. Bechgaard,
	J. Phys. Lett. (Paris) {\bf 41}, L95 (1980).
\bibitem{parkin} S. S. P. Parkin, M. Ribault, D. J\'{e}r\^{o}me, and
	K. Bechgaard, J. Phys. C {\bf 14}, 5305 (1981).
\bibitem{forro} L. Forr\'{o}, J. R. Cooper, B. Rothaemel, J. S. Schilling,
	M. Weger, and K. Bechgaard, Solid State Comm. {\bf 60}, 11 (1986).
\bibitem{pecile} R. Bozio, M. Meneghetti, D. Pedron, and C. Pecile,
	Synth. Met. {\bf 27}, B129 (1988).
\bibitem{bozio} R. Bozio, M. Meneghetti, and C. Pecile, J. Chem. Phys.
	{\bf 76}, 5785 (1982).
\bibitem{Rice} M. J. Rice, Phys. Rev. Lett. {\bf 37}, 36 (1976).
\bibitem{NMR}
	C. Bourbonnais, F. Creuzet, D. J\'{e}rome, K. Bechgaard, and
	A. Moradpour, J. Phys (Paris) Lett. {\bf 45}, L-755 (1984).
\bibitem{wzie}
	P. Wzietek, F. Creuzet, C. Bourbonnais, D. J\'{e}r\^{o}me, and A. 
	Moradpour, J. Phys. (Paris) I {\bf 3}, 171 (1993).
\bibitem{clamag} C. Bourbonnais, in Ref.~\cite{magog}.
\bibitem{claude} C. Bourbonnais, J. Phys. (Paris) I {\bf 3}, 143 (1993).
\bibitem{ducasse} L.~Ducasse, M.~Abderraba, J.~Hoarau, M.~Pesquer,
	B.~Gallois, and J.~Gaultier, J. Phys. C {\bf 19}, 3805 (1986).
\bibitem{sossc} G. Soda, D. J\'{e}r\^{o}me, M. Weger, J. M. Fabre, and
	L. Giral, Solid State Comm. {\bf 18}, 1417 (1976).
\bibitem{soda} G.~Soda, D.~J\'{e}r\^{o}me, M.~Weger, J.~Alizon, J.~Gallice,
	H.~Robert, J.~M.~Fabre, and 
	L.~Giral, J. Phys. (Paris) {\bf 38}, 931 (1977).
\bibitem{stein} P. C. Stein, A. Moradpour, and D. J\'{e}r\^{o}me,
	J. Phys. Lett. (Paris) {\bf 46}, 241 (1985).
\bibitem{photem} B. Dardel, D. Malterre, M. Grioni, P. Weibel, Y. Baer, 
	J. Voit, and D. J\'{e}r\^{o}me, Europhys. Lett. {\bf 24}, 687 (1993).
\bibitem{milseoul} F. Mila and K. Penc, Synth. Met. {\bf 70}, 997 (1995).
\bibitem{CDW} \em Electronic Properties of Inorganic
	Quasi-One-Dimensional Compounds, \rm vol.~1 and 2, edited by P. Monceau,
	D.~Reidel Publ. Comp., Dordrecht (1985).
\bibitem{lra} P.~A.~Lee, T.~M.~Rice, and P.~W.~Anderson, Phys. Rev. Lett.
	{\bf 31}, 462 (1973).
\bibitem{bluebr} C. Schlenker and J. Dumas, in \em Crystal Chemistry and
	Properties of Materials with Quasi-One-Dimensional Structures, \rm
	ed. by J. Rouxel, D. Reidel Publ. Comp., Dordrecht, 1986, p. 135.
\bibitem{ne} B. Dardel, D. Malterre, M. Grioni, P. Weibel, Y. Baer, and
	F. L\'{e}vy, Phys. Rev. Lett. {\bf 67}, 3144 (1991); 
	J.-Y. Veuillen, R. C. Cinti, and E. Al Khoury Nemeh, Europhys. Lett.
	{\bf 3}, 355 (1987).
\bibitem{hwu} Y. Hwu, P. Alm\'{e}ras, M. Marsi, H. Berger, F. L\'{e}vy,
	M. Grioni, D. Malterre, and G. Margaritondo, Phys. Rev. B {\bf 46},
	13624 (1992)
\bibitem{smith} K. E. Smith, K. Breuer, M. Greenblatt, and W. McCarrol,
	Phys. Rev. Lett. {\bf 70}, 3772 (1993).
\bibitem{degior} L. Degiorgi, G. Gr\"{u}ner, K. Kim, R. H. McKenzie,
	and P. Wachter, Phys. Rev. B {\bf 49}, 14754 (1994).
\bibitem{mcken} K. Kim, R. H. McKenzie, and J. W. Wilkins, Phys. Rev.
	Lett. {\bf 71}, 4015 (1993).
\bibitem{fujimori} N. Nakamura, A. Sekiyama, H. Namatame, A. Fujimori,
	H. Yoshihara, T. Ohtani, A. Misu, and M. Takano, Phys. Rev. B
	{\bf 49}, 16191 (1994).
\bibitem{hetero} U. Meirav, M. A. Kastner, M. Heiblum, and S. J. Wind,
	Phys. Rev. B {\bf 40}, 5871 (1989).
\bibitem{webb} F. P. Milliken, C. P. Umbach, and R. A. Webb, unpublished;
	also reported by B. Gross Levi, Physics Today {\bf 47}, 21 (1994).





\end{thebibliography}
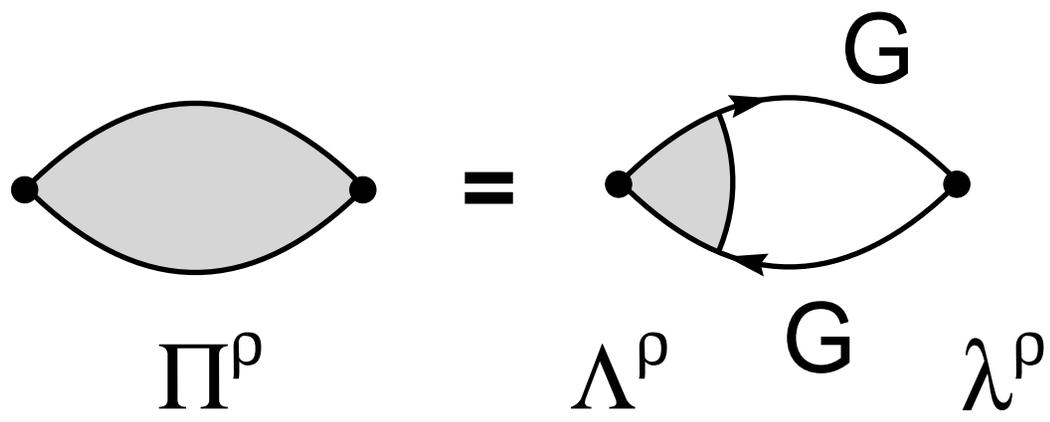
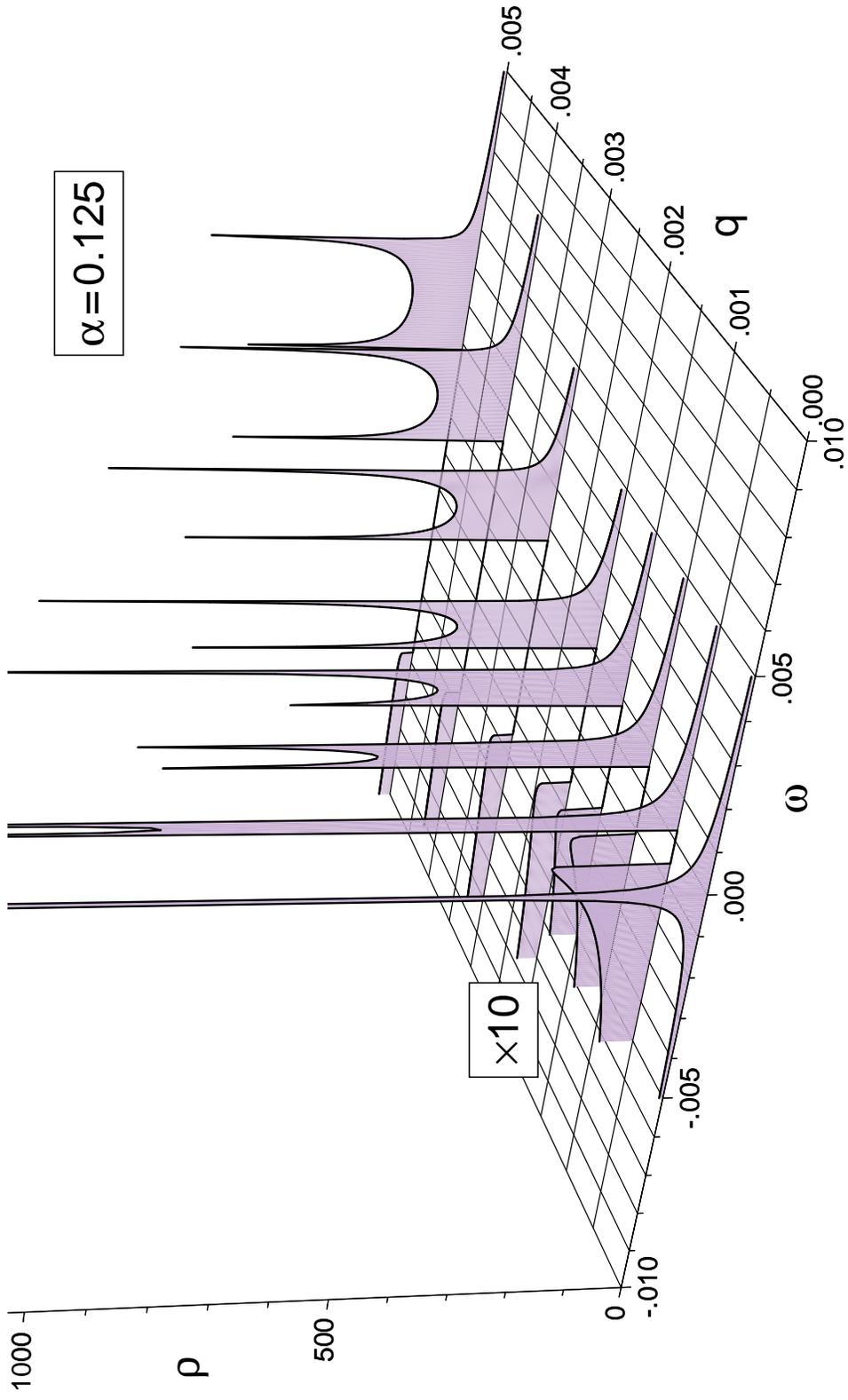
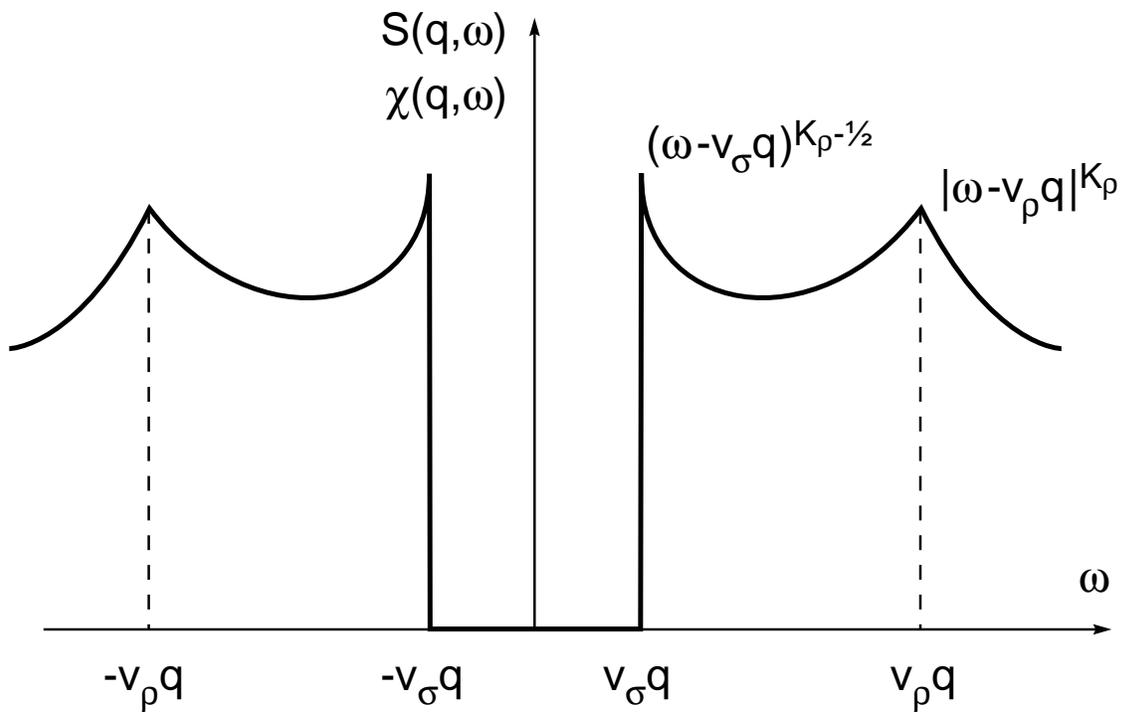
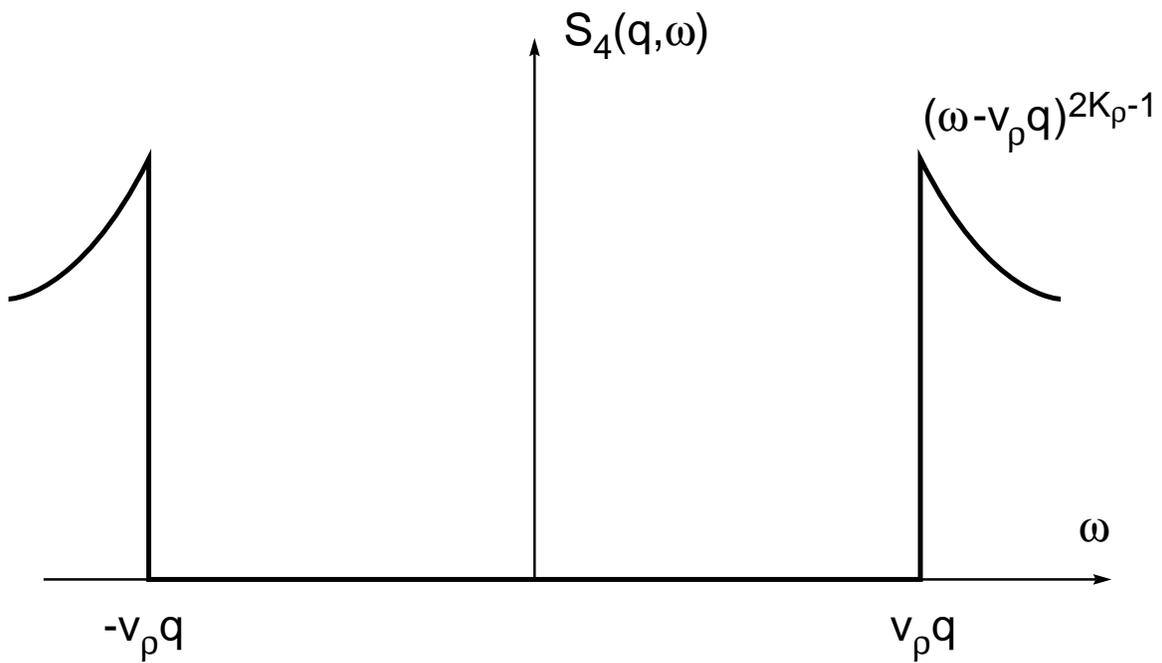
\end{document}